\documentclass[10pt,twoside]{amundi_article}

\usepackage[table]{xcolor}
\usepackage[latin9]{inputenc}
\usepackage{amssymb}
\usepackage{amsfonts}
\usepackage{amsmath}
\usepackage{fancybox}
\usepackage{fancyhdr}
\usepackage{graphicx}
\usepackage{subcaption}
\usepackage[hyphens]{url}
\usepackage{arydshln}
\usepackage{multirow}
\usepackage{pdflscape}
\usepackage{tikz}
\usepackage{comment}
\usepackage{nicefrac}
\usepackage{subcaption}
\usepackage{dsfont}
\usepackage{natbib}
\usepackage[colorlinks = true, linkcolor = blue, urlcolor = blue,
            citecolor = blue, anchorcolor = blue]{hyperref}
\usepackage{adjustbox}
\usepackage{array}
\usepackage{lipsum}
\usepackage{eurosym}
\usepackage{morefloats}

\newcolumntype{R}[2]{%
    >{\adjustbox{angle=#1,lap=\width-(#2)}\bgroup}%
    l%
    <{\egroup}%
}

\newlength{\figurewidth}
\newlength{\figureheight}
\def\figureskip{\vskip 10pt plus 2pt minus 2pt\relax}

\newtheorem{remark}{Remark}

\newtheorem{example}{Example}
\def\limfunc#1{\mathop{\rm #1}}
\def\func#1{\mathop{\rm #1}}
\newcommand{\mr}[1]{\multirow{2}{*}{#1}}
\newcommand{\mrm}[2]{\multirow{#1}{*}{#2}}

\graphicspath{}

\newcommand{\TsV}{\hspace{5pt}}

\newcommand{\TsX}{\hspace{10pt}}

\newcommand{\TsXV}{\hspace{15pt}}
\newcommand{\TsXX}{\hspace{20pt}}

\DeclareMathAlphabet{\mathpzc}{OT1}{pzc}{m}{it}
\newcommand{\RedemptionRate}{\ensuremath{\boldsymbol{\mathpzc{R}}}}
\newcommand{\AssetRate}{\ensuremath{\boldsymbol{\mathpzc{A}}}}
\newcommand{\RedemptionShock}{\ensuremath{\mathbb{R}}}
\newcommand{\AssetShock}{\ensuremath{\mathbb{A}}}

\newcommand{\rstRS}{\ensuremath{\mathbb{R}^{\mathrm{RST}}}}
\newcommand{\rstRR}{\ensuremath{\boldsymbol{\mathpzc{R}}^{\mathrm{RST}}}}

\newcommand{\spread}{\ensuremath{\boldsymbol{\mathpzc{s}}}}
\newcommand{\cashRate}{\ensuremath{\boldsymbol{\mathpzc{c}}}}

\newcommand{\cost}{\pmb{c}}
\newcommand{\impact}{\pmb{\pi}}
\newcommand{\TC}{\mathcal{TC}}

\usetikzlibrary{positioning}
\usetikzlibrary{patterns,arrows,decorations.pathreplacing}
\usetikzlibrary{backgrounds}
\usetikzlibrary{shapes.symbols}

\newcommand{\vphOne}{\vphantom{\beta_{\impact}\left( x-\kappa x^{+}\right)}}

\newif\ifResearchVersion
\ResearchVersiontrue

\begin{document}

\ifResearchVersion

\title{\textbf{\color{amundi_blue}Liquidity Stress Testing in Asset Management\\Part 3. Managing the Asset-Liability\\Liquidity Risk}%
\footnote{I am grateful to Amina Cherief, Alexis Clerc, Gr\'egoire Pacreau and
Jiali Xu for their helpful comments. This research has also benefited
from the support of Amundi Asset Management, which has
provided the data. Many thanks to Amina Cherief and Pasquale Galassi for having tested the framework
developed here on real data and Amundi mutual funds.
However, the opinions expressed in this article are those
of the author and are not meant to represent the opinions or official
positions of Amundi Asset Management.}}
\author{
{\color{amundi_dark_blue} Thierry Roncalli} \\
Quantitative Research \\
Amundi Asset Management, Paris \\
\texttt{thierry.roncalli@amundi.com}}

\date{\color{amundi_dark_blue}September 2021}

\maketitle

\begin{abstract}
This article is part of a comprehensive research project on liquidity risk
in asset management, which can be divided into three dimensions. The first
dimension covers the modeling of the liability liquidity risk (or funding
liquidity), the second dimension is dedicated to the modeling of the asset
liquidity risk (or market liquidity), whereas the third dimension considers
the management of the asset-liability liquidity risk (or asset-liability
matching). The purpose of this research is to propose a methodological
and practical framework in order to perform liquidity stress testing
programs, which comply with regulatory guidelines \citep{ESMA-2019a,
ESMA-2020a} and are useful for fund managers. The review of the academic
literature and professional research studies shows that there is a lack of
standardized and analytical models. The aim of this research project is
then to fill the gap with the goal of developing mathematical and statistical
approaches, and providing appropriate answers.\smallskip

In this third and last research paper focused on managing the
asset-liability liquidity risk, we explore the ALM tools that can be put in
place to control the liquidity gap. These ALM tools can be split into
three categories: measurement tools, management tools and monitoring tools.
In terms of measurement tools, we focus on the computation of the
redemption coverage ratio (RCR), which is the central instrument of
liquidity stress testing programs. We also study the redemption liquidation
policy and the different implementation methodologies, and we show how
reverse stress testing can be developed. In terms of liquidity management
tools, we study the calibration of liquidity buffers, the pros and cons of
special arrangements (redemption suspensions, gates, side pockets and
in-kind redemptions) and the effectiveness of swing pricing. In terms of
liquidity monitoring tools, we compare the macro- and micro-approaches to
liquidity monitoring in order to identify the
transmission channels of liquidity risk.
\end{abstract}

\medskip


\clearpage

\else

\setcounter{page}{7}

\fi

\section{Introduction}

The guidelines on liquidity stress testing in UCITS and AIFs produced by
\citet{ESMA-2020a} are rooted in the banking regulation defined by the
Basel Committee on Banking Supervision \citep{BCBS-2010,BCBS-2013}. For
instance, the redemption coverage ratio, which is the key instrument of LST
programs, is a copy-paste of the liquidity coverage ratio (LCR) in the Basel
III Accord. According to \citet{BCBS-2008}, liquidity risk management in the
banking industry must be structured around three pillars: measurement,
management and monitoring. Beyond the redemption coverage ratio, which is
typically a measurement tool, \citet{ESMA-2020a} adopt a similar approach by
mixing the three Ms.\smallskip

Liquidity risk is an important topic for the banking sector
because it concerns systemic risk. We face similar issues for the
asset management industry because it can generate big market risks.
Since liquidity risk is an ALM risk \citep[Chapter 7]{Roncalli-2020}, it  concerns both liabilities and assets. As mentioned by \citet{Brunnermeier-2009}, the interconnectedness between funding liquidity and market liquidity amplifies the liquidity risk. This is obvious in stress periods, but this is even the case in normal periods when we consider the asset management industry. The reason is that redeeming investors impose negative externalities on the remaining investors:
\begin{quote}
\textquotedblleft \textsl{Strategic interaction is a key determinant of
investors' behavior in financial markets and institutions. When choosing
their investment strategy, investors have to consider, not only the
expected fundamentals of the investment, but also the expected behavior of
other investors, which could have a first-order effect on investment
returns. Particularly interesting are situations with payoff
complementarities, where investors' incentives to take a certain action
increase if they expect that more investors will take such an action.
Payoff complementarities are expected to generate a multiplier effect, by
which they amplify the impact that shocks to fundamentals have on
investors' behavior. Such amplification is often referred to as
\textit{financial fragility}}\textquotedblright\  \citep[page
239]{Chen-2010}.
\end{quote}
This \textit{financial fragility} has been documented in several asset classes \citep{Bouveret-2021, Chernenko-2020, Fricke-2021, Fricke-2020, Rohleder-2017, Goldstein-2017b}. The negative externalities and their major impact when considering stress periods explain that financial regulators have recently paid more attention to liquidity management in the asset management industry \citep{AMF-2017, BaFin-2017, EFAMA-2020, ESRB-2017}, while the regulation of asset managers in terms of liquidity management was light in the 2000s. Nevertheless, introducing more stringent regulations in the asset management industry is not a new concept and dates back to the roadmap of the Financial Stability Board (FSB) when it was created in April 2009 after the 2008 Global Financial Crisis to monitor the stability of the financial system and manage systemic risk \citep[page 453]{Roncalli-2020}.\smallskip

However, the lack of maturity and benchmarking is an obstacle for the development of liquidity stress testing in the asset management industry. One of the big challenges for regulators is standardizing models and practices. In the case of the banking industry, the Basel Committee has been successful in proposing statistical frameworks for market and credit risks. This is not the case in the asset management industry, where academic research is relatively invisible on the liability side. As such, most solutions are in-house and not published, implying limited distribution of best practices and, generally simplistic and naive methods being developed. Against this backdrop, it is not surprising that mathematical and statistical models are completely absent from regulatory publications, especially in the case of the ESMA guidelines on liquidity stress testing in asset management.\smallskip

This paper completes a research project that began in April 2020
and was organized into three streams. The first stream covered the liability side and funding liquidity modeling. In \citet{Roncalli-lst1}, we introduced
two statistical approaches that can be used to define a
redemption shock scenario. The first one is the historical approach and
considers non-parametric risk measures such as the historical or
conditional value-at-risk. The second approach deals with frequency-severity
models, which produces parametric risk measures and stress scenarios.
Three of these probabilistic models are particularly interesting: the zero-inflated (or population-based) statistical model, the behavioral (or individual-based) model and the factor-based model. The second stream focused on the asset side and transaction cost modeling. In \citet{Roncalli-lst2}, we proposed a two-regime model to estimate ex-ante transaction costs and market impacts. This model is an extension of the square-root model and considers trading limits in order to comply with the practices of asset managers.
Based on proprietary and industry data, we were able to perform the calibration for large cap stocks, small cap stocks, sovereign bonds and corporate bonds. Moreover, we have detailed the analytics of liquidation rate, time to liquidation and liquidation shortfall to assess the liquidity risk profile of investment funds. The third stream corresponds to this research paper. The aim is to combine liability and asset risks in order to define the ALM tools. Therefore, this paper extensively mixes the previous models. For instance, a stress scenario may originate from the liabilities
or the assets or both. Synthetic measures such as the funding gap or funding ratio are essential for asset-liability management. These measures are particularly exploited for the purpose of defining appropriate liquidation policies and the management tools that can be put in place. Besides traditional management methods, asset managers are paying more and more attention to liquidity buffers. The widespread use of cash buffers for the purpose of liquidity stress testing may have some significant impacts in terms of reducing or increasing systemic risk. The recent debate on cash buffering versus cash hoarding and the \textquotedblleft \textsl{dash for cash}\textquotedblright\ episode during the Covid-19 crisis in March 2020 demonstrate that the liquidity issue in asset management remains as before. This implies that asset managers must continue to develop the required tools and adopt more responsive tools. This is especially true for monitoring tools that must use higher frequency data.\smallskip

The rest of the paper is organized as follows. Section 2 presents the liquidity measurement tools. We introduce the redemption coverage ratio (RCR) and the two computational approaches (time to liquidation and high-quality liquid assets). We also focus on the redemption liquidation policy and the differences between vertical and horizontal slicing. Compared to banks,
reverse stress testing (RST) is more complex because two dimensions can be chosen, implying that we can define a liability or an asset RST scenario.
Section 3 is dedicated to liquidity managements tools (LMTs). Besides swing pricing and special arrangements (redemption suspensions, gates, side pockets and in-kind redemptions), we extensively study the set-up of a liquidity buffer. We propose an optimization model that considers the costs and benefits of implementing a cash buffer and derive the optimal solution that depends on the risk premium of assets, the tracking error risk and the liquidation gain. Using the square-root transaction cost model, we obtain analytical formulas and test the impact of the different parameters. The liquidity monitoring tools are discussed in Section 4. We distinguish the macro-economic and micro-economic approaches. The macro-economic approach helps to define overall liquidity and is related to central bank liquidity and the economic outlook. This approach is extensively used by financial regulators and international bodies. In a liquidity stress testing framework, it must be complemented by a micro-economic approach that considers the daily liquidity at the asset class, security and issuer levels. Data collection from order books, market infrastructure and the trading desk of the asset manager is the key to successfully building a suitable monitoring system. Finally, Section 5 concludes the paper.

\section{Liquidity measurement tools}

Among the three Ms, measurement is certainly the most important and
difficult step of liquidity stress testing programs. Indeed, it encompasses
two sources of uncertainty: liability risk and asset risk. As shown by
\citet{Roncalli-lst1}, there are two main approaches for measuring the
liability risk. We can use an historical approach or a frequency-severity
framework. For this latter, we also have the choice between three models:
the zero-inflated statistical model, the behavioral model or the factor-based model. On the asset risk side, things are simpler since we generally consider
the power-law model as a standard approach. However, calibrating the
parameters remains a fragile exercise that is highly dependent on the historical data of the asset manager \citep{Roncalli-lst2}.\smallskip

As explained in the introduction, benchmarking will be a key factor for
improving these measures. Nevertheless, there is certainly another issue that
is even more detrimental. Indeed, the definition of the concepts is not
always precise, and the regulators of the asset management industry are less
prolific than the regulators of the banking industry. However, the devil is
in the details. This is why we define the different measurement concepts more precisely in this section. First, we present the redemption coverage ratio and the two approaches for computing it. Then, we focus on the redemption liquidation policy, which must specify the appropriate decision in the case of a liquidity crisis. Finally, the regulation requires that the asset manager defines reverse stress testing scenarios and explores circumstances that might cause them to occur.

\subsection{Redemption coverage ratio}

According to \citet{ESMA-2020a}, the redemption coverage ratio (RCR) is
\textquotedblleft \textit{a measurement of the ability of a fund's assets to
meet funding obligations arising from the liabilities side of the balance
sheet, such as a redemption shock}\textquotedblright. Except for
this definition\footnote{It can be found on page 7 of the ESMA guidelines
\citep{ESMA-2020a}.}, there are no other references to this concept in the
ESMA guidelines. Therefore, we must explore other resources to
clarify it, but they are few in number \citep{Bouveret-2017, IMF-2017,
ESMA-2020b}.\smallskip

The redemption coverage ratio was introduced by \citet{Bouveret-2017},
who defines it as follows:
\begin{equation}
\limfunc{RCR}=\frac{\text{Liquid assets}}{\text{Net outflows}}
\label{eq:RCR1}
\end{equation}%
where net outflows and liquid assets correspond respectively to redemption
shocks and the amount of the portfolio that can be liquidated over a given time horizon. There are two possible cases:
\begin{itemize}
\item if the RCR is above 1, then the fund's portfolio is sufficiently  liquid to cope with the redemption scenario;
\item if the RCR is below 1, then the liquidity profile of the fund may be
    worsened when the redemption scenario occurs.
\end{itemize}
In this second case, the outcome will depend largely on the market liquidity
conditions. Indeed, there is a pricing risk on the NAV because the fund will
have to sell illiquid assets in an illiquid market. The amount of additional
assets to be sold is called the liquidity shortfall (LS):
\begin{equation}
\limfunc{LS}=\max \left( 0,\text{Net outflows}-\text{Liquid assets}\right)
\label{eq:LS1}
\end{equation}%
In order to compare the liquidity profile of several funds, the measure
$\limfunc{LS}$ is expressed as a percentage of the fund's total net assets
(TNA).\smallskip

\begin{remark}
The RCR and LS measures refer to banking ALM concepts. Indeed, asset-liability management is based on two risk measures: the funding ratio
and the funding gap \citep[Chapter 7, page 376]{Roncalli-2020}. When the ALM
is applied to liquidity risk, we refer to liquidity ratio and liquidity
gap. It is obvious that the redemption coverage ratio is related to the
liquidity (coverage) ratio, while the liquidity shortfall is equivalent to
the liquidity gap.
\end{remark}

\begin{figure}[tbph]
\centering
\caption{LS and RCR for selected investment funds}
\label{fig:fsap_lux}
\figureskip
\includegraphics[scale = 0.50]{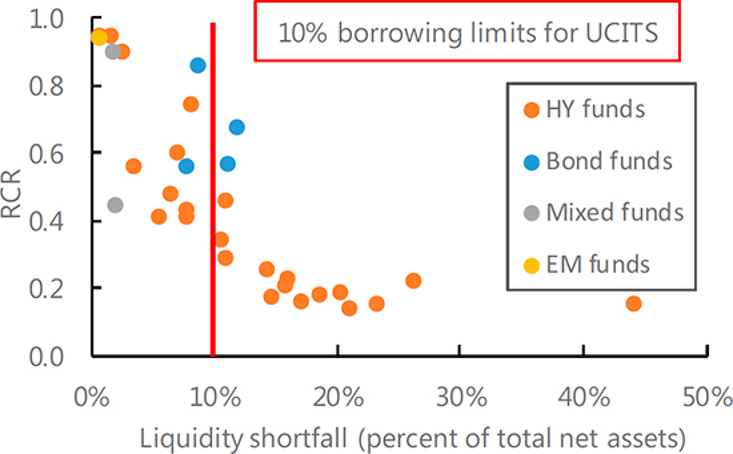}
\begin{flushleft}
{\small \textit{Source}: \citet[Figure 19, page 59]{IMF-2017}.}
\end{flushleft}
\vspace*{-10pt}
\end{figure}

The International Monetary Fund has used the redemption coverage ratio in the
case of its financial sector assessment program (FSAP) for two countries:
Luxembourg in 2017 and the United States in 2020. These two FSAP exercises
showed that a significant proportion of the funds would have enough liquid
assets to meet redemption shocks. However, the IMF found that the most vulnerable categories are HY and EM bond funds in Luxembourg \citep{IMF-2017} and HY and loan mutual funds in the US \citep{IMF-2020}. In the case of Luxembourg funds, Figure \ref{fig:fsap_lux} shows that about 30 bond funds have an RCR below 1, and 50\% of them have a liquidity shortfall greater than 10\%, which is the borrowing limit for UCITS funds.

\subsubsection{Time to liquidation approach}

\paragraph{Mathematical framework}

We consider a fund, whose asset structure is given by the vector $\omega
=\left( \omega _{1},\ldots ,\omega _{n}\right) $\ where $\omega _{i}$ is the
number of shares of security $i$ and $n$ is the number of securities that
make up the asset portfolio. By construction, the fund's total net assets are
equal to:
\begin{equation}
\limfunc{TNA}=\sum_{i=1}^{n}\omega _{i}\cdot P_{i}  \label{eq:TNA}
\end{equation}%
where $P_{i}$ is the current price of security $i$. The mathematical
expressions of Equations (\ref{eq:RCR1}) and (\ref{eq:LS1}) are:
\begin{equation}
\limfunc{RCR}=\frac{\AssetRate}{\RedemptionRate}  \label{eq:RCR2}
\end{equation}%
and:%
\begin{equation}
\limfunc{LS}=\max \left( 0,\RedemptionRate-\AssetRate\right)   \label{eq:LS2}
\end{equation}%
where $\AssetRate$ is the ratio of liquid assets in the fund and
$\RedemptionRate$ is the redemption shock expressed in \%. Following
\citet{Roncalli-lst2}, the redemption shock expressed in dollars is equal to
$\RedemptionShock=\RedemptionRate\cdot \limfunc{TNA}$. Let $q=\left(
q_{1},\ldots ,q_{n}\right) $ be a redemption portfolio and $q_{i}\left(
h\right) $ be the number of shares liquidated after $h$ trading
days\footnote{We recall that $q_{i}\left( h\right) $ is equal to:
\begin{equation}
q_{i}\left( h\right) =\min \left( \left( q_{i}-\sum_{k=0}^{h-1}q_{i}\left(
k\right) \right) ^{+},q_{i}^{+}\right)   \label{eq:asset2}
\end{equation}%
where $q_{i}\left( 0\right) =0$ and $q_{i}^{+}$ denotes the maximum number of
shares that can be sold during a trading day for the asset $i$ \citep[Section
3.2, page 14]{Roncalli-lst2}.}. The amount of liquid assets is equal to the
amount of assets that can be sold:
\begin{equation}
\AssetShock\left( h\right) =\sum_{i=1}^{n}\sum_{k=1}^{h}q_{i}\left( k\right)
\cdot P_{i}  \label{eq:asset1}
\end{equation}%
By definition, we have $\AssetShock\left( h\right) =\AssetRate\left(
h\right) \cdot \limfunc{TNA}$. We notice that asset liquidation requires a parameter $h$ to be defined, which is the time horizon. Therefore, it is better
to define RCR and LS measures as follows:
\begin{equation}
\limfunc{RCR}\left( h\right) =\frac{\AssetShock\left( h\right) }{%
\RedemptionShock}=\frac{\AssetRate\left( h\right) }{\RedemptionRate}
\label{eq:RCR3}
\end{equation}%
and:%
\begin{equation}
\limfunc{LS}\left( h\right) =\frac{\max \left( 0,\RedemptionShock-\AssetShock%
\left( h\right) \right) }{\limfunc{TNA}}=\RedemptionRate\cdot \max \left(
0,1-\limfunc{RCR}\left( h\right) \right)   \label{eq:LS3}
\end{equation}%
Since $h$ is a liquidation time horizon, the previous computation method is
called the time to liquidation (TTL) approach \citep{Bouveret-2017}.

\paragraph{Relationship with the liquidation ratio}

As its name suggests, the time to liquidation approach is related to the
liquidation ratio. Following \citet{Roncalli-lst2}, the liquidation ratio
$\mathcal{LR}\left( q;h\right) $ is the proportion of the redemption scenario
$q$ that is liquidated after $h$ trading days:
\begin{equation}
\mathcal{LR}\left( q;h\right) =\frac{\sum_{i=1}^{n}\sum_{k=1}^{h}q_{i}\left(
k\right) \cdot P_{i}}{\sum_{i=1}^{n}q_{i}\cdot P_{i}}  \label{eq:LR1}
\end{equation}%
By definition, $\mathcal{LR}\left( q;h\right) $ is between $0$ and $1$
whereas $\limfunc{RCR}\left( h\right) \geq 0$. Using Equation
(\ref{eq:asset1}), we deduce that:
\begin{equation}
\AssetShock\left( h\right) =\mathcal{LR}\left( q;h\right) \cdot \mathbb{V}%
\left( q\right)   \label{eq:asset3}
\end{equation}%
where $\mathbb{V}\left( q\right) =\sum_{i=1}^{n}q_{i}\cdot P_{i}$ is the
value function of the portfolio $q$. It follows that:%
\begin{equation}
\limfunc{RCR}\left( h\right) =\frac{\mathbb{V}\left( q\right) }{%
\RedemptionShock}\cdot \mathcal{LR}\left( q;h\right)   \label{eq:RCR4}
\end{equation}%
The redemption coverage ratio can be seen as an extension of the concept of
the liquidation ratio when the liquidation portfolio $q$ corresponds to the pool of liquid assets and the redemption shock is defined without any reference to $q$. \citet{Roncalli-lst2} define the liquidation period $h^{+}=\left\{ \inf h:\mathcal{LR}\left( q;h\right) =1\right\} $ as the number of trading days we need to liquidate the portfolio $q$. We can then have three cases:
\begin{enumerate}
\item The redemption coverage ratio is equal to the liquidation ratio if and only if the redemption scenario is equal to the value of the liquidation portfolio:%
\begin{equation}
\limfunc{RCR}\left( h\right) =\mathcal{LR}\left( q;h\right) \Leftrightarrow
\RedemptionShock=\mathbb{V}\left( q\right)
\end{equation}%
Since $\mathcal{LR}\left( q;h\right) $ is an increasing function of $h$ and $\mathcal{LR}\left( q;h^{+}\right) =1$, we have:%
\begin{equation}
\left\{
\begin{array}{ll}
\limfunc{RCR}\left( h\right) <1 & \text{if }h<h^{+} \\
\limfunc{RCR}\left( h\right) =1 & \text{if }h\geq h^{+}%
\end{array}%
\right.
\end{equation}

\item If $\mathbb{V}\left( q\right) >\RedemptionShock$, we have $\limfunc{RCR%
}\left( h\right) >\mathcal{LR}\left( q;h\right) $ and:%
\begin{equation}
\limfunc{RCR}\left( h\right) =\frac{\mathbb{V}\left( q\right) }{%
\RedemptionShock}>1\qquad \forall \,h\geq h^{+}
\end{equation}

\item If $\mathbb{V}\left( q\right) <\RedemptionShock$, we have $\limfunc{RCR%
}\left( h\right) <\mathcal{LR}\left( q;h\right) $ and:%
\begin{equation}
\limfunc{RCR}\left( h\right) <1\qquad \forall \,h\geq 0
\end{equation}
\end{enumerate}
Equation (\ref{eq:RCR4}) shows that the redemption coverage ratio is an
increasing function of $h$. From a risk management perspective, the RCR is
below one if the value $\mathbb{V}\left( q\right) $ of liquid assets is lower
than the redemption shock $\RedemptionShock$ or if the time to liquidation is
not acceptable. Let $h^{\star }=\left\{ \inf h:\limfunc{RCR}\left( h\right)
>1\right\} $ be the number of trading days we need to absorb the redemption
shock. The shorter the period $h^{\star }$ is, the better the liquidity profile. Indeed, if the period $h^{\star }$ is too long and even if $\limfunc{RCR}\left( h^{\star }\right) >1$, we cannot consider that the criterion is satisfied. This is why the risk management department must define an acceptable time to liquidation $\tau _{h}$. In this case, the liquidity profile of the fund is appropriate if and only if $\limfunc{RCR}\left( \tau _{h}\right) >1$. By definition, $\tau _{h}$ depends on the asset class. In the case of public equities, $\tau _{h}$ is equal to a few days, whereas $\tau _{h}$ may range from a few weeks to several months for private equities, depending on the liquidity objective of the investment fund.\smallskip

Similarly, the liquidity shortfall $\limfunc{LS}\left( h\right) $ can
be seen as an extension of the liquidation shortfall, which is defined as
\textquotedblleft \textsl{the remaining redemption that cannot be fulfilled
after one trading day}\textquotedblright\ \citep[Section 3.2.3, page
18]{Roncalli-lst2}:
\begin{equation}
\mathcal{LS}\left( q\right) =1-\mathcal{LR}\left( q;1\right)
\end{equation}%
Indeed, we have:%
\begin{equation}
\limfunc{LS}\left( h\right) =\RedemptionRate\cdot \max \left( 0,1-\frac{%
\mathbb{V}\left( q\right) }{\RedemptionShock}\cdot \mathcal{LR}\left(
q;h\right) \right)
\end{equation}%
In the case where $\mathbb{V}\left( q\right) =\RedemptionShock$, we obtain:
\begin{eqnarray}
\limfunc{LS}\left( h\right)  &=&\RedemptionRate\cdot \max \left( 0,1-%
\mathcal{LR}\left( q;h\right) \right)   \notag \\
&=&\RedemptionRate\cdot \left( 1-\mathcal{LR}\left( q;h\right) \right)
\notag \\
&=&\RedemptionRate\cdot \mathcal{LS}\left( q;h\right)
\end{eqnarray}%
where $\mathcal{LS}\left( q;h\right) =1-\mathcal{LR}\left( q;h\right) $ is
the \textit{generalized} liquidation shortfall, that is the remaining
redemption that cannot be fulfilled after $h$ trading days. While the
liquidation shortfall is calculated with one trading day, the liquidity
shortfall can be calculated with $h\leq \tau _{h}$. In the other cases, the
liquidity shortfall is not equal to the product of the redemption rate
$\RedemptionRate$ and the generalized liquidation shortfall because we have:
\begin{equation}
\limfunc{LS}\left( h\right) =\RedemptionRate\cdot \max \left( 0,1-\frac{%
\mathbb{V}\left( q\right) }{\RedemptionShock}\cdot \left( 1-\mathcal{LS}%
\left( q;h\right) \right) \right)
\end{equation}%
Nevertheless, we always verify that:
\begin{equation}
0\leq \limfunc{LS}\left( h\right) \leq \RedemptionRate
\end{equation}%
By construction, the liquidity shortfall cannot exceed the redemption rate.

\paragraph{Portfolio distortion}

Since the asset structure of the fund is given by the portfolio $\omega
=\left( \omega _{1},\ldots ,\omega _{n}\right) $, the portfolio weights are
equal to $w\left( \omega \right) =\left( w_{1}\left( \omega \right) ,\ldots
,w_{n}\left( \omega \right) \right) $ where:
\begin{equation}
w_{i}\left( \omega \right) =\frac{\omega _{i}\cdot P_{i}}{%
\sum_{j=1}^{n}\omega _{j}\cdot P_{j}}  \label{eq:w-omega}
\end{equation}%
Let $q=\left( q_{1},\ldots ,q_{n}\right) $ be the redemption scenario. It
follows that the redemption weights are given by:
\begin{equation}
w_{i}\left( q\right) =\frac{q_{i}\cdot P_{i}}{\sum_{j=1}^{n}q_{j}\cdot P_{j}}
\label{eq:w-q}
\end{equation}%
After the liquidation of $q$, the new asset structure is equal to $\omega
-q$, and the new weights of the portfolio become:
\begin{equation}
w_{i}\left( \omega -q\right) =\frac{\left( \omega _{i}-q_{i}\right) \cdot
P_{i}}{\sum_{j=1}^{n}\left( \omega _{j}-q_{j}\right) \cdot P_{j}}
\label{eq:w-omega_q}
\end{equation}%
Except in the case of the proportional rule $q_{i}\propto \omega _{i}$, there
is no reason that $w_{i}\left( \omega -q\right) =w_{i}\left( \omega \right)
$. In fact, we have\footnote{The weight difference $\Delta w_{i}\left( \omega
\mid q\right) $ is equal to:
\begin{eqnarray*}
\Delta w_{i}\left( \omega \mid q\right)  &=&w_{i}\left( \omega -q\right)
-w_{i}\left( \omega \right)  \\
&=&\frac{\left( \omega _{i}-q_{i}\right) \cdot P_{i}}{\mathbb{V}\left(
\omega \right) -\mathbb{V}\left( q\right) }-\frac{\omega _{i}\cdot P_{i}}{%
\mathbb{V}\left( \omega \right) } \\
&=&\frac{\mathbb{V}\left( \omega \right) \cdot \left( \omega
_{i}-q_{i}\right) \cdot P_{i}-\left( \mathbb{V}\left( \omega \right) -%
\mathbb{V}\left( q\right) \right) \cdot \omega _{i}\cdot P_{i}}{\left(
\mathbb{V}\left( \omega \right) -\mathbb{V}\left( q\right) \right) \cdot
\mathbb{V}\left( \omega \right) } \\
&=&\frac{\mathbb{V}\left( q\right) \cdot \omega _{i}\cdot P_{i}-\mathbb{V}%
\left( \omega \right) \cdot q_{i}\cdot P_{i}}{\left( \mathbb{V}\left( \omega
\right) -\mathbb{V}\left( q\right) \right) \cdot \mathbb{V}\left( \omega
\right) } \\
&=&\frac{\mathbb{V}\left( q\right) \cdot w_{i}\left( \omega \right) \cdot
\mathbb{V}\left( \omega \right) -\mathbb{V}\left( \omega \right) \cdot
w_{i}\left( q\right) \cdot \mathbb{V}\left( q\right) }{\left( \mathbb{V}%
\left( \omega \right) -\mathbb{V}\left( q\right) \right) \cdot \mathbb{V}%
\left( \omega \right) } \\
&=&\frac{\mathbb{V}\left( q\right) }{\left( \mathbb{V}\left( \omega \right) -%
\mathbb{V}\left( q\right) \right) }\left( w_{i}\left( \omega \right)
-w_{i}\left( q\right) \right)
\end{eqnarray*}%
}:%
\begin{eqnarray}
w_{i}\left( \omega -q\right)  &=&w_{i}\left( \omega \right) +\Delta
w_{i}\left( \omega \mid q\right)   \notag \\
&=&w_{i}\left( \omega \right) +\frac{\mathbb{V}\left( q\right) }{\left(
\mathbb{V}\left( \omega \right) -\mathbb{V}\left( q\right) \right) }\left(
w_{i}\left( \omega \right) -w_{i}\left( q\right) \right)
\end{eqnarray}
The previous analysis can be extended to the case $h<h^{+}$. Indeed, it
assumes that the liquidation is fully executed. Again, we can have $h^{+}\gg
\tau _{h}$, meaning the redemption shock cannot be perfectly absorbed. In
this case, we can compute $w_{i}\left( q;h\right) $ and $w_{i}\left( \omega
-q;h\right) $ by replacing $q_{i}$ with $\sum_{k=1}^{h}q_{i}\left( k\right)$.

\paragraph{Examples}

We consider a fund, whose asset structure $\omega $ is given in Table \ref{ex:rcr0}.
The investment universe is made up of $7$ assets. We also indicate the
current price $P_{i}$ and the trading limit $q_{i}^{+}$ of each asset. The
fund's total net assets are equal to $\$141.734$ mn. We assume that the
redemption shock is equal to $20\%$ or $\$28.347$ mn.

\begin{table}[tbph]
\centering
\caption{Fund's asset structure and liquidation policy}
\label{ex:rcr0}
\begin{tabular}{crrrrrrr}
\hline
Asset     & \multicolumn{1}{c}{$1$} & \multicolumn{1}{c}{$2$} &
\multicolumn{1}{c}{$3$} & \multicolumn{1}{c}{$4$} & \multicolumn{1}{c}{$5$} &
\multicolumn{1}{c}{$6$} & \multicolumn{1}{c}{$7$} \\
\hline
$\omega_i$ &  $435\,100$ & $300\,100$ & $50\,400$ & $200\,500$ & $75\,500$ & $17\,500$ & $1\,800$ \\
$w_i\left(\omega\right)$ &
               $27.32\%$ &  $26.04\%$ & $17.35\%$ &  $14.43\%$ &  $8.90\%$ &  $3.94\%$ & $2.02\%$ \\
$P_i$      &        $89$ &      $123$ &     $488$ &      $102$ &     $167$ &     $319$ & $1\,589$ \\
$q_i^{+}$  &   $20\,000$ &  $20\,000$ & $10\,000$ &  $20\,000$ & $20\,000$ &  $2\,000$ & $1\,000$ \\
\hline
\end{tabular}
\end{table}

\begin{example}[naive pro-rata liquidation]
\label{ex:rcr1} We first consider the pro-rata liquidation (also called the
proportional rule or the vertical slicing approach). In this case, the
liquidation portfolio is equal to $q=\RedemptionRate \cdot \omega = 0.20\cdot
\omega $.
\end{example}

We first determine the number of liquidated shares $q_{i}\left( h\right) $
for $h=1,2,\ldots $ (see Table \ref{tab:rcr1-2} on page \pageref{tab:rcr1-2})
in order to compute the value of $\AssetShock\left( h\right) $ and the
associated redemption coverage ratio $\limfunc{RCR}\left( h\right) $. Results
are given below in Table \ref{tab:rcr1-1}. We notice that $\limfunc{RCR}\left( 1\right) =52.53\%$ and $\limfunc{LS}\left( 1\right) =9.49\%$. If the time horizon $\tau _{h}$ to absorb the redemption shock is equal to one day, then there are not enough liquid assets since the redemption coverage ratio is less than $1$. Indeed, we need a week (or five trading days) to perfectly absorb the redemption shock. In this case, we have $\limfunc{RCR}\left( 5\right) =100\%$ and $\limfunc{LS}\left( 5\right) =0\%$. In Table \ref{tab:rcr1-1}, we also verify that $\limfunc{RCR}\left( h\right) =\mathcal{LR}\left( q;h\right)$. Moreover, we notice the convergence of the portfolio weights after the liquidation to the current portfolio weights (see Table \ref{tab:rcr1-4} on page \pageref{tab:rcr1-4}). Nevertheless, the matching of the two portfolios $\omega -q$ and $\omega $ is only valid when $h\geq h^{+}=5$.

\begin{table}[tbph]
\centering
\caption{Computation of the RCR (Example \ref{ex:rcr1}, naive pro-rata liquidation)}
\label{tab:rcr1-1}
\begin{tabular}{ccccc}
\hline
\multirow{2}{* }{$h$} & $\mathcal{LR}\left( q;h\right) $ & $\AssetShock\left( h\right) $ & $%
\limfunc{RCR}\left( h\right) $ & $\limfunc{LS}\left( h\right) $ \\
& (in \%) & (in \$ mn) & (in \%) & (in \%) \\ \hline
$1$  &  ${\TsV}52.53$  &  $14.892$  &  ${\TsV}52.53$  &  $9.49$ \\
$2$  &  ${\TsV}76.51$  &  $21.689$  &  ${\TsV}76.51$  &  $4.70$ \\
$3$  &  ${\TsV}91.51$  &  $25.939$  &  ${\TsV}91.51$  &  $1.70$ \\
$4$  &  ${\TsV}97.80$  &  $27.722$  &  ${\TsV}97.80$  &  $0.44$ \\
$5$  &       $100.00$  &  $28.347$  &       $100.00$  &  $0.00$ \\
$6$  &       $100.00$  &  $28.347$  &       $100.00$  &  $0.00$ \\
\hline
\end{tabular}
\end{table}

In this example, we assume that $q=\RedemptionRate\cdot \omega $, implying
that $\mathbb{V}\left( q\right) =\RedemptionShock$. This scheme is not
optimal because we have demonstrated that $\limfunc{RCR}\left( h^{+}\right)
=1$ and $\limfunc{RCR}\left( \tau _{h}\right) \leq 1$. The best case is then
obtained if $\tau _{h}=h^{+}$, implying the following constraints:
\begin{equation}
\limfunc{RCR}\left( \tau _{h}\right) =1\Leftrightarrow \left\{ \forall
\,i=1,\ldots ,n:q_{i}=\RedemptionRate\cdot \omega _{i}\leq \tau _{h}\cdot
q_{i}^{+}\right\}
\end{equation}%
If we set $\tau _{h}<h^{+}$, we necessarily have $\limfunc{RCR}\left( \tau
_{h}\right) <1$, meaning that there are not enough liquid assets to fulfill
the redemption scenario. Moreover, we are not sure that $q=\RedemptionRate
\cdot \omega $ is the optimal solution to maximize the redemption coverage
ratio $\limfunc{RCR}\left( \tau _{h}\right) $. Indeed, the previous analysis
suggests that $\mathbb{V}\left( q\right) >\RedemptionShock$ is a better
choice when it is possible. However, this constraint is not always satisfied
and is highly dependent on the value $\tau _{h}$ of the time horizon. In fact, the optimal solution necessarily depends on $\tau _{h}$ and is given by the following optimization problem:
\begin{eqnarray}
q^{\star }\left( \tau _{h}\right)  &=&\arg \max \limfunc{RCR}\left( \tau
_{h}\right)   \notag \\
&\text{s.t.}&\left\{
\begin{array}{l}
q\propto \omega  \\
q\geq \mathbf{0}_{n}%
\end{array}%
\right.
\end{eqnarray}%
By construction, the solution is independent from the value $\RedemptionShock
$ of the redemption shock since we have:
\begin{equation}
\arg \max \limfunc{RCR}\left( \tau _{h}\right) :=\arg \max \AssetShock\left(
\tau _{h}\right)
\end{equation}%
We obtain a trivial combinatorial problem. Indeed, the solution must satisfy
the following set of constraints:
\begin{equation}
\left\{
\begin{array}{l}
q\propto \omega  \\
q_{i}\leq \min \left( \tau _{h}\cdot q_{i}^{+},\omega _{i}\right)
\end{array}%
\right.
\end{equation}%
We deduce that:%
\begin{equation}
q^{\star }\left( \tau _{h}\right) =\varphi \left( \tau _{h}\right) \cdot
\omega
\end{equation}%
where:%
\begin{equation}
\varphi \left( \tau _{h}\right) =\inf_{i=1,\ldots ,n}\min \left( \tau
_{h}\cdot \dfrac{q_{i}^{+}}{\omega _{i}},1\right)   \label{eq:RCR-optimal1}
\end{equation}%
Moreover, we have:%
\begin{eqnarray}
\AssetShock\left( \tau _{h}\right)  &=&\sum_{i=1}^{n}\left(
\sum_{k=1}^{h}q_{i}\left( k\right) \right) P_{i}  \notag \\
&=&\sum_{i=1}^{n}q_{i}^{\star }\left( \tau _{h}\right) \cdot P_{i}  \notag \\
&=&\varphi \left( \tau _{h}\right) \left( \sum_{i=1}^{n}\omega _{i}\cdot P_{i} \right)
\notag \\
&=&\varphi \left( \tau _{h}\right) \cdot \limfunc{TNA}
\end{eqnarray}%
We conclude that the redemption coverage rate is equal to the ratio between $\varphi \left( \tau _{h}\right) $ and $\RedemptionRate$:
\begin{equation}
\limfunc{RCR}\left( \tau _{h}\right) =\frac{\varphi \left( \tau _{h}\right)
}{\RedemptionRate}  \label{eq:RCR-optimal2}
\end{equation}

\begin{example}[optimal pro-rata liquidation]
\label{ex:rcr2} We consider the optimal pro-rata liquidation when the
redemption shock $\RedemptionRate$ is equal to $20\%$ and the time horizon
$\tau_h$ varies from one trading day to one trading week.
\end{example}

In Table \ref{tab:rcr2-1}, we indicate the optimal value $\varphi \left( \tau
_{h}\right) $ for each time horizon $\tau _{h}$. We also report\footnote{We
don't need to report the statistics for $h\geq \tau _{h}$ because we have
$\mathcal{LR}\left( q;h\right) =\mathcal{LR}\left( q;\tau _{h}\right) $,
$\AssetShock\left( h\right) =\AssetShock\left( \tau _{h}\right) $,
$\limfunc{RCR}\left( h\right) =\limfunc{RCR}\left( \tau _{h}\right) $ and
$\limfunc{LS}\left( h\right) =\limfunc{LS}\left( \tau _{h}\right) $.}
$\mathcal{LR}\left( q;h\right) $, $\AssetShock\left( h\right) $,
$\limfunc{RCR}\left( h\right) $ and $\limfunc{LS}\left( h\right) $ for $h\leq
\tau _{h}$. When $\tau _{h}=1$, the optimal liquidation portfolio is equal to
$\left( 20\,000,13\,795,2\,317,9\,216,3\,470,804\right) $. The redemption
coverage ratio is equal to $22.98\%$, implying a high liquidity shortfall
representing $15.40\%$ of the total net assets. When $\tau _{h}=2$, the
optimal portfolio $q^{\star}$ becomes
$\left(40\,000,27\,589,4\,633,18\,433,6\,941, 1\,609\right) $. The redemption
coverage ratio is then equal to $45.97\%$ whereas the liquidity shortfall
represents $10.81\%$ of the total net assets. In Exercise \ref{ex:rcr1}, the
liquidation period $h^{+}$ was equal to five trading days, and we obtained
$\limfunc{RCR}\left( 5\right) =100\%$. We notice that we achieve a better
redemption coverage ratio with the optimal pro-rata liquidation rule. Indeed,
we have $\limfunc{RCR}\left( 5\right) =114.92\%$.

\begin{table}[tbph]
\centering
\caption{Computation of the RCR (Example \ref{ex:rcr2}, optimal pro-rata liquidation)}
\label{tab:rcr2-1}
\begin{tabular}{ccccccc}
\hline
\multirow{2}{*}{$\tau_h$} & $\varphi \left( \tau _{h}\right)$ &
\multirow{2}{*}{$h$} & $\mathcal{LR}\left( q;h\right) $ & $\AssetShock\left( h\right) $ & $%
\limfunc{RCR}\left( h\right) $ & $\limfunc{LS}\left( h\right) $ \\
             &               (in \%) &       &       (in \%) &     (in \$ mn) &       (in \%) &      (in \%) \\ \hline
        $1$  &         ${\TsV}4.60$  &  $1$  &      $100.00$ & ${\TsV}6.515$ & ${\TsV}22.98$ &      $15.40$ \\  \hline
\mrm{2}{$2$} & \mrm{2}{${\TsV}9.19$} &  $1$  & ${\TsV}79.18$ &      $10.317$ & ${\TsV}36.39$ &      $12.72$ \\
             &                       &  $2$  &      $100.00$ &      $13.030$ & ${\TsV}45.97$ &      $10.81$ \\  \hline
\mrm{3}{$3$} &      \mrm{3}{$13.79$} &  $1$  & ${\TsV}63.66$ &      $12.443$ & ${\TsV}43.89$ &      $11.22$ \\
             &                       &  $2$  & ${\TsV}90.02$ &      $17.595$ & ${\TsV}62.07$ & ${\TsV}7.59$ \\
             &                       &  $3$  &      $100.00$ &      $19.545$ & ${\TsV}68.95$ & ${\TsV}6.21$ \\   \hline
\mrm{4}{$4$} &      \mrm{4}{$18.39$} &  $1$  & ${\TsV}54.81$ &      $14.284$ & ${\TsV}50.39$ & ${\TsV}9.92$ \\
             &                       &  $2$  & ${\TsV}79.18$ &      $20.633$ & ${\TsV}72.79$ & ${\TsV}5.44$ \\
             &                       &  $3$  & ${\TsV}93.17$ &      $24.280$ & ${\TsV}85.65$ & ${\TsV}2.87$ \\
             &                       &  $4$  &      $100.00$ &      $26.060$ & ${\TsV}91.93$ & ${\TsV}1.61$ \\  \hline
\mrm{5}{$5$} &      \mrm{5}{$22.98$} &  $1$  & ${\TsV}47.13$ &      $15.353$ & ${\TsV}54.16$ & ${\TsV}9.17$ \\
             &                       &  $2$  & ${\TsV}70.74$ &      $23.044$ & ${\TsV}81.29$ & ${\TsV}3.74$ \\
             &                       &  $3$  & ${\TsV}85.68$ &      $27.911$ & ${\TsV}98.46$ & ${\TsV}0.31$ \\
             &                       &  $4$  & ${\TsV}94.54$ &      $30.795$ &      $108.64$ & ${\TsV}0.00$ \\
             &                       &  $5$  &      $100.00$ &      $32.575$ &      $114.92$ & ${\TsV}0.00$ \\  \hline
\end{tabular}
\vspace*{-15pt}
\end{table}

\begin{figure}[h!]
\centering
\caption{Maximum admissible redemption shock in \% (Example \ref{ex:rcr2}, optimal pro-rata liquidation)}
\label{fig:rcr2b}
\figureskip
\includegraphics[width = \figurewidth, height = \figureheight]{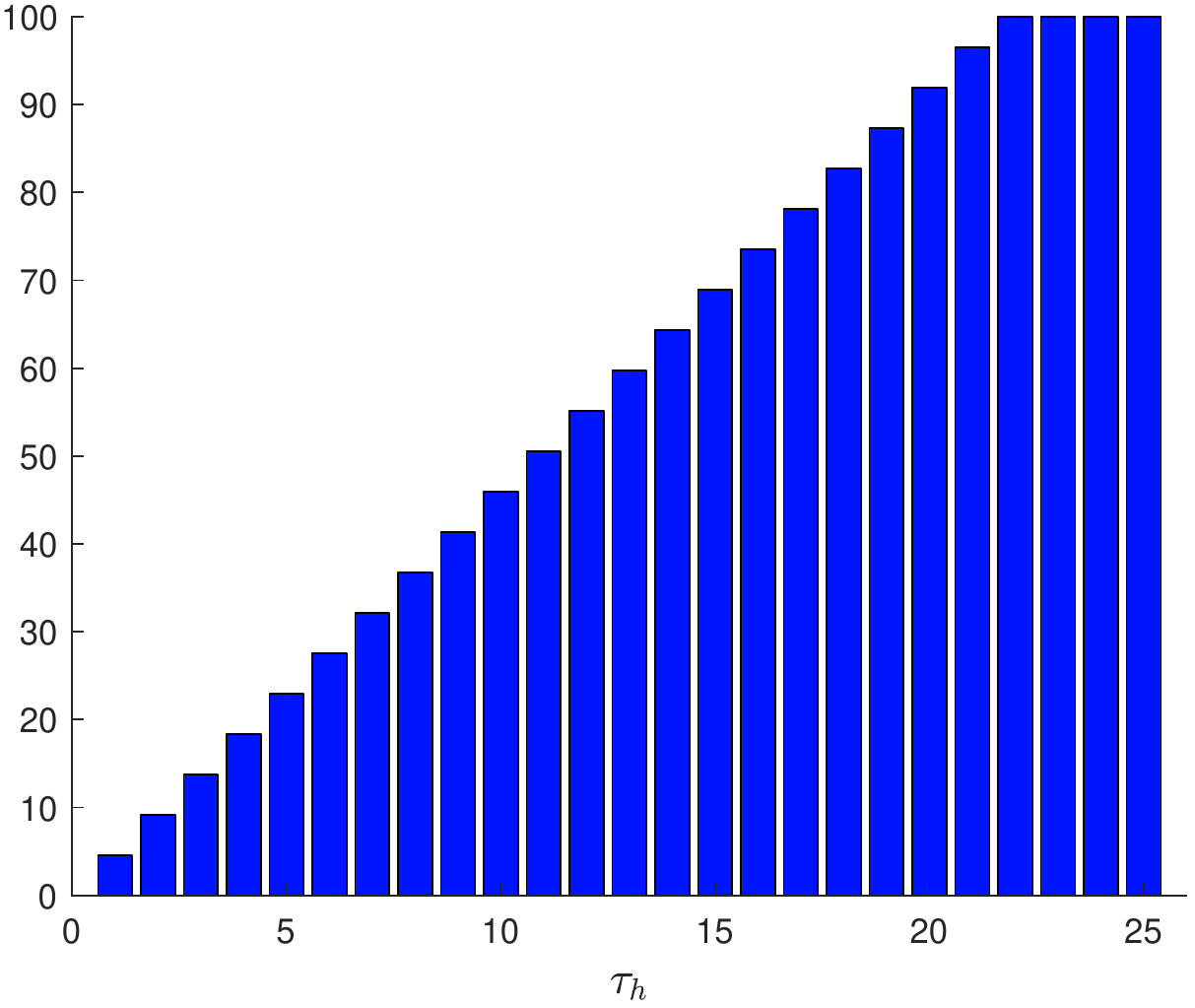}
\end{figure}

\begin{remark}
Since the optimal portfolio $q^{\star }\left( \tau _{h}\right) $ does not
depend on the redemption shock $\RedemptionShock$, $\AssetShock\left( \tau
_{h}\right) $ indicates the maximum redemption shock that can be absorbed,
implying that:
\begin{equation*}
\RedemptionShock\leq \AssetShock\left( \tau _{h}\right) \Rightarrow \limfunc{%
RCR}\left( \tau _{h}\right) \geq 1
\end{equation*}%
By definition, the maximum admissible redemption shock is equal to
$\RedemptionShock\left( \tau _{h}\right) =\AssetShock \left( \tau _{h}\right) $ or
$\RedemptionRate\left( \tau _{h}\right) =\varphi \left( \tau _{h}\right) $.
For instance, the maximum admissible redemption shock is equal to $\$6.515$
mn (or $4.60\%$ of the TNA) when the time horizon is set to one trading day.
Figure \ref{fig:rcr2b} shows the evolution of $\RedemptionRate\left(
\tau_{h}\right)$ with respect to $\tau _{h}$.
\end{remark}

\begin{example}[waterfall liquidation]
\label{ex:rcr3} We now consider the waterfall liquidation. In this case, the
fund manager liquidates assets in order of their liquidity starting from the
most liquid ones. The redemption shock is still equal to $20\%$.
\end{example}

\begin{table}[tbph]
\centering
\caption{Computation of the RCR (Example \ref{ex:rcr3}, waterfall liquidation)}
\label{tab:rcr3-1}
\begin{tabular}{ccccc}
\hline
\multirow{2}{* }{$h$} & $\mathcal{LR}\left( q;h\right) $ & $\AssetShock\left( h\right) $ & $%
\limfunc{RCR}\left( h\right) $ & $\limfunc{LS}\left( h\right) $ \\
& (in \%) & (in \$ mn) & (in \%) & (in \%) \\ \hline
$1$  &  $11.80$  &  $16.727$  &  $ 59.01$  &  $8.20$ \\
$2$  &  $23.38$  &  $33.136$  &  $116.90$  &  $0.00$ \\
$3$  &  $34.06$  &  $48.274$  &  $170.30$  &  $0.00$ \\
$4$  &  $44.21$  &  $62.661$  &  $221.05$  &  $0.00$ \\
$5$  &  $52.53$  &  $74.459$  &  $262.67$  &  $0.00$ \\
$6$  &  $57.55$  &  $81.572$  &  $287.76$  &  $0.00$ \\
\hline
\end{tabular}
\end{table}

\begin{figure}[tbph]
\centering
\caption{Maximum admissible redemption shock in \% (pro-rata vs. waterfall liquidation)}
\label{fig:rcr3b}
\figureskip
\includegraphics[width = \figurewidth, height = \figureheight]{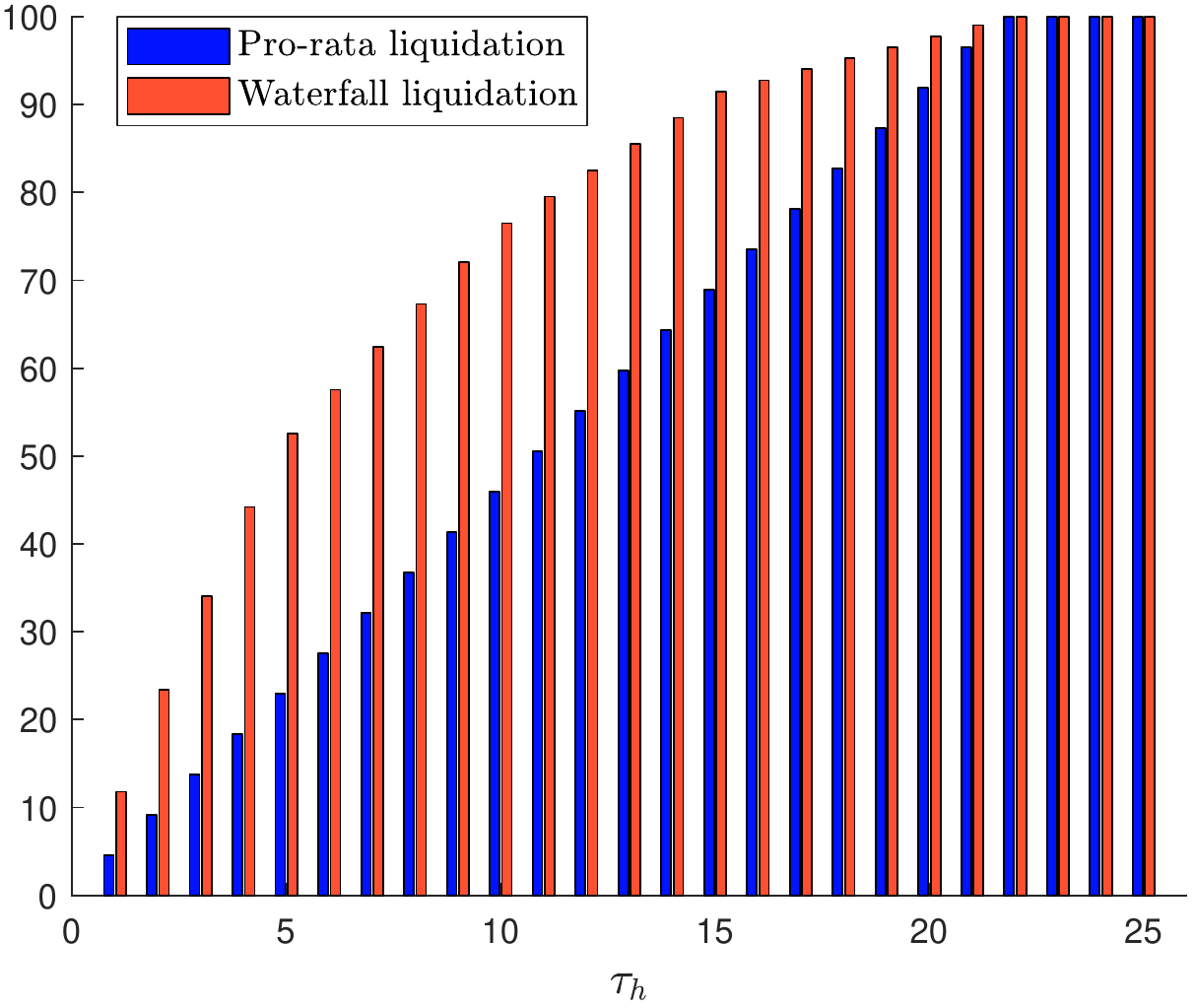}
\end{figure}

In the waterfall approach, there are no constraints on the liquidation
portfolio $q$, which is equal to the fund's portfolio $\omega $. In this
case, the redemption coverage ratio is entirely determined by the trading
limits $q^{+}$ and the current portfolio $\omega $. Every day, we sell $%
q_{i}^{+}$ shares of security $i$ until there is nothing left -- $%
q_{i}\left( h\right) =0$. Results are given in Table \ref{tab:rcr3-1}. Since
there are no constraints on the asset structure of the portfolio $\omega -q$,
we obtain higher values of the redemption coverage ratio compared to the
naive or optimal pro-rata liquidation approach. Indeed, we have $\limfunc{RCR%
}\left( 1\right) =59.01\%$, but $\limfunc{RCR}\left( 2\right) =116.90\%$. In
this example, we have $\limfunc{RCR}\left( \tau _{h}\right) >1$ when $\tau
_{h}\geq 2$. By construction, the waterfall approach will always give higher
redemption coverage ratios than the pro-rata approach. To illustrate this
property, we compare the maximum admissible redemption shock in Figure
\ref{fig:rcr3b} for the two approaches.

\subsubsection{High-quality liquid assets approach}

\paragraph{Mathematical framework}

In the high-quality liquid assets (HQLA) method, the amount of liquid
assets is estimated by splitting securities by HQLA classes and applying
liquidity weights. We assume that we have $m$ HQLA classes. Let $\func{ccf}_{k}$
denote the liquidity weight or the cash conversion factor (CCF) of
the $k^{\mathrm{th}}$ HQLA class. The ratio of liquid assets in the fund is
defined by:
\begin{equation}
\AssetRate=\sum_{i=1}^{n}w_{i}\left( \omega \right) \cdot \limfunc{CCF}\nolimits_{\ell
\left( i\right) }
\end{equation}%
where $\ell \left( i\right) $ indicates the HQLA class $k$ of security $i$.
We have:
\begin{eqnarray}
\AssetRate &=&\sum_{i=1}^{n}w_{i}\left( \omega \right) \cdot \left(
\sum_{k=1}^{m}\mathds{1}\left\{ i\in k\right\} \cdot \limfunc{CCF}\nolimits_{k}\right)
\notag \\
&=&\sum_{k=1}^{m}\left( \sum_{i=1}^{n}\mathds{1}\left\{ i\in k\right\} \cdot
w_{i}\left( \omega \right) \right) \cdot \limfunc{CCF}\nolimits_{k}  \notag \\
&=&\sum_{k=1}^{m}w_{k}\cdot \limfunc{CCF}\nolimits_{k}
\end{eqnarray}%
where $w_{k}$ is the weight of the $k^{\mathrm{th}}$ HQLA class\footnote{We also have:%
\begin{eqnarray}
\AssetShock &=&\AssetRate\cdot \func{TNA}  \notag \\
&=&\sum_{k=1}^{m}\left( w_{k}\cdot \func{TNA}\right) \cdot \limfunc{CCF}\nolimits_{k}
\notag \\
&=&\sum_{k=1}^{m}\func{TNA}\nolimits_{k}\cdot \limfunc{CCF}\nolimits_{k}
\end{eqnarray}%
where $\func{TNA}\nolimits_{k}$ is the dollar amount of the $k^{\mathrm{th}}$ HQLA
class.}. We deduce that:%
\begin{equation}
\func{RCR}=\frac{\sum_{k=1}^{m}w_{k}\cdot \limfunc{CCF}\nolimits_{k}}{\RedemptionRate}
\end{equation}%
and:%
\begin{equation}
\func{LS}=\RedemptionRate\cdot \max \left( 0,1-\frac{\sum_{k=1}^{m}w_{k}%
\cdot \limfunc{CCF}\nolimits_{k}}{\RedemptionRate}\right)
\end{equation}

\paragraph{Definition of HQLA classes}

The term HQLA refers to the liquidity coverage ratio (LCR) introduced in the
Basel III framework \citep{BCBS-2010, BCBS-2013}. An asset is considered to be a high-quality liquid asset if it can be easily converted into cash.
Therefore, the concept of HQLA is related to asset quality and asset
liquidity. The first property indicates if the asset can be sold without
discount, while the second property indicates if the asset can be easily and
quickly sold \citep{Roncalli-2020}. Thus, the LCR ratio measures whether or not the bank has the necessary assets to face a one-month stressed period of outflows. The stock of HQLA is computed by defining eligible assets and applying haircut values. For instance, corporate debt securities rated above
\textsf{BBB$-$} are eligible, implying that high yield bonds are not. Then, a
haircut of $15\%$ (resp. $50\%$) is applied to corporate bonds rated
\textsf{AA$-$} or higher (resp. between \textsf{A$+$} and \textsf{BBB$-$}).
Since the time horizon of the LCR is one month, the underlying idea is that
(1) high yield bonds can be illiquid for one month, (2) investment grade
corporate bonds can be sold during the month but with a discount, (3)
corporate bonds rated \textsf{AA$-$} or higher can lose $15\%$ of their
value in the month and (4) corporate bonds rated between \textsf{A$+$}
and \textsf{BBB$-$} can lose $50\%$ of their value in the
month.\smallskip

In Table \ref{tab:hqla1}, we report the HQLA matrix given by
\citet{Bouveret-2017} and \citet{IMF-2017}, which corresponds to the HQLA
matrix of the Basel III Accord using the following rule:
\begin{equation}
\limfunc{CCF}\nolimits_{k}=1-H_{k}
\end{equation}%
where $H_{k}$ is the haircut value. By construction, the CCF value is equal
to $100\%$ for cash. For equities, it is equal to $50\%$. Although common
equity shares are highly liquid, we can face a price drop before the
liquidation. Therefore, this value of $50\%$ mainly reflects a discount
risk. Sovereign bonds are assumed to be a perfect substitute for the cash if
the credit rating of the issuer is \textsf{AA$-$} or higher. Otherwise, the
CCF is equal to $85\%$ and $50\%$ for other IG sovereign bonds and $0\%$ for
HY sovereign bonds. In the case of corporate bonds, securities rated below
\textsf{BBB$-$} receive a CCF of $0\%$, while the CCF is respectively equal to $50\%$ and $85\%$ for \textsf{BBB$-$} to \textsf{A$+$} and \textsf{AA$-$} to \textsf{AAA}. For securitization, the CCFs are the same as for corporate bonds, except the category \textsf{BBB$-$} to \textsf{BBB$+$} for which the CCF is set to zero.

\begin{table}[tbph]
\centering
\caption{Cash conversion factors}
\label{tab:hqla1}
\begin{tabular}{cccccc}
\hline
Credit                             & \mr{Cash}        & Sovereign    & Corporate   & \mr{Securitization} & \mr{Equities}   \\
Rating                             &                  & bonds        & bonds       &                     &                 \\ \hline
\textsf{AA$-$} to \textsf{AAA}     & \mrm{4}{$100\%$} &      $100\%$ &      $85\%$ &     $85\%$          & \mrm{4}{$50\%$} \\
\textsf{A$-$} to \textsf{A$+$}     &                  & ${\TsV}85\%$ &      $50\%$ &     $50\%$          &                 \\
\textsf{BBB$-$} to \textsf{BBB$+$} &                  & ${\TsV}50\%$ &      $50\%$ & ${\TsV}0\%$         &                 \\
Below \textsf{BBB$-$}              &                  &  ${\TsX}0\%$ & ${\TsV}0\%$ & ${\TsV}0\%$         &                 \\ \hline
\end{tabular}
\begin{flushleft}
{\small \textit{Source}: \citet[Table 6, page 14]{Bouveret-2017} and \citet[Box 2, page 56]{IMF-2017}.}
\end{flushleft}
\vspace*{-10pt}
\end{table}

\begin{remark}
\citet[Exhibit 38, page 26]{ESMA-2019b} uses the same HQLA matrix, except
for securitization products. In this case, the CCFs are between $65\%$ and $%
93\%$ if the credit rating of the structure is between \textsf{AA-} and
\textsf{AAA}, and $0\%$ otherwise.
\end{remark}

\begin{figure}[tbph]
\centering
\caption{Redemption coverage ratio in \% with the HQLA approach}
\label{fig:hqla1}
\figureskip
\includegraphics[width = \figurewidth, height = \figureheight]{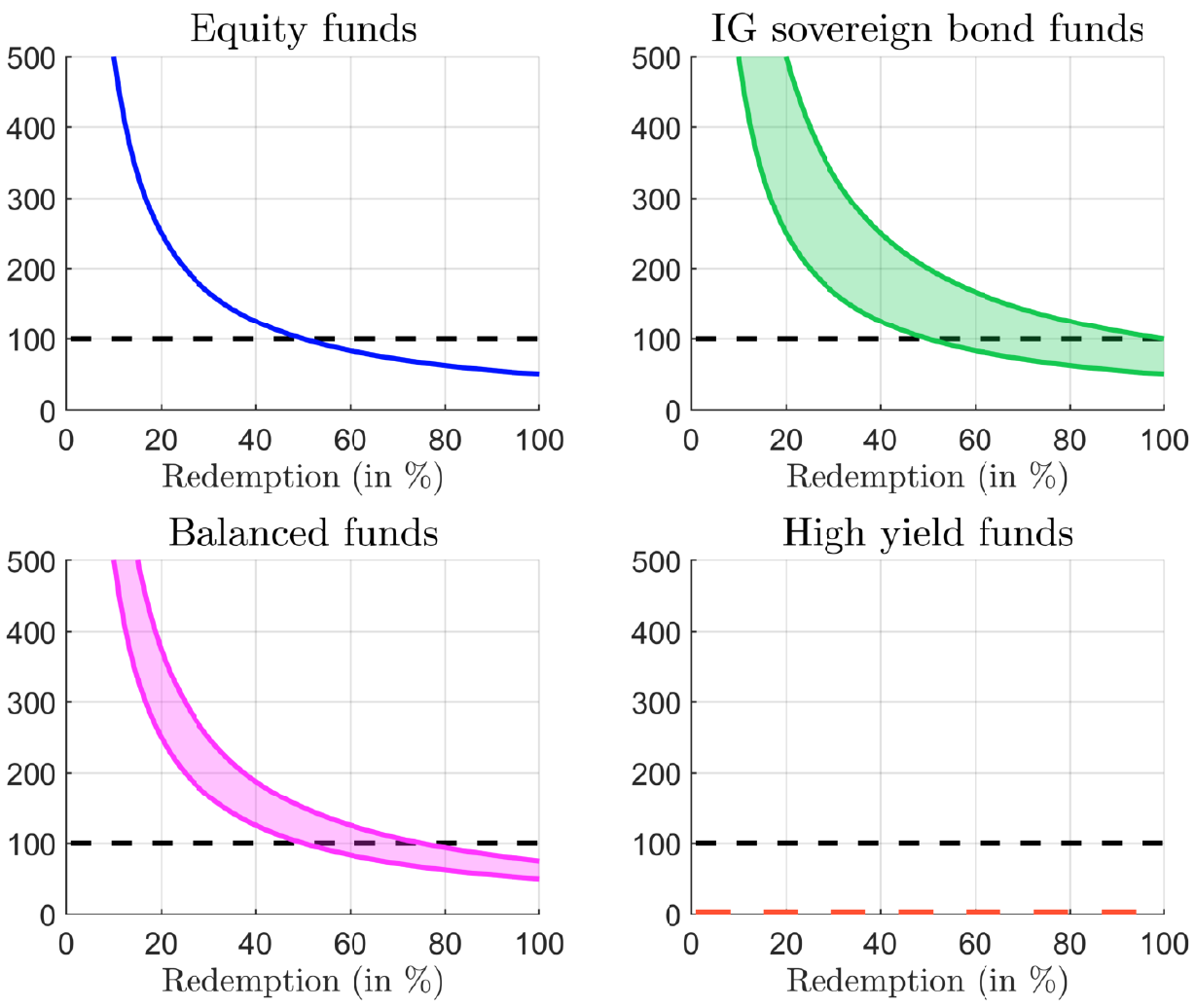}
\end{figure}

As noticed by \citet{ESMA-2019b}, \textquotedblleft \textsl{the HQLA approach
is very attractive from an operational point of view since it is easy to
compute and interpret}\textquotedblright. However, this approach has three
drawbacks. First, the HQLA matrix proposed by the IMF and ESMA is a copy/paste of
the HQLA matrix proposed by the Basel Committee, suggesting that the implicit
time horizon $\tau _{h}$ is one month or 21 trading days. However, the time
horizon is never mentioned, implying that there is a doubt about the IMF and ESMA's true intentions. Second, the granularity of the HQLA matrix is
quite coarse. For instance, there is no distinction between large cap and
small cap stocks. In the case of sovereign bonds, the CCR only depends on the
credit rating. However, we know that some bonds are more liquid than others
even if they belong to the same category of credit rating. For example,
sovereign bonds issued by France, Germany, the UK and the US are more liquid than sovereign bonds issued by Belgium, Denmark, Finland, Ireland, Japan,
Netherlands and Sweden\footnote{This can be measured by the turnover ratio.}.
We observe the same issue with peripheral debt securities (Greece, Italy,
Portugal, Spain) and EM bonds. In the case of corporate bonds, this problem is even more serious, because liquidity is not only an issuer-related question. For instance, the maturity impacts the liquidity of the bonds issued by the same company. The last drawback concerns the absence of the portfolio structure in the computation of the RCR. Indeed, the RCR depends neither on the portfolio holdings nor on the portfolio concentration. Therefore, the HQLA method is a specific top-down approach, which only focuses on asset classes. Two equity funds will have the same redemption coverage ratio for the same redemption shock (top left-hand panel in Figure \ref{fig:hqla1}). For example, we have $\func{RCR}=2.5$ if $\RedemptionRate=20\%$. The RCR is below one if the redemption shock is greater than $50\%$. For a high yield fund, the RCR is equal to zero whatever the value of the redemption shock (bottom right-hand panel
in Figure \ref{fig:hqla1}). For a balanced fund, comprised of $50\%$
IG bonds and $50\%$ public equities, we obtain the following bounds:
\begin{equation}
\frac{50\%}{\RedemptionRate}\leq \func{RCR}\leq \frac{75\%}{\RedemptionRate}
\end{equation}%
Therefore, it is obvious that the HQLA method is a macro-economic approach, that can make sense for regulators to monitor the liquidity risk at the industry level, but it is not adapted for comparing the liquidity risk of two funds.

\paragraph{Implementation of the HQLA approach}

Because of the previous comments, asset managers that would like to
implement the HQLA approach must take into account the following
considerations:
\begin{itemize}
\item The HQLA matrix must be more granular.
\item The asset manager must use different time horizons.
\item The calibration of the cash conversion factor mixes two factors\footnote{See
Appendix \ref{appendix:hqla} on page \pageref{appendix:hqla}
for the derivation of this result. A more conservative formula is
$\limfunc{CCF}\nolimits_{k}\left( \tau _{h}\right) =\limfunc{LF}%
\nolimits_{k}\left( \tau _{h}\right) \cdot \left( 1-\limfunc{DF}%
\nolimits_{k}\left( \tau _{h}\right) \right)$.}:
\begin{equation}
\limfunc{CCF}\nolimits_{k}\left( \tau _{h}\right) =\limfunc{LF}%
\nolimits_{k}\left( \tau _{h}\right) \cdot \left( 1-\limfunc{DF}%
\nolimits_{k}\left( \frac{\tau _{h}}{2}\right) \right)
\label{eq:hqla2}
\end{equation}%
where $\limfunc{LF}_{k}\left( \tau _{h}\right) $ is the (pure) liquidity
factor and $\limfunc{DF}\nolimits_{k}\left( \tau _{h}\right) $ is the
discount (or drawdown) factor.
\item The liquidity factor $\limfunc{LF}_{k}\left( \tau _{h}\right) $ is an
increasing function of $\tau _{h}$. It indicates the proportion of the HQLA
bucket that can be sold in $\tau _{h}$ trading days. By definition, we have
$\limfunc{LF}\nolimits_{k}\left( 0\right) =0$ and
$\limfunc{LF}\nolimits_{k}\left( \infty \right) =1$.

\item The drawdown factor $\limfunc{DF}\nolimits_{k}\left( \tau _{h}\right)$ is an increasing function of $\tau _{h}$. It indicates the loss value of the HQLA bucket in a worst-case scenario of a price drop after $\tau _{h}$ trading days. By definition, we have $\limfunc{DF}\nolimits_{k}\left( 0\right) =0$ and $\limfunc{DF}\nolimits_{k}\left( \infty \right) \leq 1$.
\end{itemize}
Concerning the HQLA classes, we can consider more granularity concerning the
asset class. For example, we can distinguish DM vs. EM equities, LC vs. SC
equities, etc. Moreover, we can introduce the specific risk factor of the
fund, which encompasses two main dimensions: the fund's size and its
portfolio structure. For instance, liquidating a fund of $\$100$
mn is different to liquidating a fund of $\$10$ bn. Similarly, the
liquidation of two funds with the same size can differ because of the weight
concentration difference. Indeed, liquidating a S\&P 500 index fund of $\$1$
bn is different to liquidating an active fund of $\$1$ bn that is
concentrated on $10$ American stocks. Therefore, the cash conversion factor
becomes:
\begin{equation}
\limfunc{CCF}\nolimits_{k,j}\left( \tau _{h}\right) =\limfunc{LF}%
\nolimits_{k}\left( \tau _{h}\right) \cdot \left( 1-\limfunc{DF}%
\nolimits_{k}\left( \frac{\tau _{h}}{2}\right) \right) \cdot \left( 1-%
\limfunc{SF}\nolimits_{k}\left( \limfunc{TNA}\nolimits_{j},\mathcal{H}%
_{j}\right) \right)   \label{eq:hqla3}
\end{equation}%
where $\limfunc{SF}\nolimits_{k}\in \left[ 0,1\right] $ is the specific risk
factor associated to the fund $j$. This is a decreasing function of the fund
size $\limfunc{TNA}\nolimits_{j}$ and the Herfindahl index $\mathcal{H}_{j}$
of the portfolio. Concerning the time horizon, $\tau _{h}$ can be one day,
two days, one week, two weeks or one month. Finally, the three functions
$\limfunc{LF}_{k}\left( \tau _{h}\right) $, $\limfunc{DF}\nolimits_{k}\left( \tau _{h}\right) $
and $\limfunc{SF}\nolimits_{k}\left( \limfunc{TNA}\nolimits_{j},\mathcal{H}_{j}\right)$
can be calibrated using standard econometric procedures.\smallskip

\begin{figure}[tbph]
\centering
\caption{Specification of the cash conversion factor}
\label{fig:hqla2}
\figureskip
\includegraphics[width = \figurewidth, height = \figureheight]{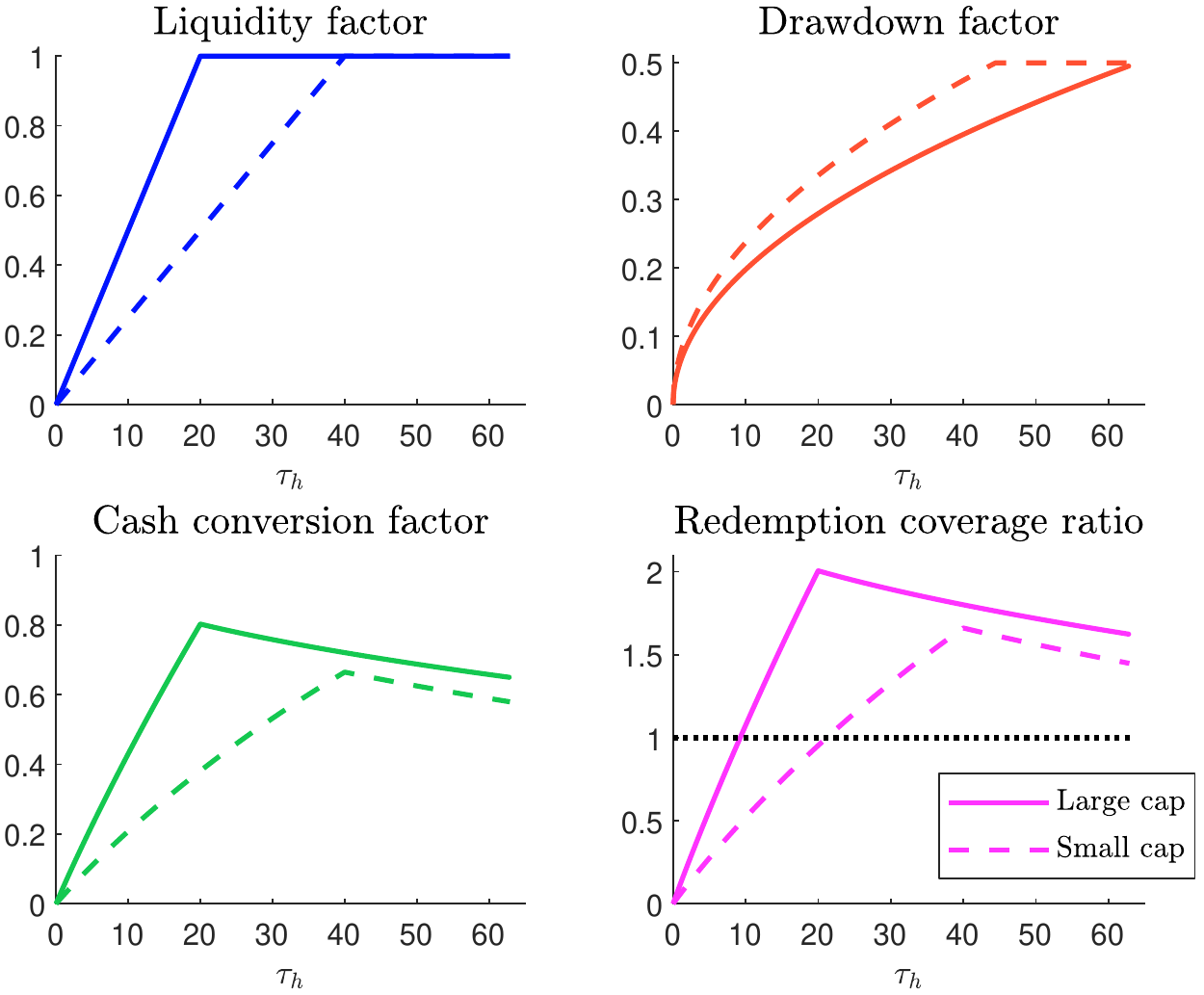}
\end{figure}

A basic specification of the liquidity factor is:
\begin{equation}
\limfunc{LF}\nolimits_{k}\left( \tau _{h}\right) =\min \left( 1.0,\lambda
_{k}\cdot \tau _{h}\right)
\end{equation}%
where $\lambda _{k}$ is the selling intensity. For the drawdown factor, it
is better to use a square root function\footnote{This is what we observe when we compute the value-at-risk of equity indices. For instance, we have reported the historical value-at-risk of the S\&P 500 index in Figure \ref{fig:hqla3} on page \pageref{fig:hqla3} for different confidence levels $\alpha $. We obtain a square-root shape. In risk management, the square-root-of-time rule is very popular and is widely used for modeling drawdown functions \citep[page 46]{Roncalli-2020}.}:
\begin{equation}
\limfunc{DF}\nolimits_{k}\left( \tau _{h}\right) =\min \left( \limfunc{MDD}%
\nolimits_{k},\eta _{k}\cdot \sqrt{\tau _{h}}\right)
\end{equation}%
where $\limfunc{MDD}\nolimits_{k}$ is the maximum drawdown and $\eta _{k}$ is
the loss intensity of the HQLA class. Let us consider the example of a large
cap equity fund, whose total net assets are equal to $\$1$ bn. The redemption
shock is set to $\$400$ mn. We assume that $\lambda _{k}=5\%$ per day, $\eta
_{k}=6.25\%$ and $\limfunc{MDD}\nolimits_{k}=50\%$. Results are reported in
Figure \ref{fig:hqla2}. We notice that the RCR depends on the value of $\tau
_{h}$. For small values of $\tau _{h}$ (less than 10 days), the RCR is below
1. For large values of $\tau _{h}$ (greater than 10 days), the RCR is above 1
because the liquidation factor overtakes the drawdown factor. Finally, we
observe that the CCF and RCR functions are increasing and then decreasing
with respect to the time horizon\footnote{This is normal since we combine an
increasing linear function with a decreasing square-root function.}. We now
consider a second fund with the same assets under management, which is
invested in small cap stocks. In this case, we assume that $\lambda _{k}$ is
reduced by a factor of two and $\eta _{k}$ is increased by 20\%. Results are
given in Figure \ref{fig:hqla2}. We verify that the small cap fund has a
lower RCR than the large cap fund.\smallskip

\begin{figure}[tbph]
\centering
\caption{Specification of the specific risk factor}
\label{fig:hqla4}
\figureskip
\includegraphics[width = \figurewidth, height = \figureheight]{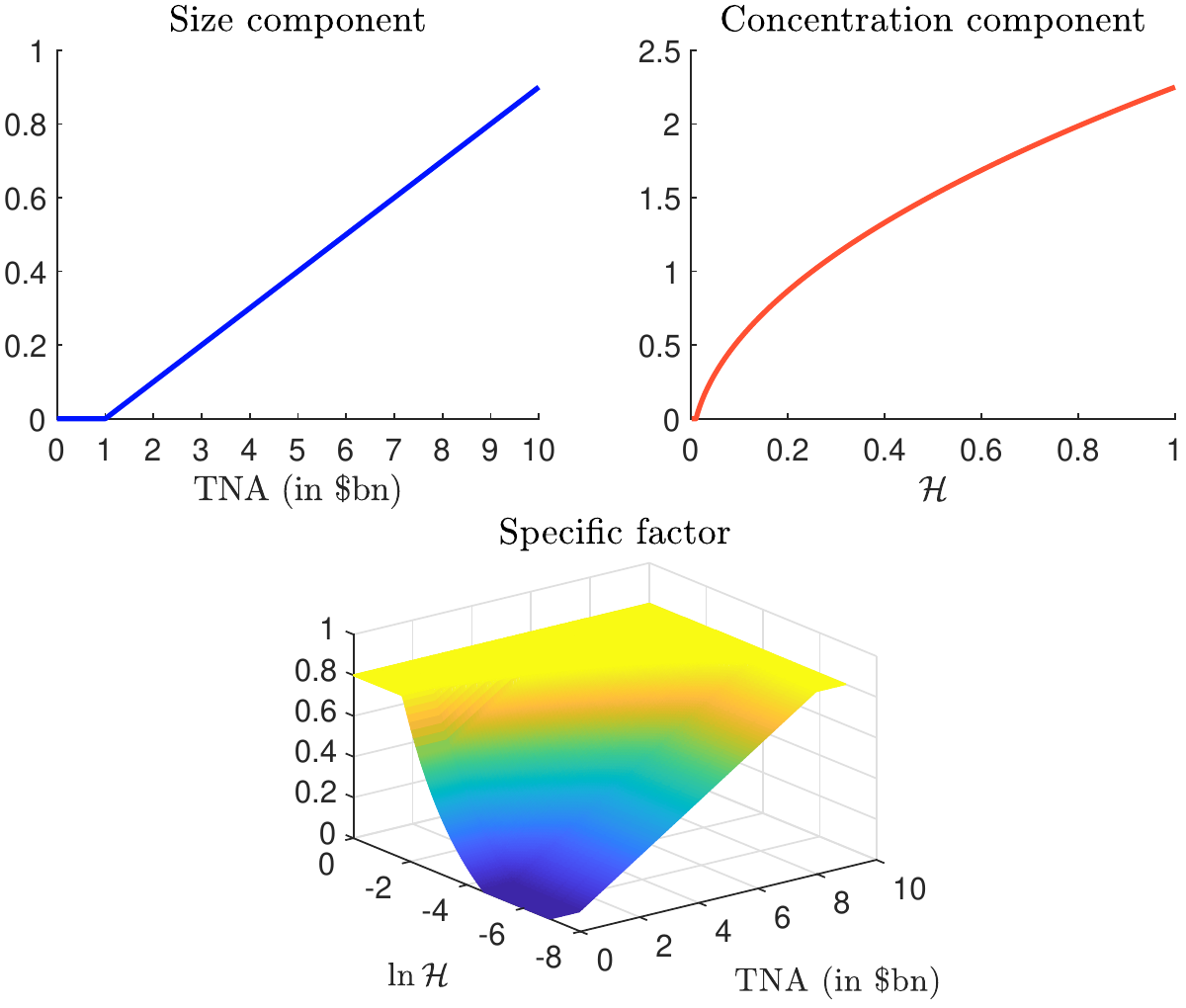}
\end{figure}

As explained previously, we should consider the specific risk of the
fund. We propose the following formula:
\begin{equation}
\limfunc{SF}\nolimits_{k}\left( \limfunc{TNA},\mathcal{H}\right) =\min
\left( \xi _{k}^{\mathrm{size}}\left( \frac{\limfunc{TNA}}{\limfunc{TNA}%
\nolimits^{\star }}-1\right) ^{+}+\xi _{k}^{\mathrm{concentration}}\left(
\sqrt{\frac{\mathcal{H}}{\mathcal{H}^{\star }}}-1\right) ^{+},\limfunc{SF}%
\nolimits^{+}\right)
\end{equation}%
where $\limfunc{TNA}$ and $\mathcal{H}$ are the total net assets and the
Herfindahl index of the fund, which is computed as
$\mathcal{H}=\sum_{i=1}^{n}w_{i}^{2}\left( \omega \right) $. By definition,
we have $n^{-1}\leq \mathcal{H}\leq 1$. $\limfunc{TNA}\nolimits^{\star }$ and
$\mathcal{H}^{\star }$ are two thresholds. Below these two limits,
$\limfunc{SF}\nolimits_{k}\left( \limfunc{TNA},\mathcal{H}\right) $ is equal
to zero. $\xi _{k}^{\mathrm{size}}$ and $\xi _{k}^{\mathrm{concentration}}$
are two coefficients that control the importance of the size and
concentration risks. Moreover, $\limfunc{SF}\nolimits^{+}$ indicates the
maximum value that can be taken by the specific risk since we have the
following inequalities:
\begin{equation}
\left\{
\begin{array}{l}
0\leq \limfunc{SF}\nolimits^{+}\leq 1 \\
0\leq \limfunc{SF}\nolimits_{k}\left( \limfunc{TNA},\mathcal{H}\right) \leq
\limfunc{SF}\nolimits^{+}%
\end{array}%
\right.
\end{equation}%
Figure \ref{fig:hqla4} illustrates the specific risk of the fund when
$\limfunc{TNA}\nolimits^{\star }=\$1$ bn, $\mathcal{H}^{\star }=1/100$, $\xi
_{k}^{\mathrm{size}}=10\%$, $\xi _{k}^{\mathrm{concentration}}=25\%$ and
$\limfunc{SF}\nolimits^{+}=0.80$. We have also reported the two components
$\limfunc{SF}\nolimits_{k}^{\mathrm{size}}\left( \limfunc{TNA}\right) $ and
$\limfunc{SF}\nolimits_{k}^{\mathrm{concentration}}\left( \mathcal{H}\right)
$:
\begin{equation}
\limfunc{SF}\nolimits_{k}\left( \limfunc{TNA},\mathcal{H}\right) =\min
\left( \limfunc{SF}\nolimits_{k}^{\mathrm{size}}\left( \limfunc{TNA}\right) +
\limfunc{SF}\nolimits_{k}^{\mathrm{concentration}}\left( \mathcal{H}\right) ,
\limfunc{SF}\nolimits^{+}\right)
\end{equation}%
It is better to use additive components than multiplicative components,
because the specific risk tends quickly to the cap value
$\limfunc{SF}\nolimits^{+}$ in this last case.

\begin{example}
\label{ex:hqla5} We assume that $\lambda _{k}=5\%$ per day, $\eta
_{k}=6.25\%$, $\limfunc{MDD}\nolimits_{k}=50\%$,
$\limfunc{TNA}\nolimits^{\star }=\$1$ bn, $\mathcal{H}^{\star }=1/100$,
$\xi_{k}^{\mathrm{size}}=10\%$, $\xi _{k}^{\mathrm{concentration}}=25\%$ and
$\limfunc{SF}\nolimits^{+}=0.80$. We consider four mutual funds, whose TNA are respectively equal to $\$1$, $\$5$, $\$7$ and $\$10$ bn. The redemption shock is equal to 40\% of the total net assets.
\end{example}

Results are given in Table \ref{tab:hqla5} with respect to the horizon time
$\tau_h$ and the fund size. We consider two concentration indices:
$\mathcal{H} = 0.01$ and $\mathcal{H} = 0.04$. We notice the impact of the
fund size on the RCR. For instance, when $\tau_h$ is set to $10$ days
and the concentration index is equal to $1\%$, $\func{RCR}$ is respectively equal to $1.08$, $0.65$, $0.43$ and $0.22$ for a fund size of $\$1$ bn, $\$5$ bn, $\$7$ bn, and $\$10$ bn. Therefore, the RCR is above one only when
the fund size is $\$1$ bn. If we increase the concentration
index, the RCR can be below one even if the fund size is small.
For instance, when $\tau_h$ is set to $10$ days and $\mathcal{H}$ is equal to $4\%$, $\func{RCR}$ is equal to $0.81$ for a fund size of $\$1$ bn. To summarize, the redemption coverage ratio is an increasing function of the time to liquidation $\tau_h$, but a decreasing function of the concentration index $\mathcal{H}$ and the fund size $\limfunc{TNA}$.

\begin{table}[tbph]
\centering
\caption{Computation of the RCR in the HQLA approach}
\label{tab:hqla5}
\begin{tabular}{ccccccccc}
\hline
$\tau_h$     & \multicolumn{4}{c}{$\mathcal{H} = 0.01$} & \multicolumn{4}{c}{$\mathcal{H} = 0.04$} \\
             & $\$1$ bn & $\$5$ bn & $\$7$ bn & $\$10$ bn & $\$1$ bn & $\$5$ bn & $\$7$ bn & $\$10$ bn \\ \hline
${\TsV}1$    & $0.12$ & $0.07$ & $0.05$ & $0.02$ & $0.09$ & $0.04$ & $0.02$ & $0.02$ \\
${\TsV}5$    & $0.56$ & $0.34$ & $0.23$ & $0.11$ & $0.42$ & $0.20$ & $0.11$ & $0.11$ \\
     $10$    & $1.08$ & $0.65$ & $0.43$ & $0.22$ & $0.81$ & $0.38$ & $0.22$ & $0.22$ \\
     $20$    & $2.01$ & $1.20$ & $0.80$ & $0.40$ & $1.50$ & $0.70$ & $0.40$ & $0.40$ \\
     $60$    & $1.64$ & $0.99$ & $0.66$ & $0.33$ & $1.23$ & $0.58$ & $0.33$ & $0.33$ \\
\hline
\end{tabular}
\end{table}

\subsection{Redemption liquidation policy}

The previous analysis demonstrates that the redemption coverage ratio is highly dependent on the redemption portfolio $q=\left( q_{1},\ldots ,q_{n}\right) $. Generally, the redemption shock is expressed as a percentage. $\RedemptionRate$ represents the proportion of the fund size that can be redeemed. Then, we can convert the redemption shock is nominal value by using the identity formula:
\begin{equation}
\RedemptionShock=\RedemptionRate\cdot \func{TNA}
\end{equation}%
For instance, if the redemption rate $\RedemptionRate$ is set to $10\%$ and
the fund size $\func{TNA}$ is equal to $\$1$ bn, the redemption shock
$\RedemptionShock$ is $\$100$ mn. However, the computation of $\func{RCR}$
requires defining the liquidation policy or the portfolio $q$. Two main
approaches are generally considered: the pro-rata liquidation and the
waterfall liquidation. The first one ensures that the asset structure of the
fund is the same before and after the liquidation. The second one minimizes
the time to liquidation. In practice, fund managers can mix the two schemes.
In this case, it is important to define the objective function in order to
understand the trade-off between portfolio distortion and liquidation time.

\subsubsection{The standard approaches}

\paragraph{Vertical slicing}

The pro-rata liquidation uses the proportional rule, implying that each
asset is liquidated such that the structure of the asset portfolio is the
same before and after the liquidation. This rule is also called the vertical
slicing approach. From a mathematical point of view, we have:
\begin{equation}
q=\RedemptionRate\cdot \omega
\end{equation}%
where $\omega $ is the fund's asset portfolio (before the liquidation).
In practice, $q_{i}$ is not necessarily an integer and must be rounded\footnote{This is why the waterfall slicing approach is also called the near proportional rule.}. For instance, if $\omega =\left( 1000,514,17\right) $ and $\RedemptionRate=10\%$, we obtain $q=\left( 100,51.4,1.7\right) $. Since we cannot sell a fraction of an asset, we can choose $q=\left(100,51,2\right) $.\smallskip

We recall that the tracking error due to the liquidation is equal to:
\begin{eqnarray}
\sigma \left( \omega \mid q\right)  &=&\sqrt{\left( w\left( \omega -q\right)
-w\left( \omega \right) \right) ^{\top }\Sigma \left( w\left( \omega
-q\right) -w\left( \omega \right) \right) }  \notag \\
&=&\sqrt{\Delta w\left( \omega \mid q\right) ^{\top }\Sigma\, \Delta w\left( \omega
\mid q\right) }
\end{eqnarray}%
where $\Sigma $ is the covariance matrix of asset returns, $w\left( \omega
\right) $ is the weight vector of portfolio $\omega $ (before liquidation)
and $w\left( \omega -q\right) $ is the weight vector of portfolio $\omega - q$ (after liquidation). The proportional rule ensures that the asset
composition does not change because of the redemption. Since the weights are
the same --- $\Delta w\left( \omega \mid q\right) =\mathbf{0}_{n}$, the
tracking error is equal to zero:
\begin{equation}
\sigma \left( \omega \mid q\right) =0
\end{equation}%
This property is important because there is no portfolio distortion with the
pro-rata liquidation rule.\smallskip

We have seen that the redemption coverage ratio is highly dependent on the time to liquidation $\tau _{h}$. In \citet[Section 3.2.2, page 18]{Roncalli-lst2}, we have defined the liquidation time as the inverse function of the liquidation ratio:%
\begin{equation}
\mathcal{LT}\left( q,p\right) =\mathcal{LR}^{-1}\left( q;p\right) =\inf
\left\{ h:\mathcal{LR}\left( q;h\right) \geq p\right\}
\end{equation}%
We now define the liquidity time (or time to liquidity) as follows:
\begin{equation}
\limfunc{TTL}\left( p\right) =\limfunc{RCR}\nolimits^{-1}\left( p\right) =\inf \left\{ h:%
\limfunc{RCR}\left( h\right) \geq p\right\}
\end{equation}%
It measures the required number of days to have a redemption coverage ratio
larger than $p$. As we have seen that $\limfunc{RCR}\left( h\right) $ and
$\limfunc{LS}\left( h\right) $ are related to $\mathcal{LR}\left( q;h\right)$
and $\mathcal{LS}\left( q;h\right) $, $\limfunc{TTL}\left( p\right) $ is also
related to $\mathcal{LT}\left( q,p\right) $. In the case where the
redemption portfolio satisfies $\RedemptionShock=\mathbb{V}\left( q\right) $,
we verify that $\limfunc{TTL}\left( p\right) =\mathcal{LT}\left( q,p\right) $
because we have $\limfunc{RCR}\left( h\right) =\mathcal{LR}\left( q;h\right)$
and $\limfunc{LS}\left( h\right) =\mathcal{LS}\left( q;h\right) $. In the
general case, we have:
\begin{eqnarray}
\limfunc{TTL}\left( p\right)  &=&\inf \left\{ h:\frac{\mathbb{V}\left( q\right)
}{\RedemptionShock}\cdot \mathcal{LR}\left( q;h\right) \geq p\right\}
\notag \\
&=&\left\{
\begin{array}{ll}
\mathcal{LT}\left( q,\dfrac{\RedemptionShock}{\mathbb{V}\left( q\right) }
\cdot p\right)  & \text{if }p\leq \dfrac{\mathbb{V}\left( q\right) }{\RedemptionShock} \\
+\infty  & \text{otherwise}
\end{array}%
\right.
\end{eqnarray}
\smallskip

While vertical slicing is optimal to minimize the tracking risk, the
liquidation of the redemption portfolio can however take a lot of time.
Indeed, the maximum we can liquidate each day is bounded by the liquidation
policy limit $q_{i}^{+}$. We have:
\begin{equation}
\sum_{h=1}^{\tau _{h}}q_{i}\left( h\right) \leq \tau _{h}\cdot q_{i}^{+}
\end{equation}%
In the case of the pro-rata liquidation rule, we have $q_{i}=\RedemptionRate
\cdot \omega _{i}$. We deduce that the redemption portfolio can be fully liquidated after $\func{TTL}\left( 1\right) =\left\lfloor \tau _{h}^{+}\right\rfloor $ days where:
\begin{equation}
\tau _{h}^{+}=\RedemptionRate\cdot \sup_{i=1,\ldots ,n}\frac{\omega _{i}}{%
q_{i}^{+}}
\end{equation}%
It may be difficult to sell some assets, because the value of $q_{i}^{+}$ is
low. Nevertheless, the remaining redemption value may be very small. This is
why fund managers generally consider in practice that the portfolio is liquidated when the proportion $p$ is set to $99\%$.

\paragraph{Horizontal slicing}

Horizontal slicing is the technical term to define waterfall liquidation. In
this approach, the portfolio is liquidated by selling the most liquid
assets first. Contrary to vertical slicing, the fund manager accepts
that the portfolio composition will be disturbed and his investment
strategy has te be modified, implying a tracking error risk:
\begin{equation}
\sigma \left( \omega \mid q\right) >0
\end{equation}

\begin{figure}[tbph]
\centering
\caption{Liquidity time in days (pro-rata versus waterfall liquidation, illiquid exposure, $p = 99\%$)}
\label{fig:ttl3c}
\figureskip
\includegraphics[width = \figurewidth, height = \figureheight]{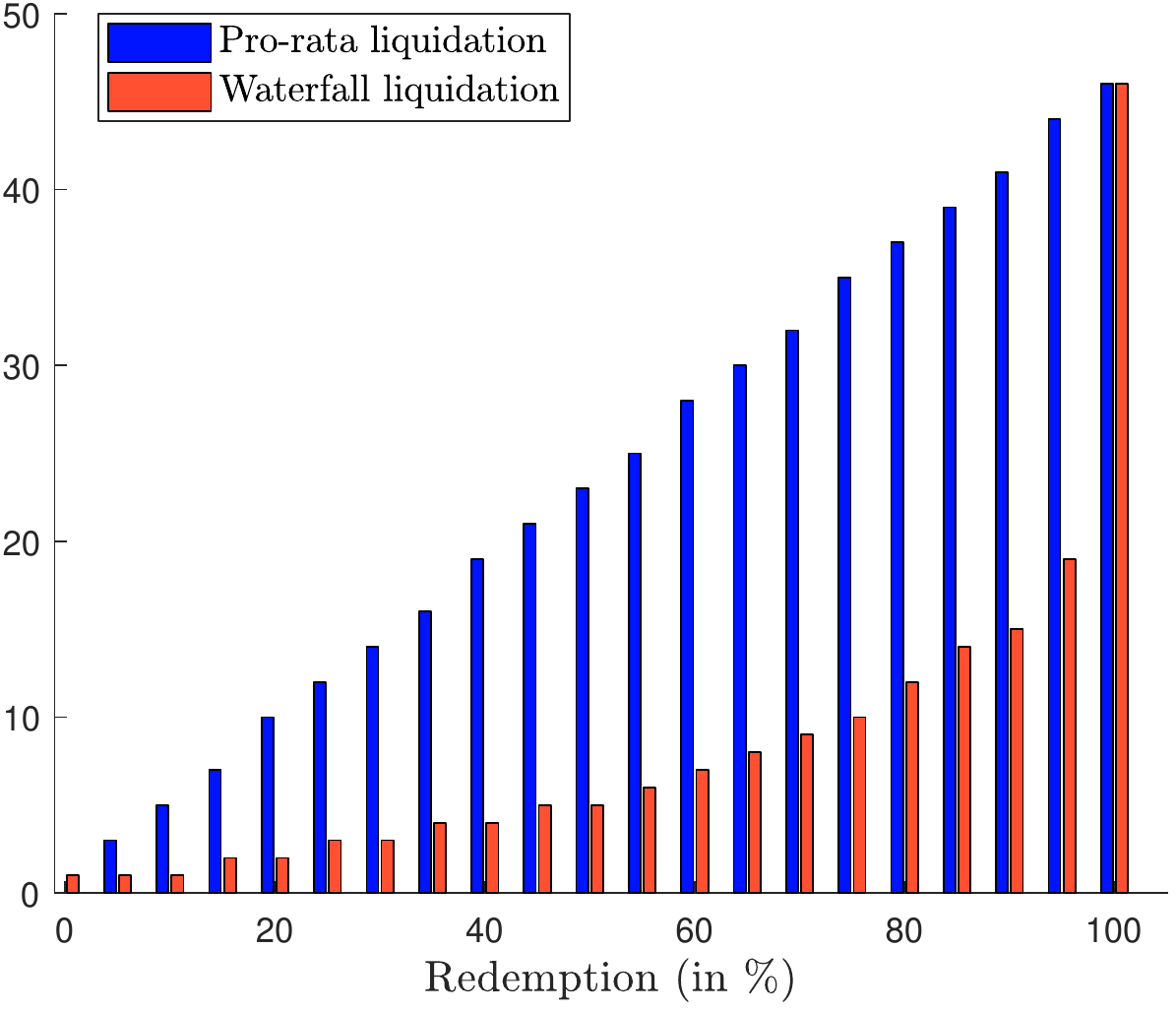}
\end{figure}

It is obvious that the waterfall approach minimizes the liquidity risk when
it is measured by the liquidity shortfall. Let us illustrate this property
with the example described in Table \ref{ex:rcr0} on page \pageref{ex:rcr0}.
If we consider the naive pro-rata liquidation rule, we obtain the liquidity
times given in Figure \ref{fig:ttl1b} on page \pageref{fig:ttl1b}. We notice
that they are very similar for $p=95\%$, $99\%$ and $100\%$. We now assume
that $q_{7}^{+}=20$, meaning that the seventh asset is not very liquid.
Therefore, we have a huge position on this asset ($\omega _{7}=1\,800$)
compared to the daily liquidation limit. If we would like to liquidate the
full exposure on this asset, it will take $90$ trading days versus $2$
trading days previously. The consequence of this illiquid exposure is that
the liquidity times are very different for $p=95\%$, $99\%$ and $100\%$ (see
Figure \ref{fig:ttl2b} on page \pageref{fig:ttl2b}). For instance, the
maximum liquidity time\footnote{It is obtained by considering the case
$\RedemptionRate=100\%$.} is respectively equal to $20$, $46$ and $90$
trading days for $p=95\%$, $99\%$ and $100\%$. Previously, the maximum
liquidity time was equal to $18$, $21$ and $22$ trading days when $q_{7}^{+}$
was equal to $1\,000$. Having some illiquid assets in the portfolio may then
dramatically increase the liquidity time when we choose the pro-rata
liquidation rule. We have also computed the liquidity time when we consider
the waterfall liquidation rule. Results are reported in Figures
\ref{fig:ttl3a} and \ref{fig:ttl3b} on page \pageref{fig:ttl3a}. We observe
two phenomena. First, if we compare Figures \ref{fig:ttl1b} and
\ref{fig:ttl3a}, we notice the higher convexity of the waterfall approach
when we increase the redemption shock. Second, we retrieve the similarity
pattern for $p=95\%$, $99\%$ and $100\%$ except for very large redemption
shocks when we have illiquid assets. The reason is that the part of illiquid
assets is much lower than the remaining value of the portfolio. Figure
\ref{fig:ttl3c} summarizes the two phenomena by comparing the pro-rata and
waterfall approaches when $q_{7}^{+}=20$.\smallskip

\begin{figure}[tbph]
\centering
\caption{Daily liquidation}
\label{fig:ttl4}
\figureskip
\includegraphics[width = \figurewidth, height = \figureheight]{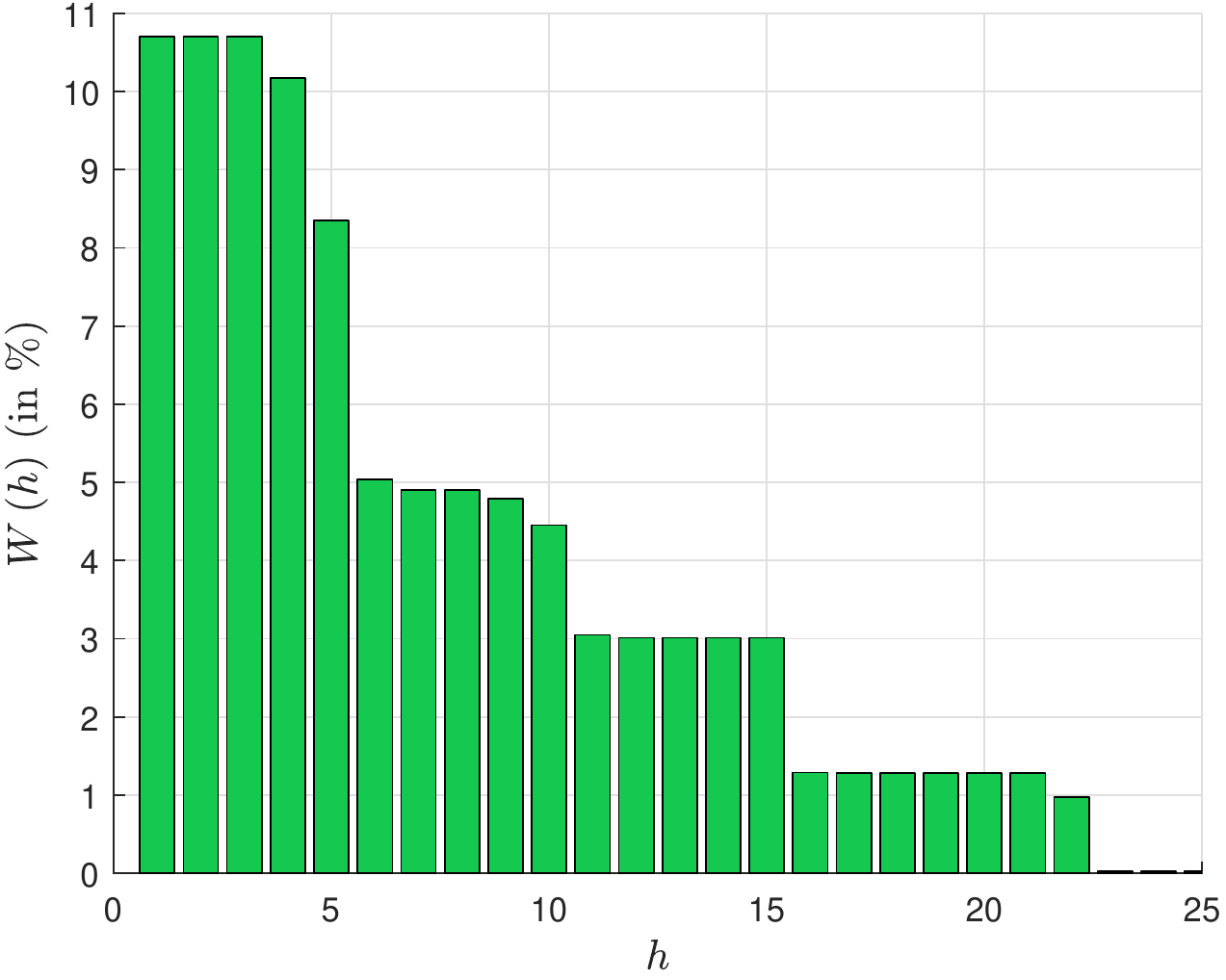}
\end{figure}

In order to determine the proportion of non-liquidated assets in the case of
the waterfall approach, we consider an analysis in terms of weights. We
recall that the portfolio weight of Asset $i$ is given by:
\begin{equation}
w_{i}\left( \omega \right) =\frac{\omega _{i}\cdot P_{i}}{\limfunc{TNA}}
\end{equation}%
Since the number of required trading days to liquidate the exposure to Asset
$i$ is equal to:
\begin{equation}
\tau _{i}\left( \omega \right) =\frac{\omega _{i}}{q_{i}^{+}}
\end{equation}%
the portfolio weight of Asset $i$ that can be liquidated with a trading day
is given by the following formula:
\begin{equation}
\psi _{i}\left( \omega \right) =\frac{w_{i}\left( \omega \right) }{\tau
_{i}\left( \omega \right) }=\frac{q_{i}^{+}\cdot P_{i}}{\limfunc{TNA}}
\end{equation}%
Using Equation (\ref{eq:LS3}) on page \pageref{eq:LS3}, we deduce that
the liquidity shortfall of a full redemption scenario under the waterfall
approach is equal to\footnote{We have $\func{LS}\left( 0\right) =100\%$.}:
\begin{equation}
\func{LS}\left( h\right) =1-\sum_{i=1}^{n}\min \left( h\cdot \psi _{i}\left(
\omega \right) ,w_{i}\left( \omega \right) \right)
\end{equation}%
The relative weight of the portfolio that can be liquidated at time $h$ is
then equal to $W\left( h\right) =\func{LS}\left( h-1\right) -\func{LS}\left(
h\right) $. $W\left( h\right)$ is the daily liquidation expressed
in \%. In Figure \ref{fig:ttl4}, we have reported the values taken by $W\left( h\right) $ for the previous example. We notice that significant liquidation occurs over the first $22$ days. After this period, the amount liquidated decreases substantially because it concerns illiquid assets.

\begin{remark}
We can use the previous analysis to determine the amount of \textquotedblleft
\textsl{illiquid assets}\textquotedblright\ in the portfolio. For that, we
choose a threshold $w^{\star }$ below which the amount liquidated is too
small\footnote{$w^{\star }$ is generally set to $0.5\%$.}:
\begin{equation}
h^{\star }=\inf \left\{ h:W\left( h\right) \leq w^{\star }\right\}
\end{equation}%
Alternatively, we can directly set the value of $h^{\star }$ above which we
assume it corresponds to an illiquid time. The amount of illiquid assets is
then equal to $\sum_{k\geq h^{\star }}W\left( h\right) $ or equivalently
$\func{LS}\left( h^{\star }-1\right) =1-\sum_{i=1}^{n}\min \left( \left(
h^{\star }-1\right) \cdot \psi _{i}\left( \omega \right) ,w_{i}\left( \omega
\right) \right) $. In the previous example, it is equal to $2.50\%$ if
$w^{\star }=1\%$ and $1.52\%$ if $w^{\star }=0.5\%$.
\end{remark}

\subsubsection{The mixing approach}

So far, the analysis of the redemption coverage ratio and the redemption
liquidation policy has been focused on the trading limits and the daily
amounts that can be liquidated. This volume-based approach is not enough and
may lead to misleading conclusions. Indeed, the previous analysis
completely omits the transaction costs. This is obviously the case of the
vertical slicing approach, where the fund manager is forced to sell exposures
that are not liquid. Therefore, no cost analysis is done in the pro-rata
liquidation rule. This is also the case in the above presentation of the
horizontal slicing approach, since the liquidation policy only considers the
daily trading limits through the variable $q^{+}$. Nevertheless, the practice
of the waterfall approach is a little bit different, because it is not
limited to the liquidity depth. Indeed, the ultimate goal of this approach is
to liquidate the exposures at the lowest cost. Therefore, it includes a cost
analysis. However, as seen previously, the waterfall approach implies a
tracking risk that is not controlled. This is not acceptable in
practice.\smallskip

The optimal liquidation approach consists in defining a maximum acceptable
level $\mathcal{TR}^{+}$ of tracking risk and to minimize the transaction
cost $\mathcal{TC}\left( q\right) $ of the liquidation portfolio:
\begin{eqnarray}
q^{\star } &=&\underset{q}{\arg \min }\mathcal{TC}\,\left( q\right)
\label{eq:liquidation1} \\
&\text{s.t.}&\left\{
\begin{array}{l}
\mathcal{TR}\left( \omega \mid q\right) \leq \mathcal{TR}^{+} \\
\mathcal{LS}\left( q;h\right) \leq \mathcal{LS}^{+} \\
\mathbf{1}_{n}^{\top }w\left( \omega -q\right) =0 \\
w\left( \omega -q\right) \geq \mathbf{0}_{n}%
\end{array}%
\right.   \notag
\end{eqnarray}%
In the case of an equity portfolio, the tracking risk is equal to the
tracking error volatility:
\begin{equation}
\mathcal{TR}\left( \omega \mid q\right) =\sigma \left( \omega \mid q\right) =
\sqrt{\Delta w\left( \omega \mid q\right) ^{\top }\Sigma \Delta w\left(
\omega \mid q\right) }
\end{equation}%
In the case of a bond portfolio, it is more difficult to define the tracking
risk because the volatility is not the right approach to measure the risk of
fixed-income instruments \citep{Roncalli-2020}. Moreover, there are several
risk dimensions to take into account. For instance, \citet{BenSlimane-2021}
considers three dimensions\footnote{In fact, \citet{BenSlimane-2021} adds two
liquidity components: the first one concerns the liquidity costs whereas the
second one concerns the liquidity depth (or the axis component of market
makers).}: sectorial risk, duration risk and credit risk. Following
\citet{BenSlimane-2021}, we can define the tracking risk as the sum of three
risk measures:
\begin{equation}
\mathcal{TR}\left( \omega \mid q\right) =\mathcal{R}_{w}\left( \omega \mid
q\right) +\mathcal{R}_{\mathrm{MD}}\left( \omega \mid q\right) +
\mathcal{R}_{\mathrm{DTS}}\left( \omega \mid q\right)
\end{equation}%
The weight risk measure $\mathcal{R}_{w}\left( \omega \mid q\right) $ is the
weight difference between Portfolio $\omega -q$ and Portfolio $\omega $
within the sector $s$:
\begin{equation}
\mathcal{R}_{w}\left( \omega \mid q\right) =\sum_{s=1}^{n_{\mathcal{S}ector}}
\left\vert \sum_{i\in \mathcal{S}ector\left( s\right) }\Delta
w_{i}\left( \omega \mid q\right) \right\vert
\end{equation}%
where $n_{\mathcal{S}ector}$ is the number of sectors and $\Delta w_{i}\left(
\omega \mid q\right) =w_{i}\left( \omega -q\right) -w_{i}\left( \omega
\right) $ is the weight distortion of Bond $i$ because of the liquidation. We
define $\mathcal{R}_{\mathrm{MD}}\left( \omega \mid q\right) $ as the
modified duration risk of $\omega -q$ with respect to $\omega $ within the
sector $s$:
\begin{equation}
\mathcal{R}_{\mathrm{MD}}\left( \omega \mid q\right) =\sum_{s=1}^{n_{%
\mathcal{S}ector}}\sum_{j=1}^{n_{\mathcal{B}ucket}}\left\vert \sum_{i\in
\mathcal{S}ector\left( s\right) }\Delta w_{i}\left( \omega \mid q\right)
\cdot \limfunc{MD}\nolimits_{i}\left( \mathcal{B}ucket_{j}\right)
\right\vert
\end{equation}%
where $n_{\mathcal{B}ucket}$ is the number of maturity buckets and
$\func{MD}_{i}\left( \mathcal{B}ucket_{j}\right) $ is the modified duration
contribution of Bond $i$ to the maturity bucket $j$. The rationale of this
definition is to track the difference in modified duration per bucket.
Finally, we define the DTS risk measure $\mathcal{R}_{\mathrm{DTS}}\left(
\omega \mid q\right) $ as the weighted DTS difference between $\omega -q$ and
$\omega $:
\begin{equation}
\mathcal{R}_{\mathrm{DTS}}\left( \omega \mid q\right) =\sum_{s=1}^{n_{%
\mathcal{S}ector}}\left\vert \sum_{i\in \mathcal{S}ector\left( s\right)
}\Delta w_{i}\left( \omega \mid q\right) \cdot \limfunc{DTS}%
\nolimits_{i}\right\vert
\end{equation}%
where $\limfunc{DTS}_{i}$ is the duration-times-spread of Bond $i$. Regarding
the transaction cost function, we recall that it is defined as follows
\citep[Equation (26), page 25]{Roncalli-lst2}:
\begin{equation}
\TC\left( q\right) =\sum_{i=1}^{n}\sum_{h=1}^{h^{+}}\mathds{1}\left\{
q_{i}\left( h\right) >0\right\} \cdot q_{i}\left( h\right) \cdot P_{i}\cdot %
\cost_{i}\left( \frac{q_{i}\left( h\right) }{v_{i}}\right)
\label{eq:liq-tc2}
\end{equation}%
where $\cost_{i}\left( x\right) $ is the unit transaction cost function
associated with Asset $i$. In \citet{Roncalli-lst2}, $\cost_{i}\left( x\right)$ follows a two-regime power-law model. We also notice that the optimization problem (\ref{eq:liquidation1}) includes a constraint related to the liquidation shortfall. Without this constraint, the solution consists in
liquidating each day an amount $q_{i}\left( h\right) $ much smaller than the
trading limit $q_{i}^{+}$ in order to minimize the transaction costs due to
the market impact. Of course, the idea is not to indefinitely delay the
liquidation. Therefore, this constraint is very important to ensure that a
significant portion of the redemption portfolio has been sold before $h$. It
follows that the optimization problem (\ref{eq:liquidation1}) can be tricky
to solve from a numerical point of view, in particular for bond funds.
Nevertheless, it perfectly illustrates the trade-off between the three risk
dimensions: the transaction cost risk $\mathcal{TC}\left( q\right) $, the
tracking risk $\mathcal{TR}\left( \omega \mid q\right) $ and the liquidation
shortfall risk $\mathcal{LS}\left( q;h\right) $.\smallskip

Once again, we consider the example described in Table \ref{ex:rcr0} on page \pageref{ex:rcr0}. We assume that the volatility of the assets is respectively equal to\footnote{The correlation matrix of asset returns is given by:
\begin{equation*}
\rho =\left(
\begin{array}{rrrrrrr}
100\% &  &  &  &  &  &  \\
10\% & 100\% &  &  &  &  &  \\
40\% & 70\% & 100\% &  &  &  &  \\
50\% & 40\% & 80\% & 100\% &  &  &  \\
30\% & 30\% & 50\% & 50\% & 100\% &  &  \\
30\% & 30\% & 50\% & 50\% &  70\% & 100\% &  \\
30\% & 30\% & 50\% & 50\% &  70\% &  70\% & 100\% \\
\end{array}%
\right)
\end{equation*}}
$20\%$, $18\%$, $15\%$, $15\%$, $22\%$, $30\%$ and $35\%$
whereas the bid-ask spread is equal to $5$, $3$, $5$, $8$, $12$, $15$ and $15$ bps. The transaction cost function corresponds to the SQRL model defined by \citet{Roncalli-lst2} with $\varphi_1 = 0.4$, $\tilde{x} = 5\%$ and $x^{+} = 10\%$. We deduce that the daily volume $v_i$ of each asset is equal to $10 \times q_i^{+}$. In Table \ref{tab:mixing1d}, we define five liquidation portfolios where the redemption rate $\RedemptionRate $ is set to $10\%$. Portfolio \#1 satisfies the pro-rata liquidation rule. We verify that the tracking risk (measured by the tracking error volatility) is equal to zero. The total transaction cost is equal to $22.4$ bps with the following break-down: $6.1$ bps for the bid-ask spread component and $16.2$ bps for the market impact component. This is a low tracking error. However, if the fund manager's objective is to liquidate the redemption in one trading day, we notice that the liquidation shortfall is equal to $23.5\%$. In Portfolio \#2, the liquidation is concentrated in the second and third assets. Because these assets are more liquid than the others, the transaction cost is lower and  equal to $20.4$ bps. Nevertheless, this portfolio leads to a high tracking error risk of $79.6$ bps. Portfolio \#3 is made up of the less liquid assets.
Therefore, it is normal to obtain a high transaction cost of $42.5$ bps.
Again, this portfolio presents a high tracking risk since we have
$\mathcal{TR}\left( \omega \mid q\right) \approx 2\%$! If the objective function is to fulfill the redemption in one day, Portfolio \#4 is a good candidate since we have $\mathcal{LS}\left( q;1\right) =0$ and the transaction cost is moderate\footnote{It is a little bit higher than the transaction cost of the vertical slicing approach.} ($\mathcal{TC}\left( q\right) =25.6$ bps). However, the tracking risk is high and is equal to $35.4\%$. Portfolio \#5 is a compromise between tracking risk and liquidity shortfall\footnote{Portfolio \#5 is equal to $40\%$ of Portfolio \#1 and $60\%$ of Portfolio \#4.}, because we have $\mathcal{TR}\left( \omega \mid q\right) =21.2$ bps, $\mathcal{TC}\left( q\right) =22.6$ bps but $\mathcal{LS}\left( q;1\right) =9.4\%$. If the objective is to find the optimal liquidation policy with the constraints $\mathcal{LS}\left( q;1\right) \leq 10\%$ and $\mathcal{TR}\left( \omega \mid q\right) =20\%$, Portfolio \#5 is a good starting point.

\begin{table}[tbph]
\centering
\caption{Comparison of five redemption portfolios}
\label{tab:mixing1d}
\begin{tabular}{ccrrrrr}
\hline
\multicolumn{2}{c}{Liquidation portfolio} & \#1 & \#2 & \#3 & \#4 & \#5 \\
\hline
$q_{1}$                                   &          &      $43\,510$ & ${\TsXX}\,0$ &   ${\TsXX}\,0$ &    $20\,000$ &      $29\,404$ \\
$q_{2}$                                   &          &      $30\,010$ &    $27\,000$ &   ${\TsXX}\,0$ &    $20\,000$ &      $24\,004$ \\
$q_{3}$                                   &          & ${\TsV}5\,040$ &    $22\,238$ &   ${\TsXX}\,0$ &    $10\,000$ & ${\TsV}8\,016$ \\
$q_{4}$                                   &          &      $20\,050$ & ${\TsXX}\,0$ &   ${\TsXX}\,0$ &    $20\,000$ &      $20\,020$ \\
$q_{5}$                                   &          & ${\TsV}7\,550$ & ${\TsXX}\,0$ &      $34\,315$ &    $18\,044$ &      $13\,846$ \\
$q_{6}$                                   &          & ${\TsV}1\,750$ & ${\TsXX}\,0$ &      $17\,500$ & ${\TsXX}\,0$ &  ${\TsX}\,700$ \\
$q_{7}$                                   &          &  ${\TsX}\,180$ & ${\TsXX}\,0$ & ${\TsV}1\,800$ & ${\TsXX}\,0$ &  ${\TsXV}\,72$ \\ \hline
$\mathcal{TR}\left( \omega \mid q\right)$ & (in bps) &    ${\TsV}0.0$ &       $79.6$ &        $201.0$ &       $35.4$ &         $21.2$ \\
$\mathcal{TC}\left( q\right)$             & (in bps) &         $22.4$ &       $20.4$ &   ${\TsV}42.5$ &       $25.6$ &         $22.6$ \\
$\mathcal{TC}_{\spread}\left( q\right)$   & (in bps) &    ${\TsV}6.1$ &  ${\TsV}4.5$ &   ${\TsV}13.8$ &  ${\TsV}6.6$ &    ${\TsV}6.4$ \\
$\mathcal{TC}_{\impact}\left( q\right)$   & (in bps) &         $16.2$ &       $15.9$ &   ${\TsV}28.7$ &       $19.1$ &         $16.2$ \\
$\mathcal{LS}\left( q;1\right)$           &  (in \%) &         $23.5$ &       $48.2$ &   ${\TsV}60.7$ &  ${\TsV}0.0$ &    ${\TsV}9.4$ \\
\hline
\end{tabular}
\end{table}

\subsection{Reverse stress testing}

Reverse stress testing is a \textquotedblleft \textsl{fund-level stress test
which starts from the identification of the pre-defined outcome with regards
to fund liquidity (e.g. the point at which the fund would no longer be liquid
enough to honor requests to redeem units) and then explores scenarios and
circumstances that might cause this to occur}\textquotedblright\ \citep[page
6]{ESMA-2020a}. Following \citet{Roncalli-2020}, reverse stress testing
consists in identifying stress scenarios that could bankrupt the fund.
Therefore, reverse stress testing can be viewed as an inverse problem.
Indeed, liquidity stress testing starts with a liability liquidity scenario
and an asset liquidity scenario in order to compute the redemption coverage
ratio. The liability liquidity scenario is defined by the redemption shock
$\RedemptionShock$ (or the redemption rate $\RedemptionRate$), while the
asset liquidity scenario is given by the stressed trading limits $q^{+}$ or
the HQLA classification. Given a time horizon $\tau _{h}$, the outcome is
$\limfunc{RCR}\left( \tau _{h}\right) $. From a theoretical point of view,
the bankruptcy of the fund depends on whether the condition $\limfunc{RCR}\left( \tau _{h}\right) \geq 1$ is satisfied or not. The underlying idea is that the fund is not viable if $\limfunc{RCR}\left( \tau _{h}\right) <1$. In practice, the fund can continue to exist because it can use short-term borrowing or
other liquidity management tools such as gates or side pockets\footnote{These
different tools will be explored in the next section on page
\pageref{section:lmt}.}. In fact, the fund's survival depends on many
parameters. However, we can consider that a too small value of
$\limfunc{RCR}\left( \tau_{h}\right) $ is critical and can produce the
collapse of the fund. Let $\limfunc{RCR}^{-}$ be the minimum acceptable level
of the redemption coverage ratio. Then, reverse stress testing consists in
finding the liability liquidity scenario and/or the asset liquidity scenario
such that $\limfunc{RCR}\left( \tau _{h}\right) =\limfunc{RCR}^{-}$.

\subsubsection{The liability RST scenario}

From a liability perspective, reverse stress testing consists in finding
the redemption shock above which the redemption coverage ratio is lower
than the minimum acceptable level:
\begin{equation}
\limfunc{RCR}\left( \tau _{h}\right) \leq \limfunc{RCR}\nolimits^{-}%
\Longrightarrow \left\{
\begin{array}{c}
\RedemptionShock\geq \rstRS\left(\tau_h\right) = \dfrac{\AssetShock\left( \tau _{h}\right) }{\limfunc{RCR%
}\nolimits^{-}} \\
\text{or} \\
\RedemptionRate\geq \rstRR\left(\tau_h\right) = \dfrac{\AssetRate\left( \tau _{h}\right) }{\limfunc{RCR}%
\nolimits^{-}}%
\end{array}%
\right.
\label{eq:rst1}
\end{equation}%
$\rstRS\left(\tau_h\right)$ (or $\rstRR\left(\tau_h\right)$) is called the
liability reverse stress testing scenario. At first sight, computing the
liability RST scenario seems to be easy since the calculation of
$\AssetShock\left( \tau _{h}\right) $ is straightforward. However, it is a
little bit more complicated since $\AssetShock\left( \tau _{h}\right) $
depends on the liquidation portfolio $q$. Therefore, we have to define $q$.
This is the hard task of reverse stress testing. Indeed, the underlying idea
is to analyze each asset exposure individually and decide the quantity of
each asset that can be sold in the market during a stress period.\smallskip

The simplest way to define $q$ is to use the multiplicative approach with
respect to the portfolio $\omega $:
\begin{equation}
q_{i}^{\mathrm{RST}}=\alpha _{i}\cdot \omega _{i}
\end{equation}%
where $\alpha _{i}$ represents the proportion of the asset $i$ than can be
sold during a liquidity stress event. In particular, $\alpha _{i}=0$
indicates that the asset is illiquid during this period. $\alpha _{i}$ also
depends on the size $\omega _{i}$. For instance, a large exposure on an asset
can lead to a small value of $\alpha _{i}$ because it can be difficult to
liquidate such exposure.\smallskip

\begin{table}[tbph]
\centering
\caption{Computation of the liability RST scenario}
\label{tab:rst1}
\begin{tabular}{c|cccc|cccc}
\hline
& & & & & & & & \\[-1em]
& \multicolumn{4}{c|}{$\rstRS\left(\tau_h\right)$ (in \$ mn)} &
\multicolumn{4}{c}{$\rstRR\left(\tau_h\right)$ (in \%)} \\
$\limfunc{RCR}\nolimits^{-}$ & $25\%$ & $75\%$ & $50\%$ & $100\%$
& $25\%$ & $75\%$ & $50\%$ & $100\%$ \\ \hline
    $\tau_h =1$ & $25.1$ & $12.6$ & ${\TsV}8.4$ & ${\TsV}6.3$ & $17.7$ & ${\TsV}8.9$ & ${\TsV}5.9$ & ${\TsV}4.4$ \\
    $\tau_h =2$ & $46.2$ & $23.1$ &      $15.4$ &      $11.5$ & $32.6$ &      $16.3$ &      $10.9$ & ${\TsV}8.1$ \\
    $\tau_h =3$ & $63.2$ & $31.6$ &      $21.1$ &      $15.8$ & $44.6$ &      $22.3$ &      $14.9$ &      $11.1$ \\
    $\tau_h =4$ & $80.1$ & $40.1$ &      $26.7$ &      $20.0$ & $56.5$ &      $28.3$ &      $18.8$ &      $14.4$ \\
$\tau_h \geq 5$ & $87.5$ & $43.8$ &      $29.2$ &      $21.9$ & $61.8$ &      $30.9$ &      $20.6$ &      $15.4$ \\
\hline
\end{tabular}
\end{table}

Let us consider again the example described in Table \ref{ex:rcr0} on page
\pageref{ex:rcr0}. We assume that the third, fifth, sixth and seventh assets
are illiquid in a stress period. For the other assets, we set $\alpha_1 =
20\%$, $\alpha_2 = 30\%$ and $\alpha_4 = 15\%$. Results are given in Table
\ref{tab:rst1}. For instance, if the minimum acceptable level of the
redemption coverage ratio is equal to $25\%$, we obtain $\rstRR\left(1\right)
= 17.7\%$. This means that the fund may support a redemption shock below
$17.7\%$, whereas the RCR limit of $25\%$ is broken if the fund experiences a
redemption shock above $17.7\%$. If the minimum acceptable level is set to
$100\%$, which is the regulatory requirement, the liability RST scenario
corresponds to $\rstRR\left(1\right) = 4.4\%$.

\begin{remark}
We don't always have a solution to Problem (\ref{eq:rst1}). Nevertheless, we
notice that:
\begin{equation}
\limfunc{RCR}\left( \infty \right) =\frac{\sum_{i=1}^{n}q_{i}^{\mathrm{RST}%
}\cdot P_{i}}{\sum_{i=1}^{n}\omega _{i}\cdot P_{i}}=\sum_{i=1}^{n}\alpha
_{i}\cdot w_{i}\left( \omega \right)
\end{equation}%
A condition to obtain a solution such that $\RedemptionShock\leq \func{TNA}$
and $\RedemptionRate\leq 1$ is to impose the constraint $\limfunc{RCR}%
\nolimits^{-}\geq \sum_{i=1}^{n}\alpha _{i}\cdot w_{i}\left( \omega \right)$.
\end{remark}

\subsubsection{The asset RST scenario}

The asset RST scenario consists in finding the asset liquidity shock above
which the redemption coverage ratio is lower than the minimum acceptable
level. Contrary to the liability RST scenario, for which the liquidity shock
is measured by the redemption rate, it is not easy to define what a liquidity
shock is when we consider the asset side. For that, we recall that the stress
testing of the assets consists in defining three multiplicative (or additive)
shocks for the bid-ask spread, the volatility and the daily volume
\cite[Section 5.4, page 51]{Roncalli-lst2}. Let $x_{i}$ be the participation
rate. We have:
\begin{equation}
x_{i}=\frac{q_{i}}{v_{i}}
\end{equation}%
where $v_{i}$ is the daily volume. The trading limit $x_{i}^{+}$ (expressed
in participation rate) is supposed to be fixed, implying that it is the same
in normal and stress periods. However, the stress period generally faces a
reduction in the daily volume, meaning that the trading limit $q_{i}^{+}$
(expressed in number of shares) is not the same:
\begin{equation}
q_{i}^{+}=\left\{
\begin{array}{ll}
v_{i}\cdot x_{i}^{+} & \text{in a normal period} \\
m_{v}\cdot v_{i}\cdot x_{i}^{+} & \text{in a stressed period}%
\end{array}%
\right.
\end{equation}%
where $m_{v}<1$ is the multiplicative shock of the daily volume. The
underlying idea of the asset RST scenario is then to define the upper limit
$m_{v}^{\mathrm{RST}}$ below which the redemption coverage ratio is lower
than the minimum acceptable level:
\begin{equation}
\limfunc{RCR}\left( \tau _{h}\right) \leq \limfunc{RCR}\nolimits^{-}%
\Longrightarrow m_{v}\leq m_{v}^{\mathrm{RST}}\left( \tau _{h}\right) <1
\end{equation}%
Nevertheless, the computation of $m_{v}^{\mathrm{RST}}\left( \tau _{h}\right)
$ requires defining a liquidation portfolio. For that, we can use the
vertical slicing approach where $q_{i}=\RedemptionRate^{\star }\cdot
\omega _{i}$ and $\RedemptionRate^{\star }$ is a standard redemption rate%
\footnote{%
A typical value of $\RedemptionRate^{\star }$ is $10\%$. It is important to
use a low value for $\RedemptionRate^{\star }$ because the asset RST scenario
measures the liquidity stress from the asset perspective, not from the
liability perspective.}. As in the case of the liability RST problem, the
solution may not exist if $\limfunc{RCR}\left( \tau _{h}\right) \leq
\limfunc{RCR}\nolimits^{-}$ when $m_{v}$ is set to one.

\begin{remark}
In the liability RST problem, a low value of $\rstRR$ indicates that the fund
is highly vulnerable. Indeed, this means that a small redemption shock may
produce a funding liquidity stress on the investment fund. In the asset RST problem, the fund is vulnerable if the value of $m_{v}^{\mathrm{RST}}$ is high. In this case, a slight deterioration of the market depth induces a market liquidity stress on the investment fund even if it faces a small redemption. To summarize, fund managers would prefer to have low values of $\rstRR$ and high values of $m_{v}^{\mathrm{RST}}$.
\end{remark}

\begin{figure}[tbph]
\centering
\caption{Computation of the asset RST scenario}
\label{fig:rst2}
\figureskip
\includegraphics[width = \figurewidth, height = \figureheight]{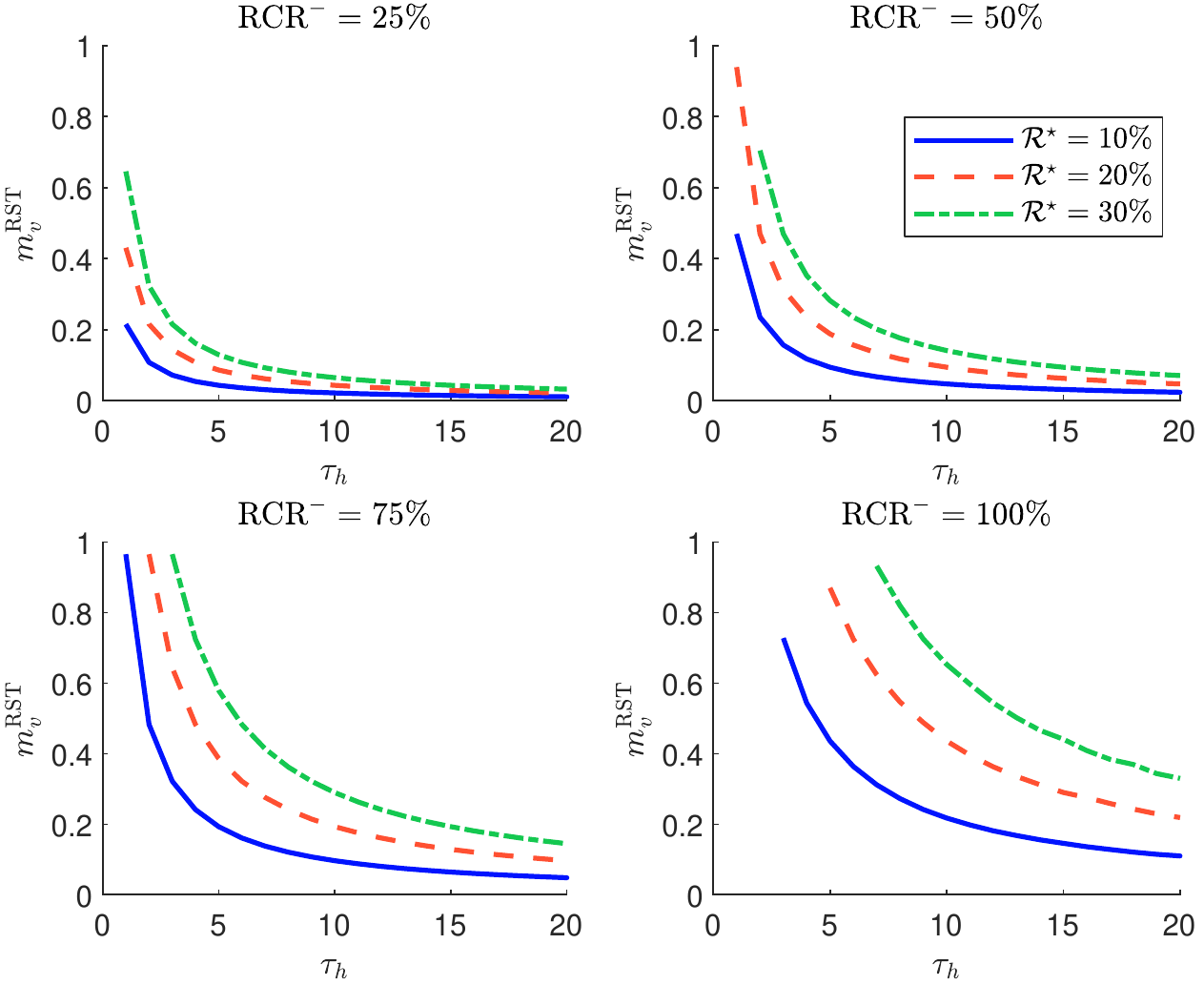}
\vspace*{-10pt}
\end{figure}

The computation of $m_{v}^{\mathrm{RST}}$ for the previous example is
reported in Figure \ref{fig:rst2}. We first notice that the solution cannot exist because there is no value of $m_{v}$ such that $\limfunc{RCR}\left( \tau _{h}\right) \leq \limfunc{RCR}\nolimits^{-}$. For instance, this is the case of $\tau _{h}\leq 6$ when $\RedemptionRate^{\star }$ is set to $30\%$ (bottom right-hand panel). By construction, $m_{v}^{\mathrm{RST}}\left( \tau _{h}\right) $ is a decreasing function of $\tau _{h}$. Indeed, the reverse stress testing scenario is more severe for short time windows than for long time windows. We also verify that $m_{v}^{\mathrm{RST}}\left( \tau _{h}\right) $ is an increasing function of $\limfunc{RCR}\nolimits^{-}$, because the constraint is tighter.

\begin{remark}
Reverse stress testing does not reduce to the computation of $\rstRR\left( \tau _{h}\right)$ or $m_{v}^{\mathrm{RST}}\left( \tau _{h}\right) $. This step must be completed by the economic analysis to understand what market or financial scenario can imply $\RedemptionRate\geq \rstRR\left(\tau_h\right)$ or $m_{v}\leq m_{v}^{\mathrm{RST}}\left( \tau _{h}\right) $.
\end{remark}

\section{Liquidity management tools}
\label{section:lmt}

Liquidity management tools are measures applied by fund managers in
exceptional circumstances to control or limit dealing in fund units
\citep{ESMA-2020a}. According to \citet{Darpeix-2020}, the main LMTs are
anti-dilution levies, gates, liquidity buffers, redemption fees,
in-kind redemptions, redemption suspensions, short-term borrowing, side
pockets and swing pricing. They can be grouped into three categories
(Table \ref{tab:esma-lmt}). First, we have liquidity buffers that may or not be mandatory, and short-term borrowing. The underlying idea is to invest a portion of assets in cash and to use it in the case of a liquidity stress. As such, this category has an impact on the structure of the asset portfolio. Second, we have special arrangements that include gates, in-kind redemptions, redemption suspensions and side pockets. The objective of this second group is to limit or delay the redemptions. Finally, we have swing pricing mechanisms\footnote{They include anti-dilution levies.}, the purpose of which is clearly to protect the remaining investors.

\begin{table}[tbph]
\centering
\caption{LMTs available to European corporate debt funds (June 2020)}
\label{tab:esma-lmt}
\scalebox{0.975}{
\begin{tabular}{llcc}
\hline
                     &                     &         AIF & UCITS  \\ \hline
Short-term borrowing &                     &      $78\%$ & $91\%$ \\ \hdashline
                     & Gates               &      $23\%$ & $73\%$ \\
Special arrangements & Side pockets        &      $10\%$ & $10\%$ \\
                     & In-kind redemptions &      $34\%$ & $77\%$ \\ \hdashline
Swing pricing        &                     & ${\TsV}7\%$ & $57\%$ \\
Anti-dilution levies &                     &      $11\%$ & $17\%$ \\ \hline
\end{tabular}}
\begin{flushleft}
{\small \textit{Source}: \citet[page 38]{ESMA-2020b}.}
\end{flushleft}
\end{table}

\subsection{Liquidity buffer and cash holding}

As noticed by \citet{Yan-2006}, cash is a critical component of mutual funds'
portfolios for three reasons. First, cash is generally used to manage the
inflows and outflows of the fund. For instance, in the case of a
subscription, the fund manager may decide to delay the investment in order to
find better investment opportunities later. In the case of a redemption, cash
can be used to liquidate a part of the portfolio without selling the risky
assets. Second, cash is important for the day-to-day management of the fund
for paying management fees, managing collateral risk, investing in
derivatives, etc. Third, cash is a financial instrument of market timing
\citep{Simutin-2010, Simutin-2014}. This explains that cash holding is an old
practice of mutual funds.\smallskip

Since the 2008 Global Financial Crisis, the importance of cash management has
increased due to liquidity policies of asset managers, and liquidity (or
cash) buffers have become a central concept in liquidity risk management.
Nevertheless, implementing a cash buffer has a cost in terms of expected
return. Therefore, cash buffer policies are increasingly integrated into
investment policies.

\subsubsection{Definition}

A liquidity buffer refers to the stock of cash instruments held by the fund
manager in order to manage the future redemptions of investors. This suggests
the intentionality of the fund manager to use the buffer only for liquidity
purposes. Because it is difficult to know whether cash is used for other
purposes (e.g. tactical allocation, supply/demand imbalance), the cash
holding of the investment fund is considered as a measurement proxy of its
liquidity buffer. \citet{Chernenko-2016} go further and suggest that cash holding is
\textquotedblleft \textsl{a good measure of a fund's liquidity transformation
activities}\textquotedblright.\smallskip

Since we use a strict definition, we consider that a liquidity buffer
corresponds to the following instruments:

\begin{itemize}
\item Cash

\begin{itemize}
\item Cash at hand

\item Deposits
\end{itemize}

\item Cash equivalents

\begin{itemize}
\item Repurchase agreements (repo)

\item Money market funds

\item Short-term debt securities
\end{itemize}
\end{itemize}

\noindent Generally, we assume that short-term debt securities have a
maturity less than one year. We notice that cash and cash equivalents do not
exactly coincide with liquid assets. Indeed, liquid assets may include stocks
and government bonds that can be liquidated the next day. Therefore, our
definition of the liquidity buffer is in fact the definition of a cash
buffer.

\subsubsection{Cost-benefit analysis}

Maintaining a cash buffer has the advantage of reducing the cost of redemption liquidation and mitigating funding risk. However, it also induces some costs in terms of return, tracking error, beta exposure, etc. Since a cash buffer corresponds to a deleverage of the risky assets, it may breach the fiduciary duties of the fund manager. Indeed, the investors pay management and performance fees in order to be fully exposed to a given asset class. Therefore, all these dimensions make the cost-benefit analysis difficult and complex, and computing an \textquotedblleft
\textit{optimal}\textquotedblright\ level of cash buffer is a difficult task from a professional point of view.

\paragraph{Cash buffer analytics}
In what follows, we define the different concepts that are necessary to
conduct a cost-benefit analysis.

\subparagraph{Cash-to-assets ratio}

We assume that a cash buffer is implemented in the fund, and we note $w_{\mathrm{cash}}$ as the cash-to-assets ratio:%
\begin{equation}
w_{\mathrm{cash}}=\frac{\mathrm{cash}}{\func{TNA}}
\end{equation}%
$w_{\mathrm{cash}}$ indicates the proportion of cash held for liquidity
purposes, whereas $w_{\mathrm{asset}}=1-w_{\mathrm{cash}}$ measures the risky
exposure to the assets. Traditionally, the fund is fully exposed to the
assets, meaning that $w_{\mathrm{cash}}=0\%$ and $w_{\mathrm{asset}}=100\%$.
Implementing a cash buffer implies that $w_{\mathrm{cash}}>0$. Nevertheless,
it is difficult to give an order of magnitude in terms of policies and
practices by asset managers. Using a sample of US funds regulated by the SEC,
\citet{Chernenko-2016} found that $w_{\mathrm{cash}}$ is equal to $7.5\%$ and
$7.9\%$ for equity and bond funds on average. However, the dispersion is very
high because $\sigma \left( w_{\mathrm{cash}}\right) $ is approximately equal
to $8\%$. Moreover, this high dispersion is observed in both the cross
section and the time series. Using the percentile statistics, we can
estimate that the common practice is to have a cash buffer between $0\%$ and $15\%$.

\subparagraph{Mean-variance analysis}

In Appendix \ref{appendix:cash} on page \pageref{appendix:cash},
we derive several statistics by comparing a fund
that is fully exposed to the assets and a fund that implements a cash
buffer. Let $R$ be the random return of this latter. We have:
\begin{equation}
\mathbb{E}\left[ R\right] =\mu _{\mathrm{asset}}-w_{\mathrm{cash}}\cdot
\left( \mu _{\mathrm{asset}}-\mu _{\mathrm{cash}}\right)
\end{equation}%
and:%
\begin{equation}
\sigma \left( R\right) =\sqrt{w_{\mathrm{cash}}^{2}\cdot \sigma _{\mathrm{%
cash}}^{2}+w_{\mathrm{asset}}^{2}\cdot \sigma _{\mathrm{asset}}^{2}+2w_{%
\mathrm{cash}}\cdot w_{\mathrm{asset}}\cdot \rho _{\func{cash},\mathrm{asset}%
}\cdot \sigma _{\mathrm{cash}}\cdot \sigma _{\mathrm{asset}}}
\end{equation}%
where $\mu _{\mathrm{cash}}$ and $\mu _{\mathrm{asset}}$ are the expected
returns of the cash and asset components, $\sigma _{\mathrm{cash}}$ and
$\sigma _{\mathrm{asset}}$ are the corresponding volatilities, and
$\rho _{\mathrm{cash},\mathrm{asset}}$ is the correlation between the cash and the assets. Since the volatility of the cash buffer is considerably lower than the volatility of the assets, we deduce that:
\begin{equation}
\sigma \left( R\right) \approx \left( 1-w_{\mathrm{cash}}\right) \cdot
\sigma _{\mathrm{asset}}
\end{equation}%
We observe that both the expected return\footnote{Because
we generally have $\mu _{\mathrm{asset}}>\mu _{\mathrm{cash}}$.} and
the volatility decrease with the introduction of the cash buffer. In
conclusion, maintaining constant liquidity consists in taking less risk
with little impact on the Sharpe ratio of the fund. Indeed, we obtain:
\begin{equation*}
\limfunc{SR}\left( R\right) \approx \limfunc{SR}\left( R_{\mathrm{asset}%
}\right)
\end{equation*}%
where $\limfunc{SR}\left( R_{\mathrm{asset}}\right) $ is the Sharpe ratio of
the assets. Therefore, the implementation of a cash buffer is equivalent to
deleveraging the asset portfolio. This result is confirmed by the portfolio's beta, which is lower than one:
\begin{equation}
\beta \left( R\mid R_{\mathrm{asset}}\right) \approx 1-w_{\mathrm{cash}}\leq 1
\end{equation}

\subparagraph{Tracking error analysis}

In this analysis, we consider that the benchmark is the asset portfolio (or the index of the corresponding asset class). On page \pageref{appendix:cash-te}, we show that the expected excess return is equal
to:
\begin{equation}
\mathbb{E}\left[ R\mid R_{\mathrm{asset}}\right] =-w_{\mathrm{cash}}\cdot
\left( \mu _{\mathrm{asset}}-\mu _{\mathrm{cash}}\right)
\end{equation}%
whereas the tracking error volatility $\sigma \left( R\mid R_{\mathrm{asset}%
}\right) $ is equal to:%
\begin{equation}
\sigma \left( R\mid R_{\mathrm{asset}}\right) \approx w_{\mathrm{cash}}\cdot
\sigma _{\mathrm{asset}}
\end{equation}%
In a normal situation where $\mu _{\mathrm{asset}}>\mu _{\mathrm{cash}}$, the
expected excess return is negative whereas the tracking error volatility is
proportional to the cash-to-assets ratio. An important result is that the
information ratio is the opposite of the Sharpe ratio of the assets:
\begin{equation}
\limfunc{IR}\left( R\mid R_{\mathrm{asset}}\right) \approx -\limfunc{SR}%
\left( R_{\mathrm{asset}}\right)
\end{equation}%
Again, this implies that the information ratio is generally negative.

\subparagraph{Liquidation gain}

The previous analysis shows that there is a cost associated to the cash
buffer. Nevertheless, there are also some benefits. The most important is the
liquidation gain, which is related to the difference of the transaction costs
without and with the cash buffer:
\begin{equation}
\mathcal{LG}\left( w_{\mathrm{cash}}\right) =\mathcal{TC}_{\mathrm{without}}-%
\mathcal{TC}_{\mathrm{with}}  \label{eq:lg1}
\end{equation}%
where $\mathcal{TC}_{\mathrm{without}}$ is the transaction cost without the cash
buffer and $\mathcal{TC}_{\mathrm{with}}$ is the transaction cost with the cash
buffer. In Appendix \ref{appendix:cash-lg} on page
\pageref{appendix:cash-lg}, we show that:
\begin{eqnarray}
\mathcal{LG}\left( w_{\mathrm{cash}}\right)  &=&\mathcal{TC}_{\mathrm{asset}%
}\left( \RedemptionRate\right) -\mathcal{TC}_{\mathrm{cash}}\left( %
\RedemptionRate\right) \cdot \mathds{1}\left\{ \RedemptionRate<w_{\mathrm{%
cash}}\right\} -  \notag \\
&&\mathcal{TC}_{\mathrm{asset}}\left( \left( \RedemptionRate-w_{\mathrm{cash}%
}\right) \right) \cdot \mathds{1}\left\{ \RedemptionRate\geq w_{\mathrm{cash}%
}\right\}   \label{eq:lg2}
\end{eqnarray}%
and:%
\begin{eqnarray}
\mathbb{E}\left[ \mathcal{LG}\left( w_{\mathrm{cash}}\right) \right]
&=&\int_{0}^{w_{\mathrm{cash}}}\left( \mathcal{TC}_{\mathrm{asset}}\left( %
\RedemptionRate\right) -\mathcal{TC}_{\mathrm{cash}}\left( \RedemptionRate%
\right) \right) \,\mathrm{d}\mathbf{F}\left( \RedemptionRate\right) +  \notag
\\
&&\int_{w_{\mathrm{cash}}}^{1}\left( \mathcal{TC}_{\mathrm{asset}}\left( %
\RedemptionRate\right) -\mathcal{TC}_{\mathrm{asset}}\left(
\RedemptionRate-w_{\mathrm{cash}} \right) \right) \,\mathrm{d}\mathbf{%
F}\left( \RedemptionRate\right)   \label{eq:lg3}
\end{eqnarray}%
where $\mathcal{TC}_{\mathrm{asset}}\left( \RedemptionRate\right) $ and
$\mathcal{TC}_{\mathrm{cash}}\left( \RedemptionRate\right) $ are the asset
and cash transaction cost functions, and $\mathbf{F}\left( x\right) $ is the
distribution function of the redemption rate $\RedemptionRate$. Implementing
a cash buffer has two main effects on the liquidity gain:
\begin{itemize}
\item First, we sell cash instead of the assets if the redemption shock is
    lower than the cash buffer and we have:
\begin{equation}
\mathcal{TC}_{\mathrm{asset}}\left( \RedemptionRate\right) \gg \mathcal{TC}_{%
\mathrm{cash}}\left( \RedemptionRate\right)
\end{equation}

\item Second, we sell a lower proportion of risky assets if the redemption
    rate is greater than the cash-to-assets ratio and we have:
\begin{equation}
\mathcal{TC}_{\mathrm{asset}}\left( \RedemptionRate\right) \gg \mathcal{TC}_{%
\mathrm{asset}}\left(\RedemptionRate-w_{\mathrm{cash}}\right)
\end{equation}
\end{itemize}%
The expected liquidation gain is then made up of two terms which are
positive:
\begin{equation}
\mathbb{E}\left[ \mathcal{LG}\left( w_{\mathrm{cash}}\right) \right] =%
\mathbb{E}\left[ \mathcal{LG}_{\mathrm{cash}}\left( w_{\mathrm{cash}}\right) %
\right] +\mathbb{E}\left[ \mathcal{LG}_{\mathrm{asset}}\left( w_{\mathrm{cash%
}}\right) \right]
\end{equation}%
with the following properties\footnote{See Appendix \ref{appendix:cash-deriv}
on page \pageref{appendix:cash-deriv}. }:
\begin{itemize}
\item $\mathbb{E}\left[ \mathcal{LG}_{\mathrm{cash}}\left(
    w_{\mathrm{cash}}\right) \right] $ is an increasing function of
    $w_{\mathrm{cash}}$ with $\mathbb{E}\left[
    \mathcal{LG}_{\mathrm{cash}}\left( 0\right) \right] =0$ and a maximum
    reached at $w_{\mathrm{cash}}^{\star }=1$:
\begin{eqnarray}
\sup \mathbb{E}\left[ \mathcal{LG}_{\mathrm{cash}}\left( w_{\mathrm{cash}%
}\right) \right]  &=&\mathbb{E}\left[ \mathcal{LG}_{\mathrm{cash}}\left(
1\right) \right]   \notag \\
&=&\int_{0}^{1}\left( \mathcal{TC}_{\mathrm{asset}}\left( \RedemptionRate%
\right) -\mathcal{TC}_{\mathrm{cash}}\left( \RedemptionRate\right) \right) \,%
\mathrm{d}\mathbf{F}\left( \RedemptionRate\right)
\end{eqnarray}

\item $\mathbb{E}\left[ \mathcal{LG}_{\mathrm{asset}}\left( 0\right)
    \right] =0$ and $\mathbb{E}\left[
    \mathcal{LG}_{\mathrm{asset}}\left(1\right) \right] =0$, implying that
    $\mathbb{E}\left[\mathcal{LG}_{\mathrm{asset}}\left(
    w_{\mathrm{cash}}\right) \right] $ is not an increasing function of
    $w_{\mathrm{cash}}$. In fact, we can show that it is a bell curve,
    which is first increasing and then decreasing.
\end{itemize}%
If we combine the two effects, we can show that:%
\begin{equation}
\frac{\partial \,\mathbb{E}\left[ \mathcal{LG}\left( w_{\mathrm{cash}%
}\right) \right] }{\partial \,w_{\mathrm{cash}}}=-\mathcal{TC}_{\mathrm{cash}%
}\left( w_{\mathrm{cash}}\right) \cdot f\left( w_{\mathrm{cash}}\right)
+\int_{w_{\mathrm{cash}}}^{1}\mathcal{TC}_{\mathrm{asset}}^{\prime }\left( %
\RedemptionRate-w_{\mathrm{cash}}\right) \,\mathrm{d}\mathbf{F}\left( %
\RedemptionRate\right)   \label{eq:lg4}
\end{equation}%
where $f\left( x\right) $ is the probability density function of the
redemption rate $\RedemptionRate$ and $\mathcal{TC}_{\mathrm{asset}}^{\prime
}$ is the derivative of the transaction cost function. We deduce that
$\mathbb{E}\left[ \mathcal{LG}\left( w_{\mathrm{cash}}\right) \right] $ is an
increasing function almost everywhere, except when $w_{\mathrm{cash}}$ is
close to one. Therefore, the function $\mathbb{E}\left[ \mathcal{LG}\left(
w_{\mathrm{cash}}\right) \right] $ reaches its maximum at a point
$w_{\mathrm{cash}}^{\star }$, which is close to $1$.\smallskip

Under the assumption that liquidating cash has zero cost and the
additive property of the transaction cost function is almost satisfied, we demonstrate
that\footnote{See Equation (\ref{eq:app-lg5}) on page \pageref{eq:app-lg5}.}:
\begin{equation}
\mathbb{E}\left[ \mathcal{LG}\left( w_{\mathrm{cash}}\right)\right] =\int_{0}^{w_{\mathrm{cash}}}\mathcal{TC%
}_{\mathrm{asset}}\left( \RedemptionRate\right) \,\mathrm{d}\mathbf{F}\left( %
\RedemptionRate\right) +\mathcal{TC}_{\mathrm{asset}}\left( w_{\mathrm{cash}%
}\right) \cdot \left( 1-\mathbf{F}\left( w_{\mathrm{cash}}\right) \right)
\label{eq:lg5}
\end{equation}%
The interpretation of this formula is very simple. The first term corresponds
to the expected transaction cost of liquidating the risky assets when the
redemption rate is lower than the cash-to-assets ratio, whereas the second
term is the transaction cost of liquidating the asset amount equivalent to
the cash buffer times the probability of observing a redemption shock greater
than the cash buffer. In Appendix \ref{appendix:cash-deriv} on page
\pageref{appendix:cash-deriv}, we demonstrate that:
\begin{equation}
\frac{\partial \,\mathbb{E}\left[ \mathcal{LG}\left( w_{\mathrm{cash}%
}\right) \right] }{\partial \,w_{\mathrm{cash}}}=\mathcal{TC}_{\mathrm{asset}%
}^{\prime }\left( w_{\mathrm{cash}}\right) \cdot \left( 1-\mathbf{F}\left(
w_{\mathrm{cash}}\right) \right)   \label{eq:lg6}
\end{equation}%
If we compare Equations (\ref{eq:lg4}) and (\ref{eq:lg6}), we observe that
they are not the same. The first term has vanished because
$\mathcal{TC}_{\mathrm{asset}}\left( \RedemptionRate\right) \approx 0$. The
second term is obtained by assuming that
$\mathcal{TC}_{\mathrm{asset}}^{\prime }$ is relatively constant\footnote{The
choice of $w_{\mathrm{cash}}$ for the derivative function
$\mathcal{TC}_{\mathrm{asset}}^{\prime }\left( \RedemptionRate\right) $ is
explained later.}:
\begin{eqnarray}
\int_{w_{\mathrm{cash}}}^{1}\mathcal{TC}_{\mathrm{asset}}^{\prime }\left( %
\RedemptionRate-w_{\mathrm{cash}}\right) \,\mathrm{d}\mathbf{F}\left( %
\RedemptionRate\right)  &\approx &\mathcal{TC}_{\mathrm{asset}}^{\prime
}\left( w_{\mathrm{cash}}\right) \int_{w_{\mathrm{cash}}}^{1}\mathrm{d}%
\mathbf{F}\left( \RedemptionRate\right)   \notag \\
&=&\mathcal{TC}_{\mathrm{asset}}^{\prime }\left( w_{\mathrm{cash}}\right)
\cdot \left( 1-\mathbf{F}\left( w_{\mathrm{cash}}\right) \right)
\end{eqnarray}%
Since $\partial _{w_{\mathrm{cash}}}\,\mathbb{E}\left[ \mathcal{LG}\left(
w_{\mathrm{cash}}\right) \right] \geq 0$, the main impact of the
approximation is to eliminate the hill effect when
$w_{\mathrm{cash}}\rightarrow 1$.

\begin{example}
\label{ex:cash3} Using a square-root model, we assume that the transaction
cost of liquidating the risky assets is equal to:
\begin{equation}
\mathcal{TC}_{\mathrm{asset}}\left( x\right) =x\cdot \left( \spread+\beta _{%
\impact}\sigma \sqrt{x}\right) \label{eq:ex-cash3a}
\end{equation}%
where $\spread$ is the bid-ask spread, $\sigma $ is the daily volatility and
$\beta _{\impact}$ is the price impact coefficient. Concerning the cash, it
may be liquidated at a fixed rate $\cashRate$:
\begin{equation*}
\mathcal{TC}_{\mathrm{cash}}\left( x\right) =x\cdot \cashRate
\end{equation*}%
where $\cashRate\ll \spread$. We also consider that the redemption rate
follows a power-law distribution:%
\begin{equation}
\mathbf{F}\left( x\right) =x^{\eta }
\end{equation}%
where $\eta >0$.
\end{example}

\begin{figure}[tbph]
\centering
\caption{Transaction cost function (\ref{eq:ex-cash3a}) in bps}
\label{fig:cash3a}
\figureskip
\includegraphics[width = \figurewidth, height = \figureheight]{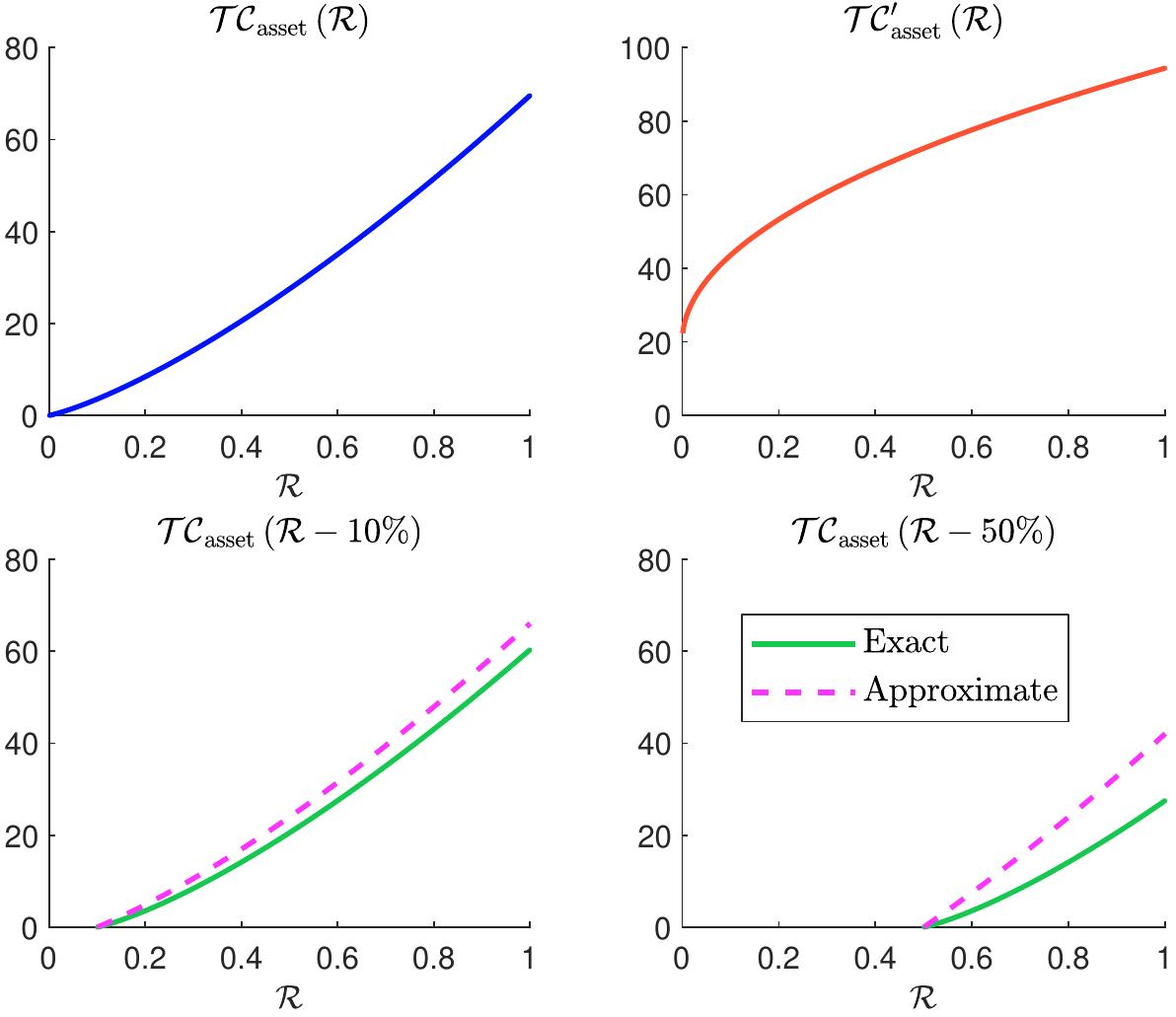}
\end{figure}

In the top left-hand panel in Figure \ref{fig:cash3a}, we have reported the
transaction cost function
$\mathcal{TC}_{\mathrm{asset}}\left(\RedemptionRate\right) $ for the
following parameters: a bid-ask spread $\spread$ of $20$ bps, a price impact
sensitivity $\beta _{\impact}$ of $0.4$ and an annualized volatility of
$20\%$. We notice that the transaction cost is between $0$ and $70$ bps.
Whereas the unit transaction cost function is concave, the total transaction
cost is convex. The first derivative $\mathcal{TC}_{\mathrm{asset}}^{\prime
}\left( \RedemptionRate\right) $ is given in the top right-hand panel in
Figure \ref{fig:cash3a}. We verify that
$\mathcal{TC}_{\mathrm{asset}}^{\prime }\left( \RedemptionRate\right) >0$,
but $\mathcal{TC}_{\mathrm{asset}}^{\prime }\left( \RedemptionRate\right) $
is far from constant. Therefore, the approximation of
$\mathcal{TC}_{\mathrm{asset}}\left( \RedemptionRate-w_{\mathrm{cash}}\right)
$ by the function $\mathcal{TC}_{\mathrm{asset}}\left( \RedemptionRate\right)
-\mathcal{TC}_{\mathrm{asset}}\left( w_{\mathrm{cash}}\right) $ is not accurate. This discrepancy is illustrated in the bottom panels in Figure
\ref{fig:cash3a} when $w_{\mathrm{cash}}$ is equal to $10\%$ and $50\%$.

\begin{figure}[tbph]
\centering
\caption{Exact vs. approximate solution of
$\mathbb{E}\left[ \mathcal{LG}\left( w_{\mathrm{cash}}\right) \right] $ in bps
(Example \ref{ex:cash3}, page \pageref{ex:cash3})}
\label{fig:cash3d}
\figureskip
\includegraphics[width = \figurewidth, height = \figureheight]{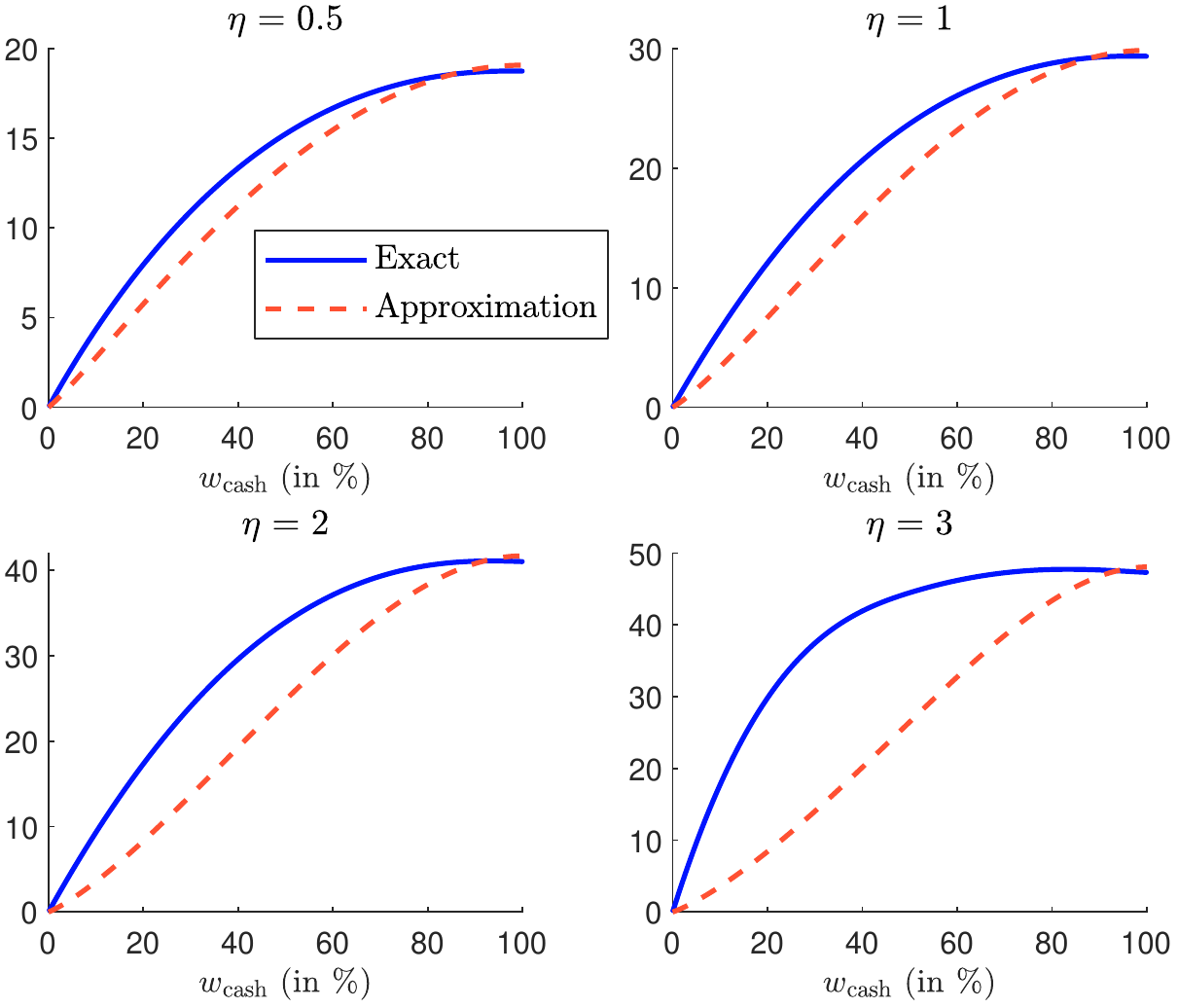}
\end{figure}

As such, this is not surprising if the exact formula of $\mathbb{E}\left[
\mathcal{LG}\left( w_{\mathrm{cash}}\right) \right] $ is:
\begin{eqnarray}
\mathbb{E}\left[ \mathcal{LG}\left( w_{\mathrm{cash}}\right) \right]  &=&%
\frac{\eta \left( \spread-\cashRate\right) }{\eta +1}\cdot w_{\mathrm{cash}%
}^{\eta +1}+\frac{2\eta \beta _{\impact}\sigma }{2\eta +3}+  \notag \\
&&\eta \spread\cdot w_{\mathrm{cash}}\left( 1-w_{\mathrm{cash}}\right) -\eta
\beta _{\impact}\sigma \cdot I\left( w_{\mathrm{cash}};\eta \right)
\label{eq:cash3b}
\end{eqnarray}%
whereas the approximate formula is very different:
\begin{equation}
\mathbb{E}\left[ \mathcal{LG}\left( w_{\mathrm{cash}}\right) \right] \approx %
\spread\cdot w_{\mathrm{cash}}+\beta _{\impact}\sigma \cdot w_{\mathrm{cash}%
}^{1.5}-\frac{\spread}{\eta +1}\cdot w_{\mathrm{cash}}^{\eta +1}-\frac{%
3\beta _{\impact}\sigma }{2\eta +3}\cdot w_{\mathrm{cash}}^{\eta +1.5}
\label{eq:cash3c}
\end{equation}%
We have reported these two functions in Figure \ref{fig:cash3d}. The
liquidation gains are expressed in bps. We observe some differences between
the exact formula (\ref{eq:cash3b}) and the approximate formula
(\ref{eq:cash3c}), but these differences tend to diminish when
$w_{\mathrm{cash}}$ tends to $1$. Moreover, the differences increase with
respect to the parameter $\eta $, which controls the shape of the redemption rate distribution function\footnote{On page
\pageref{fig:cash3b}, Figure \ref{fig:cash3b} shows the density and
distribution functions of the redemption rate. If $\eta =1$, we obtain the
uniform probability distribution. If $\eta \rightarrow 0$, the redemption
rate is located at $\RedemptionRate = 0$. If $\eta \rightarrow 1$, the
redemption rate is located at $\RedemptionRate = 1$. If $\eta <1$, the
probability that the redemption rate is lower than $50\%$ is greater than
$50\%$. If $\eta >1$, the probability that the redemption rate is lower than
$50\%$ is less than $50\%$. Therefore, $\eta $ controls the location of the
redemption rate. The greater the value of $\eta $, the greater the risk of observing a large redemption rate.}. This is normal because the probability of observing a large redemption rate increases with the parameter $\eta $. In fact, the poor approximation of $\mathbb{E}\left[ \mathcal{LG}\left(
w_{\mathrm{cash}}\right) \right] $ mainly comes from the solution of
$\mathbb{E}\left[ \mathcal{LG}_{\mathrm{asset}}\left(
w_{\mathrm{cash}}\right) \right] $ and not the solution of $\mathbb{E}\left[
\mathcal{LG}_{\mathrm{cash}}\left( w_{\mathrm{cash}}\right) \right] $ as
illustrated in Figure \ref{fig:cash3c} on page
\pageref{fig:cash3c}.\smallskip

This example allows us to verify the properties that have been demonstrated
previously. Indeed, Figure \ref{fig:cash3d} confirms that the approximate
function of $\mathbb{E}\left[ \mathcal{LG}\left( w_{\mathrm{cash}}\right)
\right] $ is increasing and reaches its maximum at $w_{\mathrm{cash}}^{\star
}=1$, whereas the exact function of $\mathbb{E}\left[ \mathcal{LG}\left(
w_{\mathrm{cash}}\right) \right] $ increases almost everywhere and only
decreases when $w_{\mathrm{cash}}$ is close to $1$. This implies that the
maximum of $\mathbb{E}\left[ \mathcal{LG}\left( w_{\mathrm{cash}}\right)
\right] $ reaches its maximum at $w_{\mathrm{cash}}^{\star }<1$. In our
example, $w_{\mathrm{cash}}^{\star }$ is equal to $97.40\%$, $96.67\%$,
$93.55\%$ and $83.37\%$ when $\eta $ is respectively equal to $0.5$, $1$, $2$
and $3$.

\begin{example}
\label{ex:cash4} We consider Example \ref{ex:cash3} on page
\pageref{ex:cash3}, but we impose a daily trading limit $x^{+}$. This example
is more realistic than the previous one, because selling $100\%$ of the
assets generally requires more than one day. This is especially true in a
liquidity stress testing framework. For example, $x^{+}=10\%$ imposes that we
can sell $10\%$ of the fund every trading day, implying that we need $10$
trading days to liquidate the fund.
\end{example}

\begin{figure}[tbph]
\centering
\caption{Transaction cost function (\ref{eq:ex-cash4a}) in bps with $x^{+} = 10\%$}
\label{fig:cash4a}
\figureskip
\includegraphics[width = \figurewidth, height = \figureheight]{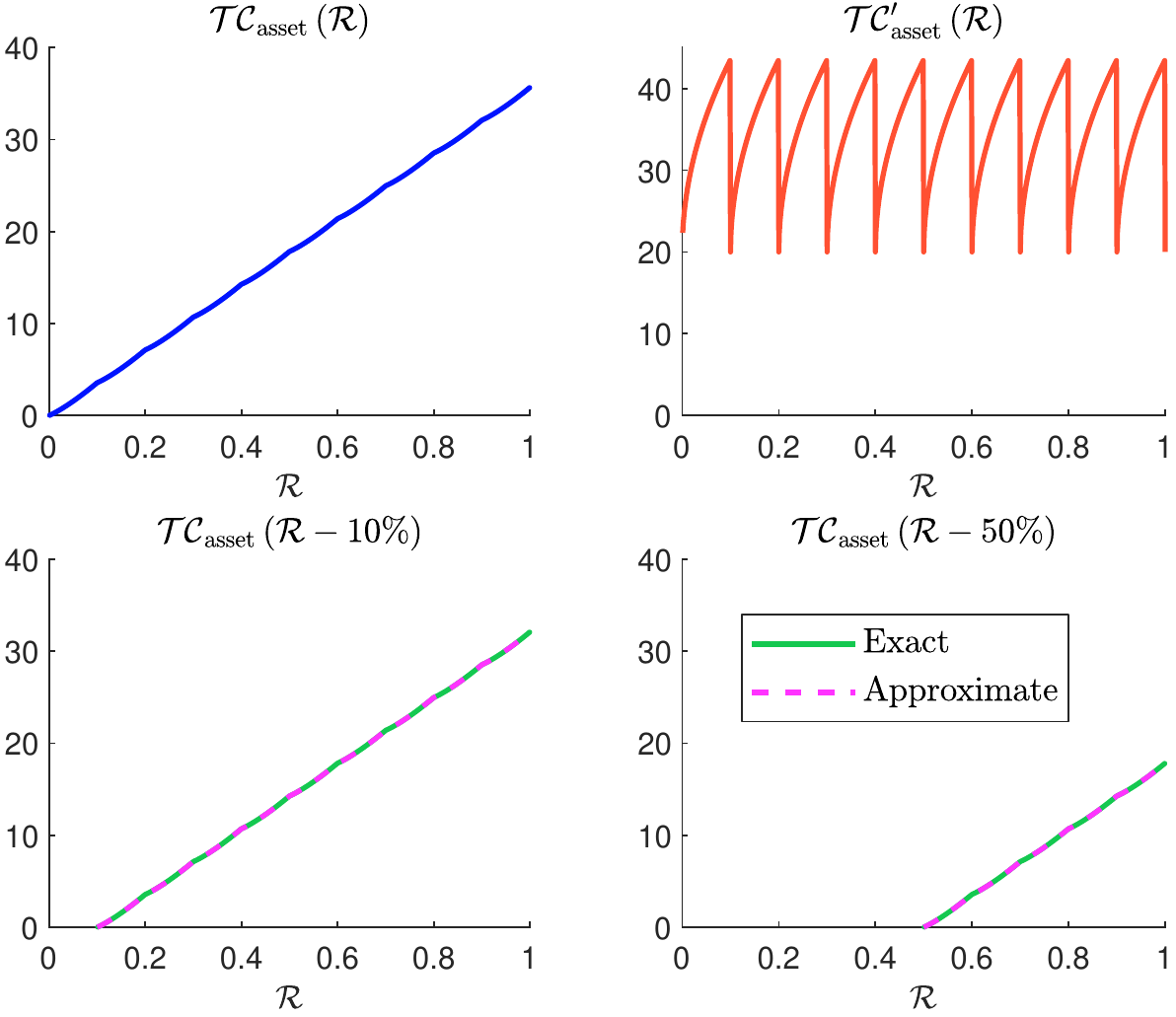}
\end{figure}

If $x\leq x^{+}$, we have:
\begin{equation}
\mathcal{TC}_{\mathrm{asset}}\left( x\right) =x\left( \spread+\beta _{\impact%
}\sigma \sqrt{x}\right)
\end{equation}%
If $x^{+}<x\leq 2x^{+}$, we need two trading days to liquidate $x$ and we
have:
\begin{eqnarray}
\mathcal{TC}_{\mathrm{asset}}\left( x\right)  &=&\underset{\text{First
trading day}}{\underbrace{x^{+}\left( \spread+\beta _{\impact}\sigma \sqrt{%
x^{+}}\right) }}+\underset{\text{Second trading day}}{\underbrace{\left(
x-x^{+}\right) \left( \spread+\beta _{\impact}\sigma \sqrt{x-x^{+}}\right) }}
\notag \\
&=&x\cdot \spread+\beta _{\impact}\sigma \cdot \left( x^{+}\sqrt{x^{+}}%
+\left( x-x^{+}\right) \sqrt{x-x^{+}}\right)
\end{eqnarray}%
More generally, if $\kappa x^{+}<x\leq \left( \kappa +1\right) x^{+}$, $x$ is
liquidated in $\kappa +1$ trading days, and we obtain:
\begin{equation}
\mathcal{TC}_{\mathrm{asset}}\left( x\right) =x\cdot \spread+\kappa \beta _{%
\impact}\sigma \cdot x^{+}\sqrt{x^{+}}+\beta _{\impact}\sigma \cdot \left(
x-\kappa x^{+}\right) \sqrt{x-\kappa x^{+}}  \label{eq:ex-cash4a}
\end{equation}%
where:%
\begin{equation}
\kappa :=\kappa \left( x;x^{+}\right) =\left\lfloor \frac{x}{x^{+}}%
\right\rfloor
\end{equation}%
Figure \ref{fig:cash4a} represents the transaction cost function
$\mathcal{TC}_{\mathrm{asset}}\left( \RedemptionRate\right) $ for the
following parameters: a bid-ask spread $\spread$ of $20$ bps, a price impact
sensitivity $\beta _{\impact}$ of $0.4$, an annualized volatility of $20\%$
and a trading limit $x^{+}=10\%$. Compared to Figure \ref{fig:cash3a}, the
transaction cost is reduced and is between $0$ and $40$ bps. This is normal
because the daily price impact is bounded in this example, and we cannot sell
more than $10\%$. The first derivative $\mathcal{TC}_{\mathrm{asset}}^{\prime
}\left( \RedemptionRate\right) $ lies in the interval $\left[ 20,43\right] $
bps and can be assumed to be constant. Therefore, the approximation of
$\mathcal{TC}_{\mathrm{asset}}\left( \RedemptionRate-w_{\mathrm{cash}}\right)
$ by the function $\mathcal{TC}_{\mathrm{asset}}\left( \RedemptionRate\right)
-\mathcal{TC}_{\mathrm{asset}}\left( w_{\mathrm{cash}}\right) $ is good as
illustrated in the bottom panels in Figure \ref{fig:cash4a} when
$w_{\mathrm{cash}}$ is equal to $10\%$ and $50\%$.\smallskip

\begin{figure}[tbph]
\centering
\caption{Maximum approximation error function
$\mathcal{M}_{\mathrm{ax}}\mathcal{E}_{\mathrm{rror}}\left( x^{+}\right) $ in bps}
\label{fig:cash4e}
\figureskip
\includegraphics[width = \figurewidth, height = \figureheight]{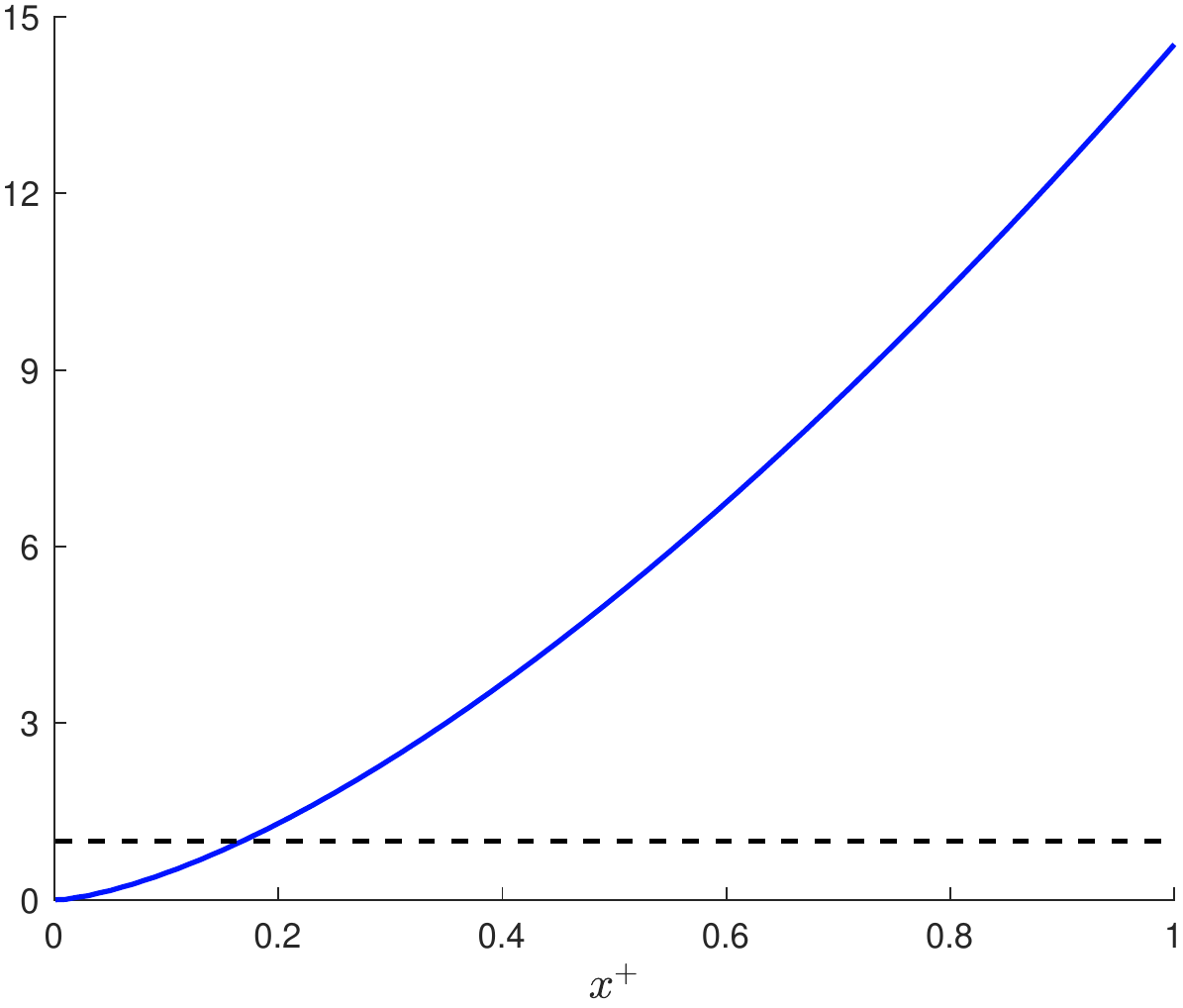}
\end{figure}

In Appendix \ref{appendix:figures} on page \pageref{fig:cash4b}, we report
the transaction cost function when the trading limit is respectively equal to
$x^{+}=30\%$ and $x^{+}=50\%$ (Figures \ref{fig:cash4b} and
\ref{fig:cash4c}). We observe that the approximation is less and less accurate when the trading limit $x^{+}$ increases. Let us define the approximation error by:
\begin{equation}
\mathcal{E}_{\mathrm{rror}}\left( w_{\mathrm{cash}};x^{+}\right) =\sup_{%
\RedemptionRate\in \left[ 0,1\right] }\left\vert \left( \mathcal{TC}_{%
\mathrm{asset}}\left( \RedemptionRate\right) -\mathcal{TC}_{\mathrm{asset}%
}\left( w_{\mathrm{cash}}\right) \right) -\mathcal{TC}_{\mathrm{asset}%
}\left( \RedemptionRate-w_{\mathrm{cash}}\right) \right\vert
\end{equation}%
This function is represented in Figure \ref{fig:cash4d} on page
\pageref{fig:cash4d} for three values of $x^{+}$: $10\%$, $20\%$ and
$30\%$. We see that the approximation error is cyclical:
\begin{equation}
\mathcal{E}_{\mathrm{rror}}\left( w_{\mathrm{cash}};x^{+}\right) =\mathcal{E}%
_{\mathrm{rror}}\left( w_{\mathrm{cash}}+k\cdot x^{+};x^{+}\right) \qquad
\text{for }k=1,2,\ldots
\end{equation}%
and we observe a modulo pattern because of the introduction of trading
limits. In Figure \ref{fig:cash4e}, we have reported the maximum
approximation error:
\begin{equation}
\mathcal{M}_{\mathrm{ax}}\mathcal{E}_{\mathrm{rror}}\left( x^{+}\right)
=\sup_{w_{\mathrm{cash}}\in \left[ 0,1\right] }\mathcal{E}\mathrm{rror}%
\left( w_{\mathrm{cash}};x^{+}\right)
\end{equation}%
The maximum error is not acceptable when we would like to trade a large amount in the market, but it is relatively low for usual trading limits. In our example, imposing $\mathcal{M}_{\mathrm{ax}}\mathcal{E}_{\mathrm{rror}}\left(
x^{+}\right) \leq 1$ bp is achieved when $x^{+}\leq 16\%$.

\begin{figure}[tbph]
\centering
\caption{Approximation of the liquidity gain
$\mathbb{E}\left[ \mathcal{LG}\left( w_{\mathrm{cash}}\right) \right] $
in bps when $x^{+} = 10\%$ (Example \ref{ex:cash4}, page \pageref{ex:cash4})}
\label{fig:cash4h}
\figureskip
\includegraphics[width = \figurewidth, height = \figureheight]{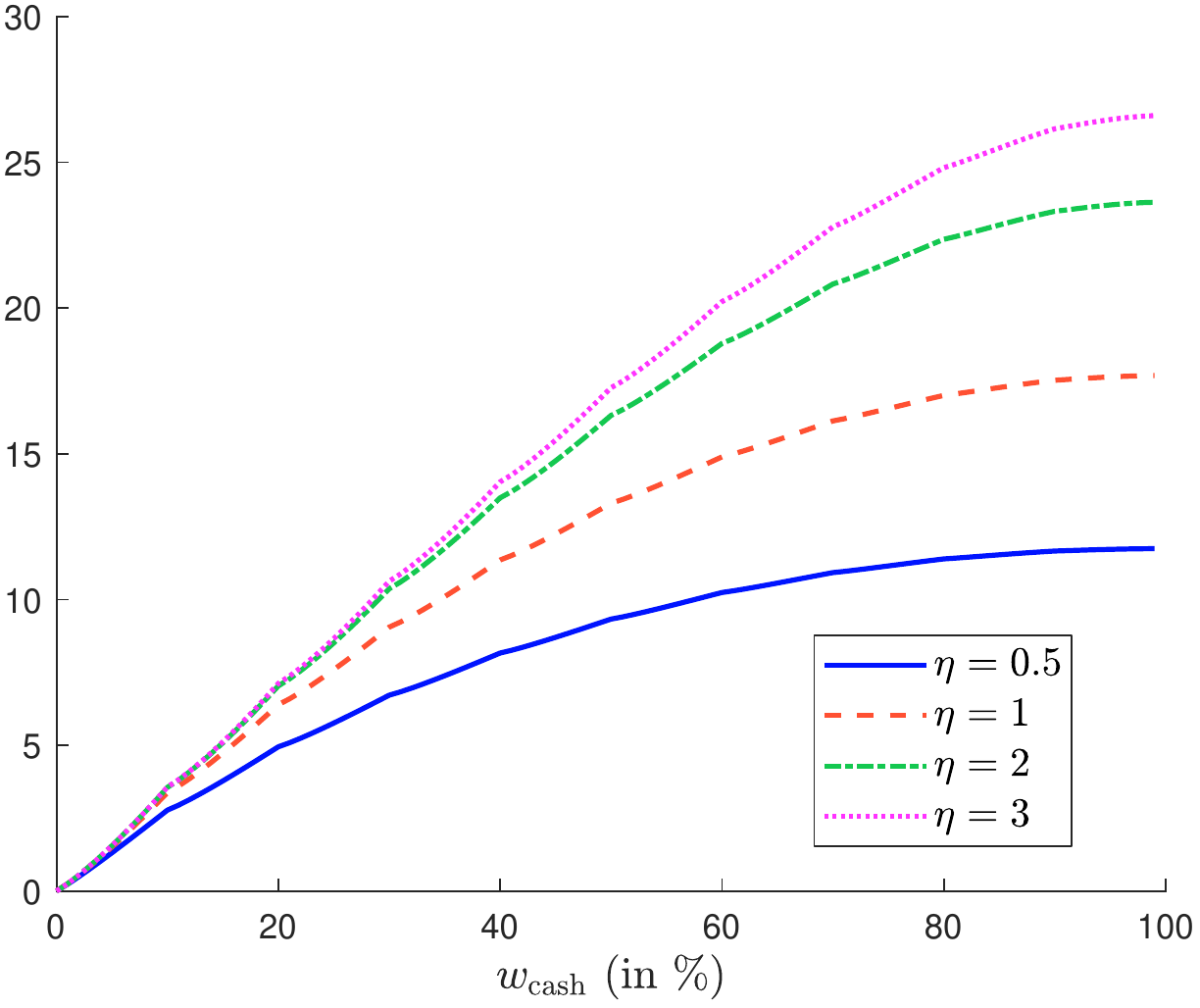}
\end{figure}

\begin{remark}
In Appendix \ref{appendix:cash4-ex} on page \pageref{appendix:cash4-ex}, we
derive the approximation of $\mathbb{E}\left[ \mathcal{LG}\left(
w_{\mathrm{cash}}\right) \right] $. Figure \ref{fig:cash4h} shows the values
of the liquidity gain when $x^{+}=10\%$. The two components are reported in
Figure \ref{fig:cash4g} on page \pageref{fig:cash4g}. Moreover, the
comparison between the exact formulas (computed with numerical integration)
and the approximation formulas is given in Figure \ref{fig:cash4i} on page
\pageref{fig:cash4i}. We verify that the approximation is very good.
\end{remark}

\paragraph{Optimal cash buffer}

We can now formulate the fund manager's optimization program. Its
objective is to minimize the expected cost of the buffer $\mathcal{BC}\left(
w_{\mathrm{cash}}\right) $ and maximize its expected gain $\mathcal{BG}\left(
w_{\mathrm{cash}}\right) $:
\begin{equation}
w_{\mathrm{cash}}^{\star }=\arg \min_{w\in \left[ 0,1\right] }\underset{%
\text{Net cost }\mathcal{NC}\left( w_{\mathrm{cash}}\right) }{\underbrace{%
\mathcal{BC}\left( w_{\mathrm{cash}}\right) -\mathcal{BG}\left( w_{\mathrm{%
cash}}\right) }}
\end{equation}%
Since the buffer cost and the buffer gain are two increasing functions, the
minimum of $\mathcal{BC}\left( w_{\mathrm{cash}}\right) $ is reached at
$w_{\mathrm{cash}}^{\star }=0$ while the maximum of $\mathcal{BG}\left(
w_{\mathrm{cash}}\right) $ is obtained for $w_{\mathrm{cash}}^{\star }=1$.
Therefore, there is a trade-off between these two functions. For instance, if
we consider that the expected cost of the cash buffer corresponds to the
opposite of the expected excess return penalized by the tracking error
variance, we obtain:
\begin{eqnarray}
\mathcal{BC}\left( w_{\mathrm{cash}}\right)  &=&-\mathbb{E}\left[ R\mid R_{%
\mathrm{asset}}\right] +\frac{\lambda }{2}\sigma ^{2}\left( R\mid R_{\mathrm{%
asset}}\right)  \\
&=&w_{\mathrm{cash}}\left( \mu _{\mathrm{asset}}-\mu _{\mathrm{cash}}\right)
+\frac{\lambda }{2}w_{\mathrm{cash}}^{2}\left( \sigma _{\mathrm{cash}%
}^{2}+\sigma _{\mathrm{asset}}^{2}-2\rho _{\mathrm{cash},\mathrm{asset}%
}\sigma _{\mathrm{cash}}\sigma _{\mathrm{asset}}\right) \notag
\end{eqnarray}%
where $\lambda \geq 0$ represents the aversion parameter to the tracking
error risk. For the specification of the buffer gain, we can choose the
expected liquidation gain:
\begin{equation}
\mathcal{BG}\left( w_{\mathrm{cash}}\right) =\mathbb{E}\left[ \mathcal{LG}%
\left( w_{\mathrm{cash}}\right) \right]
\end{equation}%
We deduce the expression of the net buffer cost $\mathcal{NBC}\left( w_{%
\mathrm{cash}}\right) $:%
\begin{eqnarray}
\mathcal{NBC}\left( w_{\mathrm{cash}}\right)  &=&w_{\mathrm{cash}}\left( \mu
_{\mathrm{asset}}-\mu _{\mathrm{cash}}\right) +  \notag \\
&&\frac{\lambda }{2}w_{\mathrm{cash}}^{2}\left( \sigma _{\mathrm{cash}%
}^{2}+\sigma _{\mathrm{asset}}^{2}-2\rho _{\mathrm{cash},\mathrm{asset}%
}\sigma _{\mathrm{cash}}\sigma _{\mathrm{asset}}\right) -  \notag \\
&&\mathbb{E}\left[ \mathcal{LG}\left( w_{\mathrm{cash}}\right) \right]
\end{eqnarray}%
It is made up of three components:
\begin{enumerate}
\item the return component that compares the expected asset return
    and the cash return;
\item the tracking error risk that measures the discrepancy of the fund's
    behavior with respect to the expected behavior;
\item the liquidity gain.
\end{enumerate}
In order to find the solution to the optimization problem, we compute the
derivative of the net buffer cost:
\begin{eqnarray}
\frac{\partial \,\mathcal{NBC}\left( w_{\mathrm{cash}}\right) }{\partial
\,w_{\mathrm{cash}}} &=&\mu _{\mathrm{asset}}-\mu _{\mathrm{cash}}+  \notag
\\
&&\lambda w_{\mathrm{cash}}\left( \sigma _{\mathrm{cash}}^{2}+\sigma _{%
\mathrm{asset}}^{2}-2\rho _{\mathrm{cash},\mathrm{asset}}\sigma _{\mathrm{%
cash}}\sigma _{\mathrm{asset}}\right) -  \notag \\
&&\frac{\partial \,\mathbb{E}\left[ \mathcal{LG}\left( w_{\mathrm{cash}%
}\right) \right] }{\partial \,w_{\mathrm{cash}}}
\end{eqnarray}%
Finally, we conclude that:
\begin{equation}
w_{\mathrm{cash}}^{\star }\in \left\{
\begin{array}{ll}
\left\{ 0\right\}  & \text{if }\partial _{w_{\mathrm{cash}}}\mathcal{NBC}%
\left( w_{\mathrm{cash}}\right) \geq 0 \\
\left\{ 1\right\}  & \text{if }\partial _{w_{\mathrm{cash}}}\mathcal{NBC}%
\left( w_{\mathrm{cash}}\right) \leq 0 \\
\left] 0,1\right[  & \text{otherwise}%
\end{array}%
\right.
\end{equation}%
The optimal value is equal to $w_{\mathrm{cash}}^{\star }=0$ in particular
when the expected return difference between the assets and the cash is
greater than the marginal expected liquidation gain:
\begin{equation}
\mu _{\mathrm{asset}}-\mu _{\mathrm{cash}}\geq \frac{\partial \,\mathbb{E}%
\left[ \mathcal{LG}\left( w_{\mathrm{cash}}\right) \right] }{\partial \,w_{%
\mathrm{cash}}}\Rightarrow \frac{\partial \,\mathcal{NBC}\left( w_{\mathrm{%
cash}}\right) }{\partial \,w_{\mathrm{cash}}}\geq 0
\end{equation}%
If the fund manager is not sensitive to the tracking error risk ($\lambda
=0$), we have:
\begin{equation}
\mu _{\mathrm{asset}}\leq 0\Longrightarrow w_{\mathrm{cash}}^{\star }=1
\end{equation}%
The two extreme solutions are easy to interpret. The first extreme case
$w_{\mathrm{cash}}^{\star }=0$ is obtained because the liquidation gain does
not compensate the (large) risk premium $\mu _{\mathrm{asset}}-\mu
_{\mathrm{cash}}$ of the assets, whereas the second extreme case
$w_{\mathrm{cash}}^{\star }=100\%$ is achieved because the fund manager
anticipates that the assets will generate a negative return. In the first case, it is inefficient to implement a cash buffer because we expect the assets to perform very well. Therefore, implementing a cash buffer will dramatically
reduce the fund's return and the cost of the liquidity stress is not
sufficient to offset this later. In the second case, it is better to
implement a $100\%$ cash buffer because we anticipate that the assets will
face a drawdown. However, if the fund manager and the investors are sensitive
to the tracking error risk, this result no longer holds. Indeed, if $\mu
_{\mathrm{asset}}\leq 0$, the sign of the derivative depends on the value of
$\lambda $:
\begin{eqnarray}
\frac{\partial \,\mathcal{NBC}\left( w_{\mathrm{cash}}\right) }{\partial
\,w_{\mathrm{cash}}} &\approx &\underset{\text{negative}}{\underbrace{\mu _{%
\mathrm{asset}}-\frac{\partial \,\mathbb{E}\left[ \mathcal{LG}\left( w_{%
\mathrm{cash}}\right) \right] }{\partial \,w_{\mathrm{cash}}}}}+  \notag \\
&&\underset{\mathrm{positive}}{\lambda \cdot \underbrace{w_{\mathrm{cash}%
}\left( \sigma _{\mathrm{cash}}^{2}+\sigma _{\mathrm{asset}}^{2}-2\rho _{%
\mathrm{cash},\mathrm{asset}}\sigma _{\mathrm{cash}}\sigma _{\mathrm{asset}%
}\right) }}
\end{eqnarray}%
For a large value of $\lambda $, $w_{\mathrm{cash}}^{\star }=100\%$ is not
optimal because it induces a high tracking error risk. This is especially true if the asset volatility $\sigma _{\mathrm{asset}}$ is large. Nevertheless, the tracking error risk vanishes if $\rho _{\mathrm{cash},\mathrm{asset}}=1$ and $%
\sigma _{\mathrm{cash}}=\sigma _{\mathrm{asset}}$, which corresponds to a
pure cash fund, but this case is obvious. All these results indicate that
the optimal cash buffer is generally equal to $0\%$ or $100\%$, whereas the
probability of obtaining an intermediate value is low.\smallskip

\begin{figure}[tbph]
\centering
\caption{Net buffer cost
($\mu _{\mathrm{asset}}-\mu _{\mathrm{cash}}=1\%$ and $\lambda =0$)}
\label{fig:cash6a}
\figureskip
\includegraphics[width = \figurewidth, height = \figureheight]{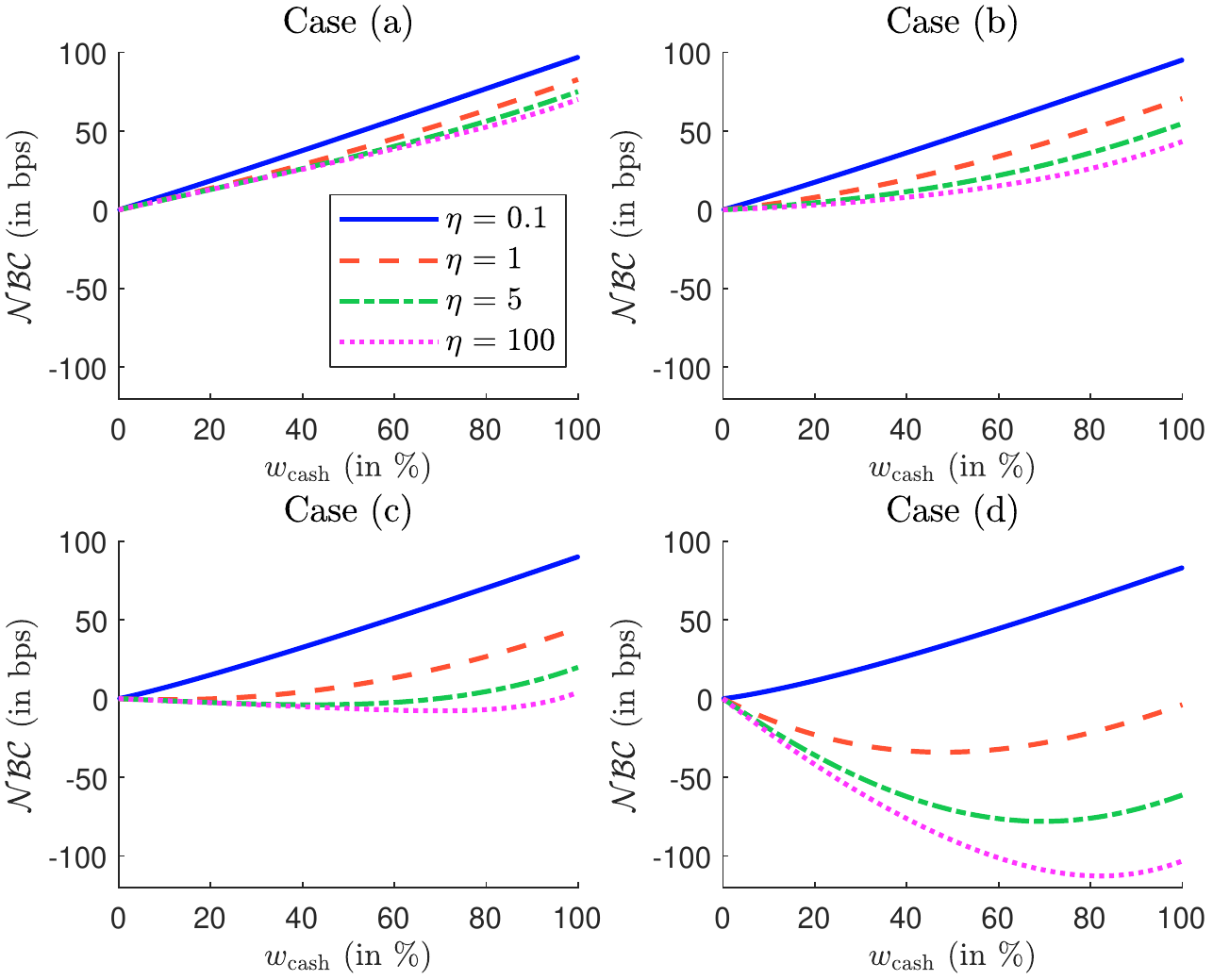}
\end{figure}

\begin{figure}[tbph]
\centering
\caption{Optimal cash buffer
($\mu _{\mathrm{asset}}-\mu _{\mathrm{cash}}=1\%$ and $\lambda =0$)}
\label{fig:cash6b}
\figureskip
\includegraphics[width = \figurewidth, height = \figureheight]{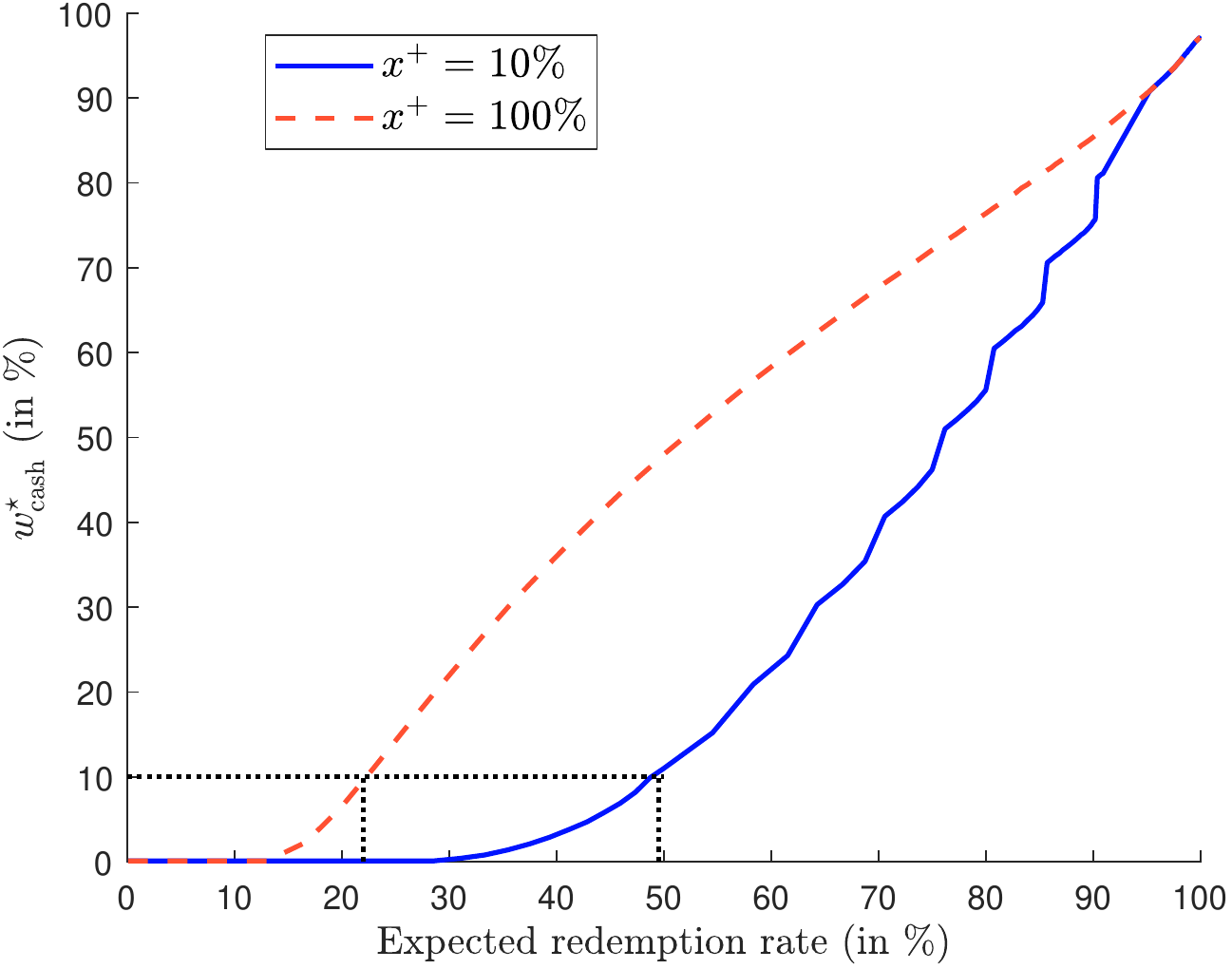}
\end{figure}

Let us illustrate the previous analysis. For the transaction cost function,
we consider the square-root model with several sets of parameters:

\begin{itemize}
\item[(a)] $\spread=20$ bps, $\cashRate=1$ bps, $\beta _{\impact}=0.40$, $%
\sigma =20\%$ and $x^{+}=10\%$

\item[(b)] $\spread=20$ bps, $\cashRate=1$ bps, $\beta _{\impact}=0.40$, $%
\sigma =20\%$ and $x^{+}=100\%$

\item[(c)] $\spread=50$ bps, $\cashRate=1$ bps, $\beta _{\impact}=0.40$, $%
\sigma =80\%$ and $x^{+}=10\%$

\item[(d)] $\spread=50$ bps, $\cashRate=1$ bps, $\beta _{\impact}=0.40$, $%
\sigma =80\%$ and $x^{+}=100\%$
\end{itemize}
The only difference between cases (a) and (b) (resp. cases (c) and
(d)) is the trading limit. There is no trading limit for cases (b) and (d),
whereas we cannot sell more than $10\%$ of total net assets in cases (a) and
(c). Cases (a) and (b) correspond to a normal period, whereas cases (c) and
(d) are more suitable for a liquidity stress period. Indeed, the bid-ask
spread is larger ($50$ bps vs. $20$ bps), and we observe a higher volatility
($80\%$ versus $20\%$). In Figure \ref{fig:cash6a}, we report the net buffer
cost $\mathcal{NBC}\left( w_{\mathrm{cash}}\right) $ when $\mu _{\mathrm{asset}}
-\mu _{\mathrm{cash}}$\ is set to $1\%$ and $\lambda $ is equal to
zero. Each plot corresponds to a different value of the parameter $\eta $.
We notice that the function $\mathcal{NBC}\left( w_{\mathrm{cash}}\right) $
is strictly increasing in cases (a) and (b), implying that the optimal cash
buffer is $w_{\mathrm{cash}}^{\star }=0$. If we consider a normal
transaction cost function, there is no interest to implement a liquidity
buffer. Cases (c) and (d) are more interesting, because the function
$\mathcal{NBC}\left( w_{\mathrm{cash}}\right) $ may be decreasing and then
increasing, meaning that $w_{\mathrm{cash}}^{\star }>0$. Therefore, it is
more interesting to use a \textquotedblleft \textit{stressed}\textquotedblright\
transaction cost function when we would like to
calculate cash buffer analytics. This is why we only focus on cases (c) and
(d) in what follows. Figure \ref{fig:cash6b} shows the optimal value
$w_{\mathrm{cash}}^{\star }$ of the cash buffer with respect to the expected
redemption rate\footnote{We have $\mathbb{E}\left[ \RedemptionRate\right] =
\frac{\eta }{\eta +1}$ when $\mathbf{F}\left( x\right) =x^{\eta }$.}. We
verify that $w_{\mathrm{cash}}^{\star }$ increases with the trading limit $x^{+}$ and the expected redemption rate. For instance, the optimal cash buffer is equal to $10\%$ if $\mathbb{E}\left[ \RedemptionRate\right] =50\%$ and $x^{+}=10\%$. If there is no trading limit, $w_{\mathrm{cash}}^{\star }=10\%$ if $\mathbb{E}\left[ \RedemptionRate\right] =23\%$. Of course, these results are extremely sensitive to the values of $\mu _{\mathrm{asset}}-\mu _{\mathrm{cash}}$, $\lambda $ and $\sigma _{\mathrm{asset}}$. For example, we obtain Figure \ref{fig:cash6c} on page \pageref{fig:cash6c} when $\mu _{\mathrm{asset}}- \mu _{\mathrm{cash}}$ is equal to $2.5\%$. $w_{\mathrm{cash}}^{\star }$ is dramatically reduced, and there is no liquidity buffer when $x^{+}=10\%$. There is also no implementation when $x^{+}=100\%$ and $\mathbb{E}\left[ \RedemptionRate\right] \leq 50\%$. Therefore, the value of $w_{\mathrm{cash}}^{\star }$ is very sensitive to $\mu _{\mathrm{asset}}- \mu _{\mathrm{cash}}$. We observe the same phenomenon with the parameter $\lambda $. Indeed, when we take into account the tracking error risk, the optimal value $w_{\mathrm{cash}}^{\star }$ is reduced\footnote{See Figures \ref{fig:cash6d} and \ref{fig:cash6e} on page \pageref{fig:cash6d}.}.\smallskip

\begin{figure}[p]
\centering
\caption{Implementation of a cash buffer when $x^{+} = 10\%$}
\label{fig:cash7c}
\figureskip
\includegraphics[width = \figurewidth, height = \figureheight]{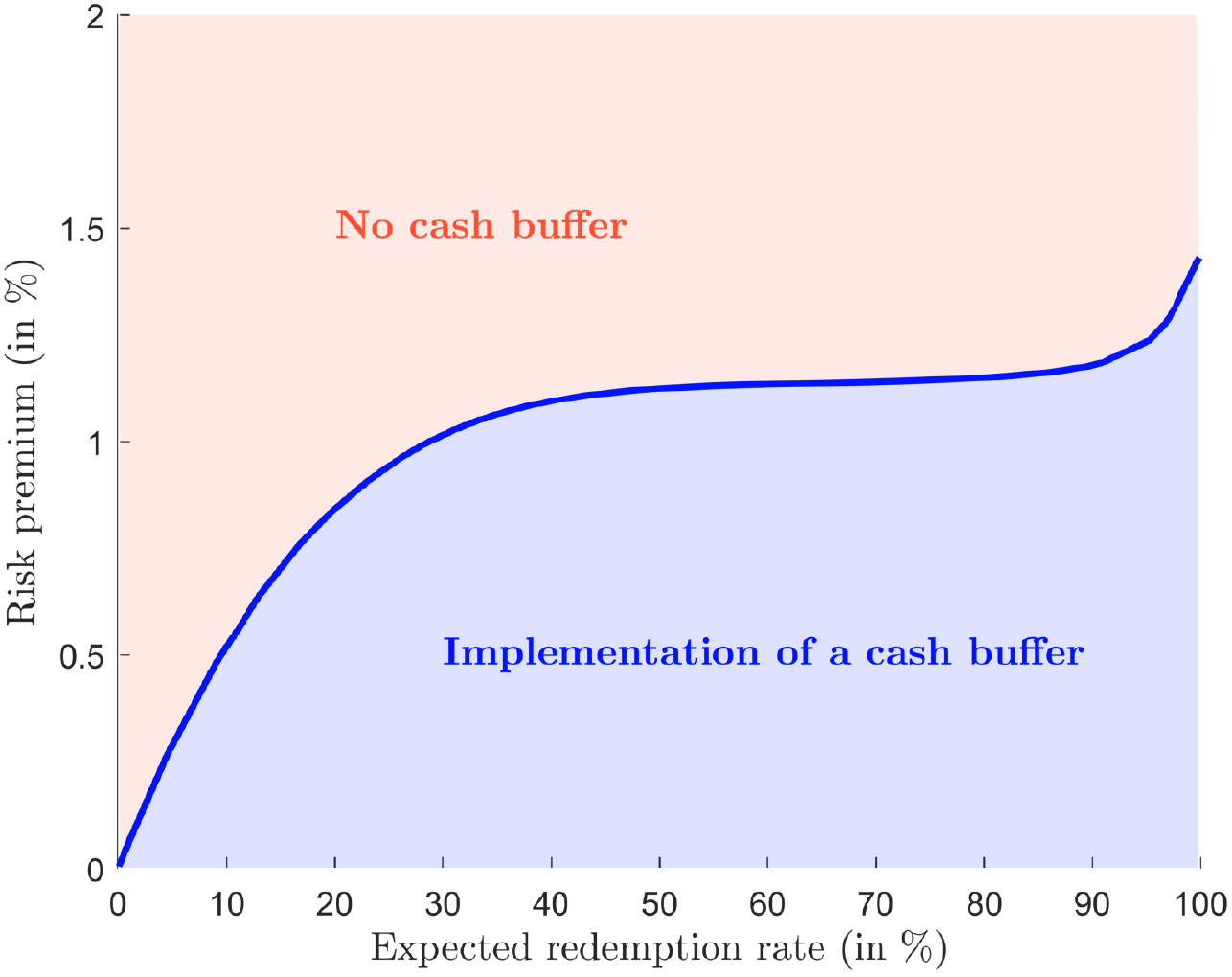}
\end{figure}

\begin{figure}[p]
\centering
\caption{Implementation of a cash buffer when $x^{+} = 100\%$}
\label{fig:cash7d}
\figureskip
\includegraphics[width = \figurewidth, height = \figureheight]{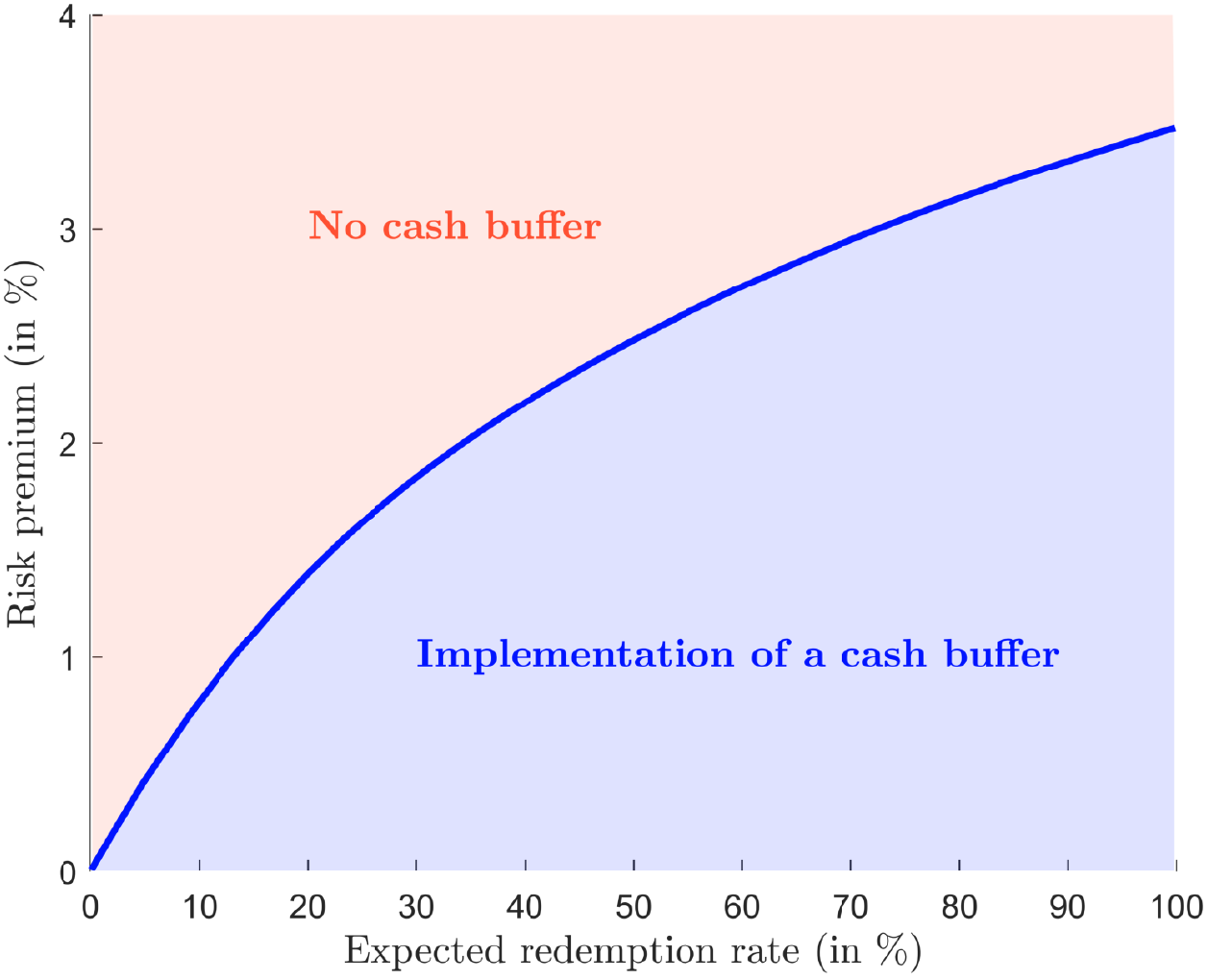}
\end{figure}

Given $w_{\mathrm{cash}}$, we define the break-even risk premium as the
value of $\mu _{\mathrm{asset}}-\mu _{\mathrm{cash}}$ such that the net cost
function is minimum. It is equal to:
\begin{equation}
\varrho \left( w_{\mathrm{cash}}\right) =\frac{\partial \,\mathbb{E}\left[
\mathcal{LG}\left( w_{\mathrm{cash}}\right) \right] }{\partial \,w_{\mathrm{%
cash}}}-\lambda w_{\mathrm{cash}}\left( \sigma _{\mathrm{cash}}^{2}+\sigma _{%
\mathrm{asset}}^{2}-2\rho _{\mathrm{cash},\mathrm{asset}}\sigma _{\mathrm{%
cash}}\sigma _{\mathrm{asset}}\right)
\end{equation}%
In Figures \ref{fig:cash7a} and \ref{fig:cash7b} on page \pageref{fig:cash6c}, we have reported the value of $\varrho \left( w_{\mathrm{cash}}\right) $ for the previous example. Once $\varrho \left( w_{\mathrm{cash}}\right) $ is computed, we obtain the following rules\footnote{For instance, Figures \ref{fig:cash7e} and \ref{fig:cash7f} on page \pageref{fig:cash7e} illustrate this set of rules for a liquidity buffer of $10\%$.}:%
\begin{equation}
\left\{
\begin{array}{l}
\mu _{\mathrm{asset}}-\mu _{\mathrm{cash}}<\varrho \left( w_{\mathrm{cash}%
}\right) \Rightarrow w_{\mathrm{cash}}^{\star }>w_{\mathrm{cash}} \\
\mu _{\mathrm{asset}}-\mu _{\mathrm{cash}}=\varrho \left( w_{\mathrm{cash}%
}\right) \Rightarrow w_{\mathrm{cash}}^{\star }=w_{\mathrm{cash}} \\
\mu _{\mathrm{asset}}-\mu _{\mathrm{cash}}>\varrho \left( w_{\mathrm{cash}%
}\right) \Rightarrow w_{\mathrm{cash}}^{\star }<w_{\mathrm{cash}}%
\end{array}%
\right.
\end{equation}%
In particular, a cash buffer must be implemented if the risk premium of the
asset is below the threshold $\varrho \left( 0\right) $:
\begin{equation}
w_{\mathrm{cash}}^{\star }>0\Leftrightarrow \mu _{\mathrm{asset}}-\mu _{%
\mathrm{cash}}<\varrho \left( 0\right) =\frac{\partial \,\mathbb{E}\left[
\mathcal{LG}\left( 0\right) \right] }{\partial \,w_{\mathrm{cash}}}
\end{equation}%
We notice that $\varrho \left( 0\right) $ does not depend on
the tracking error risk. Figures \ref{fig:cash7c} and \ref{fig:cash7d} show
when a liquidity buffer is implemented with respect to the risk premium
$\mu_{\mathrm{asset}}-\mu _{\mathrm{cash}}$ and the expected redemption rate
$\mathbb{E}\left[ \RedemptionRate\right] $.

\subsubsection{The debate on cash hoarding}

We cannot finish this section without saying a few words about the debate on
cash hoarding. Indeed, the underlying idea of the previous analysis is to
implement a cash buffer before the redemption occurs, and to help
the liquidation process during the liquidity stress period
\citep{Chernenko-2016, Goldstein-2017a, Ma-2021}. However, \citet{Morris-2017}
found that asset managers can hoard cash during redemption
periods, because they anticipate worst days. Instead of liquidating the cash
buffer to meet investor redemptions, asset managers can preserve the
liquidity of their portfolios \citep{Jiang-2021} or even increase the proportion of cash during the stress period \citep{Schrimpf-2021}. In this case, cash hoarding may amplify fire sales and seems to be contradictory with the implementation of a cash buffer. However, cash hoarding is easy to understand in our framework. Indeed, during a stress period, asset managers may anticipate a very pessimistic scenario, meaning that they dramatically reduce the expected risk premium $\mu _{\mathrm{asset}}-\mu _{\mathrm{cash}}$. This implies increasing the level of the optimal cash buffer $w_{\mathrm{cash}}^{\star }$. Therefore, the previous framework explains that cash buffering and cash hoarding are compatible if we consider that asset managers have a dynamic view of the risk premium of assets.

\subsection{Special arrangements}

Special arrangements are used extensively by the hedge fund industry. In
particular, gates and side pockets were extensively implemented during
the 2008 Global Financial Crisis after the Lehman Brothers collapse
\citep{Aiken-2015, Teo-2011}. Nevertheless, mutual funds are increasingly
familiar with these tools and are allowed in many European countries
\citep[Table 4.3.A, page 33]{Darpeix-2020}. For instance, gates, in-kind
redemptions, side pockets and redemption suspensions are active in France, Italy, Spain and the Netherlands. In Germany, gates and side pockets are not
permitted whereas side pockets are prohibited in the United Kingdom.

\subsubsection{Redemption suspension and gate}

When implementing a gate, the fund manager temporarily limits the amount of
redemptions from the fund. In this case, the gate forces the redeeming
investors to wait until the next regular withdrawal dates to receive the
balance of their withdrawal request. For instance, the fund manager can impose that the daily amount of withdrawals do not exceed $2\%$ of the fund's net assets. Let us assume a redemption rate of $5\%$ at time $t$ (investors
$A$) and $2\%$ at time $t+1$ (investors $B$). Because we have a daily gate of
$2\%$, only $40\%$ of the withdrawal of investors $A$ may be executed at time
$t$. The next $60\%$ are executed at time $t+1$ and $t+2$. Investors $B$ who
would like to redeem at time $t+1$ must wait until time $t+2$, because redeeming investors $A$ take precedence. Finally, we obtain the redemption schedule reported in Table~\ref{tab:lmt-gate1}. We notice that the last redeeming investors may be greatly penalized because of the queuing system. If there are many redemptions, the remaining investors have no incentive to redeem because they face two risks. The risk of time redemption depends on the frequency of withdrawal dates. In the case of monthly withdrawals, investors can wait several months before obtaining their cash. For instance, we observed this situation during the hedge fund crisis at the end of 2008. The second risk concerns the valuation. Indeed, the unit price can change
dramatically during the redemption gate period. This is why regulators
generally impose a maximum period for mutual funds that would like to impose
a redemption gate.\smallskip

\begin{table}[tbph]
\centering
\caption{Stress scenarios of the participation rate}
\label{tab:lmt-gate1}
\begin{tabular}{cccccc}
\hline
Redemption & Redeeming & \multicolumn{4}{c}{Time} \\
Gate & Investors         &     $t$ &   $t+1$   &   $t+2$  &  $t+3$   \\ \hline
\multirow{4}{*}{No gate}
& \multirow{2}{*}{$A$}
                         &    $5\%$ &           &          &          \\
             &           & ($100\%$)&           &          &          \\ \cline{2-6}
& \multirow{2}{*}{$B$}
                         &          &    $2\%$  &          &          \\
             &           &          & ($100\%$) &          &          \\ \hline
\multirow{4}{*}{$2\%$}
& \multirow{2}{*}{$A$}
                         &    $2\%$ &    $2\%$  &   $1\%$  &          \\
             &           &  ($40\%$)&  ($40\%$) & ($20\%$) &          \\ \cline{2-6}
& \multirow{2}{*}{$B$}
                         &          &           &   $1\%$  &   $1\%$  \\
             &           &          &           & ($50\%$) & ($50\%$) \\ \hline
\end{tabular}
\end{table}

An extreme case of a redemption gate is when the manager completely suspends redemptions from his fund. A redemption suspension is rare and was originally used by hedge funds\footnote{See for instance the famous suspension of redemptions decided by GAM after its top manager in charge of absolute return strategies was the subject of a disciplinary procedure \citep{GAM-2018}.}. However, it is now part of the liquidity management tools that can be used by mutual funds. For instance, it is the only mechanism that is available in all European jurisdictions \citep{ESRB-2017, Darpeix-2020}. It was used by at least $215$ European investment funds (with net assets totaling \euro 73.4 bn) during the coronavirus crisis in February and March 2020 \citep{Grill-2021}. The authors found that \textquotedblleft \textsl{many of those funds had invested in illiquid assets, were leveraged or had lower cash holdings than funds that were not suspended}\textquotedblright.\smallskip

At first sight, a suspension of redemptions seems to be a tougher decision
than a redemption gate. Indeed, in this last case, redemptions continue to be
accepted, but they are delayed. However, it is not certain that a redemption
gate will have less impact than a redemption suspension. In a period of fire
sales, gates can also exacerbate the liquidity crisis because of the asset
liquidation/market transmission channel of systemic risk \citep{Roncalli-2015a}. On the contrary, redemption suspensions do not
directly contribute to the asset liquidation from a theoretical point of
view. However, we generally observe higher redemptions when suspensions are stopped. This means that we can have an ex-post overreaction of investors.
In fact, it seems that a suspension of redemptions is preferable when the fund manager faces a temporary liquidity crisis such that many securities can not be priced. In the absence of price valuation, it may be good to wait until normal conditions are restored. Of course, it is not always possible and depends on the nature of the liquidity crisis.\smallskip

The impact of gates has received little attention from academics.
Nevertheless, the theoretical study of \citet{Cipriani-2014} showed that
there can be preemptive runs when a fund manager is able to impose a gate,
although it can be ex-post optimal for the fund's investors.
This illustrates the issue of strategic interaction and payoff complementarities described by \citet{Chen-2010}. Moreover, imposing a gate generally leads to a reputational risk for the fund and a negative externality for the corresponding asset class and the other similar investment funds. More generally, \citet{Voellmy-2021} showed that redemption gates are less efficient than redemption fees, which are described on page \pageref{section:redemption-fees}.

\subsubsection{Side pocket}

When a side pocket is created, the fund separates illiquid assets from
liquid assets. Therefore, the fund is split into two funds: the mirror fund,
which is made up of the liquid assets and the side pocket of illiquid
assets. Each investor in the initial fund receives the same number of units
of the mirror fund and the side pocket. The mirror fund inherits the
properties of the original fund. Therefore, the mirror fund can continue to
be subscribed or redeemed. On the contrary, the side pocket fund becomes a
closed-end fund \citep{Opromolla-2009}. The fund manager's objective is then to liquidate the assets of the side pocket fund. However, he is not forced to liquidate them immediately and can wait until market conditions improve. For instance,it took many months (and sometimes one or two years) for hedge fund to manage the side pockets created in October 2008 and for investors to retrieve their cash.\smallskip

To the best of our knowledge, the only academic study on side pocketing
is the research work conducted by \citet{Aiken-2015}, who analyzed
the behavior of $740$ hedge funds between 2006 and 2011.
The authors found that side pockets and gates are positively correlated, meaning that hedge funds both gated investors and placed assets into a side pocket during the 2008 Global Financial Crisis.
This result suggests that gates and side pockets are not mutually exclusive.
This explains the bad reputation of side pockets. Indeed, investors
generally have the feeling of facing a double sentence. A part of their
investment is segregated, and they don't know when and how much of their
capital they will retrieve. And the remainder of their investment is gated.
This is not the original objective of side pocketing, since the underlying
idea is to separate the original fund into a healthy portfolio and a bad
portfolio. But generally, the healthy fund is also gated.\smallskip

Certainly, side pocketing is a last-resort discretionary liquidity restriction because of the reputational risk. First, the fund
manager gives a strong signal to the market that the liquidity crisis is not temporary but will persist for a long time. Therefore, side pocketing indirectly contributes to strengthening the spillover effect of the liquidity crisis because market sentiment is getting worse. Second, if we restrict our analysis to the fund level, the effect of side pocketing is ambiguous.
It is obvious that it eliminates the first-mover advantage, but it is also a sign that the liquidity calibration of the original fund was worse.
Moreover, side pockets can be used to protect management fees on
the more liquid assets or to hide a poor risk management process.
This explains that side pocketing is generally followed by the collapse of the fund, which generally suffers from existing investors' withdrawing while it is not able to attract new investors.

\subsubsection{In-kind redemptions}

In-kind redemptions are non-monetary payments. In this case, the fund
manager offers a basket of securities to the redeeming investor, generally
the asset portfolio of the fund on a pro-rata basis. Since the beginning of
the 2000s, in-kind redemptions have been used extensively in order to improve the tax efficiency of US exchange traded funds \citep{Poterba-2002}. Even though they are less common in the mutual fund industry, in-kind redemptions have become increasingly popular to manage liquidity runs. For instance, according to \citet{ESRB-2017}, in-kind redemptions are the most common available tool in the European Union, just after the suspension of redemptions.\smallskip

In-kind redemptions are generally considered as an efficient tool for
managing liquidity runs since they transfer the liquidation issue to
redeeming investors. As showed by \citet{Agarwal-2020}, redemption-in-kind funds tend to deliver more illiquid securities. Moreover, these funds
\textquotedblleft \textsl{experience less flow subsequently because
investors avoid such funds where they are unable to benefit from liquidity
transformation function of funds}\textquotedblright\ \citep[page 30]{Agarwal-2020}.\smallskip

Normally, in-kind redemptions solve the valuation problem of the redemption portfolio when it corresponds to the pro-rata asset portfolio\footnote{Indeed, the valuation problem is transferred to the redeeming investors.}. This property is appealing in a period of liquidity stress. However, the pro-rata rule only concerns large redemptions in order to be sure that the rounding effect and the decimalization impact are small.
From a technical point of view, redemption-in-kind is certainly more difficult to manage than gating the fund. This certainly explains why there are few mutual funds that have applied in-kind redemptions in Europe.

\subsection{Swing pricing}

The objective of swing pricing is to protect existing investors from
dilution\footnote{This means a reduction in the fund's value.} caused by
large trading costs and market impacts due to subscriptions and/or
redemptions. Since this mechanism is relatively new, there are few research
studies on its benefit. From a theoretical and empirical point of view, it
seems that swing pricing can eliminate the first-mover advantage
\citep{Jin-2019, Capponi-2020} and mitigate the systemic risk
\citep{Malik-2017, Jin-2019}. Nevertheless, these results must be challenged
as shown by the works of \citet{Lewrick-2017a, Lewrick-2017b}:
\begin{quote}
[...] \textquotedblleft \textsl{we show that, within our theoretical
framework, swing pricing can prevent self-fulfilling runs on the fund.
However, in practice, the scope for swing pricing to prevent
self-fulfilling runs is more limited, primarily because the share of
liquidity-constrained investors is difficult to assess}\textquotedblright\
\citep{Lewrick-2017a}.\smallskip

[...] \textquotedblleft \textsl{we show that swing pricing dampens outflows
in reaction to weak fund performance, but has a limited effect during
stress episodes. Furthermore, swing pricing supports fund returns, while
raising accounting volatility, and may lead to lower cash
buffers}\textquotedblright\ \citep{Lewrick-2017b}.
\end{quote}

\subsubsection{Investor dilution}

Following \citet{Roncalli-lst1}, the total net assets (TNA) equal the total
value of assets $A\left( t\right) $ less the current or accrued liabilities
$D\left( t\right) $:
\begin{equation*}
\limfunc{TNA}\left( t\right) =A\left( t\right) -D\left( t\right)
\end{equation*}%
The net asset value (NAV) represents the share price or the unit price:
\begin{equation*}
\limfunc{NAV}\left( t\right) =\frac{\limfunc{TNA}\left( t\right) }{N\left(
t\right) }
\end{equation*}%
where the total number $N\left( t\right) $ of shares or units in issue is the
sum of all units owned by all unitholders. In the sequel, we assume that the
debits are negligible: $D\left( t\right) \ll A\left( t\right) $. This implies
that:
\begin{equation*}
\limfunc{NAV}\left( t+1\right) \approx \frac{A\left( t+1\right) }{N\left(
t+1\right) }
\end{equation*}%
$R_{A}\left( t+1\right) $ denotes the return of the assets. We can then
face three situations:
\begin{enumerate}
\item There is no net subscription or redemption flows, meaning that $
    N\left( t+1\right) =N\left( t\right) $ and $A\left( t+1\right) =\left(
    1+R\left( t+1\right) \right) \cdot A\left( t\right) $. In this case, we
    have:
\begin{eqnarray}
\limfunc{NAV}\left( t+1\right)  &=&\left( 1+R_{A}\left( t+1\right) \right)
\frac{A\left( t\right) }{N\left( t\right) }  \notag \\
&=&\left( 1+R_{A}\left( t+1\right) \right) \cdot \limfunc{NAV}\left(
t\right)   \label{eq:dilution1}
\end{eqnarray}%
The growth of the net asset value is exactly equal to the return of the
assets:%
\begin{equation*}
R_{\limfunc{NAV}}\left( t+1\right) =\frac{\limfunc{NAV}\left( t+1\right) }{%
\limfunc{NAV}\left( t\right) }-1=R_{A}\left( t+1\right)
\end{equation*}

\item If the investment fund experiences some net subscription flows, the
    number of units becomes:
\begin{equation*}
N\left( t+1\right) =N\left( t\right) +\Delta N\left( t+1\right)
\end{equation*}%
where $\Delta N\left( t+1\right) =N^{+}\left( t+1\right) $ is the number of
units to be created. At time $t+1$, we have\footnote{$\Delta N\left(
t+1\right) \cdot \limfunc{NAV}\left( t\right) $ is the amount invested in
the new assets at time $t$.}:%
\begin{eqnarray*}
A\left( t+1\right)  &=&\left( 1+R_{A}\left( t+1\right) \right) \cdot \left(
A\left( t\right) +\Delta N\left( t+1\right) \cdot \limfunc{NAV}\left(
t\right) \right)  \\
&=&\left( 1+R_{A}\left( t+1\right) \right) \cdot \left( N\left( t\right)
\cdot \limfunc{NAV}\left( t\right) +\Delta N\left( t+1\right) \cdot \limfunc{%
NAV}\left( t\right) \right)  \\
&=&\left( 1+R_{A}\left( t+1\right) \right) \cdot N\left( t+1\right) \cdot
\limfunc{NAV}\left( t\right)
\end{eqnarray*}%
and:%
\begin{equation*}
\limfunc{TNA}\left( t+1\right) =A\left( t+1\right) -\mathcal{TC}\left(
t+1\right)
\end{equation*}%
where $\mathcal{TC}\left( t+1\right) $ is the transaction cost of buying
the new assets. We deduce that:
\begin{eqnarray}
\limfunc{NAV}\left( t+1\right)  &=&\frac{A\left( t+1\right) -\mathcal{TC}%
\left( t+1\right) }{N\left( t+1\right) }  \notag \\
&=&\left( 1+R_{A}\left( t+1\right) \right) \cdot \limfunc{NAV}\left(
t\right) -\frac{\mathcal{TC}\left( t+1\right) }{N\left( t+1\right) }
\label{eq:dilution2}
\end{eqnarray}%
In this case, the growth of the net asset value is less than the return of
the assets:%
\begin{equation*}
R_{\limfunc{NAV}}\left( t+1\right) =R_{A}\left( t+1\right) -\frac{\mathcal{TC%
}\left( t+1\right) }{N\left( t+1\right) \cdot \limfunc{NAV}\left( t\right) }%
\leq R_{A}\left( t+1\right)
\end{equation*}

\item If the investment fund experiences some net redemption flows, the
    number of units becomes:
\begin{equation*}
N\left( t+1\right) =N\left( t\right) +\Delta N\left( t+1\right)
\end{equation*}%
where $\Delta N\left( t+1\right) =-N^{-}\left( t+1\right) $ and
$N^{-}\left(
t+1\right) $ is the number of units to be redeemed. At time $t+1$, we have:%
\begin{eqnarray}
\limfunc{NAV}\left( t+1\right)  &=&\frac{\left( 1+R_{A}\left( t+1\right)
\right) \cdot N\left( t\right) \cdot \limfunc{NAV}\left( t\right) -\mathcal{%
TC}\left( t+1\right) }{N\left( t\right) }  \notag \\
&=&\left( 1+R_{A}\left( t+1\right) \right) \cdot \limfunc{NAV}\left(
t\right) -\frac{\mathcal{TC}\left( t+1\right) }{N\left( t\right) }
\label{eq:dilution3}
\end{eqnarray}%
In this case, the growth of the net asset value is less than the return of
the assets:%
\begin{equation*}
R_{\limfunc{NAV}}\left( t+1\right) =R_{A}\left( t+1\right) -\frac{\mathcal{TC%
}\left( t+1\right) }{N\left( t\right) \cdot \limfunc{NAV}\left( t\right) }%
\leq R_{A}\left( t+1\right)
\end{equation*}
\end{enumerate}
When comparing Equations (\ref{eq:dilution1}), (\ref{eq:dilution2}) and
(\ref{eq:dilution3}), we notice that subscription/redemption flows may
penalize existing/remaining investors, because the net asset value is reduced
by the transaction costs that are borne by all investors in the fund:
\begin{equation*}
\limfunc{NAV}\left( t+1\mid \Delta N\left( t+1\right) =0\right) -\limfunc{NAV%
}\left( t+1\mid \Delta N\left( t+1\right) \neq 0\right) =\frac{\mathcal{TC}%
\left( t+1\right) }{\max \left( N\left( t\right) ,N\left( t+1\right) \right)}
\end{equation*}%
The decline in the net asset value is referred to as \textquotedblleft
\textit{investor dilution}\textquotedblright.\smallskip

In order to illustrate the dilution, we consider a fund with the following
characteristics: $\limfunc{NAV}\left( t\right) =\$100$, $N\left( t\right)
=10$ and $R_{A}\left( t+1\right) =5\%$. In the absence of
subscriptions/redemptions, we have:
\begin{equation*}
\limfunc{NAV}\left( t+1\right) =\left( 1+5\%\right) \times 100=105
\end{equation*}%
We assume that creating/redeeming $5$ shares induces a transaction cost of
$\$30$. In the case of a net subscription of $\$500$, we have $N\left(
t+1\right) =15$ and:
\begin{equation*}
\limfunc{NAV}\left( t+1\right) =\left( 1+5\%\right) \times 100-\frac{30}{15}%
=103
\end{equation*}%
In the case of a net redemption of $\$500$, we have $N\left( t+1\right) =5$
and:%
\begin{equation*}
\limfunc{NAV}\left( t+1\right) =\left( 1+5\%\right) \times 100-\frac{30}{10}%
=102
\end{equation*}%
The transaction cost therefore reduces the NAV and impacts all investors in the fund. Moreover, we notice that the dilution is greater for redemptions than subscriptions. The reason is that the number of shares increases in the case of a subscription, implying that the transaction cost by share is lower than in the case of a redemption.\smallskip

This asymmetry property between subscriptions and redemptions is an important
issue when considering a liquidity stress testing program. Another factor is
that the unit transaction cost is an increasing function of the size of the
subscription/redemption amount. This is particularly true in a stress market
when it is difficult to sell assets because of the low demand. If we consider
the previous example, we can assume that selling $\$500$ in a stress period
may induce a transaction cost of $\$50$. In this case, we obtain:
\begin{equation*}
\limfunc{NAV}\left( t+1\right) =\left( 1+5\%\right) \times 100-\frac{50}{10}%
=100
\end{equation*}%
This example illustrates how investor dilution is an important issue when
the fund faces redemptions in a stress period.

\subsubsection{The swing pricing principle}

The swing pricing principle means that the NAV is adjusted for net
subscriptions/redemptions. Therefore, transaction costs are only borne by the
subscribing/redeeming investors. In the case of a net redemption, the NAV
must be reduced by the transaction costs divided by the number of net
redeeming shares:
\begin{equation*}
\limfunc{NAV}\nolimits_{\mathrm{swing}}\left( t+1\right) =\limfunc{NAV}%
\nolimits_{\mathrm{gross}}\left( t+1\right) -\frac{\mathcal{TC}\left(
t+1\right) }{N^{-}\left( t+1\right) -N^{+}\left( t+1\right) }
\end{equation*}%
where $\limfunc{NAV}\nolimits_{\mathrm{gross}}$ is the \textquotedblleft
\textit{gross}\textquotedblright\ net asset value calculated before swing pricing is
applied \citep{AFG-2016}. In the case of a net subscription, the NAV becomes:
\begin{equation*}
\limfunc{NAV}\nolimits_{\mathrm{swing}}\left( t+1\right) =\limfunc{NAV}%
\nolimits_{\mathrm{gross}}\left( t+1\right) +\frac{\mathcal{TC}\left(
t+1\right) }{N^{+}\left( t+1\right) -N^{-}\left( t+1\right) }
\end{equation*}%
Therefore, the NAV\ is increased if $N^{+}-N^{-}>0$. Finally, we obtain the
following compact formula:%
\begin{equation*}
\limfunc{NAV}\nolimits_{\mathrm{swing}}\left( t+1\right) =\limfunc{NAV}%
\nolimits_{\mathrm{gross}}\left( t+1\right) +\frac{\mathcal{TC}\left(
t+1\right) }{\Delta N\left( t+1\right) }
\end{equation*}%
The adjustment only impacts investors that trade on that day, since existing
investors are not affected by this adjustment. Indeed, the total net asset is
equal to:
\begin{eqnarray*}
\limfunc{TNA}\left( t+1\right)  &=&A\left( t+1\right) -\mathcal{TC}\left(
t+1\right)  \\
&=&N\left( t\right) \cdot \limfunc{NAV}\nolimits_{\mathrm{gross}}\left(
t+1\right) +\Delta N\left( t+1\right) \cdot \limfunc{NAV}\nolimits_{\mathrm{%
swing}}\left( t+1\right) -\mathcal{TC}\left( t+1\right)  \\
&=&N\left( t\right) \cdot \limfunc{NAV}\nolimits_{\mathrm{gross}}\left(
t+1\right) +\Delta N\left( t+1\right) \cdot \limfunc{NAV}\nolimits_{\mathrm{%
gross}}\left( t+1\right)  \\
&=&N\left( t+1\right) \cdot \limfunc{NAV}\nolimits_{\mathrm{gross}}\left(
t+1\right)
\end{eqnarray*}%
meaning that it is exactly equal to the gross net asset value. If there is no redemption/subscription at time $t+2$, we obtain:
\begin{eqnarray*}
\limfunc{NAV}\left( t+2\right)  &=&\left( 1+R_{A}\left( t+2\right) \right)
\cdot \limfunc{NAV}\nolimits_{\mathrm{gross}}\left( t+1\right)  \\
&=&\left( 1+R_{A}\left( t+2\right) \right) \cdot \left( 1+R_{A}\left(
t+1\right) \right) \cdot \limfunc{NAV}\left( t+1\right)
\end{eqnarray*}%
We notice that swing pricing has protected the fund's buy-and-hold investors.\smallskip

If we consider the previous example, we have $\limfunc{NAV}\nolimits_{%
\mathrm{gross}}\left( t+1\right) =105$ and:%
\begin{equation*}
\limfunc{NAV}\nolimits_{\mathrm{swing}}\left( t+1\right) =\left\{
\begin{array}{ll}
105+\dfrac{30}{5}=111 & \text{if subscription} \\
\\
105-\dfrac{30}{5}=99 & \text{if redemption}%
\end{array}%
\right.
\end{equation*}%
We observe that swing pricing increases the fund's volatility since the NAV
adjustment with swing pricing is greater than the NAV adjustment without
swing pricing. Moreover, the adjustment is smaller for subscriptions because
the number of shares increases\footnote{Indeed, we have $\max \left( N\left(
t\right) ,N\left( t+1\right) \right) =N\left( t+1\right)
>N\left( t\right) $ in the case of net subscriptions and $\max \left( N\left(
t\right) ,N\left( t+1\right) \right) =N\left( t\right) $ in the case of net
redemptions.}. Therefore, we notice an asymmetry between subscriptions and
redemptions since the latter impact the unit price more than the former. In
the case of a liquidity crisis where there is a substantial  imbalance between demand and supply, the impact of redemptions is even stronger and the
contagion risk of a spillover effect is increased.

\subsubsection{Swing pricing in practice}

Swing pricing is regulated in Europe and the U.S. and can be used under
regulatory constraints \citep{Malik-2017}. For instance, in France, the asset
manager should inform the AMF and the fund's auditor of the implementation of
swing pricing \citep{AFG-2016}. The use of swing pricing has also been
encouraged during the Coronavirus crisis in order to manage the liquidity:
\begin{quote}
\textquotedblleft \textsl{The AMF also favors the use of swing pricing and
anti-dilution levies mechanisms during the current crisis, given the low
liquidity of certain underlying assets and the sometimes-high costs
involved in restructuring portfolios}\textquotedblright\ \citep[page 4]{AMF-2020}.
\end{quote}
According to \citet{ESMA-2020b}, swing pricing was the most used LMT in
Europe during the market stress in February and March 2020, far ahead
of redemption suspension. This follows the recommendations provided by the ESRB. Similar rules have existed in the U.S. for some years \citep{SEC-2016}, even though the use of swing pricing is less widespread than in E.U. jurisdictions.

\paragraph{Full vs. partial vs. dual pricing}

According to \citet{Jin-2019}, asset managers use three alternative pricing mechanisms:
\begin{enumerate}
\item Partial swing pricing\\
The NAV is adjusted only when the net fund flow is greater than a threshold.

\item Full swing pricing\\
The NAV is adjusted every time there is a net inflow or outflow.
Full swing pricing is a special case of partial swing pricing by considering that the threshold is equal to zero.

\item Dual pricing\\
We distinguish bid and ask NAVs, meaning that the investment fund has two NAVs. Therefore, investors purchase the fund shares at the ask price and sell at the bid price.
\end{enumerate}
Using a dataset of UK based asset managers, \citet{Jin-2019} estimated that approximately a quarter of investment funds use traditional pricing mechanisms whereas the three remaining quarters consider alternative pricing mechanisms. Within this group, the break down is the following:
$25\%$ employ full swing pricing, $50\%$ prefer partial swing pricing and $25\%$ promote dual pricing.\smallskip

Dual pricing is an extension of full swing pricing that distinguishes
between subscriptions and redemptions. However, dual pricing is more complex
to calibrate. Indeed, it is not obvious to allocate transaction costs to both
redeeming and subscribing investors because of the netting process. We have:
\begin{equation*}
\limfunc{NAV}\nolimits_{\mathrm{ask}}\left( t+1\right) =\limfunc{NAV}%
\nolimits_{\mathrm{gross}}\left( t+1\right) +\frac{\alpha \cdot \mathcal{TC}%
\left( t+1\right) }{N^{+}\left( t+1\right) }
\end{equation*}%
and:%
\begin{equation*}
\limfunc{NAV}\nolimits_{\mathrm{bid}}\left( t+1\right) =\limfunc{NAV}%
\nolimits_{\mathrm{gross}}\left( t+1\right) -\frac{\left( 1-\alpha \right)
\cdot \mathcal{TC}\left( t+1\right) }{N^{-}\left( t+1\right) }
\end{equation*}%
where $\alpha $ is the portion of the transaction costs allocated to gross
subscriptions. For instance, we can use the pro-rata rule:%
\begin{equation*}
\alpha =\frac{N^{+}\left( t+1\right) }{N^{+}\left( t+1\right) +N^{-}\left(
t+1\right) }
\end{equation*}%
but we can also penalize redeeming investors:%
\begin{equation*}
\alpha =\frac{N^{+}\left( t+1\right) }{N^{+}\left( t+1\right) +\gamma \cdot
N^{-}\left( t+1\right) }
\end{equation*}%
where $\gamma \geq 1$ is the penalization factor. Let us consider the
previous example with $N^{+}\left( t+1\right) =10$, $N^{-}\left( t+1\right)
=5$ and $\mathcal{TC}\left( t+1\right) =30$. We have $\limfunc{NAV}_{\mathrm{
swing}}\left( t+1\right) =111$. If we assume that $\gamma =1$, we have $\limfunc{NAV}\nolimits_{\mathrm{ask}}\left( t+1\right) =107$ and $\limfunc{NAV}\nolimits_{\mathrm{bid}}\left( t+1\right) =103$. If $\gamma $ is set to $2$, the previous figures become $\limfunc{NAV}\nolimits_{\mathrm{ask}}\left(
t+1\right) =106.5$ and $\limfunc{NAV}\nolimits_{\mathrm{bid}}\left(
t+1\right) =102$.

\begin{remark}
The previous example illustrates one of the drawbacks of swing pricing.
Indeed, since there are $10$ subscriptions and $5$ redemptions, the swing NAV
is greater than the gross NAV ($\$111$ vs. $\$105$). Redeeming investors benefit from the entry of new investors. In the case of dual pricing,
the unit price of redeeming investors is equal to $\$103$ (if $\gamma$ is set to $1$), which is lower than $\$111$.
\end{remark}

\paragraph{Setting the swing threshold and the swing factor}

In most cases, swing pricing is applied only when the net amount of
subscriptions and redemptions reaches a threshold\footnote{An alternative
approach is to replace $\min \left( N\left( t\right) ,N\left( t+1\right)
\right) $ with $N\left( t\right) $.}:
\begin{equation*}
\left\vert \frac{\Delta N\left( t+1\right) }{\min \left( N\left( t\right)
,N\left( t+1\right) \right) }\right\vert \geq sw_{\mathrm{threshold}}
\end{equation*}%
where $sw_{\mathrm{threshold}}$ is the swing threshold. For example,
$sw_{\mathrm{threshold}}=5\%$ implies that the swing pricing mechanism is
activated every time we observe at least $5\%$ of inflows/outflows. A swing
factor is then applied to the NAV:
\begin{equation*}
\limfunc{NAV}\nolimits_{\mathrm{swing}}\left( t+1\right) =\left\{
\begin{array}{ll}
\left( 1+sw_{\mathrm{\mathrm{factor}}}\right) \cdot \limfunc{NAV}\left(
t+1\right)  & \text{if net subscription}\geq sw_{\mathrm{threshold}} \\
\left( 1-sw_{\mathrm{\mathrm{factor}}}\right) \cdot \limfunc{NAV}\left(
t+1\right)  & \text{if net redemption}\geq sw_{\mathrm{threshold}}%
\end{array}%
\right.
\end{equation*}%
\smallskip

We can use different approaches to calibrate the parameters
$sw_{\mathrm{threshold}}$ and $sw_{\mathrm{\mathrm{factor}}}$. For instance,
we can assume that $sw_{\mathrm{threshold}}$ is constant for a family of
funds (e.g. equity funds). In this case, $sw_{\mathrm{threshold}}$ is
estimated using a historical sample of flow rates and transaction costs. The
underlying idea is to use a value of $sw_{\mathrm{threshold}}$ such that
transaction costs become significant. However, this approach may appear too
simple in a liquidity stress testing framework. Indeed, transaction costs are
larger in a stress period, meaning that $sw_{\mathrm{threshold}}$ is a
decreasing function of the stress intensity. For instance, the asset manager
can calibrate two values of $sw_{\mathrm{threshold}}$, a standard figure
which is valid for normal periods and a lower figure which is valid for
normal periods. Typical values are $5\%$ and $2\%$. The parameter
$sw_{\mathrm{\mathrm{factor}}}$ must reflect the transaction costs. Again,
two approaches are possible: ex-ante or ex-post transaction costs. In the
first case, we consider the transaction cost function calibrated to
measure the asset risk, whereas the effective cost is used in the second
case.\smallskip

By construction, the swing factor $sw_{\mathrm{\mathrm{factor}}}$ varies over
time while the swing threshold $sw_{\mathrm{threshold}}$ is more static. When
the swing pricing mechanism is applied, we can estimate the amount of
transaction costs:
\begin{equation*}
\mathcal{TC}\left( t+1\right) =sw_{\mathrm{\mathrm{factor}}}\cdot \limfunc{%
NAV}\left( t+1\right) \cdot \left\vert \Delta N\left( t+1\right) \right\vert
\end{equation*}%
We deduce that the transaction cost ratio is greater than the product of the
swing threshold and the swing factor:
\begin{eqnarray*}
\frac{\mathcal{TC}\left( t+1\right) }{\min \left( N\left( t\right) ,N\left(
t+1\right) \right) \cdot \limfunc{NAV}\left( t+1\right) } &=&sw_{\mathrm{%
\mathrm{factor}}}\cdot \left\vert \frac{\Delta N\left( t+1\right) }{\min
\left( N\left( t\right) ,N\left( t+1\right) \right) }\right\vert  \\
&\geq &sw_{\mathrm{\mathrm{factor}}}\cdot sw_{\mathrm{threshold}}
\end{eqnarray*}%
Another approach consists in fixing the value of the product:
\begin{equation*}
sw_{\mathrm{\mathrm{factor}}}\cdot sw_{\mathrm{threshold}}=sw_{\mathrm{%
product}}
\end{equation*}%
In this case, we are sure that the swing pricing is activated when the
transaction cost ratio is greater than the swing product
$sw_{\mathrm{product}}$. In the previous approaches, the swing factor is
calculated once we have verified that the fund flow is larger than
$sw_{\mathrm{threshold}}$. In this new approach, the swing factor is first
calculated in order to determine the swing threshold:
\begin{equation*}
sw_{\mathrm{threshold}}=\frac{sw_{\mathrm{product}}}{sw_{\mathrm{\mathrm{%
factor}}}}
\end{equation*}%
Therefore, the swing threshold is dynamic and changes every day.\smallskip

\begin{figure}[tbph]
\centering
\caption{Dynamic approach of swing pricing}
\label{fig:swing_pricing5}
\figureskip
\includegraphics[width = \figurewidth, height = \figureheight]{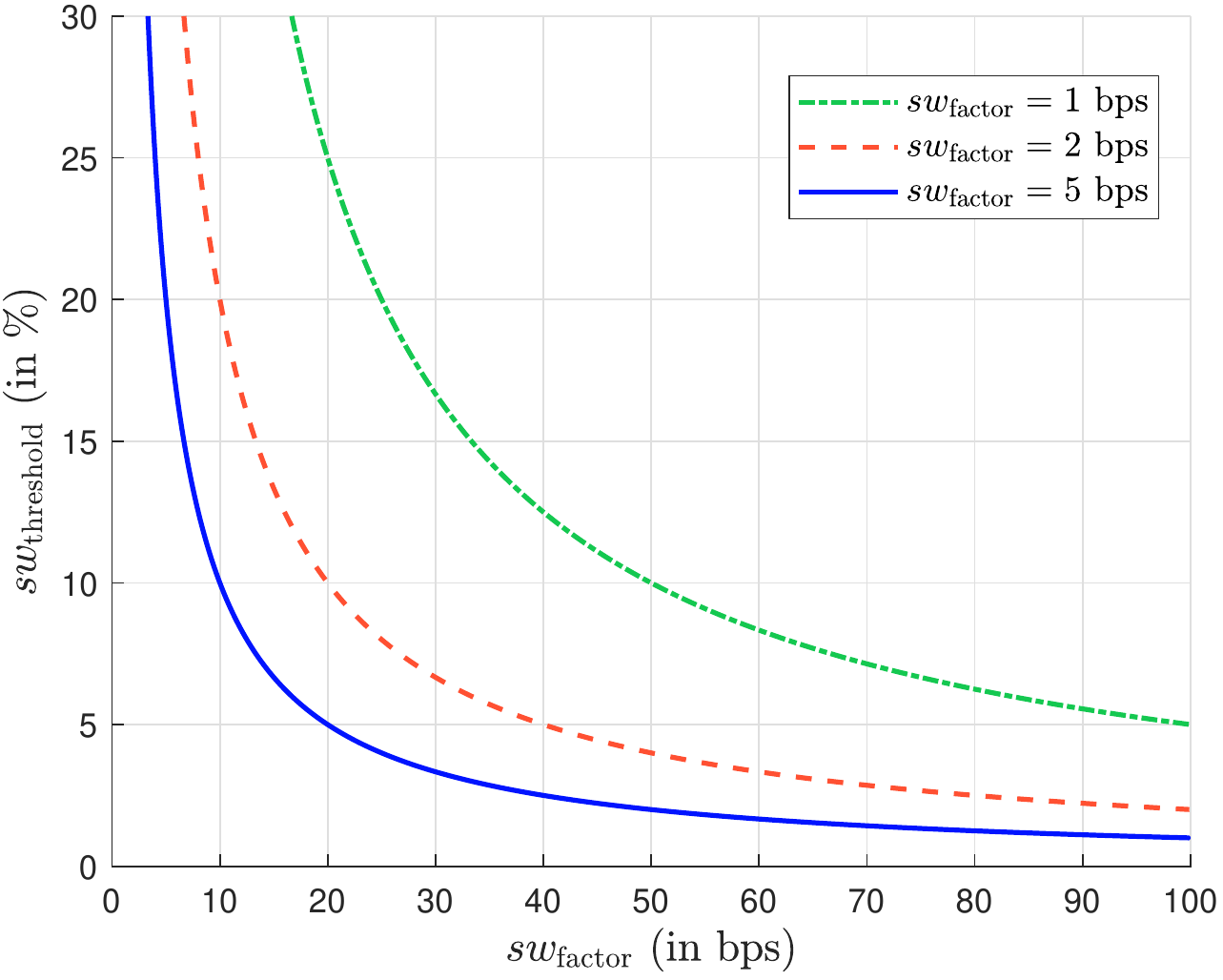}
\end{figure}

Let us see an example to illustrate the difference between the static and
dynamic approaches. We consider that $sw_{\mathrm{threshold}}=5\%$ and
$sw_{\mathrm{\mathrm{factor}}}=40$ bps. We deduce that
$sw_{\mathrm{product}}=2$ bps. In the static approach, the swing pricing
mechanism is not activated if we face a redemption rate of $4\%$ whatever the
value of the swing factor. We assume that we are in a period of stress and a
redemption rate of $4\%$ implies a swing factor of $60$ bps. In the dynamic
approach, the swing threshold is equal to $3.33\%$, implying the activation
of the swing pricing mechanism. More generally, we have a hyperbolic
relationship between $sw_{\mathrm{\mathrm{factor}}}$ and
$sw_{\mathrm{threshold}}$ as illustrated in Figure \ref{fig:swing_pricing5}.

\subsubsection{Anti-dilution levies}

Anti-dilution levies (ADL) are very close to swing pricing since the fund
manager does not use the transaction costs to adjust the NAV, but to adjust
entry and exit fees. According to \citet{AFG-2016}, these fees are equal to:
\begin{equation*}
\scalebox{0.9}{
\begin{tabular}{c|c:c|c:c|c}
\hline
& \multicolumn{2}{c|}{$N^{+}>N^{-}$} & \multicolumn{2}{c|}{$N^{+}<N^{-}$} &
Pro-rata \\ \hline
$\limfunc{ADL}\nolimits^{+}$ & $\dfrac{\mathcal{TC}\left( t+1\right) }{%
N^{+}\left( t+1\right) }$ & $\dfrac{\mathcal{TC}\left( t+1\right) }{\Delta
N\left( t+1\right) }$ & $0$ & $0$ & $\dfrac{\mathcal{TC}\left( t+1\right) }{%
N^{+}\left( t+1\right) +N^{-}\left( t+1\right) }$ \\
$\limfunc{ADL}\nolimits^{-}$ & $0$ & $0$ & $\dfrac{\mathcal{TC}\left(
t+1\right) }{N^{-}\left( t+1\right) }$ & $-\dfrac{\mathcal{TC}\left(
t+1\right) }{\Delta N\left( t+1\right) }$ & $\dfrac{\mathcal{TC}\left(
t+1\right) }{N^{+}\left( t+1\right) +N^{-}\left( t+1\right) }$ \\ \hline
\end{tabular}}
\end{equation*}
where $\limfunc{ADL}\nolimits^{+}$ is the entry fees and
$\limfunc{ADL}\nolimits^{-}$ is the exit fees. In the case of a pro-rata
rule, the transaction costs are borne by subscribing and redeeming investors.
In the other cases, the transaction costs are charged to subscribing
investors if $N^{+}>N^{-}$ or redeeming investors if $N^{+}<N^{-}$. Moreover,
anti-dilution levies may or may not recognize netting figures. This is why
we have reported two columns for the cases $N^{+}>N^{-}$ and
$N^{+}<N^{-}$.\smallskip

The main advantage of anti-dilution levies is that the NAV is not noisy and
reflects the fair value of the unit price. Remaining investors may be
sensitive because the mark-to-market is smoother. However,
subscribing/redeeming investors may prefer swing pricing, because they may
pay more attention to additional costs than to an adjusted price. Indeed,
they may have the feeling that swing pricing is fairer, because the
published NAV applies to all investors, whereas entry/exit costs only concern
them.

\begin{remark}
Sometimes there is a confusion between redemption fees\label{section:redemption-fees} and exit
fees\footnote{See for instance \citet{Greene-2007} and \citet{Lenkey-2016}.}.
Indeed, redemption fees are charged to investors in a systematic way
whatever the market conditions. They are indicated in the prospectus and their level is disclosed. Therefore, they are not a liquidity management tool for liquidity stress testing. On the contrary, swing pricing and anti-dilution levies are only charged in stress markets. Their levels are not necessarily disclosed. Table \ref{tab:swing-pricing1} summarizes the differences between these three mechanisms.
\end{remark}

\begin{table}[tbph]
\centering
\caption{Differences between redemption fees, anti-dilution levies and swing pricing}
\label{tab:swing-pricing1}
\begin{tabular}{cccc}
\hline
\multirow{2}{*}{Characteristics}
                   & Redemption                & Anti-dilution & Swing            \\
                   & fees                      & levies        & pricing          \\ \hline
\mr{Justification} & \mr{No obligation}        & \multicolumn{2}{c}{Documented and general principles}    \\
                   &                           & \multicolumn{2}{c}{externalised to fundholders}          \\ \hdashline
Requirements       & \mr{Any redemption}       & \multicolumn{2}{c}{Based on the net S/R balance}         \\
for activation     &                           & \multicolumn{2}{c}{and an activation threshold}          \\ \hdashline
Indication to      & Level of fees             & \multicolumn{2}{c}{No detail}                            \\
the level          & defined in the prospectus & \multicolumn{2}{c}{concerning the parameters}            \\
\hline
\end{tabular}
\begin{flushleft}
{\small \textit{Source}: \citet[page 15]{Darpeix-2020}.}
\end{flushleft}
\end{table}

\subsubsection{Effectiveness of swing pricing}

Based on the empirical study of US funds and their Luxembourg counterparts,
\citet{Lewrick-2017b} noticed that negative returns imply larger outflows
during normal market conditions for US funds. In stressed markets, in particular during the 2013 US taper tantrum, they found no difference. Since swing pricing is applied in Luxembourg and not in the US during the study period, they concluded that swing pricing failed to reduce the liquidity risk. For \citet{Capponi-2020}, the reason lies in the scale and application
of swing pricing. These authors consider that swing factors are too small
and must be larger to reduce the incentive to redeem immediately to
capture the first-mover advantage premium.\smallskip

We reiterate here that the purpose of swing pricing is to protect the remaining investors. In particular, during a period of market stress,
they do not pay other investors' transaction costs. The objective
of swing pricing is not to prevent a liquidity crisis, but
it may help fund managers to better rebalance their portfolio and
reduce the use of horizontal slicing.

\begin{remark}
This section dedicated to liquidity management tools demonstrates that there
is not one solution, but several approaches to managing the liquidity risk.
Nevertheless, this section also shows that the perfect tool does not exist.
In liquidity risk, the number of known and unknown unknowns is much greater than the number of known knowns. Therefore, the tools presented here give a partial answer to the problem, because the liquidity issue concerns the balance between buying and selling forces. Therefore, it makes sense to complement the liquidity framework by monitoring its level.
\end{remark}

\section{Liquidity monitoring tools}

It is obvious that monitoring liquidity is an important stage of a
liquidity stress testing program. For instance, the step is mandatory in
banking regulations. In addition to the LCR and NSFR, \citet{BCBS-2013} defines a set of liquidity risk monitoring tools, in order to \textquotedblleft \textsl{capture specific information related to a bank's cash flows, balance sheet structure, available unencumbered collateral and certain market indicators}\textquotedblright. We can classify these tools\footnote{The exhaustive list of liquidity monitoring metrics in the case of the Basel III framework is available in \citet{BCBS-2019}.} into two categories. The first category concerns the metrics that measure the liquidity at a global level. It corresponds to the macro-economic approach of liquidity monitoring, and it mainly uses market-wide information. The second category is specific to the managed portfolios. It corresponds to the micro-economic approach of liquidity monitoring, and it mainly uses security-based information.

\subsection{Macro-economic approach to liquidity monitoring}

The ESMA risk assessment uses several metrics \citep[page 4]{TRV-2021} to
monitor financial risks:
(a) risk participants (market environment, securities markets,
infrastructure and services, asset management and consumers),
(b) risk categories (liquidity risk, market risk, contagion risk, credit
risk, operational risk) and
(c) risk drivers (macro-economic environment, interest-rate environment,
sovereign and private debt markets, infrastructure disruption, political and
event risks).
Some of them are interesting when monitoring global liquidity from
an asset management viewpoint. We notice that the starting point of ESMA for
measuring financial risks is the market environment, more precisely the
economic outlook (real output, inflation risk, etc.) and the policy
responses. Therefore, central bank liquidity is an important monitoring
metric. Besides money market conditions, monitoring the banking
sector is also essential, because of its interconnectedness with asset
managers and asset owners. Therefore, statistics on the repo
market activity are important to track\footnote{Other statistics are easily available such as the one-month return of banks' stocks, CDS and credit spreads of banks, etc.}. Of course, traditional market risk metrics can be used to assess global liquidity. These include market sentiment (for instance, the levels of the VIX index and flight to liquidity), the performance of asset classes, the average level of
credit spreads (for both sovereign and corporate bonds), the high yield
premium, etc. Finally, the analysis of inflows/outflows, the number of
active LMTs, liquidity demand from investors, the dynamics of trading
volumes and the average bid-ask spreads can complement the macro-economic
approach to liquidity monitoring.

\subsection{Micro-economic approach to liquidity monitoring}

The micro-economic approach focuses on asset classes, security
instruments and issuers. For instance, the global liquidity metric gives
no information on the liquidity of US municipal bonds, the investment grade
segment of the ETF market, Italian BTP bonds, etc. The underlying idea is to
then use more specific measures, including the daily spread, the daily volume, the number of daily quotes, etc. These metrics can be computed
by asset class, security or issuer. Other information may be useful, for instance market-making activity, issuance activity, ETF liquidity, etc. Other important information flows are the metrics that can be computed from order books or the activity of trading desks. A typical example is the order imbalance proposed by \citet{Easley-2015}.

\section{Conclusion}

This article concludes a series of research studies dedicated to liquidity stress testing in asset management. As already said, academics and
professionals have so far paid little attention when asset-liability
management concerns the asset management industry. The goal of this research
project was then to fill the gap to develop mathematical and statistical
approaches and provide appropriate answers. The first part of this project
was dedicated to the liability liquidity risk \citep{Roncalli-lst1} and focused on the statistical modeling of redemption shocks. The second part concerned the asset liquidity risk \citep{Roncalli-lst2} and dealt with the modeling of the transaction cost function. Finally, this article, which constitutes the third part of the project, establishes the ALM framework of the liquidity risk in asset management \citep{Roncalli-lst3}. It is organized around the three Ms: measurement, management and monitoring.\smallskip

The primary liquidity measurement tool is the redemption coverage ratio or
RCR. Using a redemption scenario, the RCR measures the fund
manager's ability to liquidate the redemption portfolio in a stress period. Two methods exist to calculate the RCR: the time to liquidation approach and
the HQLA framework. The RCR depends on several assumptions about liability and asset risks, but also on the liquidity policy (trading limits
and liquidation method\footnote{We compare vertical slicing (naive and optimal pro-rata liquidation) and horizontal slicing (waterfall liquidation).}). We show that the latter has a big impact.
Moreover, we show how reverse stress testing can be implemented, in
particular how to define liability and asset RST scenarios.\smallskip

Liquidity management tools are many and varied. However, we can classify
them into three main categories. The first category concerns cash
management and the implementation of liquidity buffers. For that, we propose
an analytical framework that compares the costs and benefits of a cash
buffer. Therefore, we are able to define the optimal value of the cash
buffer, which depends on marginal transaction costs, the expected return of
assets and the sensitivity to the tracking error risk. In particular, we
illustrate the central role of the risk premium. This analysis enabled us to
reconcile the paradox around cash buffering and cash hoarding. In
particular, we explain cash hoarding by the dynamic implementation of a cash
buffer when the asset manager formulates negative expectations on the risk
premium. The second category of LMTs are special arrangements. It concerns
redemption suspensions, gates, side pockets and in-kind redemptions.
Finally, the last category revolves around swing pricing.\smallskip

Liquidity monitoring tools are more classical since they are not specific to
the asset management industry. Indeed, central and commercial banks,
regulators, market makers, investors, hedge funds and asset managers use
very similar approaches. This is especially true for global liquidity that
is highly dependent on central bank liquidity, economic outlook and market
sentiment. Monitoring liquidity at asset class, security or issuer level
is more challenging, but this is mainly a data management project.\smallskip

Once again, financial regulation has sped up the development of
liquidity risk management with the publication of the ESMA guidelines
on liquidity stress testing in UCITS and AIFs. Even though these guidelines
are less specific than those applied in the banking sector, they give
sufficient information about what is expected and the road
that asset managers must take in the future in terms of liquidity management.
This study has been completed with the sole aim of complying with
ESMA guidelines and asset management practices.
It can be viewed as a benchmark for asset managers and a guidebook
for academics, who want to develop practical models in this research field.

\clearpage

\bibliographystyle{apalike}

\begin{thebibliography}{99}

\bibitem[Agarwal \textsl{et al.}(2020)]{Agarwal-2020} \textsc{Agarwal}, V.,
    \textsc{Ren}, H., \textsc{Shen}, K., and \textsc{Zhao}, H. (2020),
    Redemption in Kind and Mutual Fund Liquidity Management,
    \textit{SSRN}, \url{www.ssrn.com/abstract=3527846}.

\bibitem[Aiken \textsl{et al.}(2015)]{Aiken-2015} \textsc{Aiken}, A.L.,
    \textsc{Clifford}, C.P., and \textsc{Ellis}, J.A. (2015), Hedge Funds and
    Discretionary Liquidity Restrictions, \textit{Journal of Financial
    Economics}, 116(1), pp. 197-218.

\bibitem[AFG(2016)]{AFG-2016}
    Association Fran\c{c}aise de la Gestion Financi\`ere (2016),
    Code of Conduct for Asset Managers using Swing Pricing and Variable Anti
    Dilution Levies, \textit{Professional Guide}.

\bibitem[AMF(2017)]{AMF-2017}
    Autorit\'e des March\'es Financiers (2017),
    The Use of Stress Tests as Part of Risk Management, \textit{AMF Guide}, February.

\bibitem[AMF(2020)]{AMF-2020} Autorit\'e des March\'es Financiers (2020),
    Continuity of Management Activities during the Coronavirus Crisis,
    \textit{AMF Report}, March.

\bibitem[BCBS(2008)]{BCBS-2008}
    Basel Committee on Banking Supervision (2008), \textit{Principles for Sound
    Liquidity Risk Management and Supervision},
    September 2008.

\bibitem[BCBS(2010)]{BCBS-2010}
    Basel Committee on Banking Supervision (2010), \textit{Basel III: A
    Global Regulatory Framework for More Resilient Banks and Banking Systems},
    December 2010 (revision June 2011).

\bibitem[BCBS(2013)]{BCBS-2013}
    Basel Committee on Banking Supervision (2013), \textit{Basel III:
    The Liquidity Coverage Ratio and Liquidity Risk Monitoring Tools}, January
    2013.

\bibitem[BCBS(2019)]{BCBS-2019}
    Basel Committee on Banking Supervision (2019), \textit{Supervisory
    Review Process --- SRP50: Liquidity Monitoring Metrics}, December
    2019.

\bibitem[Ben Slimane(2021)]{BenSlimane-2021}
    \textsc{Ben Slimane}, M. (2021),
    Bond Index Tracking with Genetic Algorithms,
    \textit{Amundi Working Paper}.

\bibitem[Bouveret(2017)]{Bouveret-2017}
    \textsc{Bouveret}, A. (2017), Liquidity Stress Tests for Investment
    Funds: A Practical Guide, \textit{IMF Working Paper}, 17/226.

\bibitem[Bouveret and Yu(2021)]{Bouveret-2021}
    \textsc{Bouveret}, A., and \textsc{Yu}, J. (2021), Risks and Vulnerabilities
    in the US Bond Mutual Fund Industry, \textit{IMF Working Paper}, 21/109.

\bibitem[Brunnermeier and Pedersen(2009)]{Brunnermeier-2009}
    \textsc{Brunnermeier}, M.K., and \textsc{Pedersen}, L.H. (2009),
    Market Liquidity and Funding Liquidity, \textit{Review of Financial
    Studies}, 22(6), pp. 2201-2238.

\bibitem[BaFin(2017)]{BaFin-2017} Bundesanstalt f\"{u}r
    Finanzdienstleistungsaufsicht (2017), Liquidity Stress Testing by German
    Asset Management Companies, December 2017.

\bibitem[Capponi \textsl{et al.}(2020)]{Capponi-2020}
    \textsc{Capponi}, A., \textsc{Glasserman}, P., and \textsc{Weber}, M. (2020),
    Swing Pricing for Mutual Funds: Breaking the Feedback Loop
    between Fire Sales and Fund Redemptions,
    \textit{Management Science}, 66(8), pp. 3581-3602.

\bibitem[Chen \textsl{et al.}(2010)]{Chen-2010} \textsc{Chen}, Q.,
    \textsc{Goldstein}, I., and \textsc{Jiang}, W. (2010),
    Payoff Complementarities and Financial Fragility: Evidence
    from Mutual Fund Outflows, \textit{Journal of Financial Economics},
    97(2), pp. 239-262.

\bibitem[Chernenko and Sunderam(2016)]{Chernenko-2016} \textsc{Chernenko},
    S., and \textsc{Sunderam}, A. (2016), Liquidity Transformation in Asset
    Management: Evidence from the Cash Holdings of Mutual Funds,
    \textit{NBER}, 22391.

\bibitem[Chernenko and Sunderam(2020)]{Chernenko-2020}
    \textsc{Chernenko}, S., and \textsc{Sunderam}, A. (2020), Do Fire Sales Create
    Externalities?, \textit{Journal of Financial Economics}, 135(3), pp. 602-628.

\bibitem[Cipriani \textsl{et al.}(2014)]{Cipriani-2014} \textsc{Cipriani},
    M., \textsc{Martin}, A., \textsc{McCabe}, P., and \textsc{Parigi}, B.M.
    (2014), Gates, Fees, and Preemptive Runs, Federal Reserve Bank of New
    York, \textit{Staff Reports}, 670.

\bibitem[Darpeix \textsl{et al.}(2020)]{Darpeix-2020} \textsc{Darpeix}, P-E.,
    \textsc{Le Moign}, C., \textsc{M\^{e}me}, N., and \textsc{Novakovic}, M. (2020),
    Overview and Inventory of French Funds' Liquidity Management Tools,
    Banque de France, \textit{Working Paper}, 775, pp. 1-37.

\bibitem[Easley \textsl{et al.}(2015)]{Easley-2015} \textsc{Easley}, D.,
    \textsc{de Prado}, M.L., and \textsc{O'Hara}, M. (2015),
     Optimal Execution Horizon, \textit{Mathematical Finance}, 25(3), pp. 640-672.

\bibitem[EFAMA(2020)]{EFAMA-2020}
    European Fund and Asset Management Association (2020),
    Managing Fund Liquidity Risk in Europe --- Recent Regulatory Enhancements
    \& Proposals for Further Improvements, \textit{AMIC/EFAMA Report}, January.

\bibitem[ESMA(2019a)]{ESMA-2019a}
    European Securities and Markets Authority (2019a), Guidelines on Liquidity
    Stress Testing in UCITS and AIFs, \textit{Final Report}, \textit{ESMA34-39-882}, September.

\bibitem[ESMA(2019b)]{ESMA-2019b}
    European Securities and Markets Authority (2019b), Stress Simulation for Investment Funds
    (Stresi), \textit{Economic Report}, \textit{ESMA50-164-2458}, September.

\bibitem[ESMA(2020a)]{ESMA-2020a}
    European Securities and Markets Authority (2020a), Guidelines on Liquidity
    Stress Testing in UCITS and AIFs, \textit{ESMA34-39-897}, July.

\bibitem[ESMA(2020b)]{ESMA-2020b}
    European Securities and Markets Authority (2020b),
    Recommendation of the European Systemic Risk Board (ESRB)
    on Liquidity Risk in Investment Funds,
    \textit{Report}, \textit{ESMA34-39-1119}, November.

\bibitem[ESMA TRV Report(2021)]{TRV-2021}
    European Securities and Markets Authority (2021),
    Trends, Risks and Vulnerabilities (TRV),
    \textit{Report}, 2, \textit{ESMA50-165-1842}, September.

\bibitem[ESRB(2017)]{ESRB-2017}
    European Systemic Risk Board (2017),
    Recommendation of 7 December 2017 on Liquidity and
    Leverage Risks in Investment Funds,
    \textit{Report}, \textit{ESRB/2017/6}, December.

\bibitem[Fricke and Fricke(2021)]{Fricke-2021} \textsc{Fricke}, C., and
    \textsc{Fricke}, D. (2021), Vulnerable Asset Management? The Case of
    Mutual Funds, \textit{Journal of Financial Stability}, 52,
    100800.

\bibitem[Fricke and Wilke(2020)]{Fricke-2020} \textsc{Fricke}, D., and
    \textsc{Wilke}, H. (2020), Connected Funds,
    \textit{Discussion Paper}, 48, Deutsche Bundesbank.

\bibitem[GAM(2018)]{GAM-2018} GAM (2018), GAM Fund Boards Suspend Dealing
    in Unconstrained/Absolute Return Funds, \textit{Press Release},
    August $2^{\mathrm{nd}}$ 2018, \url{www.gam.com}.

\bibitem[Goldstein(2017)]{Goldstein-2017a} \textsc{Goldstein}, I. (2017)
    Comment on \textquotedblleft \textsl{Redemption Risk and
    Cash Hoarding by Asset Managers}\textquotedblright\
    by Morris, Shim, and Shin, \textit{Journal of Monetary Economics},
    89, pp. 88-91.

\bibitem[Goldstein \textsl{et al.}(2017)]{Goldstein-2017b} \textsc{Goldstein},
    I., \textsc{Jiang}, H., and \textsc{Ng}, D.T. (2017),
    Investor Flows and Fragility in Corporate Bond Funds,
    \textit{Journal of Financial Economics}, 126(3), pp. 592-613.

\bibitem[Greene \textsl{et al.}(2007)]{Greene-2007} \textsc{Greene}, J.T.,
    \textsc{Hodges}, C.W., and \textsc{Rakowski}, D.A. (2007), Daily Mutual
    Fund Flows and Redemption Policies, \textit{Journal of Banking \&
    Finance}, 31(12), pp. 3822-3842.

\bibitem[Grill \textsl{et al.}(2021)]{Grill-2021} \textsc{Grill}, M.,
    \textsc{Vivar}, L.M., and \textsc{Wedow}, M. (2021),
    The Suspensions of Redemptions during the COVID-19 Crisis --
    A Case for Pre-emptive Liquidity Measures?, European Central Bank,
    \textit{Macroprudential Bulletin}, 12, April.

\bibitem[Jiang \textsl{et al.}(2021)]{Jiang-2021} \textsc{Jiang}, H.,
    \textsc{Li}, D., and \textsc{Wang}, A. (2021),
    Dynamic Liquidity Management by Corporate Bond Mutual Funds,
    \textit{Journal of Financial and Quantitative Analysis}, 56(5), pp. 1622-1652.

\bibitem[Jin \textsl{et al.}(2019)]{Jin-2019} \textsc{Jin}, D.,
    \textsc{Kacperczyk}, M., \textsc{Kahraman}, B., and \textsc{Suntheim}, F. (2019),
    Swing Pricing and Fragility in Open-end Mutual Funds,
    \textit{IMF Working Paper}, 19/227.

\bibitem[IMF(2017)]{IMF-2017} International Monetary Fund (2017),
    Luxembourg Financial Sector Assessment Program (FSAP): Technical Note \&
    Risk Analysis, \textit{IMF Staff Country Reports}, 17/261, August.

\bibitem[IMF(2020)]{IMF-2020} International Monetary Fund (2020), United
    States Financial Sector Assessment Program (FSAP): Technical Note, Risk Analysis
    and Stress Testing the Financial Sector,
    \textit{IMF Staff Country Reports}, 20/247, August.

\bibitem[Lenkey and Song(2016)]{Lenkey-2016} \textsc{Lenkey}, S., and
    \textsc{Song}, F. (2016), Redemption Fees and Information-based Runs,
    \textit{SSRN}, \url{www.ssrn.com/abstract=2796717}.

\bibitem[Lewrick and Schanz(2017a)]{Lewrick-2017a} \textsc{Lewrick}, U.,
    and \textsc{Schanz}, J. (2017a), Liquidity Risk in Markets with Trading Frictions:
    What can Swing Pricing Achieve, \textit{BIS Working Paper}, 663.

\bibitem[Lewrick and Schanz(2017b)]{Lewrick-2017b} \textsc{Lewrick}, U.,
    and \textsc{Schanz}, J. (2017b), Is the Price Right? Swing Pricing and
    Investor Redemptions, \textit{BIS Working Paper}, 664.

\bibitem[Ma \textsl{et al.}(2021)]{Ma-2021} \textsc{Ma}, Y.,
    \textsc{Xiao}, K., and \textsc{Zeng}, Y. (2021),
    Mutual Fund Liquidity Transformation and Reverse Flight to Liquidity,
    \textit{SSRN}, \url{www.ssrn.com/abstract=3640861}.

\bibitem[Malik and Lindner(2017)]{Malik-2017} \textsc{Malik}, S. and
    \textsc{Lindner}, P. (2017), On Swing Pricing and Systemic Risk Mitigation,
    \textit{IMF Working Paper}, 17/159.

\bibitem[Morris \textsl{et al.}(2017)]{Morris-2017} \textsc{Morris}, S.,
    \textsc{Shim}, I., and \textsc{Shin}, H.S. (2017), Redemption Risk and
    Cash Hoarding by Asset Managers, \textit{Journal of Monetary Economics},
    89, pp. 71-87.

\bibitem[Opromolla(2009)]{Opromolla-2009} \textsc{Ppromolla}, G. (2009),
    Facing the Financial Crisis: Bank of Italy's Implementing Regulation on
    Hedge Funds, \textit{Journal of Investment Compliance}, 10(2), pp. 41-44.

\bibitem[Poterba and Shoven(2002)]{Poterba-2002} \textsc{Poterba}, J.M., and \textsc{Shoven}, J.B. (2002),
    Exchange-Traded Funds: A New Investment Option for Taxable Investors,
    \textit{American Economic Review}, 92(2), pp. 422-427.

\bibitem[Rohleder \textsl{et al.}(2017)]{Rohleder-2017} \textsc{Rohleder},
    M., \textsc{Schulte}, D., and \textsc{Wilkens}, M. (2017), Management of Flow
    Risk in Mutual Funds, \textit{Review of Quantitative Finance and
    Accounting}, 48(1), pp. 31-56.

\bibitem[Roncalli(2020)]{Roncalli-2020} \textsc{Roncalli}, T. (2020),
    \textit{Handbook of Financial Risk Management}, Chapman \& Hall/CRC
    Financial Mathematics Series.

\bibitem[Roncalli \textsl{et al.}(2021a)]{Roncalli-lst1}
    \textsc{Roncalli}, T., \textsc{Karray-Meziou}, F., \textsc{Pan}, F., and \textsc{Regnault}, M. (2021a),
    Liquidity Stress Testing in Asset Management --- Part 1. Modeling the Liability Liquidity Risk,
    \textit{Amundi Working Paper}.

\bibitem[Roncalli \textsl{et al.}(2021b)]{Roncalli-lst2}
    \textsc{Roncalli}, T., \textsc{Cherief}, A., \textsc{Karray-Meziou}, F., and \textsc{Regnault}, M. (2021b),
    Liquidity Stress Testing in Asset Management --- Part 2. Modeling the Asset Liquidity Risk,
    \textit{Amundi Working Paper}.

\bibitem[Roncalli(2021c)]{Roncalli-lst3}
    \textsc{Roncalli}, T. (2021c),
    Liquidity Stress Testing in Asset Management --- Part 3. Managing the Asset-Liability Liquidity Risk,
    \textit{Amundi Working Paper}.

\bibitem[Roncalli and Weisang(2015a)]{Roncalli-2015a}
    \textsc{Roncalli}, T., and \textsc{Weisang}, G. (2015a), Asset
    Management and Systemic Risk, \textit{SSRN},
    \url{www.ssrn.com/abstract=2610174}.

\bibitem[Roncalli and Weisang(2015b)]{Roncalli-2015b} \textsc{Roncalli}, T.,
    and \textsc{Weisang}, G. (2015b), \textit{Response to FSB-IOSCO Second Consultative Document,
    Assessment Methodologies for Identifying Non-Bank Non-Insurer
    Global Systemically Important Financial Institutions}, May 28,
    \url{https://www.fsb.org/wp-content/uploads/Thierry-Roncalli-and-Guillaume-Weisang.pdf}.

\bibitem[Schrimpf \textsl{et al.}(2021)]{Schrimpf-2021} \textsc{Schrimpf}, A.,
    \textsc{Shim}, I., and \textsc{Shin}, H.S. (2021),
    Liquidity Management and Asset Sales by Bond Funds in the Face of
    Investor Redemptions in March 2020, \textit{BIS Bulletin}, 39, pp. 1-6.

\bibitem[SEC(2016)]{SEC-2016}
    Securities and Exchange Commission (2016), Investment Company Swing Pricing,
    \textit{Final Rule}, IC-32316, October.

\bibitem[Simutin(2010)]{Simutin-2010} \textsc{Simutin}, M. (2010),
    Excess Cash and Stock Returns, \textit{Financial Management}, 39(3), pp. 1197-1222.

\bibitem[Simutin(2014)]{Simutin-2014} \textsc{Simutin}, M. (2014),
    Cash Holdings and Mutual Fund Performance,
    \textit{Review of Finance}, 18(4), pp. 1425-1464.

\bibitem[Teo(2011)]{Teo-2011} \textsc{Teo}, M. (2011), The Liquidity Risk of
    Liquid Hedge Funds, \textit{Journal of Financial Economics}, 100(1), pp.
    24-44.

\bibitem[Voellmy(2021)]{Voellmy-2021} \textsc{Voellmy}, L. (2021), Preventing
    Runs with Fees and Gates, \textit{Journal of Banking \& Finance}, 125, 106065.

\bibitem[Yan(2006)]{Yan-2006} \textsc{Yan}, X. (2006),
    The Determinants and Implications of Mutual Fund Cash Holdings:
    Theory and Evidence, \textit{Financial Management}, 35(2), pp. 67-91.

\end{thebibliography}

\clearpage

\appendix

\section*{Appendix}

\section{Glossary}

\subsection*{ALMT}
\label{glossary:almt}
ALMT (or a-LMT) is the acronym of \textquotedblleft \textit{Additional
Liquidity Management Tool}\textquotedblright. They include the tools
applied by asset managers in exceptional circumstances to control or limit
dealing in fund units/shares in the interests of investors. Examples of ALMT
are suspension of dealing in units, deferral of dealing, \hyperref[gloassary:sp]{side-pocketing} and
\hyperref[glossary:sa]{special arrangements}.

\subsection*{Anti-diluation levy}
\label{glossary:adlv}
Anti-diluation levies correspond to entry and exit fees. Their levels are
calculated with respect to the transaction costs induced by subscriptions
and redemptions.

\subsection*{Cash buffer}
\label{glossary:cb}
A cash buffer is a special type of \hyperref[gloassary:lbuffer]{liquidity buffers} that is exclusively composed of cash instruments and cash equivalents.

\subsection*{Cash conversion factor}
\label{glossary:ccf}
A CCF is a multiplicative factor, which indicates how to convert \$1 of
assets into a liquid cash exposure.

\subsection*{Cash hoarding}
\label{glossary:ch}
Cash hoarding corresponds to a situation where the asset manager increases
its cash holding in a liquidity stress period.

\subsection*{Gate}
\label{glossary:gate}
When a gate is implemented, the fund manager temporarily limits the amount
of redemptions.

\subsection*{Horizontal slicing}
\label{glossary:hs}
See \hyperref[glossary:wl]{waterfall liquidation}.

\subsection*{HQLA class}
\label{glossary:hqla}
The term HQLA refers to high quality liquid asset. An HQLA class groups all
the securities that present the same ability to be converted into cash.

\subsection*{In-kind redemption}
\label{glossary:ikr}
When in-kind redemptions are implemented, the fund manager offers a basket
of securities to the redeeming investor. In-kind redemptions are also called
physical redemptions as opposed to cash redemptions that imply a monetary
payment.

\subsection*{Liquidation policy}
\label{glossary:lp}
See \hyperref[glossary:tl]{trading limit}.

\subsection*{Liquidation time}
\label{glossary:lt}
See \hyperref[glossary:ttl]{time to liquidation}.

\subsection*{Liquidity buffer}
\label{glossary:lbuffer}
A liquidity buffer refers to the stock of liquid instruments held by the
fund manager in order to manage future redemptions.

\subsection*{Liquidity management tool}
\label{glossary:lmt}
Liquidity management tools include \hyperref[glossary:lbuffer]{liquidity buffers},
\hyperref[glossary:sa]{special arrangements},
\hyperref[glossary:sp]{swing pricing} and
\hyperref[glossary:adl]{anti-dilution levies}. See also
\hyperref[glossary:almt]{ALMT}.

\subsection*{Liquidity shortfall}
\label{glossary:ls}
The liquidation shortfall is defined as the residual redemption that cannot
be fulfilled after $h$ trading days. It is expressed as a percentage of the
redemption value. If it is equal to $0\%$, this means that we can liquidate
the redemption in $h$ trading days. More generally, its mathematical
expression is:
\begin{equation*}
\func{LS}\left( h\right) =\RedemptionRate\cdot \max \left( 0,1-\limfunc{RCR}%
\left( h\right) \right)
\end{equation*}

\subsection*{Pro-rata liquidation}
\label{glossary:prl}
The pro-rata liquidation uses the proportional rule, implying that each
asset is liquidated such that the structure of the portfolio is the same
before and after the liquidation.

\subsection*{Redemption coverage ratio}
\label{glossary:rcr}
The redemption coverage ratio $\limfunc{RCR}\left( h\right) $ is the
proportion of the redemption that is liquidated after $h$ trading days. We
generally focus on daily and weekly liquidation ratios $\limfunc{RCR}\left(
1\right) $ and $\limfunc{RCR}\left( 5\right) $. The RCR is also used to
define the \hyperref[glossary:lt]{liquidation time} (or \hyperref[glossary:ttl]%
{time to liquidation}), which is an important measure for managing the
liquidity risk.

\subsection*{Redemption scenario}
\label{glossary:rs}
A redemption scenario $q$ is defined by the vector $\left( q_{1},\ldots
,q_{n}\right) $ where $q_{i}$ is the number of shares of security $i$ to
sell. This scenario can be expressed in dollars:
\begin{equation*}
Q:=\left( Q_{1},\ldots ,Q_{n}\right) =\left( q_{1}P_{1},\ldots
,q_{n}P_{n}\right)
\end{equation*}%
where $P_{i}$ is the price of security $i$. The redemption scenario may also
be defined by its dollar value $\mathbb{R}$:
\begin{equation*}
\mathbb{R}=\mathbb{V}\left( q\right) =\sum_{i=1}^{n}q_{i}P_{i}
\end{equation*}

\subsection*{Redemption suspension}
\label{glossary:rs2}
A redemption suspension is a temporary measure where the investors are
unable to withdraw their capital in the fund.

\subsection*{Reverse stress testing}
\label{glossary:rst}

\subsection*{Side pocket}
\label{glossary:sp}
A side pocket is a segregated portfolio of illiquid assets.

\subsection*{Special arrangement}
\label{glossary:sa}
Special arrangements are specific types of \hyperref[glossary:lmt]{LMT}
measures available to some AIFs and which impact investors' redemption
rights, such as \hyperref[glossary:sp]{side pockets} or
\hyperref[glossary:gate]{gates}.

\subsection*{Swing pricing}
\label{glossary:swing}
Swing pricing is a NAV adjustment process to incorporate redemption and subscription costs.

\subsection*{Time to liquidation}
\label{glossary:ttl}
The time to liquidation is the inverse function of the
\hyperref[glossary:lr]{liquidation ratio}. It indicates the minimum number of days required to liquidate the proportion $p$ of the redemption.

\subsection*{Trading limit}
\label{glossary:tl}
The trading limit $q^{+}$ is the maximum number of shares that can be sold
in one trading day.

\subsection*{Vertical slicing}
\label{glossary:vs}
See \hyperref[glossary:prl]{pro-rata liquidation}.

\subsection*{Waterfall liquidation}
\label{glossary:wl}
In this approach, the portfolio is liquidated by selling the most
liquid assets first.

\clearpage

\section{Mathematical results}

\subsection{Computation of the cash conversion factor}
\label{appendix:hqla}

We define $H\left( t\right)$ as the following integral function:%
\begin{equation}
H\left( t\right) =\int_{0}^{t}f\left( u\right) \left( 1-\int_{0}^{u}g\left(
s\right) \,\mathrm{d}s\right) \,\mathrm{d}u
\end{equation}%
where $f\left( u\right) \geq 0$ and $g\left( u\right) \geq 0$ are two
positive functions. We note:%
\begin{equation}
F\left( t\right) =\int_{0}^{t}f\left( u\right) \,\mathrm{d}u
\end{equation}%
and:%
\begin{equation}
G\left( t\right) =\int_{0}^{t}g\left( u\right) \,\mathrm{d}u
\end{equation}%
We assume that:
\begin{itemize}
\item $F\left( t\right) $ is an increasing function with $F\left( 0\right) =0
$ and $F\left( \infty \right) =1$;

\item $G\left( t\right) $ is an increasing function with $G\left( 0\right) =0
$ and $G\left( \infty \right) \leq 1$.
\end{itemize}
We deduce that
\begin{equation}
0\leq H\left( t\right) \leq 1
\end{equation}%
In the case where $f\left( u\right) =\xi $ is constant, we obtain:%
\begin{eqnarray}
H\left( t\right)  &=&\xi \int_{0}^{t}\left( 1-\int_{0}^{u}g\left( s\right) \,%
\mathrm{d}s\right) \,\mathrm{d}u  \notag \\
&=&\frac{F\left( t\right) \int_{0}^{t}\left( 1-G\left( u\right) \right) \,%
\mathrm{d}u}{t}
\end{eqnarray}%
because we have $F\left( t\right) =\xi t$. Using the integral mean value
theorem, we deduce that $\int_{0}^{t}G\left( u\right) \,\mathrm{d}u=t\left(
1-G\left( c\right) \right) $ where $c\in \left[ 0,t\right] $. If $g\left(
u\right) $ is relatively smooth, it follows that:%
\begin{eqnarray}
H\left( t\right)  &=&F\left( t\right) \left( 1-G\left( c\right) \right)
\notag \\
&\approx &F\left( t\right) \left( 1-G\left( \frac{t}{2}\right) \right)
\label{eq:hqla3}
\end{eqnarray}%
This result has been obtained by considering that $f\left( u\right) $ is
constant. Nevertheless, we assume that this result holds in the general
case.\smallskip

Let us apply the previous result to the computation of the cash conversion
factor. $f\left( u\right) $ is the instantaneous amount of the liquidation
portfolio that can be sold in the market at time $u$, whereas $F\left(
t\right) $ is the cumulated amount of the liquidation portfolio that can be
sold between $0$ and $t$. $G\left( u\right) =\int_{0}^{u}g\left( s\right) \,\mathrm{d}s$
is the drawdown during the period $\left[ 0,t\right] $. Using
the notations on page \pageref{eq:hqla2}, Equation (\ref{eq:hqla3}) becomes:
\begin{equation}
\limfunc{CCF}\left( t\right) =\limfunc{LF}\left( t\right) \left( 1-\limfunc{%
DF}\left( \frac{t}{2}\right) \right)   \label{eq:hqla4}
\end{equation}

\subsection{Analytics of the cash buffer}
\label{appendix:cash}

\subsubsection{Mean-variance analysis of the portfolio}
\label{appendix:cash-mv}

Let $w_{\mathrm{cash}}$ be the cash-to-assets ratio:%
\begin{equation}
w_{\mathrm{cash}}=\frac{\mathrm{cash}}{\func{TNA}}
\end{equation}%
The random return of the portfolio that includes the cash buffer  is equal
to:%
\begin{eqnarray}
R &=&w_{\mathrm{cash}}\cdot R_{\mathrm{cash}}+\left( 1-w_{\mathrm{cash}%
}\right) \cdot R_{\mathrm{asset}}  \notag \\
&=&R_{\mathrm{asset}}-w_{\mathrm{cash}}\cdot \left( R_{\mathrm{asset}}-R_{%
\mathrm{cash}}\right)
\end{eqnarray}%
where $R_{\mathrm{cash}}$ and $R_{\mathrm{asset}}$ are the random returns of
the cash and the assets. We deduce that:%
\begin{equation}
\mathbb{E}\left[ R\right] =\mu _{\mathrm{asset}}-w_{\mathrm{cash}}\cdot
\left( \mu _{\mathrm{asset}}-\mu _{\mathrm{cash}}\right)
\end{equation}%
and:%
\begin{eqnarray}
\sigma ^{2}\left( R\right)  &=&w_{\mathrm{cash}}^{2}\cdot \sigma _{\mathrm{%
cash}}^{2}+\left( 1-w_{\mathrm{cash}}\right) ^{2}\cdot \sigma _{\mathrm{asset%
}}^{2}+  \notag \\
&&2w_{\mathrm{cash}}\cdot \left( 1-w_{\mathrm{cash}}\right) \cdot \rho _{%
\func{cash},\mathrm{asset}}\cdot \sigma _{\mathrm{cash}}\cdot \sigma _{%
\mathrm{asset}}
\end{eqnarray}%
where $\mu _{\mathrm{cash}}$ and $\mu _{\mathrm{asset}}$ are the expected
returns of the cash and asset components, $\sigma _{\mathrm{cash}}$ and $%
\sigma _{\mathrm{asset}}$ are the corresponding volatilities, and $\rho _{%
\mathrm{cash},\mathrm{asset}}$ is the correlation between the cash and the
assets. Generally, we assume that $\sigma _{\mathrm{cash}}\approx 0$ (or $%
\sigma _{\mathrm{cash}}\ll \sigma _{\mathrm{asset}}$), implying that:%
\begin{equation}
\sigma \left( R\right) \approx \left( 1-w_{\mathrm{cash}}\right) \cdot
\sigma _{\mathrm{asset}}
\end{equation}

\subsubsection{Tracking error analysis of the portfolio}
\label{appendix:cash-te}

Since the tracking error due to the cash buffer is given by:
\begin{eqnarray}
e &=&R-R_{\mathrm{asset}}  \notag \\
&=&-w_{\mathrm{cash}}\cdot \left( R_{\mathrm{asset}}-R_{\mathrm{cash}%
}\right)
\end{eqnarray}%
we obtain the following formula for the expected excess return:%
\begin{eqnarray}
\mathbb{E}\left[ R\mid R_{\mathrm{asset}}\right]  &=&\mathbb{E}\left[ R-R_{%
\mathrm{asset}}\right]   \notag \\
&=&-w_{\mathrm{cash}}\cdot \left( \mu _{\mathrm{asset}}-\mu _{\mathrm{cash}%
}\right)
\end{eqnarray}%
whereas the tracking error volatility $\sigma \left( R\mid R_{\mathrm{asset}%
}\right) $ is equal to:%
\begin{eqnarray}
\sigma ^{2}\left( R\mid R_{\mathrm{asset}}\right)  &=&\sigma ^{2}\left( R-R_{%
\mathrm{asset}}\right)   \notag \\
&=&w_{\mathrm{cash}}^{2}\cdot \left( \sigma _{\mathrm{cash}}^{2}+\sigma _{%
\mathrm{asset}}^{2}-2\rho _{\mathrm{cash},\mathrm{asset}}\cdot \sigma _{%
\mathrm{cash}}\cdot \sigma _{\mathrm{asset}}\right)
\end{eqnarray}%
If we assume that $\sigma _{\mathrm{cash}}\approx 0$, it follows that:%
\begin{equation}
\sigma \left( R\mid R_{\mathrm{asset}}\right) \approx w_{\mathrm{cash}}\cdot
\sigma _{\mathrm{asset}}
\end{equation}

\subsubsection{Beta and correlation of the portfolio}
\label{appendix:cash-beta}

The covariance between the portfolio return and the asset return is equal to:%
\begin{eqnarray}
\func{cov}\left( R,R_{\mathrm{asset}}\right)  &=&\mathbb{E}\left[ R\cdot R_{%
\mathrm{asset}}\right] -\mathbb{E}\left[ R\right] \cdot \mathbb{E}\left[ R_{%
\mathrm{asset}}\right]   \notag \\
&=&\mathbb{E}\left[ w_{\mathrm{cash}}\cdot R_{\mathrm{cash}}\cdot R_{\mathrm{%
asset}}+\left( 1-w_{\mathrm{cash}}\right) \cdot R_{\mathrm{asset}}^{2}\right]
-  \notag \\
&&\left( \mu _{\mathrm{asset}}-w_{\mathrm{cash}}\cdot \left( \mu _{\mathrm{%
asset}}-\mu _{\mathrm{cash}}\right) \right) \cdot \mu _{\mathrm{asset}}
\notag \\
&=&w_{\mathrm{cash}}\cdot \left( \rho _{\mathrm{cash},\mathrm{asset}}\cdot
\sigma _{\mathrm{cash}}\cdot \sigma _{\mathrm{asset}}+\mu _{\mathrm{cash}%
}\cdot \mu _{\mathrm{asset}}\right) +  \notag \\
&&\left( 1-w_{\mathrm{cash}}\right) \cdot \left( \sigma _{\mathrm{asset}%
}^{2}+\mu _{\mathrm{asset}}^{2}\right) -  \notag \\
&&\left( \mu _{\mathrm{asset}}^{2}-w_{\mathrm{cash}}\cdot \left( \mu _{%
\mathrm{asset}}^{2}-\mu _{\mathrm{cash}}\cdot \mu _{\mathrm{asset}}\right)
\right)   \notag \\
&=&w_{\mathrm{cash}}\cdot \rho _{\mathrm{cash},\mathrm{asset}}\cdot \sigma _{%
\mathrm{cash}}\cdot \sigma _{\mathrm{asset}}+  \notag \\
&&\left( 1-w_{\mathrm{cash}}\right) \cdot \sigma _{\mathrm{asset}}^{2}
\end{eqnarray}%
We deduce that:%
\begin{eqnarray}
\beta \left( R\mid R_{\mathrm{asset}}\right)  &=&\frac{\func{cov}\left( R,R_{%
\mathrm{asset}}\right) }{\sigma ^{2}\left( R_{\mathrm{asset}}\right) }
\notag \\
&=&1-\frac{w_{\mathrm{cash}}}{\sigma _{\mathrm{asset}}^{2}}\left( \sigma _{%
\mathrm{asset}}^{2}-\rho _{\mathrm{cash},\mathrm{asset}}\cdot \sigma _{%
\mathrm{cash}}\cdot \sigma _{\mathrm{asset}}\right)
\end{eqnarray}%
If $\sigma _{\mathrm{cash}}\approx 0$, we obtain:%
\begin{equation}
\beta \left( R\mid R_{\mathrm{asset}}\right) \approx 1-w_{\mathrm{cash}}
\end{equation}%
and:%
\begin{eqnarray}
\rho \left( R,R_{\mathrm{asset}}\right)  &=&\frac{\func{cov}\left( R,R_{%
\mathrm{asset}}\right) }{\sigma \left( R\right) \cdot \sigma \left( R_{\func{%
asset}}\right) }  \notag \\
&\approx &1
\end{eqnarray}

\subsubsection{Sharpe and information ratios}
\label{appendix:cash-sr}

The Sharpe ratio is equal to:%
\begin{eqnarray}
\limfunc{SR}\left( R\right)  &=&\frac{\mathbb{E}\left[ R\right] -\mathbb{E}%
\left[ R_{\mathrm{cash}}\right] }{\sigma \left( R\right) }  \notag \\
&=&\frac{\left( 1-w_{\mathrm{cash}}\right) \cdot \left( \mu _{\mathrm{asset}%
}-\mu _{\mathrm{cash}}\right) }{\sigma \left( R\right) }
\end{eqnarray}%
For the information ratio, we obtain:%
\begin{eqnarray}
\limfunc{IR}\left( R\mid R_{\mathrm{asset}}\right)  &=&\frac{\mathbb{E}\left[
R\mid R_{\mathrm{asset}}\right] }{\sigma \left( R\mid R_{\mathrm{asset}%
}\right) }  \notag \\
&=&-\frac{\mu _{\mathrm{asset}}-\mu _{\mathrm{cash}}}{\sqrt{\sigma _{\mathrm{%
cash}}^{2}+\sigma _{\mathrm{asset}}^{2}-2\rho _{\mathrm{cash},\mathrm{asset}%
}\cdot \sigma _{\mathrm{cash}}\cdot \sigma _{\mathrm{asset}}}}
\end{eqnarray}%
If $\sigma _{\mathrm{cash}}\approx 0$, we deduce that:%
\begin{equation}
\limfunc{SR}\left( R\right) \approx \frac{\mu _{\mathrm{asset}}-\mu _{%
\mathrm{cash}}}{\sigma _{\mathrm{asset}}}=\limfunc{SR}\left( R_{\mathrm{asset%
}}\right)
\end{equation}%
and:%
\begin{equation}
\limfunc{IR}\left( R\mid R_{\mathrm{asset}}\right) \approx -\frac{\mu _{%
\mathrm{asset}}-\mu _{\mathrm{cash}}}{\sigma _{\mathrm{asset}}}=-\limfunc{SR}%
\left( R_{\mathrm{asset}}\right)
\end{equation}

\subsubsection{Liquidation gain}
\label{appendix:cash-lg}

Without the cash buffer, the transaction cost of the redemption shock
$\RedemptionShock=\RedemptionRate\cdot \limfunc{TNA}$ is equal to:
\begin{equation}
\mathcal{TC}_{\mathrm{without}}=\mathcal{TC}_{\mathrm{asset}}\left( %
\RedemptionRate\cdot \limfunc{TNA}\right)
\end{equation}%
where $\mathcal{TC}_{\mathrm{asset}}\left( V\right) $ is the transaction
cost\footnote{The unit of the transaction cost function is expressed in \% of
the total net assets.} when liquidating the amount $V$ of assets. With the
cash buffer, the breakdown of redemption shock is:
\begin{eqnarray}
\RedemptionShock &=&\hspace{33.58pt}\RedemptionShock_{\mathrm{cash}}\hspace{%
33.585pt}+\hspace{25pt}\RedemptionShock_{\mathrm{asset}}  \notag \\
&=&\underset{\text{Cash liquidation}}{\underbrace{\min \left( w_{\mathrm{cash%
}},\RedemptionRate\right) \cdot \limfunc{TNA}}}+\underset{\text{Asset
liquidation}}{\underbrace{\left( \RedemptionRate-w_{\mathrm{cash}}\right)
^{+}\cdot \limfunc{TNA}}}
\end{eqnarray}%
Indeed, the fund manager first sells the cash until the redemption rate
reaches the cash-to-assets ratio, and then liquidates the assets if
necessary:
\begin{equation}
\RedemptionShock=\left\{
\begin{array}{ll}
0 & \text{if }\RedemptionRate=0 \\
\RedemptionRate\cdot \limfunc{TNA} & \text{if }0<\RedemptionRate\leq w_{%
\mathrm{cash}} \\
w_{\mathrm{cash}}\cdot \limfunc{TNA}+\left( \RedemptionRate-w_{\mathrm{cash}%
}\right) \cdot \limfunc{TNA} & \text{if }\RedemptionRate>w_{\mathrm{cash}}%
\end{array}%
\right.
\end{equation}%
We deduce that the transaction cost has two components:
\begin{equation}
\mathcal{TC}_{\mathrm{with}}=\mathcal{TC}_{\mathrm{cash}}\left( \min \left(
w_{\mathrm{cash}},\RedemptionRate\right) \cdot \limfunc{TNA}\right) +%
\mathcal{TC}_{\mathrm{asset}}\left( \left( \RedemptionRate-w_{\mathrm{cash}%
}\right) ^{+}\cdot \limfunc{TNA}\right)
\end{equation}%
where $\mathcal{TC}_{\mathrm{cash}}\left( V\right) $ is the transaction cost
when liquidating the amount $V$ of cash. Another more tractable expression of
$\mathcal{TC}_{\mathrm{with}}$ is:
\begin{eqnarray}
\mathcal{TC}_{\mathrm{with}} &=&\mathcal{TC}_{\mathrm{cash}}\left( %
\RedemptionRate\cdot \limfunc{TNA}\right) \cdot \mathds{1}\left\{ %
\RedemptionRate\leq w_{\mathrm{cash}}\right\} +  \notag \\
&&\mathcal{TC}_{\mathrm{asset}}\left( \left( \RedemptionRate-w_{\mathrm{cash}%
}\right) \cdot \limfunc{TNA}\right) \cdot \mathds{1}\left\{ \RedemptionRate%
>w_{\mathrm{cash}}\right\}
\end{eqnarray}%
It follows that the liquidation gain of implementing a cash buffer is:
\begin{equation}
\mathcal{LG}=\mathcal{TC}_{\mathrm{without}}-\mathcal{TC}_{\mathrm{with}}
\end{equation}%
We deduce that:%
\begin{eqnarray}
\mathcal{LG} &=&\mathcal{TC}_{\mathrm{asset}}\left( \RedemptionRate\cdot
\limfunc{TNA}\right) -\mathcal{TC}_{\mathrm{cash}}\left( \RedemptionRate%
\cdot \limfunc{TNA}\right) \cdot \mathds{1}\left\{ \RedemptionRate\leq w_{%
\mathrm{cash}}\right\} -  \notag \\
&&\mathcal{TC}_{\mathrm{asset}}\left( \left( \RedemptionRate-w_{\mathrm{cash}%
}\right) \cdot \limfunc{TNA}\right) \cdot \mathds{1}\left\{ \RedemptionRate%
>w_{\mathrm{cash}}\right\}   \notag \\
&=&\mathcal{TC}_{\mathrm{asset}}\left( \RedemptionRate\cdot \limfunc{TNA}%
\right) \cdot \mathds{1}\left\{ \RedemptionRate\leq w_{\mathrm{cash}%
}\right\} +\mathcal{TC}_{\mathrm{asset}}\left( \RedemptionRate\cdot \limfunc{%
TNA}\right) \cdot \mathds{1}\left\{ \RedemptionRate>w_{\mathrm{cash}%
}\right\} -  \notag \\
&&\mathcal{TC}_{\mathrm{cash}}\left( \RedemptionRate\cdot \limfunc{TNA}%
\right) \cdot \mathds{1}\left\{ \RedemptionRate\leq w_{\mathrm{cash}%
}\right\} -\mathcal{TC}_{\mathrm{asset}}\left( \left( \RedemptionRate-w_{%
\mathrm{cash}}\right) \cdot \limfunc{TNA}\right) \cdot \mathds{1}\left\{ %
\RedemptionRate>w_{\mathrm{cash}}\right\}   \notag \\
&=&\left( \mathcal{TC}_{\mathrm{asset}}\left( \RedemptionRate\cdot \limfunc{%
TNA}\right) -\mathcal{TC}_{\mathrm{cash}}\left( \RedemptionRate\cdot
\limfunc{TNA}\right) \right) \cdot \mathds{1}\left\{ \RedemptionRate\leq w_{%
\mathrm{cash}}\right\} +  \notag \\
&&\left( \mathcal{TC}_{\mathrm{asset}}\left( \RedemptionRate\cdot \limfunc{%
TNA}\right) -\mathcal{TC}_{\mathrm{asset}}\left( \left( \RedemptionRate-w_{%
\mathrm{cash}}\right) \cdot \limfunc{TNA}\right) \right) \cdot \mathds{1}%
\left\{ \RedemptionRate>w_{\mathrm{cash}}\right\}   \notag \\
&=&\mathcal{LG}_{\mathrm{cash}}+\mathcal{LG}_{\mathrm{asset}}
\label{eq:app-lg1}
\end{eqnarray}%
where:
\begin{equation}
\mathcal{LG}_{\mathrm{cash}}=\left( \mathcal{TC}_{\mathrm{asset}}\left( %
\RedemptionRate\cdot \limfunc{TNA}\right) -\mathcal{TC}_{\mathrm{cash}%
}\left( \RedemptionRate\cdot \limfunc{TNA}\right) \right) \cdot \mathds{1}%
\left\{ \RedemptionRate\leq w_{\mathrm{cash}}\right\}
\end{equation}%
and:
\begin{equation}
\mathcal{LG}_{\mathrm{asset}}=\left( \mathcal{TC}_{\mathrm{asset}}\left( %
\RedemptionRate\cdot \limfunc{TNA}\right) -\mathcal{TC}_{\mathrm{asset}%
}\left( \left( \RedemptionRate-w_{\mathrm{cash}}\right) \cdot \limfunc{TNA}%
\right) \right) \cdot \mathds{1}\left\{ \RedemptionRate>w_{\mathrm{cash}%
}\right\}
\end{equation}%
Finally, we conclude that:
\begin{eqnarray}
\mathbb{E}\left[ \mathcal{LG}\right]  &=&\mathbb{E}\left[ \mathcal{LG}_{%
\mathrm{cash}}\right] +\mathbb{E}\left[ \mathcal{LG}_{\mathrm{asset}}\right]
\notag \\
&=&\int_{0}^{w_{\mathrm{cash}}}\left( \mathcal{TC}_{\mathrm{asset}}\left( %
\RedemptionRate\cdot \limfunc{TNA}\right) -\mathcal{TC}_{\mathrm{cash}%
}\left( \RedemptionRate\cdot \limfunc{TNA}\right) \right) \,\mathrm{d}%
\mathbf{F}\left( \RedemptionRate\right) +  \notag \\
&&\int_{w_{\mathrm{cash}}}^{1}\left( \mathcal{TC}_{\mathrm{asset}}\left( %
\RedemptionRate\cdot \limfunc{TNA}\right) -\mathcal{TC}_{\mathrm{asset}%
}\left( \left( \RedemptionRate-w_{\mathrm{cash}}\right) \cdot \limfunc{TNA}%
\right) \right) \,\mathrm{d}\mathbf{F}\left( \RedemptionRate\right)
\label{eq:app-lg2}
\end{eqnarray}%
where $\mathbf{F}\left( x\right) $ is the distribution function of the
redemption rate $\RedemptionRate$.\smallskip

We can simplify the previous expressions in two different ways. If we assume
that $\mathcal{TC}_{\mathrm{cash}}\left( \RedemptionShock\right) \approx 0$,
we have:
\begin{equation}
\mathcal{LG}_{\mathrm{cash}}\approx \mathcal{TC}_{\mathrm{asset}}\left( %
\RedemptionRate\cdot \limfunc{TNA}\right) \cdot \mathds{1}\left\{ %
\RedemptionRate\leq w_{\mathrm{cash}}\right\}   \label{eq:app-lg3a}
\end{equation}%
and:
\begin{equation}
\mathbb{E}\left[ \mathcal{LG}_{\mathrm{cash}}\right] =\int_{0}^{w_{\mathrm{%
cash}}}\mathcal{TC}_{\mathrm{asset}}\left( \RedemptionRate\cdot \limfunc{TNA}%
\right) \,\mathrm{d}\mathbf{F}\left( \RedemptionRate\right)
\label{eq:app-lg3b}
\end{equation}%
We can also simplify $\mathcal{LG}_{\mathrm{asset}}$ with the following
approximation:
\begin{equation}
\mathcal{TC}_{\mathrm{asset}}\left( \left( \RedemptionRate-w_{\mathrm{cash}%
}\right) \cdot \limfunc{TNA}\right) \approx \mathcal{TC}_{\mathrm{asset}%
}\left( \RedemptionRate\cdot \limfunc{TNA}\right) -\mathcal{TC}_{\mathrm{%
asset}}\left( w_{\mathrm{cash}}\cdot \limfunc{TNA}\right)
\end{equation}%
This approximation is valid if the transaction cost function is perfectly
additive. This is not the case because of the price impact. However, the
transaction cost function may be decomposed as a sum of daily transaction
costs. Because of the liquidation policy limits, the daily transaction costs
are almost the same for large redemptions and can justify the previous
approximation. To better illustrate the underlying idea, let us assume that
$\RedemptionRate=30\%$ and $w_{\mathrm{cash}}=5\%$. Moreover, we suppose that
we can liquidate $5\%$ of the total net assets every day with a total cost of
$7$ bps in the stress regime\footnote{We recall that the transaction cost
function is expressed in $\%$ of the total net assets, and not with respect
to the liquidation amount. A total cost of $7$ bps for the fund when the
redemption rate is equal to $5\%$ is then equivalent to a unit transaction
cost of $140$ bps}. Liquidating $30\%$ is performed in 6 days:
$\mathcal{TC}_{\mathrm{asset}}\left( \RedemptionRate \cdot
\limfunc{TNA}\right) =6\times 7=42$ bps. Liquidating $30\%-5\%$ is performed
in 5 days and we have:
\begin{eqnarray}
\mathcal{TC}_{\mathrm{asset}}\left( \left( 30\%-5\%\right) \cdot \limfunc{TNA%
}\right)  &=&\mathcal{TC}_{\mathrm{asset}}\left( 25\%\cdot \limfunc{TNA}%
\right)   \notag \\
&=&5\times 7=35\text{ bps}  \notag \\
&=&42-7  \notag \\
&=&\mathcal{TC}_{\mathrm{asset}}\left( 30\%\cdot \limfunc{TNA}\right) -%
\mathcal{TC}_{\mathrm{asset}}\left( 5\%\cdot \limfunc{TNA}\right)
\end{eqnarray}%
In practice, we don't verify the strict equality because of many factors, but
we can consider that the approximation is relatively valid compared to all
uncertainties of a stress testing program. Therefore, we have:
\begin{eqnarray}
\mathcal{LG}_{\mathrm{asset}} &\approx &\left( \mathcal{TC}_{\mathrm{asset}%
}\left( \RedemptionRate\cdot \limfunc{TNA}\right) -\mathcal{TC}_{\mathrm{%
asset}}\left( \RedemptionRate\cdot \limfunc{TNA}\right) +\mathcal{TC}_{%
\mathrm{asset}}\left( w_{\mathrm{cash}}\cdot \limfunc{TNA}\right) \right)
\cdot \mathds{1}\left\{ \RedemptionRate>w_{\mathrm{cash}}\right\}   \notag \\
&=&\mathcal{TC}_{\mathrm{asset}}\left( w_{\mathrm{cash}}\cdot \limfunc{TNA}%
\right) \cdot \mathds{1}\left\{ \RedemptionRate>w_{\mathrm{cash}}\right\}
\label{eq:app-lg4a}
\end{eqnarray}%
and:%
\begin{eqnarray}
\mathbb{E}\left[ \mathcal{LG}_{\mathrm{asset}}\right]  &=&\int_{w_{\mathrm{%
cash}}}^{1}\mathcal{TC}_{\mathrm{asset}}\left( w_{\mathrm{cash}}\cdot
\limfunc{TNA}\right) \,\mathrm{d}\mathbf{F}\left( \RedemptionRate\right)
\notag \\
&=&\mathcal{TC}_{\mathrm{asset}}\left( w_{\mathrm{cash}}\cdot \limfunc{TNA}%
\right) \cdot \left( 1-\mathbf{F}\left( w_{\mathrm{cash}}\right) \right)
\label{eq:app-lg4b}
\end{eqnarray}%
Finally, we obtain:
\begin{eqnarray}
\mathbb{E}\left[ \mathcal{LG}\right]  &=&\int_{0}^{w_{\mathrm{cash}}}%
\mathcal{TC}_{\mathrm{asset}}\left( \RedemptionRate\cdot \limfunc{TNA}%
\right) \,\mathrm{d}\mathbf{F}\left( \RedemptionRate\right) +  \notag \\
&&\mathcal{TC}_{\mathrm{asset}}\left( w_{\mathrm{cash}}\cdot \limfunc{TNA}%
\right) \cdot \left( 1-\mathbf{F}\left( w_{\mathrm{cash}}\right) \right)
\label{eq:app-lg5}
\end{eqnarray}

\begin{remark}
Since $\mathcal{TC}_{\mathrm{asset}}\left( \RedemptionShock\right) $ is a
function, we can replace it by the function $\mathcal{TC}_{\mathrm{asset}%
}\left( \RedemptionRate\right) $ without any impact on the previous
equations. This is equivalent to normalize the total net assets --- $%
\limfunc{TNA}=1$.
\end{remark}

\subsubsection{First derivative of $\mathbb{E}\left[ \mathcal{LG}\left( w_{%
\mathrm{cash}}\right) \right] $}
\label{appendix:cash-deriv}

\paragraph{Exact formula}

The first derivative of $\mathbb{E}\left[ \mathcal{LG}_{\mathrm{cash}}\left(
w_{\mathrm{cash}}\right) \right] $ satisfies:
\begin{eqnarray}
\frac{\partial \,\mathbb{E}\left[ \mathcal{LG}_{\mathrm{cash}}\left( w_{%
\mathrm{cash}}\right) \right] }{\partial \,w_{\mathrm{cash}}} &=&\left(
\mathcal{TC}_{\mathrm{asset}}\left( w_{\mathrm{cash}}\right) -\mathcal{TC}_{%
\mathrm{cash}}\left( w_{\mathrm{cash}}\right) \right) \cdot f\left( w_{%
\mathrm{cash}}\right)   \notag \\
&\geq &0
\end{eqnarray}%
where $f\left( x\right) $ is the probability density function of the
redemption rate $\RedemptionRate$. For $\mathbb{E}\left[
\mathcal{LG}_{\mathrm{asset}}\left( w_{\mathrm{cash}}\right) \right] $, we
use the Leibniz integral rule:
\begin{eqnarray}
\frac{\partial \,\mathbb{E}\left[ \mathcal{LG}_{\mathrm{asset}}\left( w_{%
\mathrm{cash}}\right) \right] }{\partial \,w_{\mathrm{cash}}} &=&-\mathcal{TC%
}_{\mathrm{asset}}\left( w_{\mathrm{cash}}\right) \cdot f\left( w_{\mathrm{cash}%
}\right) +  \notag \\
&&\int_{w_{\mathrm{cash}}}^{1}\mathcal{TC}_{\mathrm{asset}}^{\prime }\left( %
\RedemptionRate-w_{\mathrm{cash}}\right) \,\mathrm{d}\mathbf{F}\left( %
\RedemptionRate\right)
\end{eqnarray}%
where $\mathcal{TC}_{\mathrm{asset}}^{\prime }$ is the derivative of the
transaction cost function, which is assumed to be positive. We have:
\begin{eqnarray}
\frac{\partial \,\mathbb{E}\left[ \mathcal{LG}_{\mathrm{asset}}\left(
0\right) \right] }{\partial \,w_{\mathrm{cash}}} &=&\int_{0}^{1}\mathcal{TC}%
_{\mathrm{asset}}^{\prime }\left( \RedemptionRate\right) \,\mathrm{d}\mathbf{%
F}\left( \RedemptionRate\right)   \notag \\
&\geq &0
\end{eqnarray}%
and:%
\begin{eqnarray}
\frac{\partial \,\mathbb{E}\left[ \mathcal{LG}_{\mathrm{asset}}\left(
1\right) \right] }{\partial \,w_{\mathrm{cash}}} &=&-\mathcal{TC}_{\mathrm{%
asset}}\left( 1\right) \cdot f\left( 1\right)   \notag \\
&<&0
\end{eqnarray}%
Finally, we obtain:
\begin{equation}
\frac{\partial \,\mathbb{E}\left[ \mathcal{LG}\left( w_{\mathrm{cash}%
}\right) \right] }{\partial \,w_{\mathrm{cash}}}
= -\mathcal{TC}_{\mathrm{cash}}\left( w_{\mathrm{cash}}\right) \cdot
f\left( w_{\mathrm{cash}}\right) +
\int_{w_{\mathrm{cash}}}^{1}\mathcal{TC}_{\mathrm{asset}}^{\prime }\left( %
\RedemptionRate-w_{\mathrm{cash}}\right) \,\mathrm{d}\mathbf{F}\left( %
\RedemptionRate\right)
\end{equation}%
It follows that:%
\begin{eqnarray}
\frac{\partial \,\mathbb{E}\left[ \mathcal{LG}\left( 0\right) \right] }{%
\partial \,w_{\mathrm{cash}}} &=&\int_{0}^{1}\mathcal{TC}_{\mathrm{asset}%
}^{\prime }\left( \RedemptionRate-w_{\mathrm{cash}}\right) \,\mathrm{d}%
\mathbf{F}\left( \RedemptionRate\right)   \notag \\
&\geq &0
\end{eqnarray}%
and:%
\begin{eqnarray}
\frac{\partial \,\mathbb{E}\left[ \mathcal{LG}\left( 1\right) \right] }{%
\partial \,w_{\mathrm{cash}}} &=&-\mathcal{TC}_{\mathrm{cash}}\left(
1\right) \cdot f\left( 1\right)   \notag \\
&<&0
\end{eqnarray}

\paragraph{Approximate formula}

The first derivative of $\mathbb{E}\left[ \mathcal{LG}_{\mathrm{cash}}\left(
w_{\mathrm{cash}}\right) \right] $ satisfies:%
\begin{eqnarray}
\frac{\partial \,\mathbb{E}\left[ \mathcal{LG}_{\mathrm{cash}}\left( w_{%
\mathrm{cash}}\right) \right] }{\partial \,w_{\mathrm{cash}}} &=&\mathcal{TC}%
_{\mathrm{asset}}\left( w_{\mathrm{cash}}\right) \cdot f\left( w_{\mathrm{%
cash}}\right)   \notag \\
&\geq &0
\end{eqnarray}%
For $\mathbb{E}\left[ \mathcal{LG}_{\mathrm{asset}}\left( w_{\mathrm{cash}%
}\right) \right] $, we obtain:%
\begin{equation}
\frac{\partial \,\mathbb{E}\left[ \mathcal{LG}_{\mathrm{asset}}\left( w_{%
\mathrm{cash}}\right) \right] }{\partial \,w_{\mathrm{cash}}}=-\mathcal{TC}_{%
\mathrm{asset}}\left( w_{\mathrm{cash}}\right) \cdot f\left( w_{\mathrm{cash}%
}\right) +\mathcal{TC}_{\mathrm{asset}}^{\prime }\left( w_{\mathrm{cash}%
}\right) \cdot \left( 1-\mathbf{F}\left( w_{\mathrm{cash}}\right) \right)
\end{equation}%
We have:%
\begin{eqnarray}
\frac{\partial \,\mathbb{E}\left[ \mathcal{LG}_{\mathrm{asset}}\left(
0\right) \right] }{\partial \,w_{\mathrm{cash}}} &=&\mathcal{TC}_{\mathrm{%
asset}}^{\prime }\left( 0\right) \cdot \left( 1-\mathbf{F}\left( 0\right)
\right)   \notag \\
&\geq &0
\end{eqnarray}%
and:%
\begin{eqnarray}
\frac{\partial \,\mathbb{E}\left[ \mathcal{LG}_{\mathrm{asset}}\left(
1\right) \right] }{\partial \,w_{\mathrm{cash}}} &=&-\mathcal{TC}_{\mathrm{%
asset}}\left( 1\right) \cdot f\left( 1\right)   \notag \\
&<&0
\end{eqnarray}%
Finally, we conclude that:%
\begin{eqnarray}
\frac{\partial \,\mathbb{E}\left[ \mathcal{LG}\left( w_{\mathrm{cash}%
}\right) \right] }{\partial \,w_{\mathrm{cash}}} &=&\mathcal{TC}_{\mathrm{%
asset}}\left( w_{\mathrm{cash}}\right) \cdot f\left( w_{\mathrm{cash}%
}\right) -\mathcal{TC}_{\mathrm{asset}}\left( w_{\mathrm{cash}}\right)
f\left( w_{\mathrm{cash}}\right) +  \notag \\
&&\mathcal{TC}_{\mathrm{asset}}^{\prime }\left( w_{\mathrm{cash}}\right)
\cdot \left( 1-\mathbf{F}\left( w_{\mathrm{cash}}\right) \right)   \notag \\
&=&\mathcal{TC}_{\mathrm{asset}}^{\prime }\left( w_{\mathrm{cash}}\right)
\cdot \left( 1-\mathbf{F}\left( w_{\mathrm{cash}}\right) \right)   \notag \\
&\geq &0
\end{eqnarray}

\subsubsection{Closed-form formula of Example \protect\ref{ex:cash3} on page \protect\pageref{ex:cash3}}
\label{appendix:cash3-ex}

\paragraph{Exact formula}

If $\mathcal{TC}_{\mathrm{asset}}\left( x\right) =x\cdot \left( \spread%
+\beta _{\impact}\sigma \sqrt{x}\right) $, $\mathcal{TC}_{\mathrm{cash}%
}\left( x\right) =x\cdot \cashRate$ and $\mathbf{F}\left( x\right) =x^{\eta }
$, we have:%
\begin{eqnarray}
\mathbb{E}\left[ \mathcal{LG}_{\mathrm{cash}}\left( w_{\mathrm{cash}}\right) %
\right]  &=&\int_{0}^{w_{\mathrm{cash}}}\left( \mathcal{TC}_{\mathrm{asset}%
}\left( \RedemptionRate\right) -\mathcal{TC}_{\mathrm{cash}}\left( %
\RedemptionRate\right) \right) \,\mathrm{d}\mathbf{F}\left( \RedemptionRate%
\right)   \notag \\
&=&\eta \int_{0}^{w_{\mathrm{cash}}}x\cdot \left( \spread-\cashRate+\beta _{%
\impact}\sigma \sqrt{x}\right) \cdot x^{\eta -1}\,\mathrm{d}x  \notag \\
&=&\eta \left( \spread-\cashRate\right) \int_{0}^{w_{\mathrm{cash}}}x^{\eta
}\,\mathrm{d}x+\eta \beta _{\impact}\sigma \int_{0}^{w_{\mathrm{cash}%
}}x^{\eta +0.5}\,\mathrm{d}x  \notag \\
&=&\frac{\eta \left( \spread-\cashRate\right) }{\eta +1}\cdot w_{\mathrm{cash%
}}^{\eta +1}+\frac{2\eta \beta _{\impact}\sigma }{2\eta +3}\cdot w_{\mathrm{%
cash}}^{\eta +1.5}
\end{eqnarray}%
and:%
\begin{eqnarray}
\mathbb{E}\left[ \mathcal{LG}_{\mathrm{asset}}\left( w_{\mathrm{cash}%
}\right) \right]  &=&\int_{w_{\mathrm{cash}}}^{1}\left( \mathcal{TC}_{%
\mathrm{asset}}\left( \RedemptionRate\right) -\mathcal{TC}_{\mathrm{asset}%
}\left( \left( \RedemptionRate-w_{\mathrm{cash}}\right) \right) \right) \,%
\mathrm{d}\mathbf{F}\left( \RedemptionRate\right)   \notag \\
&=&\eta \int_{w_{\mathrm{cash}}}^{1}\left( \spread w_{\mathrm{cash}}+\beta _{%
\impact}\sigma \left( x\sqrt{x}-\left( x-w_{\mathrm{cash}}\right) \sqrt{x-w_{%
\mathrm{cash}}}\right) \right) \cdot x^{\eta -1}\,\mathrm{d}x  \notag \\
&=&\eta \spread w_{\mathrm{cash}}\int_{w_{\mathrm{cash}}}^{1}\,\mathrm{d}%
x+\eta \beta _{\impact}\sigma \int_{w_{\mathrm{cash}}}^{1}x^{\eta +0.5}\,%
\mathrm{d}x-  \notag \\
&&\eta \beta _{\impact}\sigma \int_{w_{\mathrm{cash}}}^{1}\left( x-w_{%
\mathrm{cash}}\right) ^{1.5}x^{\eta -1}\,\mathrm{d}x
\end{eqnarray}%
If we denote by $I\left( w_{\mathrm{cash}};\eta \right) $\ the integral $%
\int_{w_{\mathrm{cash}}}^{1}\left( x-w_{\mathrm{cash}}\right) ^{1.5}x^{\eta
-1}\,\mathrm{d}x$, we obtain:%
\begin{equation}
\mathbb{E}\left[ \mathcal{LG}_{\mathrm{asset}}\left( w_{\mathrm{cash}%
}\right) \right] =\eta \spread\cdot w_{\mathrm{cash}}\left( 1-w_{\mathrm{cash%
}}\right) +\frac{2\eta \beta _{\impact}\sigma }{2\eta +3}\cdot \left( 1-w_{%
\mathrm{cash}}^{\eta +1.5}\right) -\eta \beta _{\impact}\sigma \cdot I\left(
w_{\mathrm{cash}};\eta \right)
\end{equation}%
where\footnote{See Equation (\ref{eq:app-hypergeometric2}) in Appendix
\ref{appendix:hypergeometric} on page \pageref{eq:app-hypergeometric2}.}:
\begin{equation}
I\left( w_{\mathrm{cash}};\eta \right) =\frac{2}{5}\left( 1-w_{\mathrm{cash}%
}\right) ^{5/2}w_{\mathrm{cash}}^{\eta -1}\,_{2}\mathfrak{F}_{1}\left(
1-\eta ,\frac{5}{2};\frac{7}{2};\frac{w_{\mathrm{cash}}-1}{w_{\mathrm{cash}}}%
\right)
\end{equation}
We deduce that:%
\begin{eqnarray}
\mathbb{E}\left[ \mathcal{LG}\left( w_{\mathrm{cash}}\right) \right]  &=&%
\frac{\eta \left( \spread-\cashRate\right) }{\eta +1}\cdot w_{\mathrm{cash}%
}^{\eta +1}+\frac{2\eta \beta _{\impact}\sigma }{2\eta +3}+  \notag \\
&&\eta \spread\cdot w_{\mathrm{cash}}\left( 1-w_{\mathrm{cash}}\right) -\eta
\beta _{\impact}\sigma \cdot I\left( w_{\mathrm{cash}};\eta \right)
\end{eqnarray}

\paragraph{Approximate formula}

We have:%
\begin{eqnarray}
\mathbb{E}\left[ \mathcal{LG}_{\mathrm{cash}}\left( w_{\mathrm{cash}}\right) %
\right]  &=&\int_{0}^{w_{\mathrm{cash}}}\mathcal{TC}_{\mathrm{asset}}\left(
x\right) \,\mathrm{d}\mathbf{F}\left( x\right)   \notag \\
&=&\eta \int_{0}^{w_{\mathrm{cash}}}x\cdot \left( \spread+\beta _{\impact%
}\sigma \sqrt{x}\right) \cdot x^{\eta -1}\,\mathrm{d}x  \notag \\
&=&\eta \spread\int_{0}^{w_{\mathrm{cash}}}x^{\eta }\,\mathrm{d}x+\eta \beta
_{\impact}\sigma \int_{0}^{w_{\mathrm{cash}}}x^{\eta +0.5}\,\mathrm{d}x
\notag \\
&=&\frac{\eta \spread}{\eta +1}\cdot w_{\mathrm{cash}}^{\eta +1}+\frac{2\eta
\beta _{\impact}\sigma }{2\eta +3}\cdot w_{\mathrm{cash}}^{\eta +1.5}
\end{eqnarray}%
and:%
\begin{eqnarray}
\mathbb{E}\left[ \mathcal{LG}_{\mathrm{asset}}\left( w_{\mathrm{cash}%
}\right) \right]  &=&\mathcal{TC}_{\mathrm{asset}}\left( w_{\mathrm{cash}%
}\right) \cdot \left( 1-\mathbf{F}\left( w_{\mathrm{cash}}\right) \right)
\notag \\
&=&w_{\mathrm{cash}}\cdot \left( \spread+\beta _{\impact}\sigma \sqrt{w_{%
\mathrm{cash}}}\right) \cdot \left( 1-w_{\mathrm{cash}}^{\eta }\right)
\notag \\
&=&\spread\cdot w_{\mathrm{cash}}-\spread\cdot w_{\mathrm{cash}}^{\eta
+1}+\beta _{\impact}\sigma \cdot w_{\mathrm{cash}}^{1.5}-\beta _{\impact%
}\sigma \cdot w_{\mathrm{cash}}^{\eta +1.5}
\end{eqnarray}%
We deduce that:%
\begin{equation}
\mathbb{E}\left[ \mathcal{LG}\left( w_{\mathrm{cash}}\right) \right] =\spread%
\cdot w_{\mathrm{cash}}+\beta _{\impact}\sigma \cdot w_{\mathrm{cash}}^{1.5}-%
\frac{\spread}{\eta +1}\cdot w_{\mathrm{cash}}^{\eta +1}-\frac{3\beta _{%
\impact}\sigma }{2\eta +3}\cdot w_{\mathrm{cash}}^{\eta +1.5}
\end{equation}

\subsubsection{Closed-form formula of Example \protect\ref{ex:cash4} on page
\protect\pageref{ex:cash4}}
\label{appendix:cash4-ex}

We have:
\begin{eqnarray}
\mathcal{TC}_{\mathrm{asset}}\left( x\right)  &=&\underset{\text{linear}}{%
\underbrace{\vphOne x\cdot \spread}+}\underset{\text{constant}}{\underbrace{\vphOne \kappa
\beta _{\impact}\sigma \cdot x^{+}\sqrt{x^{+}}}}+\underset{\text{nonlinear}}{%
\underbrace{\beta _{\impact}\sigma \cdot \left( x-\kappa x^{+}\right) \sqrt{%
x-\kappa x^{+}}}}  \notag \\
&:=&g\left( x;\kappa ,x^{+}\right)
\end{eqnarray}%
where:%
\begin{equation}
\kappa :=\kappa \left( x;x^{+}\right) =\left\lfloor \frac{x}{x^{+}}%
\right\rfloor
\end{equation}%
We recall that $\mathcal{TC}_{\mathrm{cash}}\left( x\right) =x\cdot \cashRate
$ and $\mathbf{F}\left( x\right) =x^{\eta }$. By denoting $\kappa
_{\mathrm{cash}}=\kappa \left( w_{\mathrm{cash}};x^{+}\right) $, we obtain:
\begin{eqnarray}
\mathbb{E}\left[ \mathcal{LG}_{\mathrm{cash}}\left( w_{\mathrm{cash}}\right) %
\right]  &=&\int_{0}^{w_{\mathrm{cash}}}\mathcal{TC}_{\mathrm{asset}}\left(
x\right) \,\mathrm{d}\mathbf{F}\left( x\right)   \notag \\
&=&\int_{0}^{x^{+}}\mathcal{TC}_{\mathrm{asset}}\left( x\right) \,\mathrm{d}%
\mathbf{F}\left( x\right) +\int_{x^{+}}^{2x^{+}}\mathcal{TC}_{\mathrm{asset}%
}\left( x\right) \,\mathrm{d}\mathbf{F}\left( x\right) +\ldots + \notag \\
&&
\int_{\left(
\kappa _{\mathrm{cash}}-1\right) x^{+}}^{\kappa _{\mathrm{cash}}x^{+}}%
\mathcal{TC}_{\mathrm{asset}}\left( x\right) \,\mathrm{d}\mathbf{F}\left(
x\right) +
\int_{\kappa _{\mathrm{cash}}x^{+}}^{w_{\mathrm{cash}}}\mathcal{TC}_{%
\mathrm{asset}}\left( x\right) \,\mathrm{d}\mathbf{F}\left( x\right)   \notag
\\
&=&\sum_{k=1}^{\kappa _{\mathrm{cash}}}\int_{\left( k-1\right)
x^{+}}^{kx^{+}}\mathcal{TC}_{\mathrm{asset}}\left( x\right) \,\mathrm{d}%
\mathbf{F}\left( x\right) +\int_{\kappa _{\mathrm{cash}}x^{+}}^{w_{\mathrm{%
cash}}}\mathcal{TC}_{\mathrm{asset}}\left( x\right) \,\mathrm{d}\mathbf{F}%
\left( x\right) \notag \\
&&
\end{eqnarray}%
We have the following cases:%
\begin{equation}
\mathcal{TC}_{\mathrm{asset}}\left( x\right) =\left\{
\begin{array}{ll}
g\left( x;k-1,x^{+}\right)  & \text{if }x\in \left[ \left( k-1\right)
x^{+},kx^{+}\right]  \\
g\left( x;\kappa _{\mathrm{cash}},x^{+}\right)  & x\in \left[ \kappa _{%
\mathrm{cash}}x^{+},w_{\mathrm{cash}}\right]
\end{array}%
\right.
\end{equation}%
We deduce that:%
\begin{eqnarray}
(\ast ) &=&\int_{\left( k-1\right) x^{+}}^{kx^{+}}\mathcal{TC}_{\mathrm{asset%
}}\left( x\right) \,\mathrm{d}\mathbf{F}\left( x\right)   \notag \\
&=&\eta \spread\int_{\left( k-1\right) x^{+}}^{kx^{+}}x^{\eta }\,\mathrm{d}x+
\notag \\
&&\eta \left( k-1\right) \beta _{\impact}\sigma x^{+}\sqrt{x^{+}}%
\int_{\left( k-1\right) x^{+}}^{kx^{+}}x^{\eta -1}\,\mathrm{d}x+  \notag \\
&&\eta \beta _{\impact}\sigma \int_{\left( k-1\right) x^{+}}^{kx^{+}}\left(
x-\left( k-1\right) x^{+}\right) \sqrt{x-\left( k-1\right) x^{+}}x^{\eta
-1}\,\mathrm{d}x
\end{eqnarray}%
and:%
\begin{eqnarray}
(\ast ) &=&\int_{\kappa _{\mathrm{cash}}x^{+}}^{w_{\mathrm{cash}}}\mathcal{TC%
}_{\mathrm{asset}}\left( x\right) \,\mathrm{d}\mathbf{F}\left( x\right)
\notag \\
&=&\eta \spread\int_{\kappa _{\mathrm{cash}}x^{+}}^{w_{\mathrm{cash}%
}}x^{\eta }\,\mathrm{d}x+  \notag \\
&&\eta \kappa _{\mathrm{cash}}\beta _{\impact}\sigma x^{+}\sqrt{x^{+}}%
\int_{\kappa _{\mathrm{cash}}x^{+}}^{w_{\mathrm{cash}}}x^{\eta -1}\,\mathrm{d%
}x+  \notag \\
&&\eta \beta _{\impact}\sigma \int_{\kappa _{\mathrm{cash}}x^{+}}^{w_{%
\mathrm{cash}}}\left( x-\kappa _{\mathrm{cash}}x^{+}\right) \sqrt{x-\kappa _{%
\mathrm{cash}}x^{+}}x^{\eta -1}\,\mathrm{d}x
\end{eqnarray}%
For the first and second terms, we have:%
\begin{eqnarray}
\int_{a}^{b}x^{\eta }\,\mathrm{d}x &=&\left[ \frac{x^{\eta +1}}{\eta +1}%
\right] _{a}^{b}  \notag \\
&=&\frac{b^{\eta +1}-a^{\eta +1}}{\eta +1}
\end{eqnarray}%
and:%
\begin{equation}
\int_{a}^{b}x^{\eta -1}\,\mathrm{d}x=\frac{b^{\eta }-a^{\eta }}{\eta }
\end{equation}%
We also notice that:%
\begin{equation}
\sum_{k=1}^{\kappa _{\mathrm{cash}}}\frac{\left( kx^{+}\right) ^{\eta
+1}-\left( \left( k-1\right) x^{+}\right) ^{\eta +1}}{\eta +1}+\frac{w_{%
\mathrm{cash}}^{\eta +1}-\left( \kappa _{\mathrm{cash}}x^{+}\right) ^{\eta
+1}}{\eta +1}=\frac{w_{\mathrm{cash}}^{\eta +1}}{\eta +1}
\end{equation}%
We note:%
\begin{equation*}
\mathcal{H}\left( w_{\mathrm{cash}},\kappa _{\mathrm{cash}},x^{+}\right)
=\sum_{k=1}^{\kappa _{\mathrm{cash}}}\left( k-1\right) \left( kx^{+}\right)
^{\eta }-\left( \left( k-1\right) x^{+}\right) ^{\eta }+\kappa _{\mathrm{cash%
}}\left( w_{\mathrm{cash}}^{\eta }-\left( \kappa _{\mathrm{cash}%
}x^{+}\right) ^{\eta }\right)
\end{equation*}%
For the third term, we have:%
\begin{equation}
\int_{a}^{b}\left( x-a\right) \sqrt{x-a}x^{\eta -1}\,\mathrm{d}x=I\left(
a,b;\eta \right)
\end{equation}%
Except for some specific values\footnote{%
See Appendix \ref{appendix:sinh} on page \pageref{appendix:sinh}.} of $\eta $,
this term has no closed-form formula and we use a numerical solution. We
conclude that:
\begin{eqnarray}
\mathbb{E}\left[ \mathcal{LG}_{\mathrm{cash}}\left( w_{\mathrm{cash}}\right) %
\right]  &=&\eta \spread\frac{w_{\mathrm{cash}}^{\eta +1}}{\eta +1}+  \notag
\\
&&\beta _{\impact}\sigma x^{+}\sqrt{x^{+}}\mathcal{H}\left( w_{\mathrm{cash}%
},\kappa _{\mathrm{cash}},x^{+}\right) + \\
&&\eta \beta _{\impact}\sigma \left( \sum_{k=1}^{\kappa _{\mathrm{cash}%
}}I\left( \left( k-1\right) x^{+},kx^{+};\eta \right) +I\left( \kappa _{%
\mathrm{cash}}x^{+},w_{\mathrm{cash}};\eta \right) \right)   \notag \\
&&
\end{eqnarray}%
The computation of $\mathbb{E}\left[ \mathcal{LG}_{\mathrm{asset}}\left( w_{%
\mathrm{cash}}\right) \right] $ gives:%
\begin{eqnarray}
\mathbb{E}\left[ \mathcal{LG}_{\mathrm{asset}}\left( w_{\mathrm{cash}%
}\right) \right]  &=&\mathcal{TC}_{\mathrm{asset}}\left( w_{\mathrm{cash}%
}\right) \cdot \left( 1-\mathbf{F}\left( w_{\mathrm{cash}}\right) \right)
\notag \\
&=&\spread\left( w_{\mathrm{cash}}-w_{\mathrm{cash}}^{\eta +1}\right)
+\kappa _{\mathrm{cash}}\beta _{\impact}\sigma x^{+}\sqrt{x^{+}}\left( 1-w_{%
\mathrm{cash}}^{\eta }\right) +  \notag \\
&&\beta _{\impact}\sigma \cdot \left( w_{\mathrm{cash}}-\kappa _{\mathrm{cash%
}}x^{+}\right) \sqrt{w_{\mathrm{cash}}-\kappa _{\mathrm{cash}}x^{+}}\cdot
\left( 1-w_{\mathrm{cash}}^{\eta }\right)
\end{eqnarray}%
We conclude that:%
\begin{eqnarray}
\mathbb{E}\left[ \mathcal{LG}\left( w_{\mathrm{cash}}\right) \right]
&=&\eta \spread\frac{w_{\mathrm{cash}}^{\eta +1}}{\eta +1}+\beta _{\impact%
}\sigma x^{+}\sqrt{x^{+}}\mathcal{H}\left( w_{\mathrm{cash}},\kappa _{%
\mathrm{cash}},x^{+}\right) +  \notag \\
&&\eta \beta _{\impact}\sigma \left( \sum_{k=1}^{\kappa _{\mathrm{cash}%
}}I\left( \left( k-1\right) x^{+},kx^{+};\eta \right) +I\left( \kappa _{%
\mathrm{cash}}x^{+},w_{\mathrm{cash}};\eta \right) \right) +  \notag \\
&&\spread\left( w_{\mathrm{cash}}-w_{\mathrm{cash}}^{\eta +1}\right) +\kappa
_{\mathrm{cash}}\beta _{\impact}\sigma x^{+}\sqrt{x^{+}}\left( 1-w_{\mathrm{%
cash}}^{\eta }\right) +  \notag \\
&&\beta _{\impact}\sigma \cdot \left( w_{\mathrm{cash}}-\kappa_{\mathrm{cash}} x^{+}\right)
\sqrt{w_{\mathrm{cash}}-\kappa_{\mathrm{cash}} x^{+}}\cdot \left( 1-w_{\mathrm{cash}}^{\eta
}\right)
\end{eqnarray}

\subsection{Computation of the integral function
$I\left( w_{\mathrm{cash}};\protect\eta \right) $}
\label{appendix:hypergeometric}

We consider the following integral:
\begin{equation}
I\left( w_{\mathrm{cash}};\eta \right) =\int_{w_{\mathrm{cash}}}^{1}\left(
x-w_{\mathrm{cash}}\right) ^{3/2}x^{\eta -1}\,\mathrm{d}x
\end{equation}%
where $\eta >0$ and $w_{\mathrm{cash}}\in \left[ 0,1\right] $.

\subsubsection{Preliminary result}

Let $\mathfrak{B}\left( a,b\right) =\int_{0}^{1}x^{a-1}\left( 1-x\right)
^{b-1}\,\mathrm{d}x$ and $_{2}\mathfrak{F}_{1}\left( a,b;c;z\right) ={%
\displaystyle\sum_{n=0}^{\infty }}\dfrac{\left( a\right) _{n}\left( b\right)
_{n}}{\left( c\right) _{n}}\dfrac{z^{n}}{n!}$ be the beta function and the
ordinary hypergeometric function\footnote{$\left( a\right) _{n}$ is the
rising Pochhammer symbol.}. If $c>b>0$, we know that:
\begin{equation}
\mathfrak{B}\left( b,c-b\right) \,_{2}\mathfrak{F}_{1}\left( a,b;c;z\right)
=\int_{0}^{1}x^{b-1}\left( 1-x\right) ^{c-b-1}\left( 1-zx\right) ^{-a}\,%
\mathrm{d}x  \label{eq:app-hypergeometric1}
\end{equation}%
If $c=1+b$, we deduce that:
\begin{eqnarray}
\int_{0}^{1}x^{b-1}\left( 1-zx\right) ^{-a}\,\mathrm{d}x &=&\mathfrak{B}%
\left( b,1\right) \,_{2}\mathfrak{F}_{1}\left( a,b;c;z\right)   \notag \\
&=&\frac{_{2}\mathfrak{F}_{1}\left( a,b;c;z\right) }{b}
\end{eqnarray}%
because:%
\begin{eqnarray}
\mathfrak{B}\left( b,1\right)  &=&\int_{0}^{1}x^{b-1}\,\mathrm{d}x  \notag \\
&=&\left[ \frac{x^{b}}{b}\right] _{0}^{1}  \notag \\
&=&\frac{1}{b}
\end{eqnarray}

\subsubsection{Main result}

We consider the change of variable:%
\begin{equation}
y=\frac{x-w_{\mathrm{cash}}}{1-w_{\mathrm{cash}}}
\end{equation}%
We deduce that:%
\begin{eqnarray}
I\left( w_{\mathrm{cash}};\eta \right)  &=&\int_{0}^{1}\left( \left( 1-w_{%
\mathrm{cash}}\right) y\right) ^{3/2}\left( w_{\mathrm{cash}}+\left( 1-w_{%
\mathrm{cash}}\right) y\right) ^{\eta -1}\left( 1-w_{\mathrm{cash}}\right) \,%
\mathrm{d}y  \notag \\
&=&\left( 1-w_{\mathrm{cash}}\right) ^{5/2}w_{\mathrm{cash}}^{\eta
-1}\int_{0}^{1}y^{3/2}\left( 1+\left( \frac{1-w_{\mathrm{cash}}}{w_{\mathrm{%
cash}}}\right) y\right) ^{\eta -1}\,\mathrm{d}y  \notag \\
&=&\left( 1-w_{\mathrm{cash}}\right) ^{5/2}w_{\mathrm{cash}}^{\eta -1}%
\mathfrak{B}\left( \frac{5}{2},1\right) \,_{2}\mathfrak{F}_{1}\left( 1-\eta ,%
\frac{5}{2};\frac{7}{2};\frac{w_{\mathrm{cash}}-1}{w_{\mathrm{cash}}}\right)
\notag \\
&=&\frac{2}{5}\left( 1-w_{\mathrm{cash}}\right) ^{5/2}w_{\mathrm{cash}%
}^{\eta -1}\,_{2}\mathfrak{F}_{1}\left( 1-\eta ,\frac{5}{2};\frac{7}{2};%
\frac{w_{\mathrm{cash}}-1}{w_{\mathrm{cash}}}\right)
\label{eq:app-hypergeometric2}
\end{eqnarray}

\begin{remark}
From a theoretical point of view, Equation (\ref{eq:app-hypergeometric2}) is
only valid for $0.5<w_{\mathrm{cash}}\leq 1$ if we adopt the definition
$_{2}\mathfrak{F}_{1}\left( a,b;c;z\right) ={\displaystyle\sum_{n=0}^{\infty
}}\dfrac{\left( a\right) _{n}\left( b\right) _{n}}{\left( c\right)
_{n}}\dfrac{z^{n}}{n!}$ because we must have $\left\vert z\right\vert <1$.
Nevertheless, we can show that Equation (\ref{eq:app-hypergeometric1})
remains valid for $\left\vert z\right\vert \geq 1$ if we consider that the
hypergeometric function is the solution of Euler's hypergeometric
differential equation. In this case, we can use Equation
(\ref{eq:app-hypergeometric2}) for $0\leq w_{\mathrm{cash}}\leq 0.5$, but we
must be careful about the numerical implementation of the hypergeometric
function $_{2}\mathfrak{F}_{1}\left( a,b;c;z\right) $.
\end{remark}

\subsubsection{Special cases}

\paragraph{Specific values of $\protect\eta$}

If $\eta =0.5$, we have:%
\begin{eqnarray}
I\left( w_{\mathrm{cash}};0.5\right)  &=&\int_{w_{\mathrm{cash}}}^{1}\frac{%
\left( x-w_{\mathrm{cash}}\right) ^{3/2}}{\sqrt{x}}\,\mathrm{d}x  \notag \\
&=&\frac{\left( 4-10w_{\mathrm{cash}}\right) \sqrt{1-w_{\mathrm{cash}}}+3w_{%
\mathrm{cash}}^{2}\left( 2\ln \left( 1+\sqrt{1-w_{\mathrm{cash}}}\right)
-\ln \left( w_{\mathrm{cash}}\right) \right) }{8} \notag \\
&&
\end{eqnarray}%
\smallskip

If $\eta =1$, we obtain:
\begin{eqnarray}
I\left( w_{\mathrm{cash}};1\right) &=&\int_{w_{\mathrm{cash}}}^{1}\left(
x-w_{\mathrm{cash}}\right) ^{3/2}\,\mathrm{d}x  \notag \\
&=&\left[ \frac{\left( x-w_{\mathrm{cash}}\right) ^{5/2}}{2.5}\right] _{w_{%
\mathrm{cash}}}^{1}  \notag \\
&=&\frac{2}{5}\left( 1-w_{\mathrm{cash}}\right) ^{5/2}
\end{eqnarray}%
We verify that:
\begin{equation}
\,_{2}\mathfrak{F}_{1}\left( 1-1,\frac{5}{2},\frac{7}{2};z\right) =1
\end{equation}
\smallskip

If $\eta =2$, we have:
\begin{equation}
I\left( w_{\mathrm{cash}};2\right) =\int_{w_{\mathrm{cash}}}^{1}\left( x-w_{%
\mathrm{cash}}\right) ^{3/2}x\,\mathrm{d}x
\end{equation}%
Let $y=\left( x-w_{\mathrm{cash}}\right) ^{3/2}$, we have $x=y^{2/3}+w_{%
\mathrm{cash}}$ and $\mathrm{d}x=\dfrac{2}{3}y^{-1/3}\,\mathrm{d}y$. It
follows that:%
\begin{eqnarray}
I\left( w_{\mathrm{cash}};2\right) &=&\int_{0}^{\left( 1-w_{\mathrm{cash}%
}\right) ^{3/2}}y\left( y^{2/3}+w_{\mathrm{cash}}\right) \frac{2}{3}%
y^{-1/3}\,\mathrm{d}y  \notag \\
&=&\frac{2}{3}\int_{0}^{\left( 1-w_{\mathrm{cash}}\right) ^{3/2}}\left(
y^{4/3}+w_{\mathrm{cash}}y^{2/3}\right) \,\mathrm{d}y  \notag \\
&=&\frac{2}{3}\left[ \frac{3y^{7/3}}{7}+3w_{\mathrm{cash}}\frac{y^{5/3}}{5}%
\right] _{0}^{\left( 1-w_{\mathrm{cash}}\right) ^{3/2}}  \notag \\
&=&\frac{2\left( 1-w_{\mathrm{cash}}\right) ^{7/2}}{7}+\frac{2w_{\mathrm{cash%
}}\left( 1-w_{\mathrm{cash}}\right) ^{5/2}}{5}  \notag \\
&=&\frac{2}{5}\left( 1-w_{\mathrm{cash}}\right) ^{5/2}w_{\mathrm{cash}%
}\left( 1+\frac{5}{7}\left( \frac{1-w_{\mathrm{cash}}}{w_{\mathrm{cash}}}%
\right) \right)
\end{eqnarray}%
We verify that:
\begin{equation}
\,_{2}\mathfrak{F}_{1}\left( 1-2,\frac{5}{2},\frac{7}{2};z\right) =
1-\frac{5}{7}z
\end{equation}
\smallskip

If $\eta $ is integer, $_{2}\mathfrak{F}_{1}\left( 1-\eta
,\frac{5}{2},\frac{7}{2};z\right) $ is a polynomial function. The analytical
solution can be computed using the Wolfram's alpha
platform\footnote{\url{https://www.wolframalpha.com}} and the hypergeometric
function \texttt{Hypergeometric2F1[1-eta,5/2,7/2,z]} by replacing
\texttt{eta} with the corresponding integer. For instance, we have:
\begin{equation}
\,_{2}\mathfrak{F}_{1}\left( 1-3,\frac{5}{2},\frac{7}{2};z\right) =
\frac{35z^{2}-90z+63}{63}
\end{equation}%
and:%
\begin{equation}
\,_{2}\mathfrak{F}_{1}\left( 1-3,\frac{5}{2},\frac{7}{2};z\right) =
\frac{-105z^{3}+385z^{2}-495z+231}{231}
\end{equation}%
It is then straightforward to find the analytical solution of $I\left(
w_{\mathrm{cash}};\eta \right) $ by using Equation
(\ref{eq:app-hypergeometric2}) and replacing $z$ by
$\dfrac{w_{\mathrm{cash}}-1}{w_{\mathrm{cash}}}$.

\paragraph{Specific values of $w_{\mathrm{cash}}$}

If $w_{\mathrm{cash}}=0$, we have:
\begin{equation}
I\left( 0;\eta \right) =\frac{2}{2\eta +3}
\end{equation}%
We deduce that:
\begin{equation}
\mathbb{E}\left[ \mathcal{LG}_{\mathrm{asset}}\left( 0\right) \right] =0
\end{equation}
\smallskip

If $w_{\mathrm{cash}}=0.5$, we have:
\begin{equation}
_{2}\mathfrak{F}_{1}\left( 1-\eta ,\frac{5}{2},\frac{7}{2};1\right) =\frac{15%
\sqrt{\pi }\Gamma \left( \eta \right) }{8\Gamma \left( \eta +\frac{5}{2}%
\right) }
\end{equation}%
and:
\begin{equation}
I\left( 0.5;\eta \right) =\frac{3\sqrt{\pi }\Gamma \left( \eta \right) }{%
2^{\eta +7/2}\Gamma \left( \eta +\frac{5}{2}\right) }
\end{equation}%
We deduce that:
\begin{equation}
\mathbb{E}\left[ \mathcal{LG}_{\mathrm{asset}}\left( 0.5\right) \right] =%
\frac{\eta \spread}{4}+\eta \beta _{\impact}\sigma \left( \frac{2}{2\eta +3}%
\left( 1-\frac{1}{2}^{\eta +1.5}\right) -\frac{3\sqrt{\pi }\Gamma \left(
\eta \right) }{2^{\eta +7/2}\Gamma \left( \eta +\frac{5}{2}\right) }\right)
\end{equation}
\smallskip

If $w_{\mathrm{cash}}=1$, we have:
\begin{equation}
_{2}\mathfrak{F}_{1}\left( 1-\eta ,\frac{5}{2},\frac{7}{2};0\right) =1
\end{equation}%
and:%
\begin{equation}
I\left( 1;\eta \right) =0
\end{equation}%
We deduce that:
\begin{equation}
\mathbb{E}\left[ \mathcal{LG}_{\mathrm{asset}}\left( 1\right) \right] =0
\end{equation}

\subsection{Computation of $I\left( a,b;\protect\eta \right) $ in some
special cases} \label{appendix:sinh}

Using the results derived previously, we deduce that:%
\begin{eqnarray}
I\left( a,b;1\right)  &=&\frac{2}{5}\left( b-a\right) ^{2.5}  \notag \\
I\left( a,b;2\right)  &=&\frac{2}{35}\left( b-a\right) ^{2.5}\left(
2a+5b\right)   \notag \\
I\left( a,b;3\right)  &=&\frac{2}{315}\left( b-a\right) ^{2.5}\left(
8a^{2}+20ab+35b^{2}\right)
\end{eqnarray}%
For $\eta =0.5$, we have:%
\begin{equation}
I\left( a,b;0.5\right) =\frac{1}{4}\sqrt{b-a}\left( \left( 2b-5a\right)
\sqrt{b}-3\psi \left( a,b\right) \right)
\end{equation}%
where:%
\begin{equation}
\psi \left( a,b\right) =\left\{
\begin{array}{ll}
\func{Re}\left( a^{2}\left( a-b\right) ^{-0.5}\sin ^{-1}\sqrt{\dfrac{b}{a}}%
\right)  & \text{if }a\neq 0 \\
0 & \text{if }a=0%
\end{array}%
\right.
\end{equation}

\clearpage

\section{Additional results}

\subsection{Tables}
\label{appendix:tables}

\begin{table}[tbph]
\centering
\caption{Number of liquidated shares $q_{i}\left( h\right)$ (Example \ref{ex:rcr1}, naive pro-rata liquidation)}
\label{tab:rcr1-2}
\begin{tabular}{c|rrrrrrr}
\hline
\multirow{2}{*}{$h$} & \multicolumn{7}{c}{Asset} \\
      & \multicolumn{1}{c}{\#1} & \multicolumn{1}{c}{\#2} & \multicolumn{1}{c}{\#3}
      & \multicolumn{1}{c}{\#4} & \multicolumn{1}{c}{\#5} & \multicolumn{1}{c}{\#6}
      & \multicolumn{1}{c}{\#7} \\ \hline
$1$   & $20\,000$ & $20\,000$ & $10\,000$ & $20\,000$ & $15\,100$ & $2\,000$ & $360$ \\
$2$   & $20\,000$ & $20\,000$ &      $80$ & $20\,000$ &       $0$ & $1\,500$ &   $0$ \\
$3$   & $20\,000$ & $20\,000$ &       $0$ &     $100$ &       $0$ &      $0$ &   $0$ \\
$4$   & $20\,000$ & $     20$ &       $0$ &       $0$ &       $0$ &      $0$ &   $0$ \\
$5$   & $ 7\,020$ & $      0$ &       $0$ &       $0$ &       $0$ &      $0$ &   $0$ \\
$6$   & $      0$ & $      0$ &       $0$ &       $0$ &       $0$ &      $0$ &   $0$ \\ \hline
Total & $87\,020$ & $60\,020$ & $10\,080$ & $40\,100$ & $15\,100$ & $3\,500$ & $360$ \\ \hline
\end{tabular}
\end{table}

\begin{table}[tbph]
\centering
\caption{Weights $w_{i}\left( q; h\right)$ in \% (Example \ref{ex:rcr1}, naive pro-rata liquidation)}
\label{tab:rcr1-3}
\begin{tabular}{c|rrrrrrr}
\hline
\multirow{2}{*}{$h$} & \multicolumn{7}{c}{Asset} \\
      & \multicolumn{1}{c}{\#1} & \multicolumn{1}{c}{\#2} & \multicolumn{1}{c}{\#3}
      & \multicolumn{1}{c}{\#4} & \multicolumn{1}{c}{\#5} & \multicolumn{1}{c}{\#6}
      & \multicolumn{1}{c}{\#7} \\ \hline
$1$   & $11.95$ & $16.52$ & $32.77$ & $13.70$ &      $16.93$ & $4.28$ & $3.84$ \\
$2$   & $16.41$ & $22.68$ & $22.68$ & $18.81$ &      $11.63$ & $5.15$ & $2.64$ \\
$3$   & $20.59$ & $28.45$ & $18.96$ & $15.77$ & ${\TsV}9.72$ & $4.30$ & $2.21$ \\
$4$   & $25.68$ & $26.63$ & $17.74$ & $14.75$ & ${\TsV}9.10$ & $4.03$ & $2.06$ \\
$5$   & $27.32$ & $26.04$ & $17.35$ & $14.43$ & ${\TsV}8.90$ & $3.94$ & $2.02$ \\
$6$   & $27.32$ & $26.04$ & $17.35$ & $14.43$ & ${\TsV}8.90$ & $3.94$ & $2.02$ \\ \hline
\end{tabular}
\end{table}

\begin{table}[tbph]
\centering
\caption{Weights $w_{i}\left( \omega - q; h\right)$ in \% (Example \ref{ex:rcr1}, naive pro-rata liquidation)}
\label{tab:rcr1-4}
\begin{tabular}{c|rrrrrrr}
\hline
\multirow{2}{*}{$h$} & \multicolumn{7}{c}{Asset} \\
      & \multicolumn{1}{c}{\#1} & \multicolumn{1}{c}{\#2} & \multicolumn{1}{c}{\#3}
      & \multicolumn{1}{c}{\#4} & \multicolumn{1}{c}{\#5} & \multicolumn{1}{c}{\#6}
      & \multicolumn{1}{c}{\#7} \\ \hline
$0$   & $27.32$ & $26.04$ & $17.35$ & $14.43$ & $8.90$ & $3.94$ & $2.02$ \\
$1$   & $29.13$ & $27.16$ & $15.54$ & $14.51$ & $7.95$ & $3.90$ & $1.80$ \\
$2$   & $29.29$ & $26.65$ & $16.39$ & $13.64$ & $8.40$ & $3.72$ & $1.91$ \\
$3$   & $28.83$ & $25.50$ & $16.99$ & $14.13$ & $8.71$ & $3.86$ & $1.98$ \\
$4$   & $27.72$ & $25.90$ & $17.26$ & $14.35$ & $8.85$ & $3.92$ & $2.01$ \\
$5$   & $27.32$ & $26.04$ & $17.35$ & $14.43$ & $8.90$ & $3.94$ & $2.02$ \\
$6$   & $27.32$ & $26.04$ & $17.35$ & $14.43$ & $8.90$ & $3.94$ & $2.02$ \\ \hline
\end{tabular}
\end{table}

\begin{table}[tbph]
\centering
\caption{Number of liquidated shares $q_{i}\left( h\right)$ (Example \ref{ex:rcr3}, waterfall liquidation)}
\label{tab:rcr3-2}
\begin{tabular}{c|rrrrrrr}
\hline
\multirow{2}{*}{$h$} & \multicolumn{7}{c}{Asset} \\
      & \multicolumn{1}{c}{\#1} & \multicolumn{1}{c}{\#2} & \multicolumn{1}{c}{\#3}
      & \multicolumn{1}{c}{\#4} & \multicolumn{1}{c}{\#5} & \multicolumn{1}{c}{\#6}
      & \multicolumn{1}{c}{\#7} \\ \hline
${\TsV}1$ &    $20\,000$ &  $20\,000$ & $10\,000$ &  $20\,000$ & $20\,000$ &  $2\,000$ & $1\,000$ \\
${\TsV}2$ &    $20\,000$ &  $20\,000$ & $10\,000$ &  $20\,000$ & $20\,000$ &  $2\,000$ &    $800$ \\
${\TsV}3$ &    $20\,000$ &  $20\,000$ & $10\,000$ &  $20\,000$ & $20\,000$ &  $2\,000$ &      $0$ \\
${\TsV}4$ &    $20\,000$ &  $20\,000$ & $10\,000$ &  $20\,000$ & $15\,500$ &  $2\,000$ &      $0$ \\
${\TsV}5$ &    $20\,000$ &  $20\,000$ & $10\,000$ &  $20\,000$ &       $0$ &  $2\,000$ &      $0$ \\
${\TsV}6$ &    $20\,000$ &  $20\,000$ &     $400$ &  $20\,000$ &       $0$ &  $2\,000$ &      $0$ \\
${\TsV}7$ &    $20\,000$ &  $20\,000$ &       $0$ &  $20\,000$ &       $0$ &  $2\,000$ &      $0$ \\
${\TsV}8$ &    $20\,000$ &  $20\,000$ &       $0$ &  $20\,000$ &       $0$ &  $2\,000$ &      $0$ \\
${\TsV}9$ &    $20\,000$ &  $20\,000$ &       $0$ &  $20\,000$ &       $0$ &  $1\,500$ &      $0$ \\
     $10$ &    $20\,000$ &  $20\,000$ &       $0$ &  $20\,000$ &       $0$ &       $0$ &      $0$ \\
     $11$ &    $20\,000$ &  $20\,000$ &       $0$ &      $500$ &       $0$ &       $0$ &      $0$ \\
     $12$ &    $20\,000$ &  $20\,000$ &       $0$ &        $0$ &       $0$ &       $0$ &      $0$ \\
     $13$ &    $20\,000$ &  $20\,000$ &       $0$ &        $0$ &       $0$ &       $0$ &      $0$ \\
     $14$ &    $20\,000$ &  $20\,000$ &       $0$ &        $0$ &       $0$ &       $0$ &      $0$ \\
     $15$ &    $20\,000$ &  $20\,000$ &       $0$ &        $0$ &       $0$ &       $0$ &      $0$ \\
     $16$ &    $20\,000$ &      $100$ &       $0$ &        $0$ &       $0$ &       $0$ &      $0$ \\
     $17$ &    $20\,000$ &        $0$ &       $0$ &        $0$ &       $0$ &       $0$ &      $0$ \\
     $18$ &    $20\,000$ &        $0$ &       $0$ &        $0$ &       $0$ &       $0$ &      $0$ \\
     $19$ &    $20\,000$ &        $0$ &       $0$ &        $0$ &       $0$ &       $0$ &      $0$ \\
     $20$ &    $20\,000$ &        $0$ &       $0$ &        $0$ &       $0$ &       $0$ &      $0$ \\
     $21$ &    $20\,000$ &        $0$ &       $0$ &        $0$ &       $0$ &       $0$ &      $0$ \\
     $22$ &    $15\,100$ &        $0$ &       $0$ &        $0$ &       $0$ &       $0$ &      $0$ \\
     $23$ &          $0$ &        $0$ &       $0$ &        $0$ &       $0$ &       $0$ &      $0$ \\ \hline
Total     &   $435\,100$ & $300\,100$ & $50\,400$ & $200\,500$ & $75\,500$ & $17\,500$ & $1\,800$ \\ \hline
\end{tabular}
\end{table}

\begin{table}[tbph]
\centering
\caption{Weights $w_{i}\left( q; h\right)$ in \% (Example \ref{ex:rcr3}, waterfall liquidation)}
\label{tab:rcr3-3}
\begin{tabular}{c|rrrrrrr}
\hline
\multirow{2}{*}{$h$} & \multicolumn{7}{c}{Asset} \\
      & \multicolumn{1}{c}{\#1} & \multicolumn{1}{c}{\#2} & \multicolumn{1}{c}{\#3}
      & \multicolumn{1}{c}{\#4} & \multicolumn{1}{c}{\#5} & \multicolumn{1}{c}{\#6}
      & \multicolumn{1}{c}{\#7} \\ \hline
$1$   & $10.64$ & $14.71$ & $29.17$ & $12.20$ & $19.97$ & $3.81$ & $9.50$ \\
$2$   & $10.74$ & $14.85$ & $29.45$ & $12.31$ & $20.16$ & $3.85$ & $8.63$ \\
$3$   & $11.06$ & $15.29$ & $30.33$ & $12.68$ & $20.76$ & $3.96$ & $5.92$ \\
$4$   & $11.36$ & $15.70$ & $31.15$ & $13.02$ & $20.12$ & $4.07$ & $4.56$ \\
$5$   & $11.95$ & $16.52$ & $32.77$ & $13.70$ & $16.93$ & $4.28$ & $3.84$ \\
$6$   & $13.09$ & $18.09$ & $30.15$ & $15.01$ & $15.46$ & $4.69$ & $3.51$ \\ \hline
\end{tabular}
\end{table}

\begin{table}[tbph]
\centering
\caption{Weights $w_{i}\left( \omega - q; h\right)$ in \% (Example \ref{ex:rcr3}, waterfall liquidation)}
\label{tab:rcr3-4}
\begin{tabular}{c|rrrrrrr}
\hline
\multirow{2}{*}{$h$} & \multicolumn{7}{c}{Asset} \\
      & \multicolumn{1}{c}{\#1} & \multicolumn{1}{c}{\#2} & \multicolumn{1}{c}{\#3}
      & \multicolumn{1}{c}{\#4} & \multicolumn{1}{c}{\#5} & \multicolumn{1}{c}{\#6}
      & \multicolumn{1}{c}{\#7} \\ \hline
$0$   & $27.32$ & $26.04$ &      $17.35$ & $14.43$ & $8.90$ & $3.94$ & $2.02$ \\
$1$   & $29.55$ & $27.56$ &      $15.77$ & $14.73$ & $7.41$ & $3.96$ & $1.02$ \\
$2$   & $32.38$ & $29.46$ &      $13.66$ & $15.07$ & $5.46$ & $3.97$ & $0.00$ \\
$3$   & $35.72$ & $31.60$ &      $10.65$ & $15.33$ & $2.77$ & $3.93$ & $0.00$ \\
$4$   & $39.97$ & $34.24$ & ${\TsV}6.42$ & $15.54$ & $0.00$ & $3.83$ & $0.00$ \\
$5$   & $44.33$ & $36.58$ & ${\TsV}0.29$ & $15.24$ & $0.00$ & $3.56$ & $0.00$ \\
$6$   & $46.61$ & $36.82$ & ${\TsV}0.00$ & $13.65$ & $0.00$ & $2.92$ & $0.00$ \\ \hline
\end{tabular}
\end{table}

\clearpage

\subsection{Figures}
\label{appendix:figures}

\begin{figure}[tbph]
\centering
\caption{Calibration of the drawdown function (S\&P 500 index, 1990-2020, historical value-at-risk)}
\label{fig:hqla3}
\figureskip
\includegraphics[width = \figurewidth, height = \figureheight]{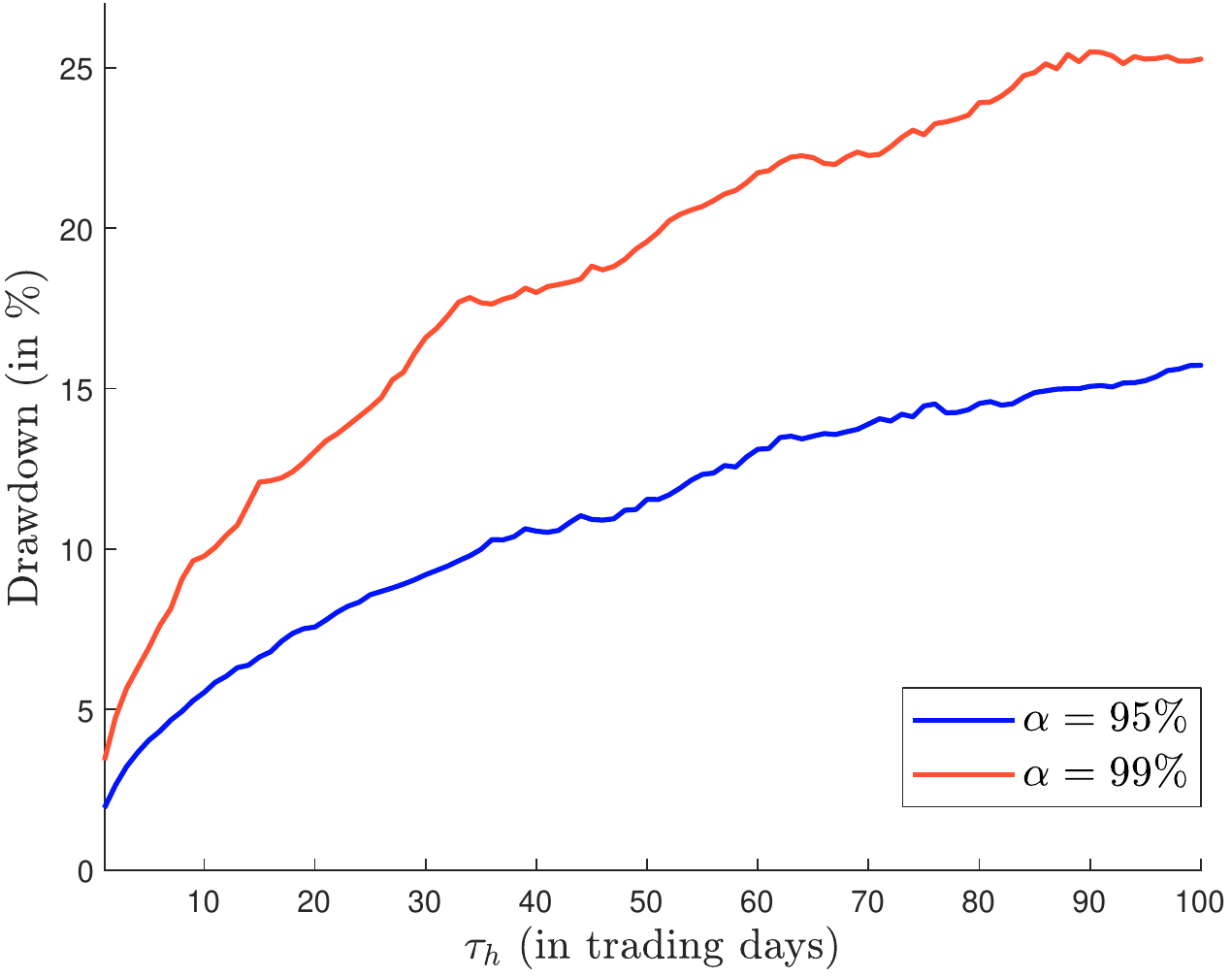}
\end{figure}

\begin{figure}[tbph]
\centering
\caption{Liquidity time in days (naive pro-rata liquidation)}
\label{fig:ttl1b}
\figureskip
\includegraphics[width = \figurewidth, height = \figureheight]{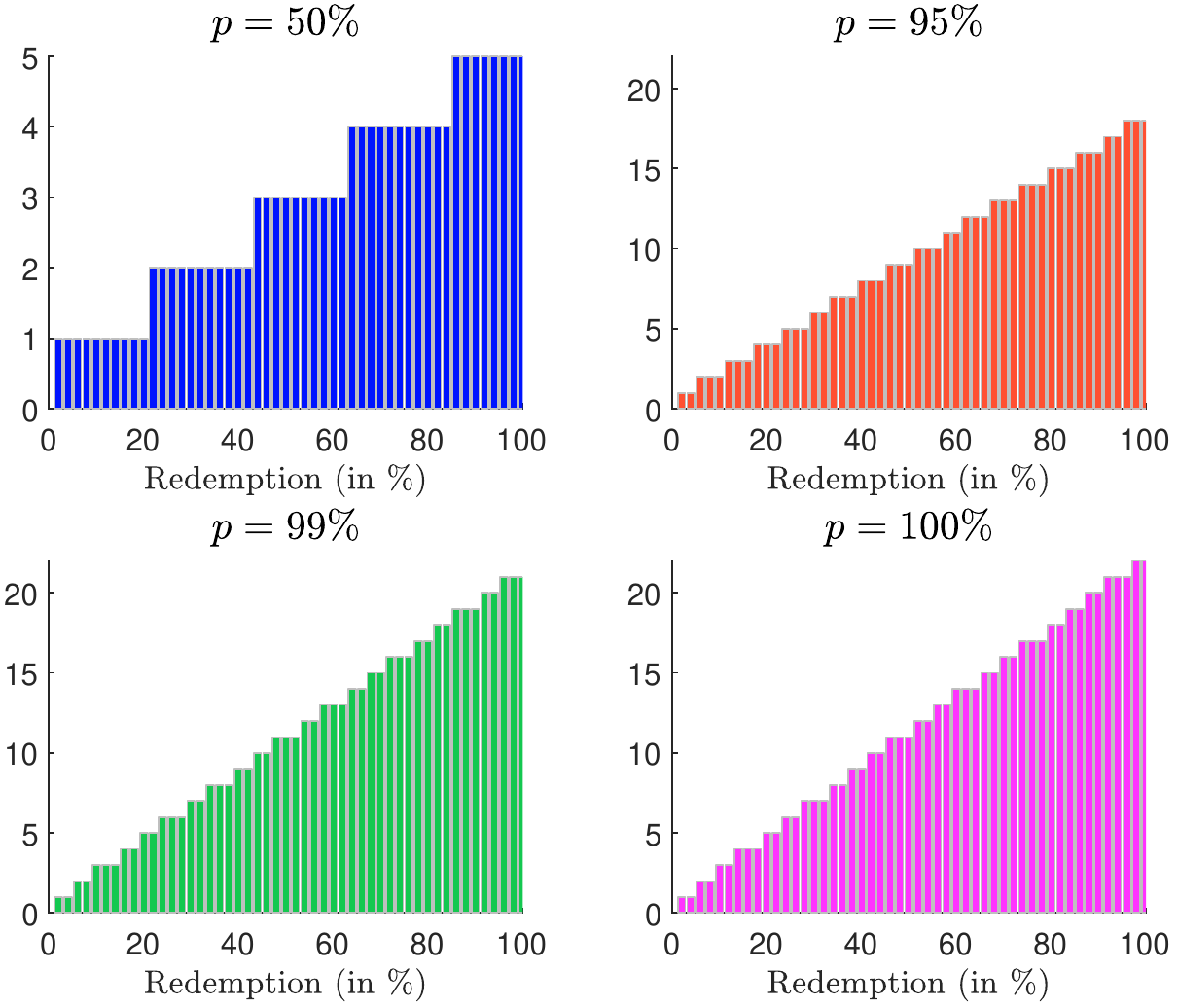}
\end{figure}

\begin{figure}[tbph]
\centering
\caption{Liquidity time in days (naive pro-rata liquidation, illiquid exposure)}
\label{fig:ttl2b}
\figureskip
\includegraphics[width = \figurewidth, height = \figureheight]{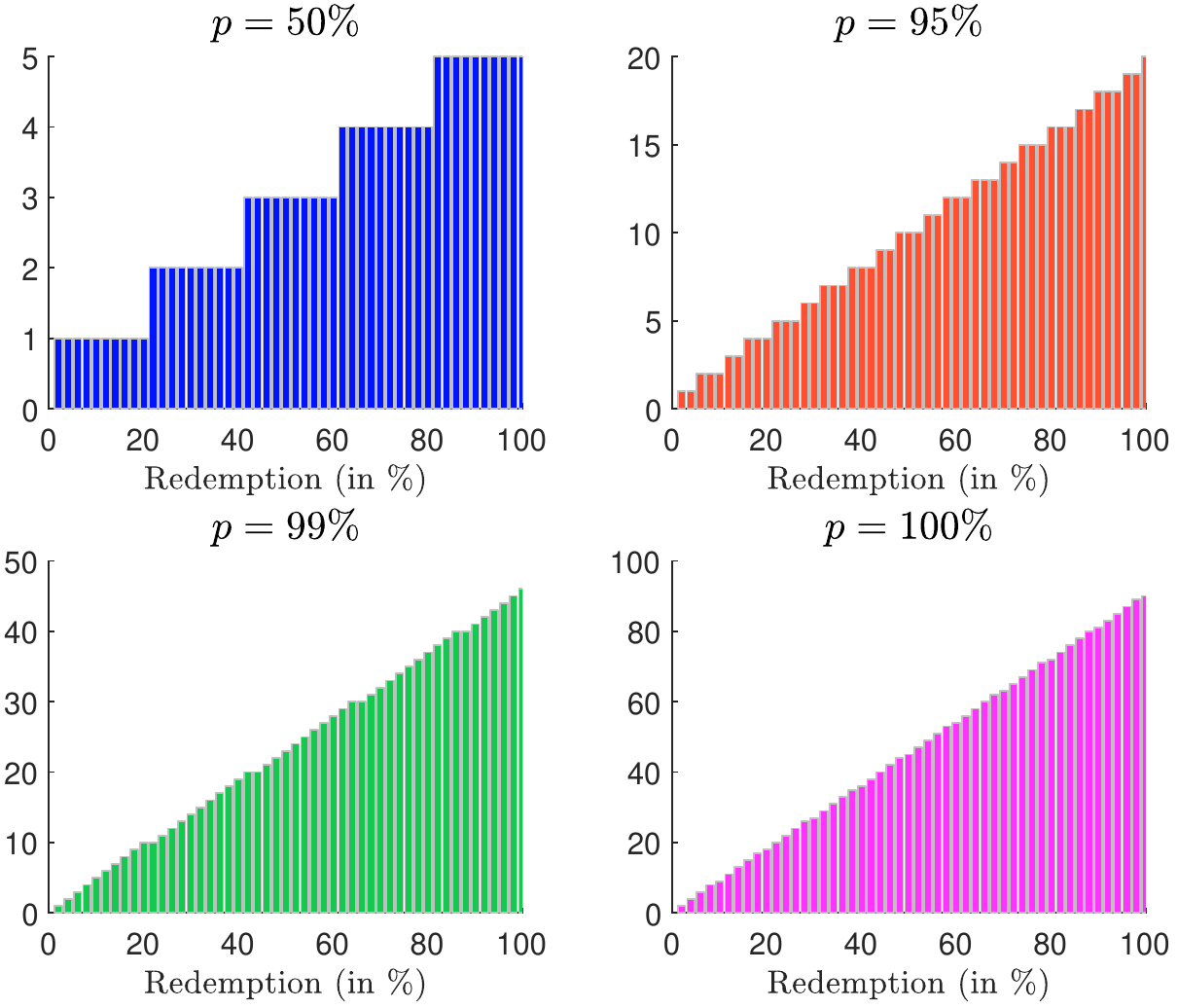}
\end{figure}

\begin{figure}[tbph]
\centering
\caption{Liquidity time in days (waterfall liquidation)}
\label{fig:ttl3a}
\figureskip
\includegraphics[width = \figurewidth, height = \figureheight]{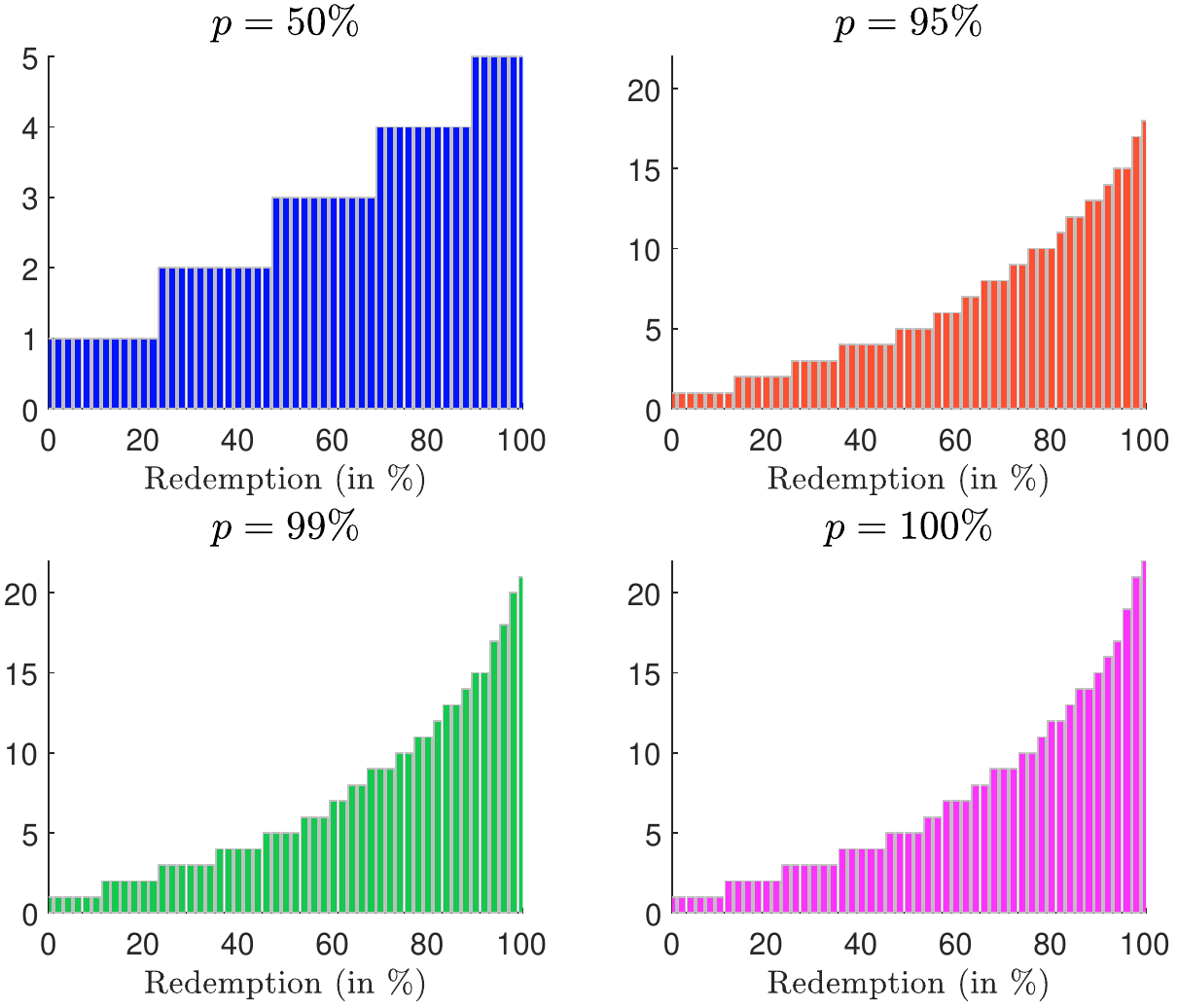}
\end{figure}

\begin{figure}[tbph]
\centering
\caption{Liquidity time in days (waterfall liquidation, illiquid exposure)}
\label{fig:ttl3b}
\figureskip
\includegraphics[width = \figurewidth, height = \figureheight]{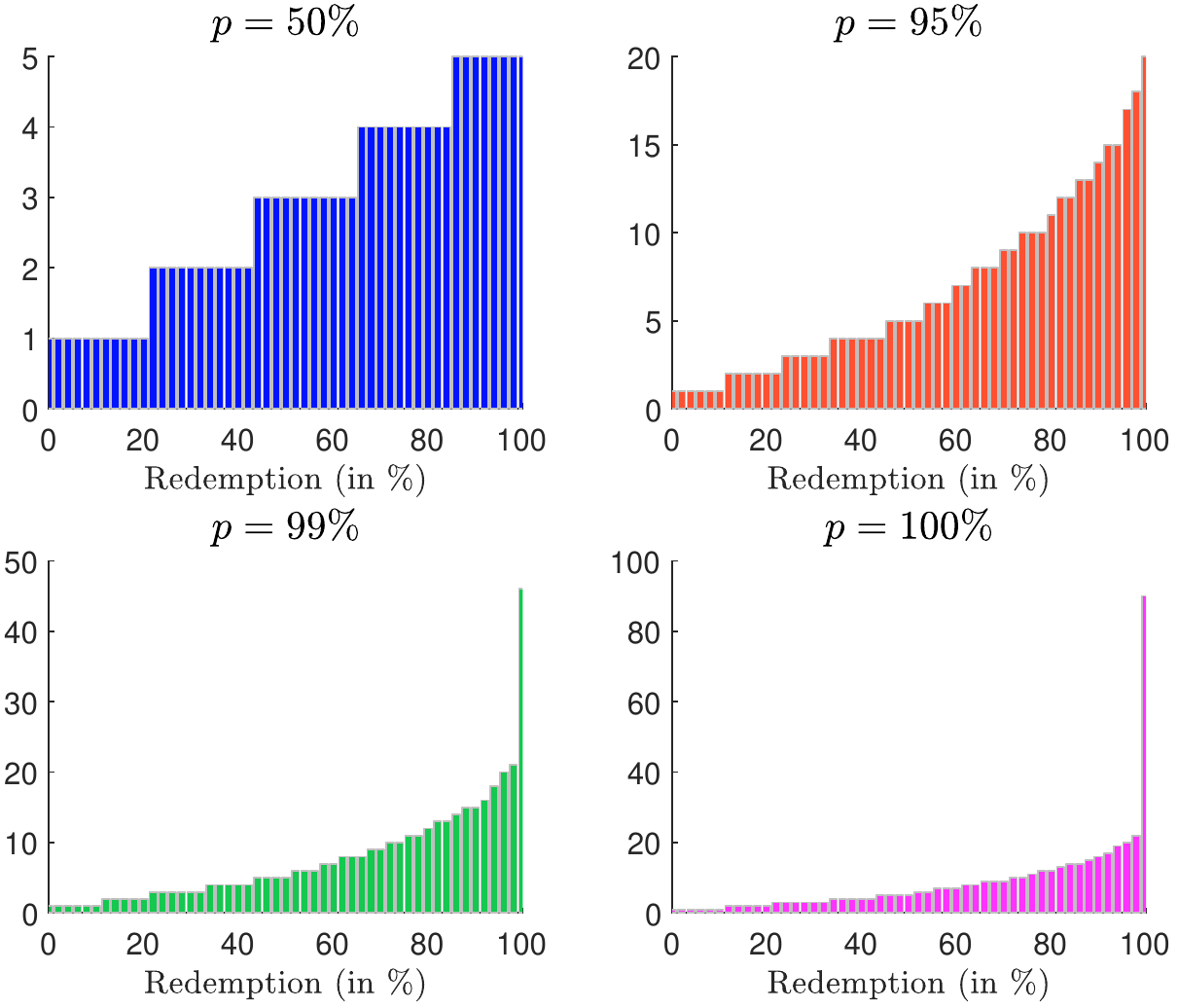}
\end{figure}

\begin{figure}[tbph]
\centering
\caption{Impact of the cash buffer on the portfolio return
($\mu _{\mathrm{asset}}=10\%$, $\sigma _{\mathrm{asset}}=20\%$ and $\rho _{\mathrm{cash},\mathrm{asset}}=0\%$)}
\label{fig:cash2a}
\figureskip
\includegraphics[width = \figurewidth, height = \figureheight]{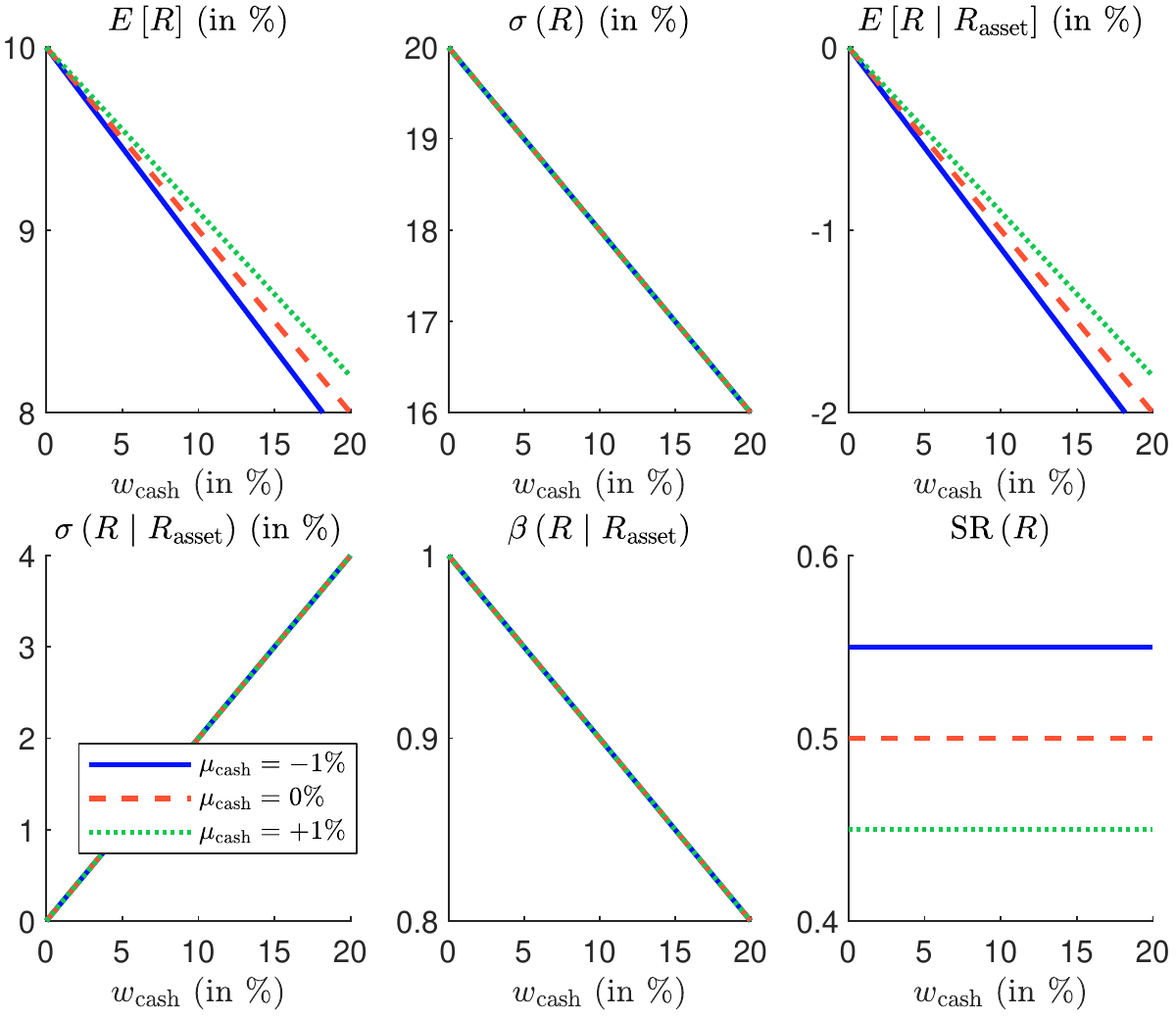}
\end{figure}

\begin{figure}[tbph]
\centering
\caption{Impact of the cash buffer on the portfolio return
($\mu _{\mathrm{asset}}=3\%$, $\sigma _{\mathrm{asset}}=5\%$ and $\rho _{\mathrm{cash},\mathrm{asset}}=20\%$)}
\label{fig:cash2b}
\figureskip
\includegraphics[width = \figurewidth, height = \figureheight]{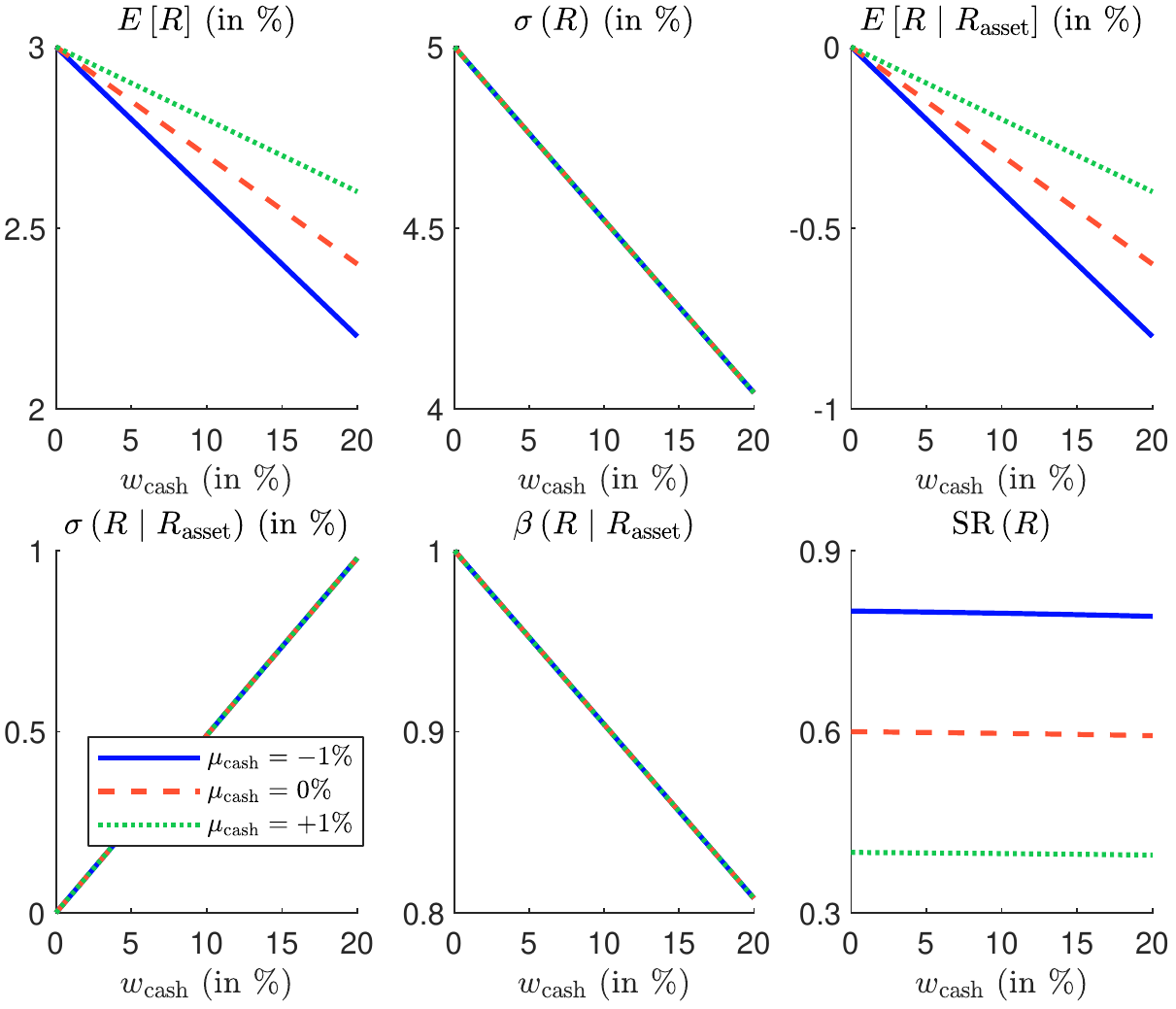}
\end{figure}

\begin{figure}[tbph]
\centering
\caption{Probability distribution function of the redemption rate $\RedemptionRate$
(Example \ref{ex:cash3}, page \pageref{ex:cash3})}
\label{fig:cash3b}
\figureskip
\includegraphics[width = \figurewidth, height = \figureheight]{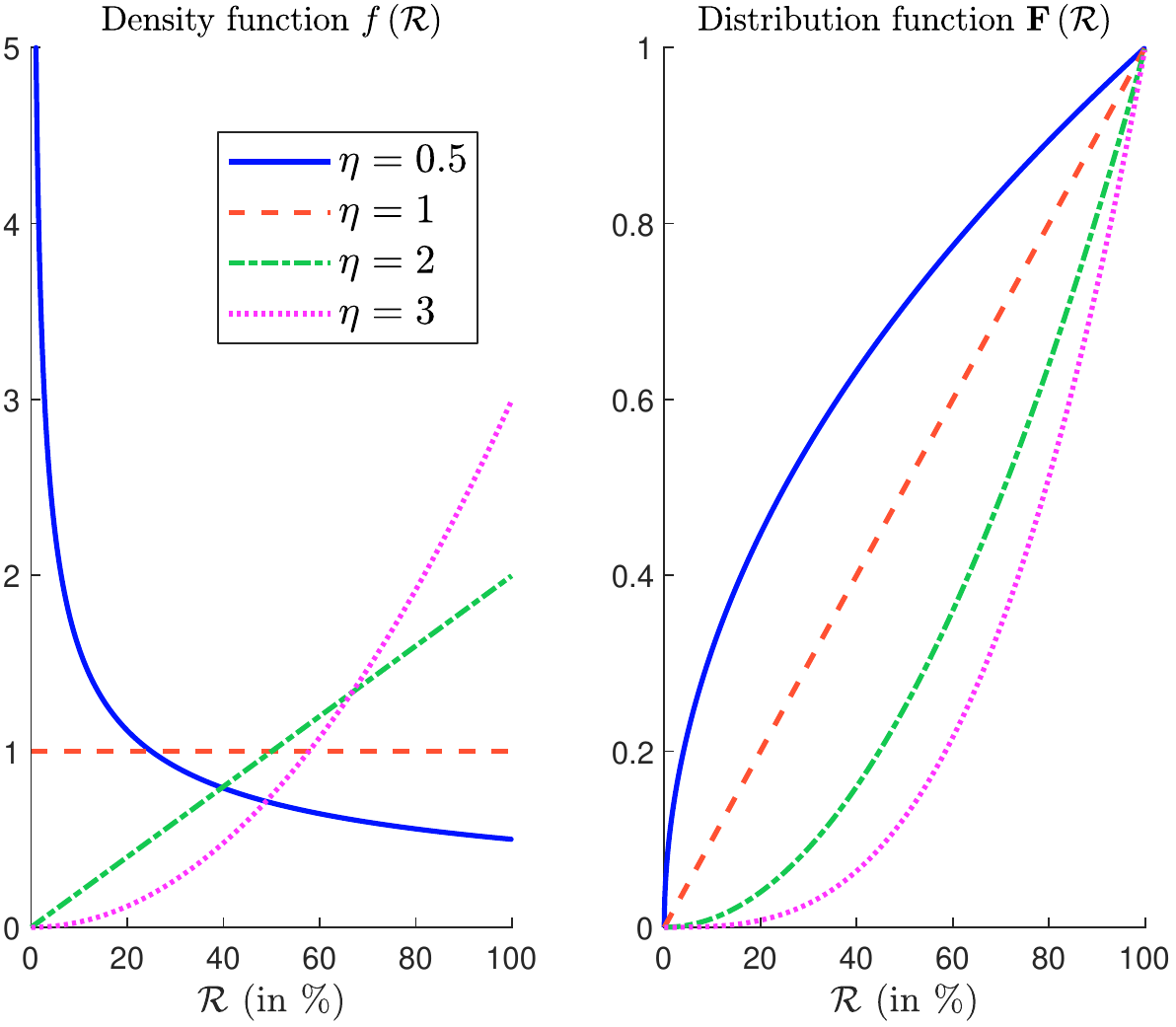}
\end{figure}

\begin{figure}[tbph]
\centering
\caption{Exact vs. approximate solution of
$\mathbb{E}\left[ \mathcal{LG}_{\mathrm{cash}}\left( w_{\mathrm{cash}}\right) \right] $
and
$\mathbb{E}\left[ \mathcal{LG}_{\mathrm{asset}}\left(w_{\mathrm{cash}}\right) \right] $ in bps
(Example \ref{ex:cash3}, page \pageref{ex:cash3})}
\label{fig:cash3c}
\figureskip
\includegraphics[width = \figurewidth, height = \figureheight]{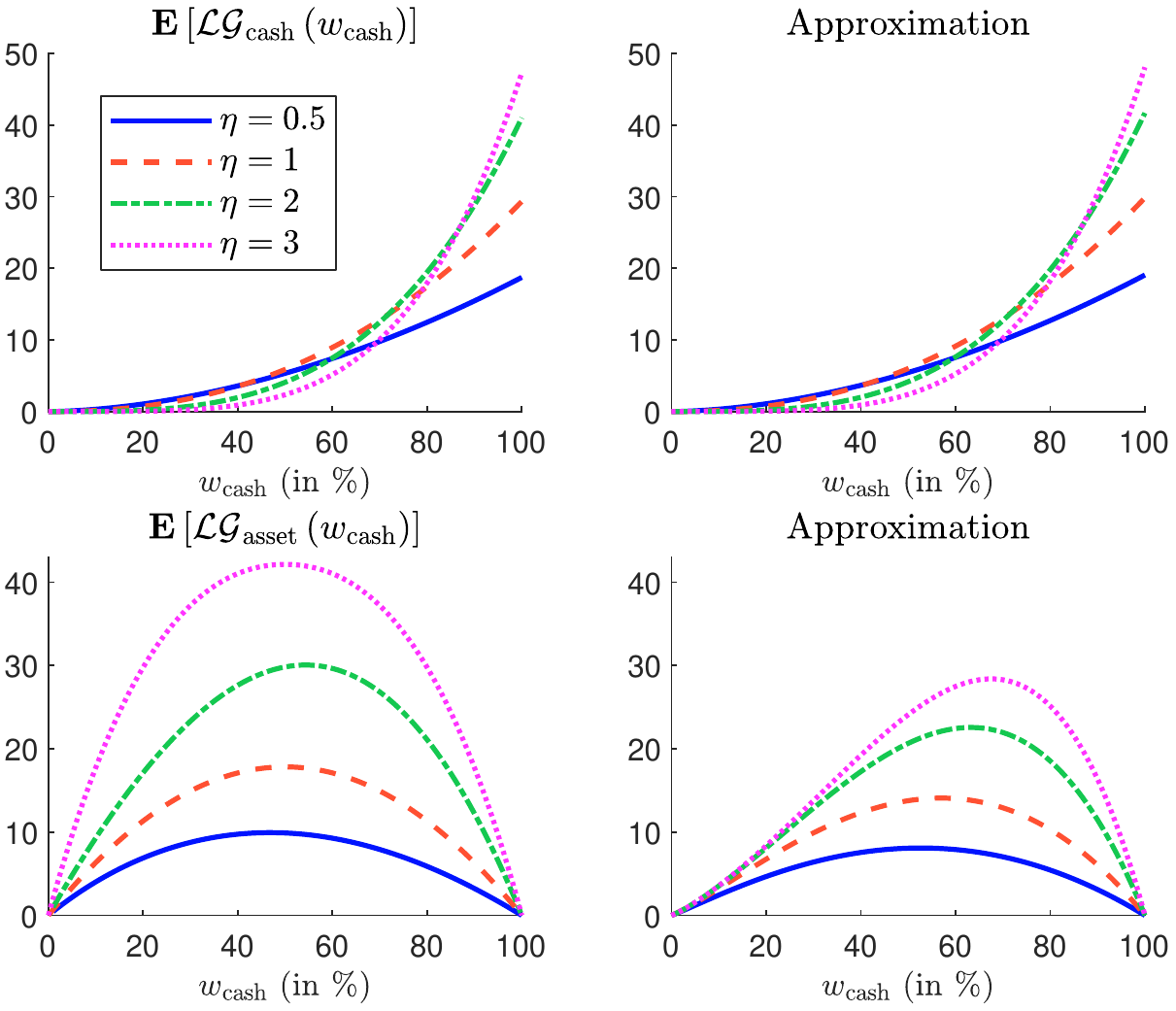}
\end{figure}

\begin{figure}[tbph]
\centering
\caption{Transaction cost function (\ref{eq:ex-cash4a}) in bps with $x^{+} = 30\%$}
\label{fig:cash4b}
\figureskip
\includegraphics[width = \figurewidth, height = \figureheight]{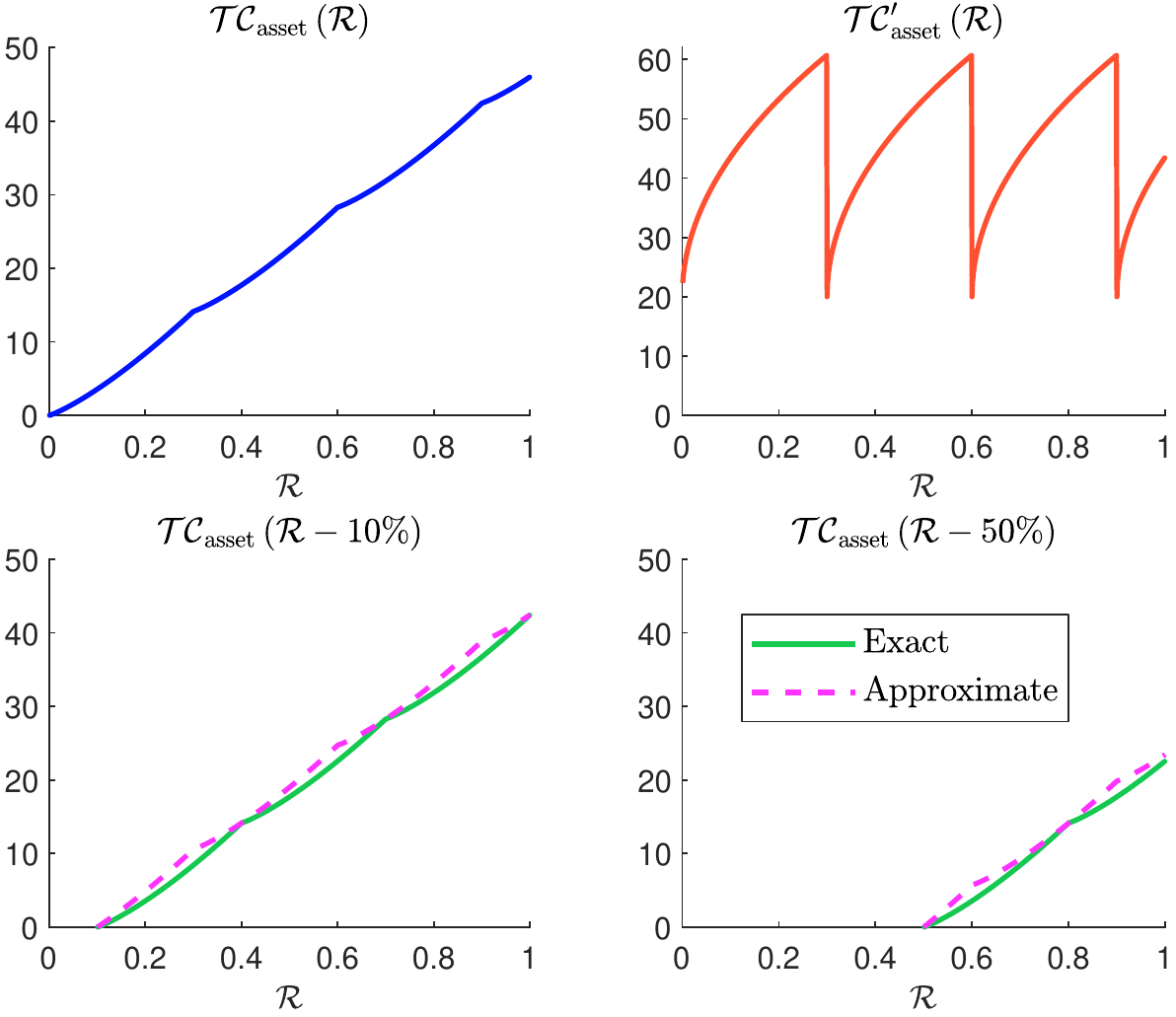}
\end{figure}

\begin{figure}[tbph]
\centering
\caption{Transaction cost function (\ref{eq:ex-cash4a}) in bps with $x^{+} = 50\%$}
\label{fig:cash4c}
\figureskip
\includegraphics[width = \figurewidth, height = \figureheight]{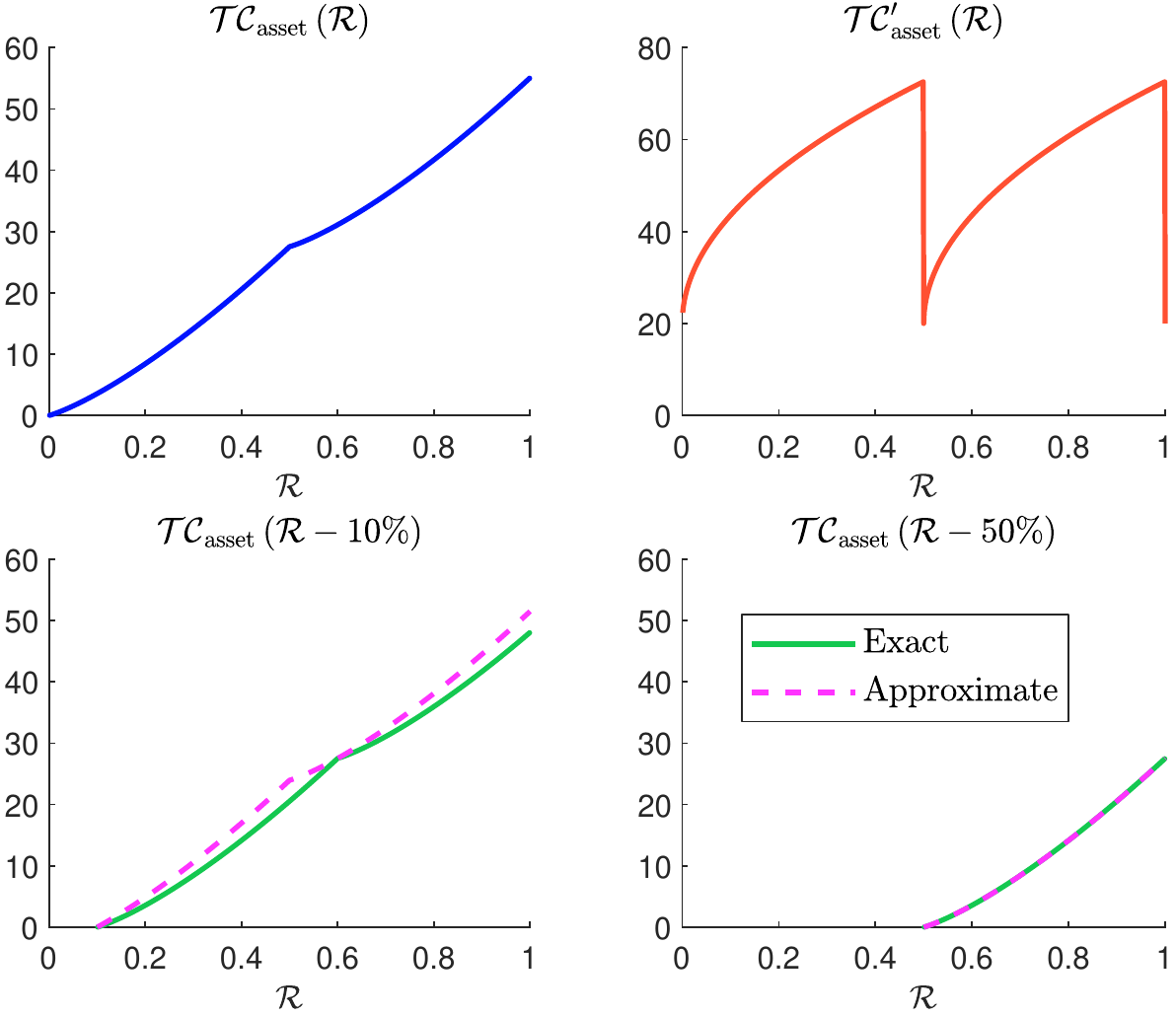}
\end{figure}

\begin{figure}[tbph]
\centering
\caption{Approximation error function
$\mathcal{E}_{\mathrm{rror}}\left( w_{\mathrm{cash}};x^{+}\right) $ in bps}
\label{fig:cash4d}
\figureskip
\includegraphics[width = \figurewidth, height = \figureheight]{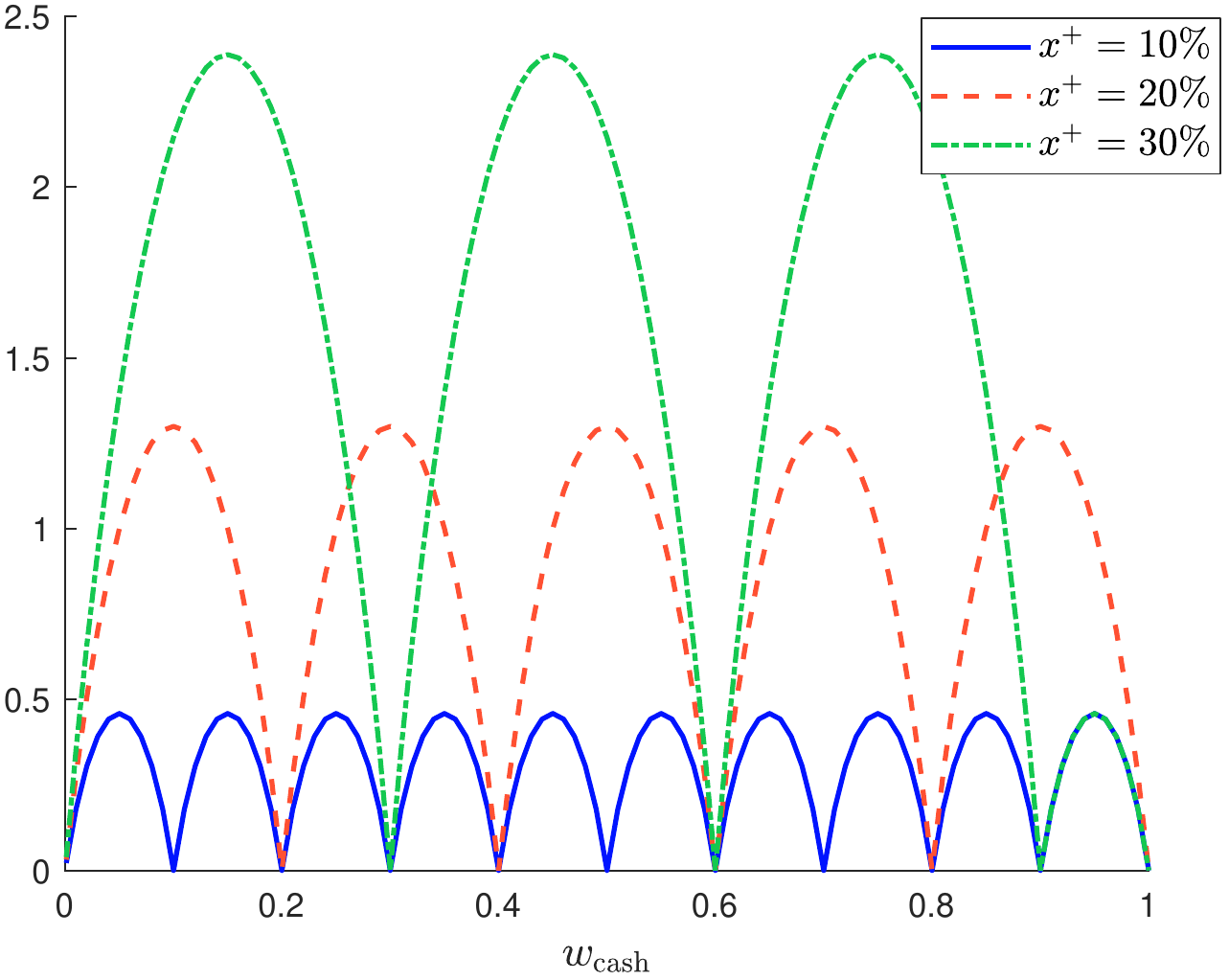}
\end{figure}

\begin{figure}[tbph]
\centering
\caption{Approximation of the liquidity gains
$\mathbb{E}\left[ \mathcal{LG}_{\mathrm{cash}}\left( w_{\mathrm{cash}}\right) \right] $
and
$\mathbb{E}\left[ \mathcal{LG}_{\mathrm{asset}}\left(w_{\mathrm{cash}}\right) \right] $ in bps
(Example \ref{ex:cash4}, page \pageref{ex:cash4})}
\label{fig:cash4g}
\figureskip
\includegraphics[width = \figurewidth, height = \figureheight]{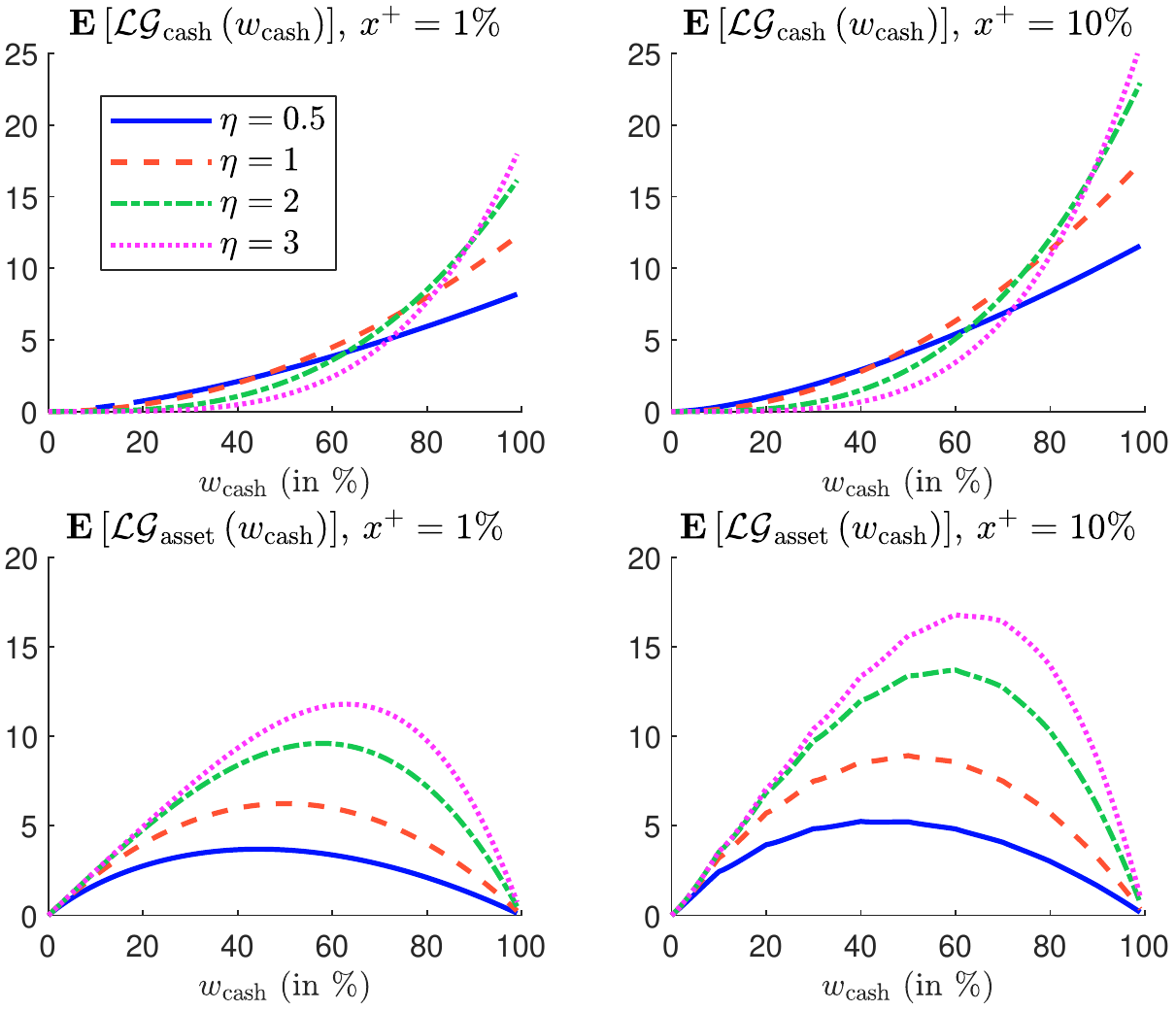}
\end{figure}

\begin{figure}[tbph]
\centering
\caption{Comparison of exact and approximate formulas
in bps when $x^{+} = 10\%$ (Example \ref{ex:cash4}, page \pageref{ex:cash4})}
\label{fig:cash4i}
\figureskip
\includegraphics[width = \figurewidth, height = \figureheight]{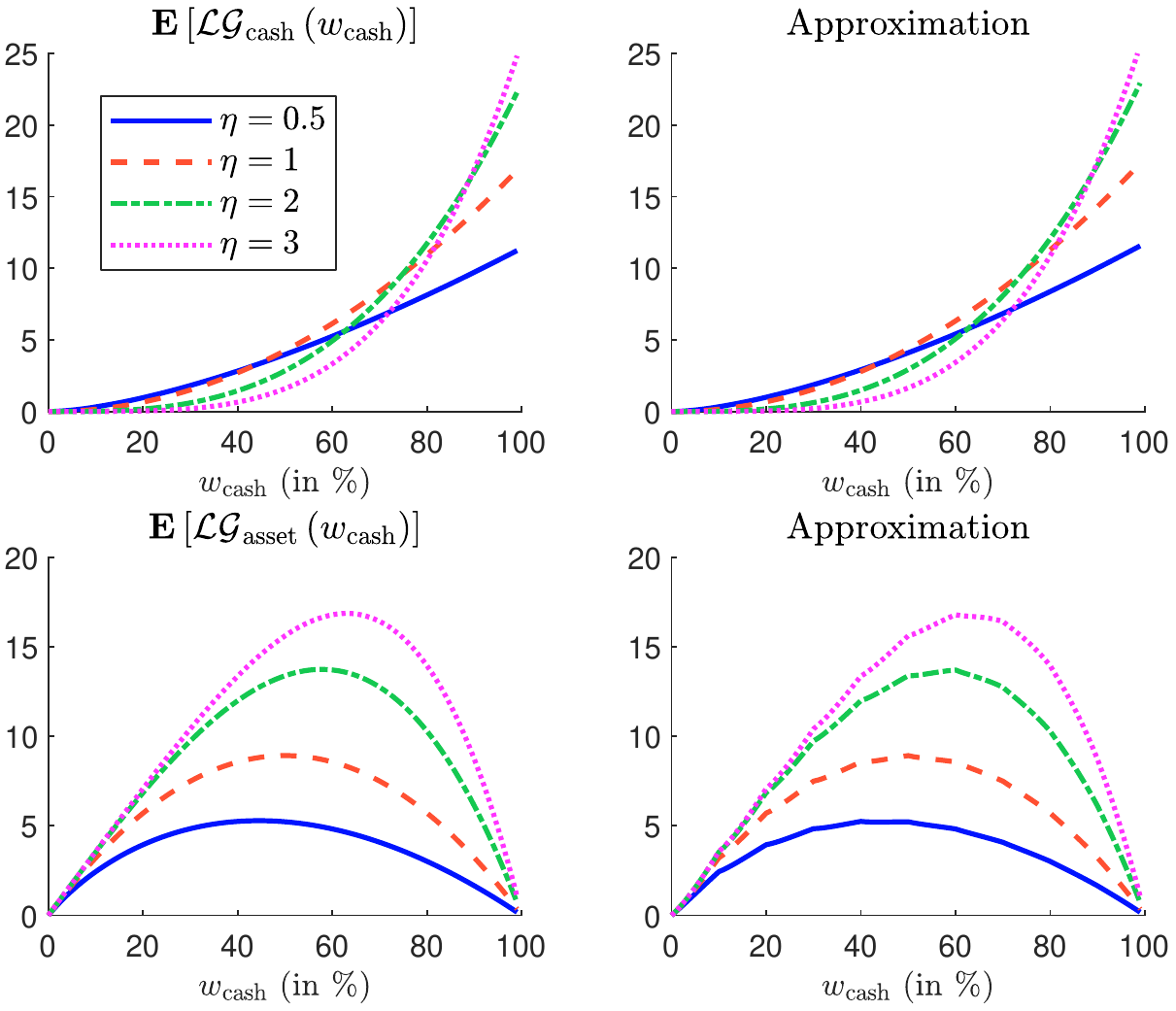}
\end{figure}

\begin{figure}[tbph]
\centering
\caption{Optimal cash buffer
($\mu _{\mathrm{asset}}-\mu _{\mathrm{cash}}=2.5\%$ and $\lambda =0$)}
\label{fig:cash6c}
\figureskip
\includegraphics[width = \figurewidth, height = \figureheight]{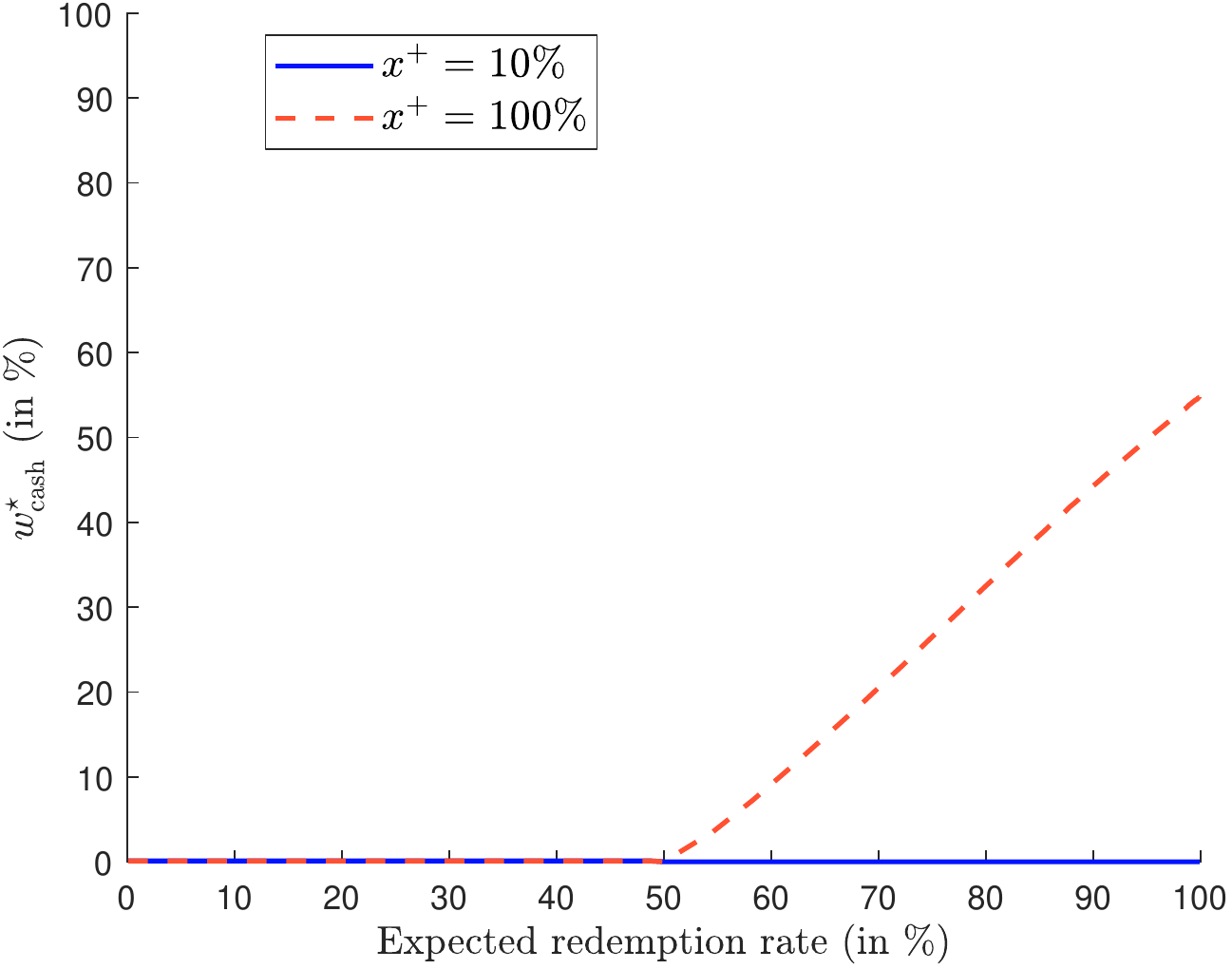}
\end{figure}

\begin{figure}[tbph]
\centering
\caption{Optimal cash buffer
($\mu _{\mathrm{asset}}-\mu _{\mathrm{cash}}=1\%$, $\lambda =0.25$ and
$\sigma _{\mathrm{asset}}=20\%$)}
\label{fig:cash6d}
\figureskip
\includegraphics[width = \figurewidth, height = \figureheight]{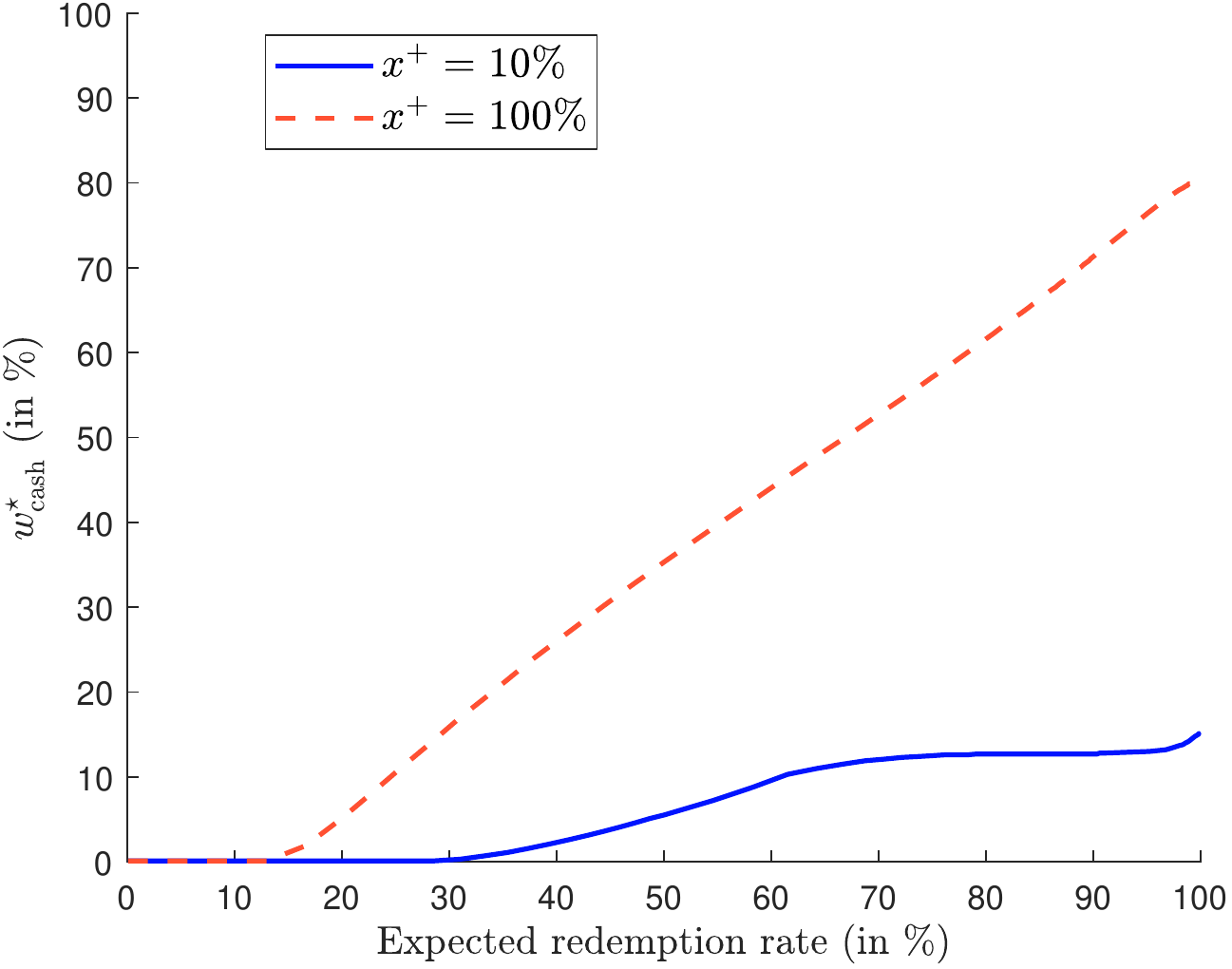}
\end{figure}

\begin{figure}[tbph]
\centering
\caption{Optimal cash buffer
($\mu _{\mathrm{asset}}-\mu _{\mathrm{cash}}=1\%$, $\lambda = 2$ and
$\sigma _{\mathrm{asset}}=20\%$)}
\label{fig:cash6e}
\figureskip
\includegraphics[width = \figurewidth, height = \figureheight]{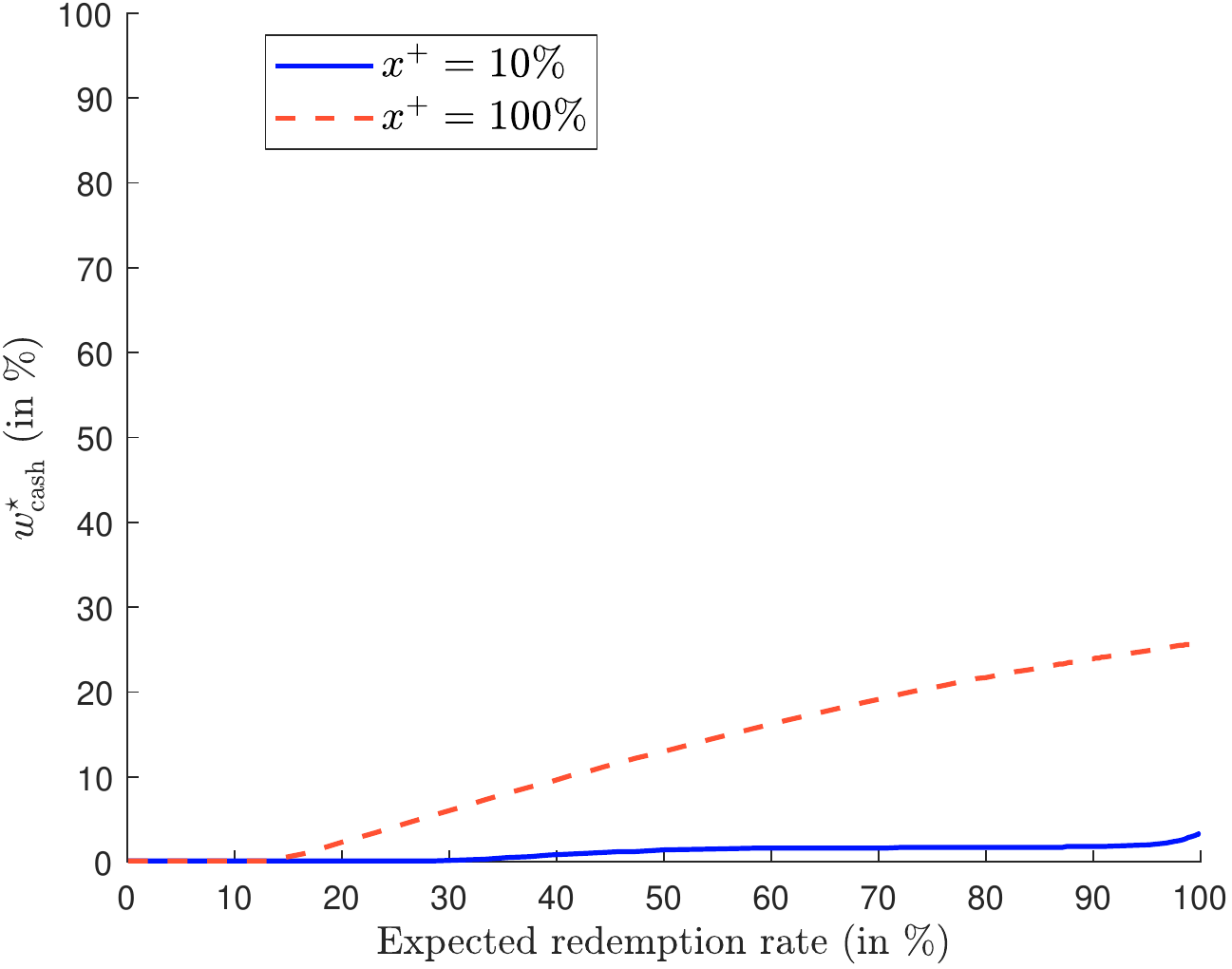}
\end{figure}

\begin{figure}[tbph]
\centering
\caption{Break-even risk premium
$\varrho \left( w_{\mathrm{cash}}\right) $ in \% ($x^{+}=10\%$, $\lambda =0$)}
\label{fig:cash7a}
\figureskip
\includegraphics[width = \figurewidth, height = \figureheight]{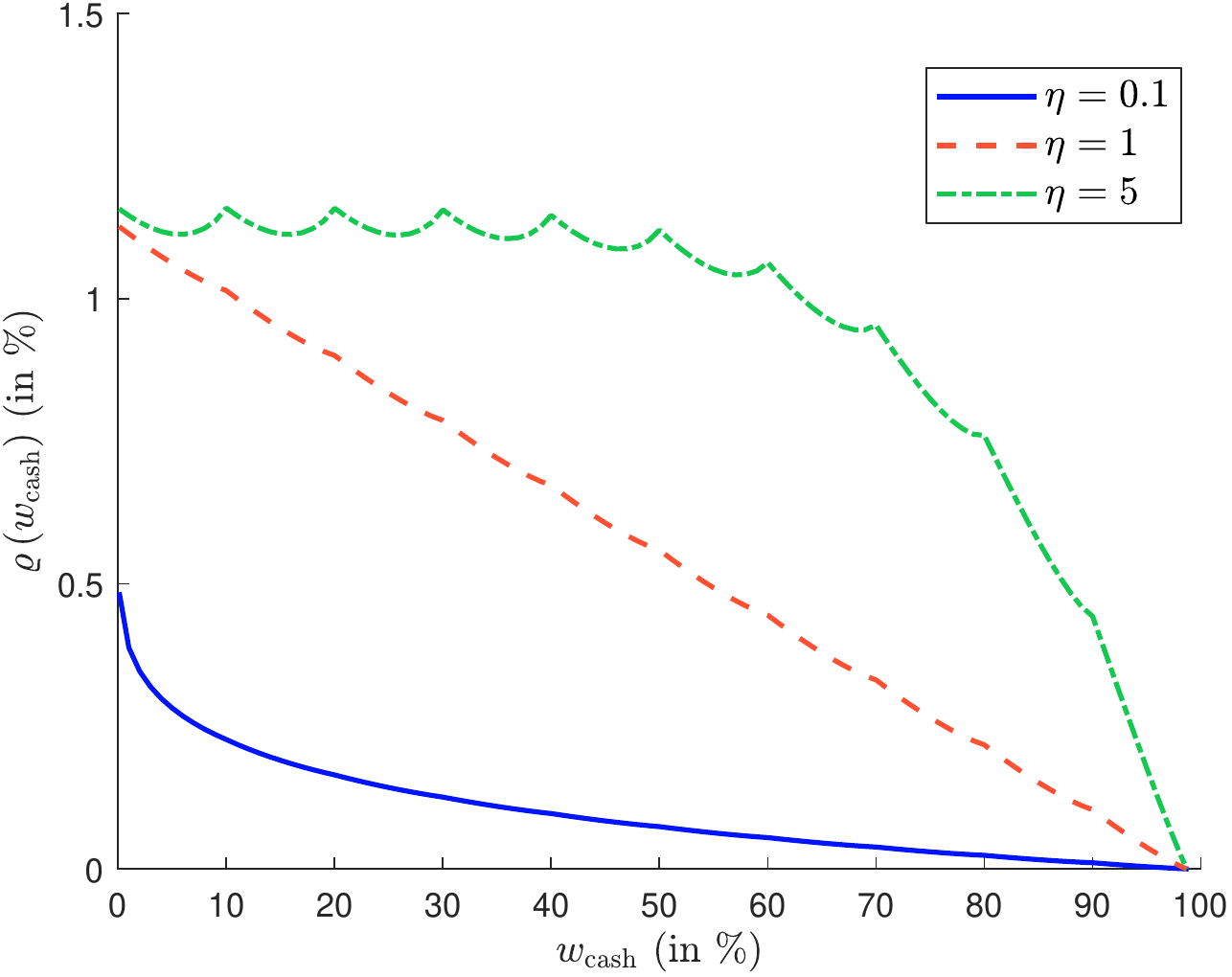}
\end{figure}

\begin{figure}[tbph]
\centering
\caption{Break-even risk premium
$\varrho \left( w_{\mathrm{cash}}\right) $ in \% ($x^{+}=100\%$, $\lambda =0$)}
\label{fig:cash7b}
\figureskip
\includegraphics[width = \figurewidth, height = \figureheight]{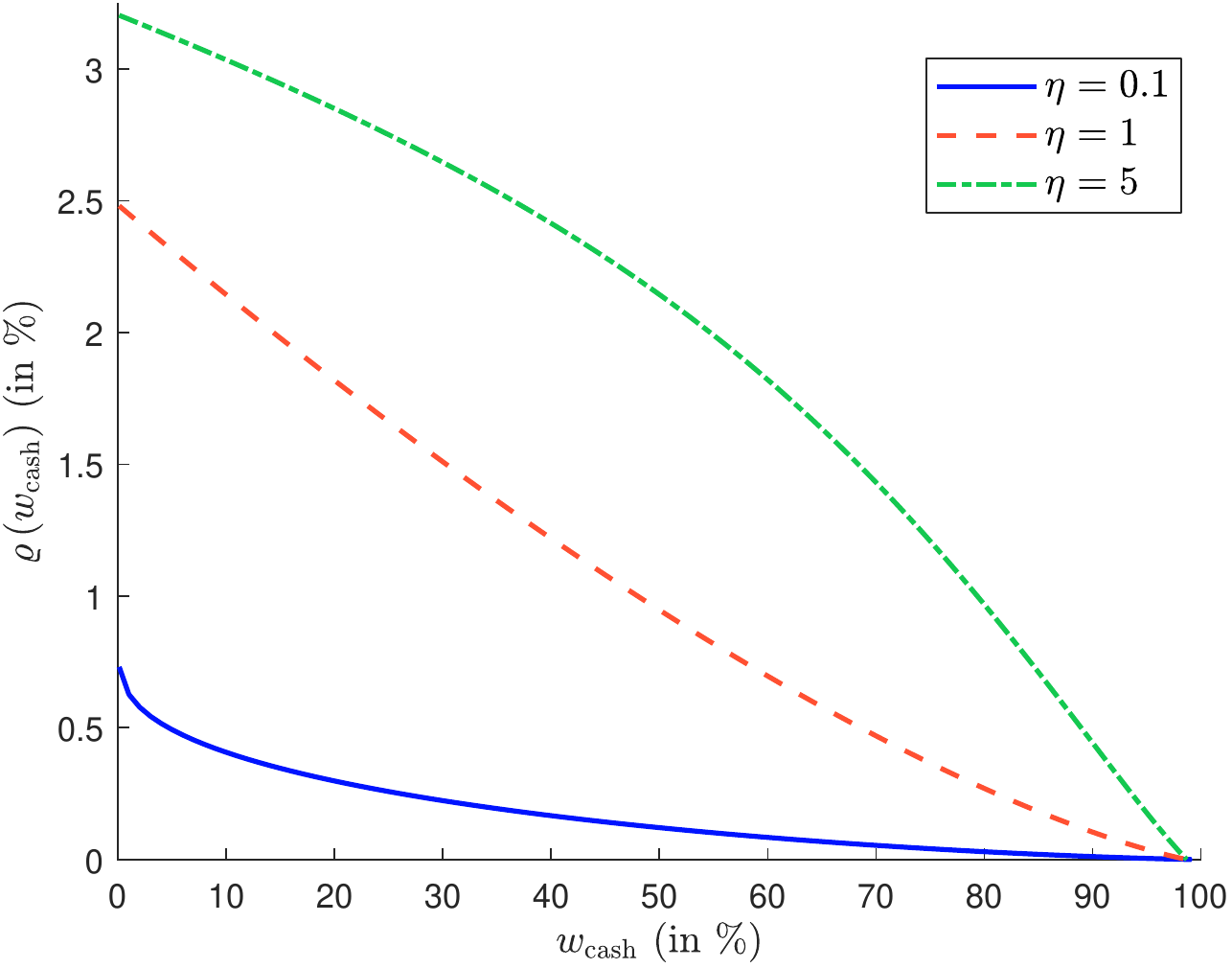}
\end{figure}

\begin{figure}[tbph]
\centering
\caption{Decision rule for implementing a cash buffer of $10\%$
($x^{+}=10\%$, $\lambda =0$)}
\label{fig:cash7e}
\figureskip
\includegraphics[width = \figurewidth, height = \figureheight]{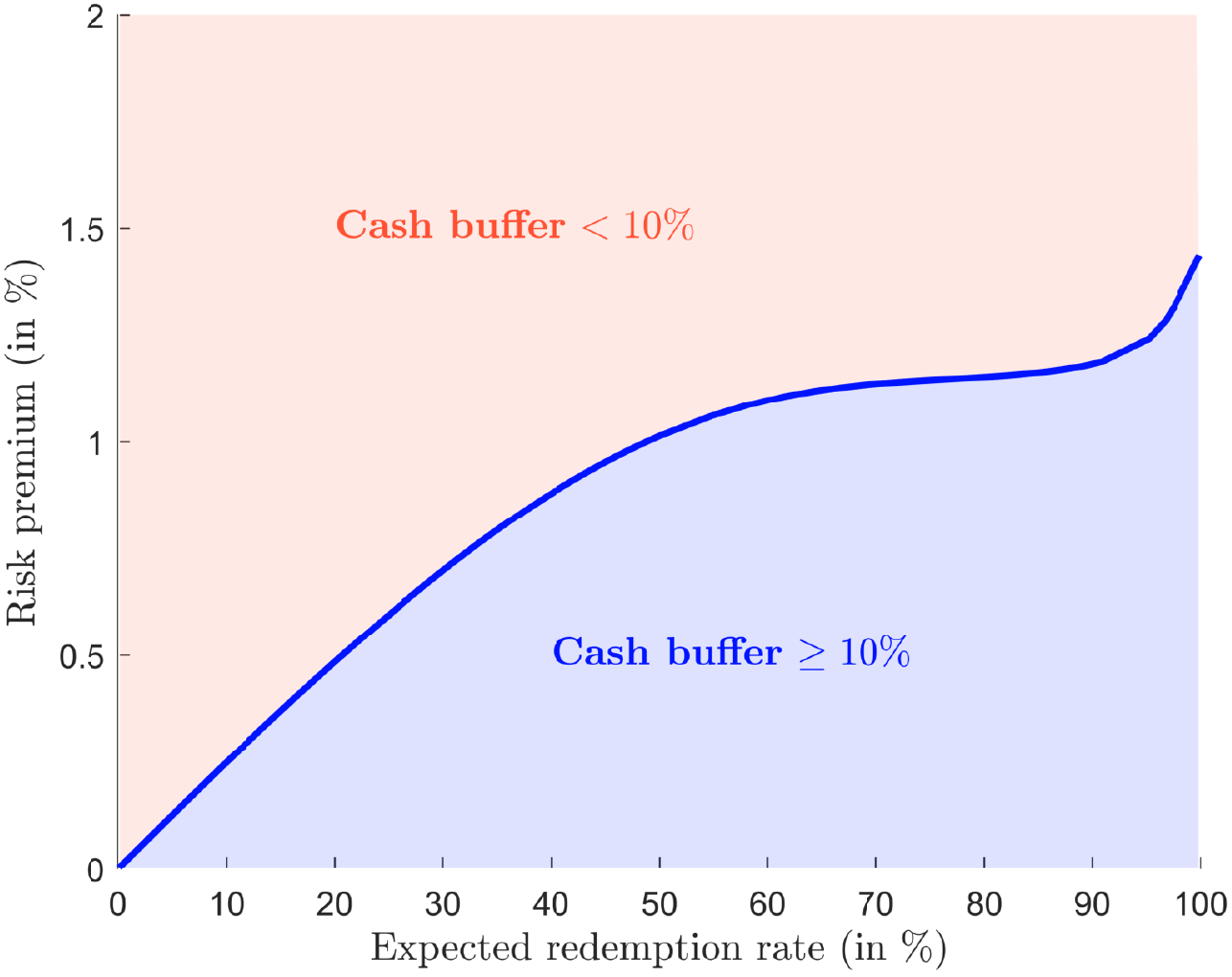}
\end{figure}

\begin{figure}[tbph]
\centering
\caption{Decision rule for implementing a cash buffer of $10\%$
($x^{+}=100\%$, $\lambda =0$)}
\label{fig:cash7f}
\figureskip
\includegraphics[width = \figurewidth, height = \figureheight]{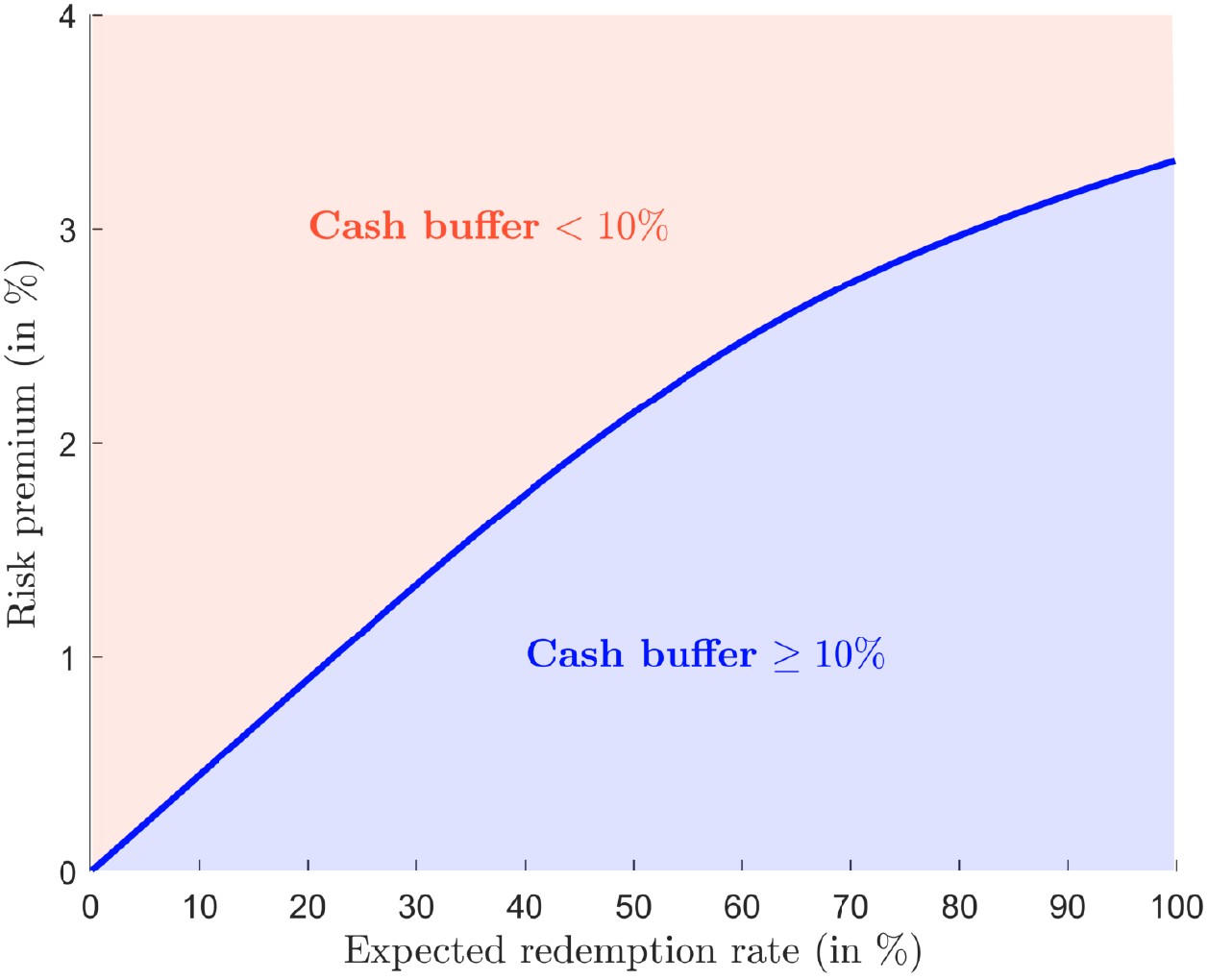}
\end{figure}

\clearpage

\tableofcontents

\end{document}